\documentclass[review]{elsarticle}

\usepackage{lineno,hyperref}
\usepackage{geometry}
\usepackage{float}
\usepackage{amsmath,amssymb,amsfonts,amsthm}    
\usepackage{graphicx}                           
\usepackage[font=small,skip=0pt]{caption}
\usepackage{subcaption}
\usepackage{mathtools}
\usepackage{multirow,makecell}
\usepackage{array}
\usepackage{accents}
\usepackage{leftidx}
\usepackage{chngcntr}
\usepackage[page]{appendix}
\usepackage{xcolor}
\usepackage{natbib}
\theoremstyle{definition}
\newtheorem{definition}
{Remark}[section]
\newcommand\munderbar[1]{%
  \underaccent{\bar}{#1}}
\makeatletter
\DeclareRobustCommand{\rvdots}{%
  \vbox{
    \baselineskip4\p@\lineskiplimit\z@
    \kern-\p@
    \hbox{.}\hbox{.}\hbox{.}
  }}
\makeatother
\usepackage{cleveref}

\modulolinenumbers[5]

\journal{Computer Methods in Applied Mechanics and Engineering}

\usepackage{etoolbox}
\patchcmd{\pprintMaketitle}
  {\fi\hrule}
  {\fi\ifvoid\extrainfobox\else\unvbox\extrainfobox\par\vskip5pt\fi\hrule}
  {}{}

\newenvironment{extrainfo}
  {\global\setbox\extrainfobox=\vbox\bgroup\parindent=0pt }
  {\egroup}
\newsavebox\extrainfobox

\usepackage{numcompress}\biboptions{authoryear}

\date{}
\begin{document}

\begin{frontmatter}

\title{An isogeometric finite element formulation for geometrically exact Timoshenko beams with extensible directors}

\author[LBB]{Myung-Jin Choi\corref{mycorrespondingauthor}}
\ead{choi@lbb.rwth-aachen.de}

\author[AICES,Gdansk,IIT]{Roger A. Sauer}
\ead{sauer@aices.rwth-aachen.de}

\author[LBB]{Sven Klinkel}
\ead{klinkel@lbb.rwth-aachen.de}

\cortext[mycorrespondingauthor]{Corresponding author}

\address[LBB]{Chair of Structural Analysis and Dynamics, RWTH Aachen University, Mies-van-der-Rohe Str.\,1,\,52074 Aachen, Germany}
\address[AICES]{Aachen Institute for Advanced Study in Computational Engineering Science (AICES), RWTH Aachen University, Templergraben 55, 52062 Aachen, Germany}
\address[Gdansk]{Faculty of Civil and Environmental Engineering, Gda\'{n}sk University of Technology, ul. Narutowicza 11/12, 80-233 Gda\'{n}sk, Poland}
\address[IIT]{Department of Mechanical Engineering, Indian Institute of Technology Kanpur, UP 208016, India}

\begin{extrainfo}
\begin{center}
{\small Published\footnote{This PDF is the personal version of an article whose final publication is available at \href{www.sciencedirect.com}{www.sciencedirect.com}} in \textit{Computer Methods in Applied Mechanics and Engineering}, \href{https://doi.org/10.1016/j.cma.2021.113993}{10.1016/j.cma.2021.113993}}\\
{\small Submitted on 25. October 2020, Revised on 14. April 2021, Accepted on 2. June 2021}
\end{center}
\end{extrainfo}

\begin{abstract}
An isogeometric finite element formulation for geometrically and materially nonlinear Timoshenko beams is presented, which incorporates in-plane deformation of the cross-section described by two extensible director vectors. Since those directors belong to the space ${\Bbb R}^3$, a configuration can be additively updated. The developed formulation allows direct application of nonlinear three-dimensional constitutive equations without zero stress conditions. Especially, the significance of considering correct surface loads rather than applying an equivalent load directly on the central axis is investigated. Incompatible linear in-plane strain components for the cross-section have been added to alleviate Poisson locking by using an enhanced assumed strain (EAS) method. In various numerical examples exhibiting large deformations, the accuracy and efficiency of the presented beam formulation is assessed in comparison to brick elements. We particularly use hyperelastic materials of the St.\,Venant-Kirchhoff and compressible Neo-Hookean types.\\
\end{abstract}


\end{frontmatter}

\linenumbers

\section{Introduction}
\begin{figure*}[htb]	
	\centering
	\begin{subfigure}[b] {0.45\textwidth} \centering
		\includegraphics[width=\linewidth]{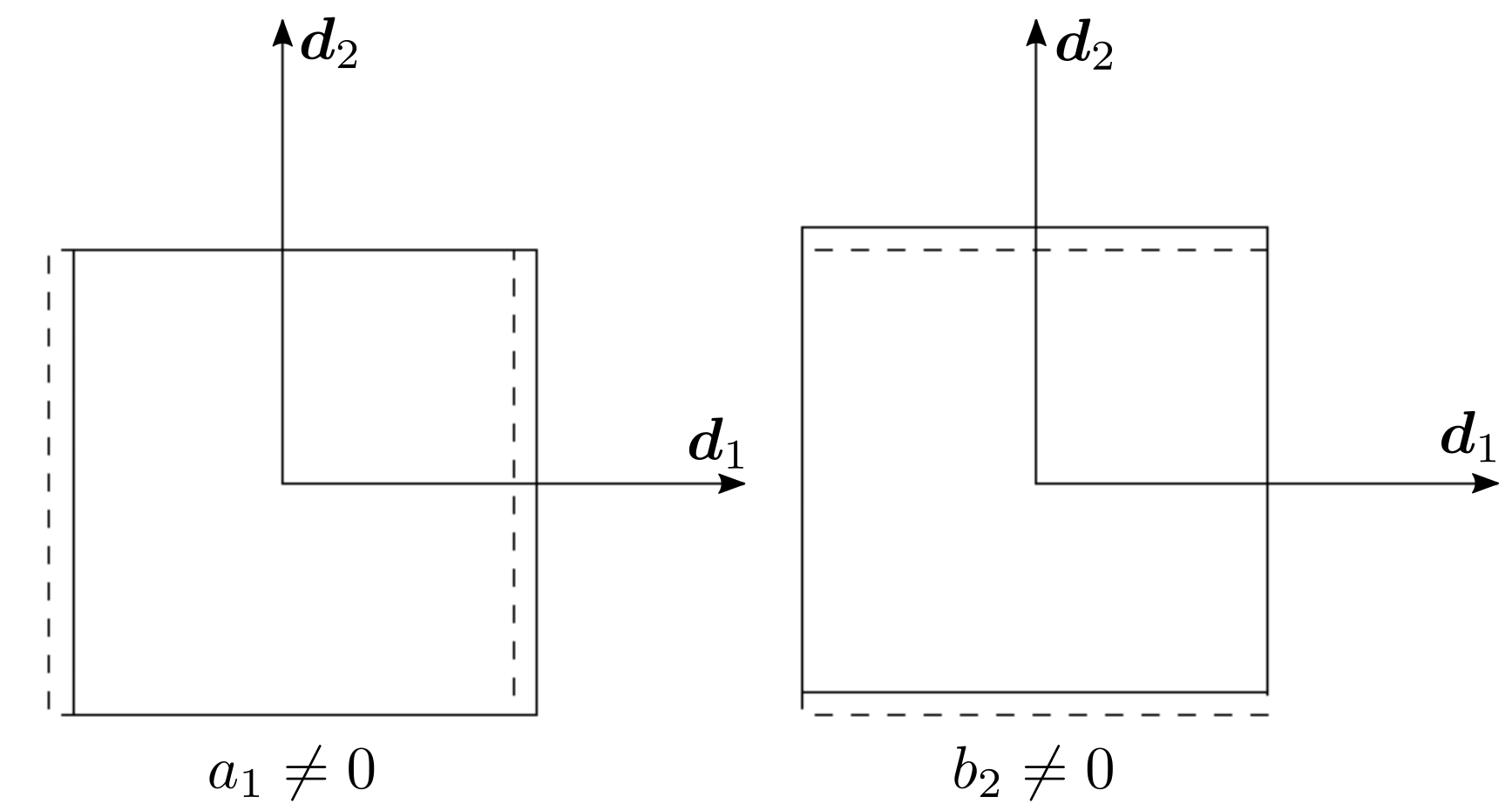}
		\caption{Linear strain due to quadratic terms}
		\label{intro_cs_deform_2nd}
	\end{subfigure}		
	\begin{subfigure}[b] {0.45\textwidth} \centering
		\includegraphics[width=\linewidth]{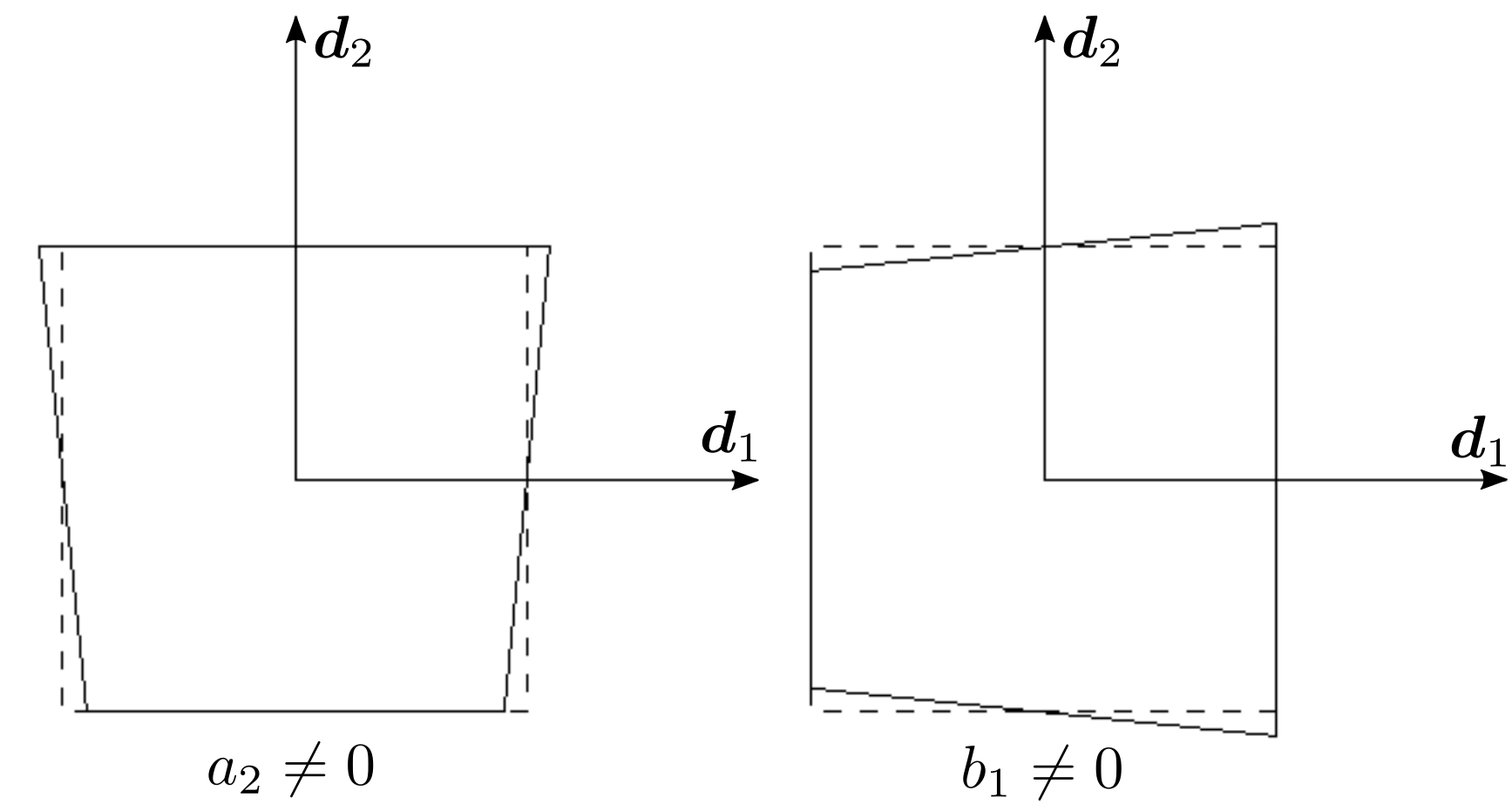}
		\caption{Linear strain due to bilinear terms}
		\label{intro_cs_deform_2nd_trapezoid}
	\end{subfigure}
\caption{{Illustration of in-plane deformations of the cross-section with linear strain field. The dashed and solid lines represent the undeformed and deformed cross-sections, respectively. (a) The through-the-thickness stretching strain is linear along the $\xi^1$ and $\xi^2$ directions in case of $a_1\ne0$ and $b_2\ne0$, respectively. (b) The through-the-thickness stretching strain is linear along the $\xi^2$ and $\xi^1$ directions in case of $a_2\ne0$ and $b_1\ne0$, respectively. Note that the deformed cross-sections have trapezoidal shapes.}}
\label{intro_cs_deform_linear}	
\end{figure*}

A rod (or rod-like body) can be regarded as a spatial curve, to which two deformable vectors, called \textit{directors} are assigned. This curve is also called \textit{directed} or \textit{Cosserat} curve. The balance laws can be stated directly in terms of the curve velocity and director velocity vectors, and their work conjugate force and director force vectors, which eventually yields the equations of motion in the one-dimensional (curve) domain \citep{green1966general}. Since we actually deal with a three-dimensional continuum, one can consistently derive the equations of motion of the rod from those of the full three-dimensional continuum. This \textit{dimensional reduction}, or \textit{degeneration} procedure is based on a suitable kinematic assumption, and this dimensionally reduced theoretical model is referred to as \textit{beam} model. An \textit{exact} expansion of the position vector of any point of the beam at time $t$ is given as \citep{antman1966dynamical}
\begin{equation}
\label{intro_beam_kin_general}
{{\boldsymbol{x}}_t} = \sum\limits_{p = 0}^\infty  {\sum\limits_{q = 0}^p {{{({\xi ^1})}^{p - q}}{{({\xi ^2})}^q}{{\boldsymbol{d}}^{(p - q,q)}}(\xi^3,t)} },
\end{equation}
where $\xi^\gamma$ ($\gamma=1,2$) denote the two coordinates in transverse (principal) directions of the cross-section plane, $\xi^3$ denotes the coordinate along the central axis, and 
\begin{equation}
\label{def_taylor_deriv_xt}
{{\boldsymbol{d}}^{(p - q,q)}}(\xi^3,t) \coloneqq \frac{1}{{(p - q)!q!}}\left( {\frac{{{\partial ^p}{{\boldsymbol{x}}_t}}}{{\partial {{({\xi^1})}^{p - q}}\partial {{({\xi^2})}^q}}}} \right).
\end{equation}
Using the full conservation laws of a three-dimensional continuum as a starting point, applying the kinematics in Eq.\,(\ref{intro_beam_kin_general}) offers an \textit{exact} reparameterization of the three-dimensional theory into the one-dimensional one \citep{antman1966dynamical,green1968rods}. However, this theory has infinite number of equations and unknowns, which makes it intractable for a finite element formulation and computation. The \textit{first order theory} assumes the position vector to be a linear function of the coordinates $\xi^{\gamma}$, i.e. \citep{volterra1956equations,antman1966dynamical}
\begin{equation}\label{intro_beam_th_str_cur_config}
{{\boldsymbol{x}}_t} = {{\boldsymbol{\varphi }}}({\xi^3},t) + \sum\limits_{\gamma  = 1}^2 {{\xi ^\gamma }{{\boldsymbol{d}}_\gamma }({\xi ^3},t)},
\end{equation}
where ${\boldsymbol{\varphi }}(\xi^3,t)\equiv{{\boldsymbol{d}}^{(0,0)}}(\xi^3,t)$ denotes the position of the beam central axis, and two directors are denoted by ${\boldsymbol{d}}_1(\xi ^3,t)\equiv{\boldsymbol{d}}_{}^{(1,0)}(\xi ^3,t)$ and ${\boldsymbol{d}}_2(\xi ^3,t)\equiv{\boldsymbol{d}}_{}^{(0,1)}(\xi ^3,t)$. This approximation simplifies the strain field; it physically implies that planar cross-sections still remain planar after deformation, but allows for constant in-plane stretching and shear deformations of the cross-section. This implies that the linear in-plane strain field in the cross-section due to the Poisson effect in bending mode cannot be accommodated in the first order theory\footnote{One can find an analytical example and discussion on this in section 6 of \citet{green1967linear}.}, which consequently increases the bending stiffness. This problem is often referred to as \textit{Poisson locking}, and the resulting error does not reduce with mesh refinement along the central axis since the displacement field in the cross-section is still linear \citep{bischoff1997shear}. One may extend the formulation in Eq.\,(\ref{intro_beam_th_str_cur_config}) to quadratic displacement field in the cross-section by adding the second order terms about the coordinates $\xi^{\gamma}$ in order to allow for a linear in-plane strain field. There are several theoretical works on this \textit{second order theory} including the work by \citet{pastrone1978dynamics} and on even higher $N$-th order theory by \citet{antman1966dynamical}. Since shell formulations have only one thickness direction, higher-order formulations are simpler than for beams. Several works including \citet{parisch1995continuum}, \citet{brank2002nonlinear}, and \citet{hokkanen2019isogeometric} employed second order theory in shell formulations. In beam formulations, several previous works considering the extensible director kinematics, which allows in-plane cross-section deformations, can be found. A theoretical study to derive balance equations and objective strain measures based on the polar decomposition of the in-plane cross-sectional deformation can be found in \citet{kumar2011geometrically}. Further extension to initially curved beams was proposed in \citet{genovese2014two}, where unconstrained quaternion parameters were utilized to represent both in-plane stretching and rotation of cross-sections. In those works, constitutive models are typically simplified to the form of quadratic strain energy density function. \citet{durville2012contact} also employed a first order theory in frictional beam-to-beam contact problems, where the constitutive law was simplified to avoid Poisson locking. \citet{coda2009solid} employed second order theory combined with an additional warping degree-of-freedom. However, it turns out that the linear in-plane strain field for the cross-section is not complete, so that the missing bilinear terms may lead to severe Poisson locking. In order to have a linear strain field in the cross-section with the increase of the number of unknowns minimized, one may extend the kinematics of Eq.\,(\ref{intro_beam_th_str_cur_config}) to
\begin{equation}\label{intro_beam_th_str_cur_config_2nd}
{\boldsymbol{x}} = {\boldsymbol{\varphi }}({\xi ^3}) + {\xi ^1}\left( {1 + {a_1}{\xi ^1} + {a_2}{\xi ^2}} \right){{\boldsymbol{d}}_1}({\xi ^3}) + {\xi ^2}\left({1 + {b_1}{\xi ^1} + {b_2}{\xi ^2}} \right){{\boldsymbol{d}}_2}({\xi ^3}),
\end{equation}
where four additional unknown coefficient functions $a_{\gamma}=a_{\gamma}({\xi^3})$ and $b_{\gamma}=b_{\gamma}({\xi^3})$ $({\gamma}=1,2)$ are introduced. Here and hereafter, the dependence of variables on time $t$ is usually omitted for brevity of expressions. This enrichment enables additional modes of the cross-sectional deformation (see Fig.\,\ref{intro_cs_deform_linear} for an illustration), which are also induced in bending deformation due to the Poisson effect.
{Eq.\,(\ref{intro_beam_th_str_cur_config_2nd}) recovers the kinematic assumption\footnote{In this paper, we focus on in-plane deformation of cross-section, although additional warping degrees-of-freedom was considered in the work of \cite{coda2009solid}. This restricts the range of application to compact convex cross-sections, where the warping effect is not pronounced.} in \cite{coda2009solid} if $a_2=b_1=0$, which means the absence of bilinear terms, so that the trapezoidal cross-section deformation, shown in Fig.\,\ref{intro_cs_deform_2nd_trapezoid}, cannot be accomodated. Therefore, Poisson locking cannot be effectively alleviated. In this paper, we employ the enhanced assumed strain (EAS) method to circumvent Poisson locking in the first order theory. In order to verify the significance of those bilinear terms in Eq.\,(\ref{intro_beam_th_str_cur_config_2nd}), in a numerical example of section \ref{ex_cant_b_end_f}, we compare two different EAS formulations based on five and nine enhanced strain parameters, respectively. The formulation of five enhanced strain parameters is obtained by ignoring the incompatible modes of trapezoidal cross-section deformation, i.e., it considers only the incompatible modes of Fig.\,\ref{intro_cs_deform_2nd}. The other one with nine enhanced strain parameters considers the whole set of incompatible linear cross-section modes, i.e., it considers both of the incompatible modes of Fig.\,\ref{intro_cs_deform_2nd} and \ref{intro_cs_deform_2nd_trapezoid}.} 

{The enhanced assumed strain (EAS) method developed in \citet{simo1990class} is based on the three-field Hu-Washizu variational principle. As the independent stress field is eliminated from the variational formulation by an orthogonality condition, it becomes a two-field variational formulation in terms of displacement and enhanced strain fields. Further, the enhanced strain parameters can be condensed out on the element level; thus the basic features of a displacement-based formulation are preserved. This method was generalized in the context of  nonlinear problems in \citet{simo1992geometrically} in which a multiplicative decomposition of the deformation gradient into compatible and incompatible parts is used. One can refer to several works including \citet{buchter1994three}, \citet{betsch19964}, \citet{bischoff1997shear}, and \cite{brank2002nonlinear} for EAS-based shell formulations. In this paper, we apply the EAS method to the beam formulation. Beyond previous beam formulations based on the kinematics of extensible directors, our work has the following highlights:}
\begin{itemize}
\item {Consistency in balance equations and boundary conditions: The director field as well as the central axis displacement field satisfy the momentum balance equations and boundary conditions consistently derived from those of the three-dimensional continuum body. In the formulation of \citet{coda2009solid} and \citet{durville2012contact}, there are no detailed expressions of balance equations, beam strains, and stress resultants. To the best of our knowledge, in those works, the finite element formulation can be obtained by substituting the beam kinematic expression of the current material point position into the deformation gradient of three-dimensional elasticity. This \textit{solid-like} formulation yields an equivalent finite element formulation through a much more simplified derivation process. However, in addition to the possibility of applying mixed variational formulations in future works, the derivation of balance equations, beam strains, and stress resultants turns out to be significant in the interpretation of coupling between different strain components (for examples, see sections \ref{ex_end_mnt_subsub_axial} and \ref{ex_end_mnt_subsub_th}.)}
\item {We employ the EAS-method, where the additional strain parameters are statically condensed out, so that the same number of nodal degrees-of-freedom is used as in the pure displacement-based formulation. Each of the enhanced in-plane transverse normal strain components is linear in both of $\xi^1$ and $\xi^2$, which is in contrast to the strains obtained from the kinematic assumption in \cite{coda2009solid}. In the numerical example of section \ref{ex_cant_b_end_f}, it is verified that this further enrichment alleviates Poisson locking more effectively.} 
\item Significance of correct surface loads: The consistently derived traction boundary condition shows that considering the correct surface load leads to an external director stress couple term that turns out to play a significant role in the accuracy of analysis.
\item Incorporation of general hyperelastic constitutive laws: As we consider the complete six stress components without any zero stress condition, our beam formulation naturally includes a straightforward interface for general three-dimensional constitutive laws.
\item Verification by comparison with brick element solution: We verify the accuracy and efficiency of our beam formulation by comparison with the results from brick elements.
\end{itemize}
It turns out that if linear shape functions are used to interpolate the director field, an artificial thickness stretch arises in bending deformations due to parasitic strain terms, and it eventually increases the bending stiffness. This effect is called \textit{curvature thickness locking}. Since the parasitic terms vanish at the nodal points, the assumed natural strain (ANS) method interpolates the transverse normal (through-the-thickness) stretch at nodes instead of evaluating it at Gauss integration points \citep*{betsch1995assumed, bischoff1997shear}. {For membrane and transverse shear locking, there are several other existing methods, for examples, selective reduced integration method in \citet{adam2014improved}, Greville quadrature method in \citet{zou2021galerkin}, and mixed variational formulation in \citet{wackerfuss2009mixed,wackerfuss2011nonlinear}. However, since curvature-thickness, membrane, and transverse shear locking issues become less significant by mesh refinement or higher-order basis functions, especially in low to moderate slenderness ratio of our interests, no special treatment is implemented in this paper (see the investigation on those locking issues in section \ref{ex_beam_end_mnt_allev_lock}). Further investigation on the application of existing method remains future work.}

If we restrict the two directors in Eq.\,(\ref{intro_beam_th_str_cur_config}) to be orthonormal, which physically means that the cross-section is rigid, large rotations of the cross-section can be described by an orthogonal transformation. In planar static problems, \citet{reissner1972one} derived the force and moment balance equations, from which the strain-displacement relation is obtained via the principle of virtual work and work conjugate relations. Since this approach poses no assumption on the magnitude of deformations, it is often called \textit{geometrically exact beam theory}. This work was extended to three-dimensional dynamic problems by \citet{simo1985finite}, which was followed by the finite element formulation of static problems in \citet{simo1986three}. An additional degree-of-freedom related to torsion-warping deformation was added in \citet{simo1991geometrically}, and this work was extended by \citet{gruttmann1998geometrical} to consider eccentricity with arbitrary cross-section shapes. There have been a number of works on the parameterization of finite rotations, and the multiplicative or additive configuration update process. One may refer to the overviews on this given by \citet{meier2014objective} and \citet{crisfield1999objectivity}. In \citet{crisfield1999objectivity}, it was pointed out that the usual spatial discretization of the strain measures in \citet{simo1986three} leads to non-invariance of the interpolated strain measures in rigid body rotation, even though the strain measures in continuum form are objective. This non-objectivity stems from the non-commutativity, i.e., non-vectorial nature of the finite rotation. To retain the objectivity of strain measures in the underlying continuum formulation, the isoparametric interpolation of director vectors is used instead of interpolating the rotational parameters (see for example \citealp{betsch2002frame, romero2002objective, eugster2014director}), and the subsequent weak form of finite element formulation is reformulated. As those beam formulations still assume rigid cross-sections, the orthonormality condition of the director vectors should be satisfied. Several methods to impose the constraint can be found in the literature, examples are the Lagrange multiplier method \citep{betsch2002frame, eugster2014director}, and the introduction of nodal rotational degrees-of-freedom \citep*{betsch2002frame, romero2002objective}. {In order to preserve the objectivity and path-independence in the rotation interpolation, several methods have been developed; for examples, orthogonal interpolation of relative rotation vectors \citep{crisfield1999objectivity,ghosh2009frame}, geodesic interpolation \citep{sander2010geodesic}, interpolation of quaternion parameters \citep{zupan2013virtual}. \citet{romero2004interpolation} compared several rotation interpolation schemes in perspective of computational accuracy and efficiency. A more comprehensive review on geometrically exact finite element beam formulations can be found in \citet{meier2019geometrically}}. In the isoparametric approximation of directors, employed in our beam formulation, the director vectors belong to $\Bbb{R}^3$, that is, no orthonormality condition is imposed. This means that the cross-section can undergo in-plane deformations like transverse normal stretch and in-plane shear deformations. {Further, it enables us to avoid the rotation group, which is a nonlinear manifold, in the configuration space of the beam, and consequently complicates the configuration and strain update process \citep{durville2012contact}. \citet{coda2009solid} and \citet{coda2011fem}, who employed an isoparametric interpolation of directors without orthonormality condition, presented several numerical examples showing the objectivity and path-independence of the finite element formulation.}

Classical beam theories introduce the zero transverse stress condition based on the assumption that the transverse stresses are much smaller than the axial and transverse shear stresses. Thus, six stress components in the three-dimensional theory reduce to three components including the transverse shear components in the Timoshenko beam theory. However, this often complicates the application of three-dimensional nonlinear material laws, and requires a computationally expensive iteration process. Global and local iteration algorithms to enforce the zero stress condition at Gauss integration points were developed in \citet{de1991zero} and \citet{klinkel2002using}, respectively. One can also refer to several recent works on Kirchhoff-Love shell formulations with general three-dimensional constitutive laws, where the transverse normal strain component can be condensed out by applying the plane stress condition in an analytical or iterative manner, for example, for hyperelasticity by \citet{kiendl2015isogeometric} and \citet{duong2017new}, and elasto-plasticity by \citet{ambati2018isogeometric}. There are several other finite element formulations to dimensionally reduce slender three-dimensional bodies and incorporate general three-dimensional constitutive laws. The so-called \textit{solid beam formulation} uses a single brick element\footnote{This is sometimes called a \textit{solid element}.} in thickness direction. To avoid severe stiffening effects typically observed in low-order elements, a brick element was developed based on the EAS method in geometrically nonlinear problems \citep*{klinkel1997geometrical}. A brick element combined with EAS, ANS, and reduced integration methods in order to alleviate locking was presented in \citet{frischkorn2013solid}. The absolute nodal coordinate (ANC) formulation uses slope vectors as nodal variables to describe the orientation of the cross-section. The \textit{fully parameterized} ANC element enables straightforward implementation of general nonlinear constitutive laws. A comprehensive review on the ANC element can be found in \citet{gerstmayr2013review}, and one can also refer to a comparison with the geometrically exact beam formulation in \citet{romero2008comparison}. \citet{wackerfuss2009mixed, wackerfuss2011nonlinear} presented a mixed variational formulation, which allows a straightforward interface to arbitrary three-dimensional constitutive laws, where each node has the common three translational and three rotational degrees-of-freedom, as the additional degrees-of-freedom are eliminated on element level via static condensation.

Isogeometric analysis (IGA) was introduced in \citet{hughes2005isogeometric} to bridge the gap between computer-aided design (CAD) and computer-aided engineering (CAE) like finite element analysis (FEA) by employing non-uniform rational B-splines (NURBS) basis functions to approximate the solution field as well as the geometry. IGA enables exact geometrical representation of initial configuration in CAD to be directly utilized in the analysis without any approximation even in coarse level of spatial discretization. Further, the high-order continuity in NURBS basis function is advantageous in describing the beam and shell kinematics under the Kirchhoff-Love constraint, which requires at least $C^1$-continuity in the displacement field. IGA was utilized for example in \citet{kiendl2015isogeometric}, \citet{duong2017new}, and \citet{ambati2018isogeometric} for Kirchhoff-Love shells, and in \citet{bauer2020weak} for Euler-Bernoulli beams. For geometrically exact Timoshenko beams, an isogeometric collocation method was presented by \citet{marino2016isogeometric}, and it was extended to a mixed formulation in \citet{marino2017locking}. An isogeometric finite element formulation and configuration design sensitivity analysis were presented in \citet{choi2019isogeometric}. Recently, \citet{vo2020total} used the Green-Lagrange strain measure with the St.\,Venant-Kirchhoff material model under the zero stress condition. There have been several works to develop optimal quadrature rules for higher order NURBS basis functions to alleviate shear and membrane locking, for examples, a selective reduced integration in \citet{adam2014improved}, and Greville quadrature in \cite{zou2021galerkin}. Since our beam formulation allows for additional cross-sectional deformations from which another type of locking due to the coupling between bending and cross-section deformations appears, it requires further investigation to apply those quadrature rules to our beam formulation, which remains future work.

{There are many applications where one may find deformable cross-sections of rods or rod-like bodies with low or moderate slenderness ratios. Although one can find many beam structures with open and thin-walled cross-sections in industrial applications, which requires to consider torsion-warping deformations, we focus on convex cross-sections in this paper, and the incorporation of out-of-plane deformations in the cross-section remains future work. Our beam formulation is useful for the analysis of beams with low to moderate slenderness ratios, where the deformation of cross-section shape is significant, for examples, due to local contact or the Poisson effect. For example, our beam formulation can be applied to the analysis of lattice or textile structures where individual ligaments or fibers have moderate slenderness ratio, and coarse-grained modeling of carbon nanotubes and DNA. Those applications are often characterized by circular or elliptical cross-section shapes. For highly slender beams, it has been shown that the assumption of undeformable cross-sections and shear-free deformations, i.e., Kirchhoff-Love theory, can be effectively and efficiently utilized \citep{meier2019geometrically}, since it enables to further reduce the number of degrees-of-freedom and avoid numerical instability due to the coupling of shear and cross-sectional deformations with bending deformation. This formulation was successfully applied to contact problems, for example, contact interactions in complex system of fibers \citep{meier2017unified}. As the slenderness ratio decreases, the analysis of local contact with cross-sectional deformations becomes significant. One example is the coupling between normal extension of the cross-section and bending deformation that can be found in the works of \citet{naghdi1989significance} and \citet{nordenholz1997steady}.} Especially, \citet{naghdi1989significance} illustrated that the difference in the transverse normal forces on the upper and lower lateral surfaces leads to flexural deformation via the Poisson effect. They also showed that the consideration of transverse normal strains plays a significant role to accurately predict a continuous surface force distribution. Another example that can lead to significant deformation of the beam cross section is local contact and adhesion of soft beams. For example, in \citet{sauer2009multiscale}, the adhesion mechanism of geckos was described by beam-to-rigid surface contact, where no deformation through the beam thickness was assumed, even though local contact can be expected to have a significant influence on beam deformation. \citet{olga2018contact} applied the Hertz theory to incorporate the effect of cross-section deformation in beam-to-beam contact, where the penalty parameter in the contact constraint was obtained as a function of the amount of penetration. Another interesting application can be found in the development of continuum models for atomistic structures like carbon nanotubes. \citet{kumar2011rod} developed a beam model for single-walled carbon nanotubes that allows for deformation of the nanotube's lateral surface in a one-dimensional framework, which can be an efficient substitute to two-dimensional shell models. 

The remainder of this paper is organized as follows. In section \ref{beam_kin}, we present the beam kinematics based on extensible directors. In section \ref{eq_motion}, we derive the momentum balance equations from the balance laws of a three-dimensional continuum, and define stress resultants and director stress couples. In section \ref{var_for_weak_form}, we derive the beam strain measures that are work conjugate to the stress resultants and director stress couples. Further, the expression of external stress resultants and director stress couples are obtained from the surface loads. In section \ref{var_form_constitutive_law} we detail the process of reducing three-dimensional hyperelastic constitutive laws to one-dimensional ones. {In section \ref{eas_formulation}, we present the enhanced assumed strain method to alleviate Poisson locking.} In section \ref{num_ex}, we verify the developed beam formulation in various numerical examples by comparing the results with those of IGA brick elements. For completeness, appendices to the beam formulation and further numerical examples are given in Appendices \ref{app_theory} and \ref{app_hypelas_conv_test}, respectively.

\section{Beam kinematics}
\label{beam_kin}
\begin{figure}[htp]
\centering	
\includegraphics[width=0.565\linewidth]{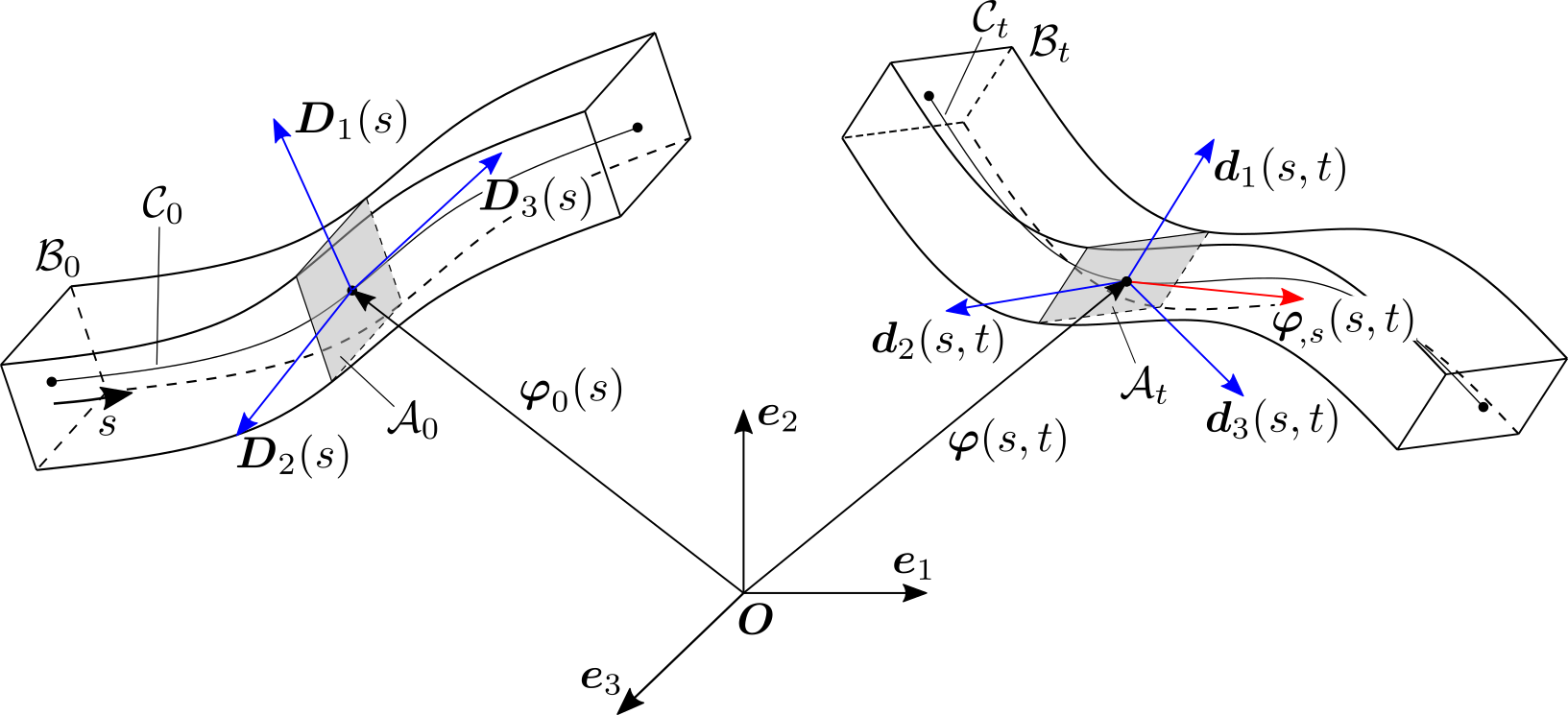}
\caption{A schematic illustration of the beam kinematics in the initial and current configurations.}
\label{beam_kin_3_domains}
\end{figure}

The configuration of a beam is described by a family of \textit{cross-sections} whose centroids\footnote{In this paper, the \textit{centroid} refers to the mass centroid. If we assume a constant mass density, it coincides with the \textit{geometrical} centroid.} are connected by a spatial curve referred to as the \textit{central axis}. An initial (undeformed) configuration of the central axis $\mathcal{C}_0$ is given by a spatial curve parameterized by a parametric coordinate $\xi\in{\Bbb{R}^1}$, i.e., ${\mathcal{C}_0}:\,{\xi} \to {{\boldsymbol{\varphi }}_0}({\xi}) \in {{\Bbb{R}}^3}$. The initial configuration of the central axis is reparameterized by the arc-length parameter $s \in \left[ {0,L} \right] \subset {{\Bbb{R}}^1}$, that is, ${\mathcal{C}_0}:\,s \to {{\boldsymbol{\varphi }}_0}(s) \in {{\Bbb{R}}^3}$. $L$ represents the length of the initial central axis. This reparameterization is advantageous to simplify the subsequent expressions due to $\left\| {{{\boldsymbol{\varphi }}_{0,s}}} \right\| = 1$. The cross-section $\mathcal{A}_0\subset {\Bbb{R}^2}$ is spanned by two orthonormal base vectors ${{\boldsymbol{D}}_{\gamma}}(s) \in {\Bbb R}^3$ ($\gamma  = 1,2$), which are called \textit{initial directors}, aligned along the principal directions of the second moment of inertia of the cross-section. Further, ${{{\boldsymbol{D}}_3}(s)}$ is defined as a unit normal vector to the initial cross-section. In this paper, it is assumed that the cross-section is orthogonal to the central axis in the initial configuration, so that we simply obtain ${{{\boldsymbol{D}}_3}(s)} \coloneqq {{\boldsymbol{\varphi }}_{0,{s}}}(s)$, which is tangent to the initial central axis. Here and hereafter, $(\bullet)_{,s}$ denotes the partial differentiation with respect to the arc-length parameter $s$. The current (deformed) configuration of the central axis is defined by the spatial curve ${{\mathcal{C}}_t}:\,s \to {{\boldsymbol{\varphi }}}(s,t) \in {{\Bbb{R}}^3}$, where $t\in{\Bbb R}^{+}$ denotes time. In the current configuration, the cross-section $\mathcal{A}_t\subset {\Bbb{R}^2}$ is defined by a plane normal to the \textit{unit vector} ${{\boldsymbol{d}_3}}(s,t) \in {{\Bbb{R}}^3}$, and the plane is spanned by two base vectors ${\boldsymbol{d}_{\gamma}}(s,t) \in {\Bbb R}^3$ ($\gamma = 1,2$), which are referred to as \textit{current directors}. In contrast to the initial configuration, those current directors are not necessarily orthogonal to each other or of unit length. Their length only needs to satisfy
\begin{equation}\label{beam_th_str_lambda_def}
{\lambda _\gamma }(s,t) \coloneqq \left\| {{{\boldsymbol{d}}_\gamma }(s,t)} \right\| > 0\,\,{\text{for}}\,\,s \in [0,L].
\end{equation}
Furthermore, in the current configuration, the cross-section remains plane but not necessarily normal to the tangent vector ${\boldsymbol{\varphi}_{\!,s}}(s,t)$, due to transverse shear deformation. ${{\boldsymbol{d}_3}}(s,t)$, which is normal to the current cross-section, can be obtained from the current directors as
\begin{equation}\label{beam_th_str_calc_d3_vec}
{{\boldsymbol{d}}_3} = \frac{{{{\boldsymbol{d}}_1} \times {{\boldsymbol{d}}_2}}}{{\left\| {{{\boldsymbol{d}}_1} \times {{\boldsymbol{d}}_2}} \right\|}}\,\,\text{where}\,\,{\left\| {{{\boldsymbol{d}}_1} \times {{\boldsymbol{d}}_2}} \right\|}\ne 0. 
\end{equation}
Note that the condition ${\left\| {{{\boldsymbol{d}}_1} \times {{\boldsymbol{d}}_2}} \right\|}\ne 0$ precludes the physically unreasonable situation of infinite in-plane shear deformation of the cross-section. We also postulate the condition
\begin{equation}\label{beam_th_str_calc_d3_vec}
{{\boldsymbol{\varphi}}_{\!,s}} \cdot ({{\boldsymbol{d}}_1} \times {{\boldsymbol{d}}_2}) > 0,
\end{equation}
which precludes the unphysical situation of infinite transverse shear deformation. We define $\left\{ {{{\boldsymbol{e}}_1},{{\boldsymbol{e}}_2},{{\boldsymbol{e}}_3}} \right\}$ as a standard Cartesian basis in ${{\Bbb R}^3}$. Fig.\,\ref{beam_kin_3_domains} schematically illustrates the above kinematic description of the initial and current beam configurations. 
\begin{figure}[htp]
\centering	
\includegraphics[width=0.4\linewidth]{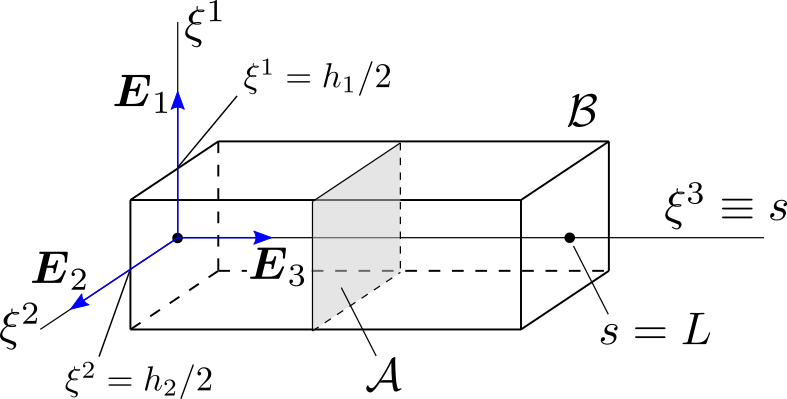}
\caption{An example of the reference domain $\mathcal{B}$ in the case of rectangular cross-section with dimension $h_1\times{h_2}$.}
\label{beam_ref_domain}
\end{figure}
\noindent We define a \textit{reference domain} ${\mathcal{B}}\coloneqq (0,L) \times {\mathcal{A}}$, where ${\mathcal{A}}$ denotes the open domain of coordinates $\xi^1$ and $\xi^2$. For example, for a rectangular cross-section with dimension $h_1\times{h_2}$ we have ${({\xi^1},{\xi^2})}\in{\mathcal{A}}\coloneqq (-{h_1}/2,{h_1}/2)\times(-{h_2}/2,{h_2}/2)$, see Fig.\,\ref{beam_ref_domain} for an illustration. The location of each point in the reference domain is expressed in terms of the coordinates ${\xi^1}$, ${\xi^2}$, and ${\xi^3}$ in the standard Cartesian basis in ${{\Bbb R}^3}$ denoted by ${{\boldsymbol{E}}_1}$, ${{\boldsymbol{E}}_2}$, and ${{\boldsymbol{E}}_3}$. We then define two mappings from the reference domain to the initial configuration $\mathcal{B}_0$ and to the current configuration $\mathcal{B}_t$ respectively by ${{\boldsymbol{x}}_0}:{\mathcal{B}} \to {{\mathcal{B}}_0}$ and ${{\boldsymbol{x}}_t}:{\mathcal{B}} \to {{\mathcal{B}}_t}$. The deformation from the initial to the current configuration is then expressed by the mapping 
\begin{equation}
{{\boldsymbol{\Phi }}_t} \coloneqq {{\boldsymbol{x}}_t} \circ {{\boldsymbol{x}}_0}^{ - 1}:{{\mathcal{B}}_0} \to {{\mathcal{B}}_t}.
\end{equation}
The initial (undeformed) configuration is expressed by
\begin{equation}\label{beam_th_str_init_config}
{{\boldsymbol{x}}_0} = {{\boldsymbol{x}}_0}({\xi ^1},{\xi ^2},{\xi ^3}) \coloneqq {{\boldsymbol{\varphi }}_0}(s) + {{\xi ^\gamma }{{\boldsymbol{D}}_\gamma }(s)},
\end{equation}
where ${\xi^3} \equiv s$. We note that the coordinates ${\xi_1},{\xi_2},{\xi_3}$ are chosen to have dimension of length, and so the director vectors ${\boldsymbol{d}_1}$ and ${\boldsymbol{d}_2}$ are dimensionless. Here and hereafter, unless stated otherwise, repeated Latin indices like $i$ and $j$ imply summation over $1$ to $3$, and repeated Greek indices like $\alpha$, $\beta$, and $\gamma$ imply summation over $1$ to $2$. Also, it is noted that the parameter $s$ is often replaced by $\xi^3$ for notational convenience. We define a covariant basis ${{\boldsymbol{G}}_{i}} \coloneqq \partial {{\boldsymbol{x}}_0}/\partial {\xi ^i}$ ($i=1,2,3)$, which then follows as
\begin{equation}\label{beam_th_str_init_cov_base}
\left\{ \begin{array}{l}
\begin{aligned}
{{\boldsymbol{G}}_1}({{\xi}^1},{{\xi}^2},{{\xi}^3}) &= {{\boldsymbol{D}}_1}(s),\\
{{\boldsymbol{G}}_2}({{\xi}^1},{{\xi}^2},{{\xi}^3}) &= {{\boldsymbol{D}}_2}(s),\\
{{\boldsymbol{G}}_3}({{\xi}^1},{{\xi}^2},{{\xi}^3}) &= {{\boldsymbol{D}}_3}(s) + {{\xi ^\gamma }{{\boldsymbol{D}}_{\gamma ,s}}(s)}.\\
\end{aligned}
\end{array} \right.
\end{equation}
\noindent The Fr{\'{e}}chet derivative of the initial configuration is then written as
\begin{align}\label{beam_th_str_frechet_deriv}
D{{\boldsymbol{x}}_0} \coloneqq {{\boldsymbol{G}}_{i}} \otimes {{{\boldsymbol{E}}^i}},
\end{align}
where ${{\boldsymbol{E}}^i} \equiv {{\boldsymbol{E}}_i}$. From the orthogonality condition ${\boldsymbol{G}_{i}} \cdot {{\boldsymbol{G}}^j} = \delta _i^j$, where the Kronecker-delta symbol is defined as
\begin{equation}
\delta _i^j = \left\{ {\begin{array}{*{20}{c}}
{0\,\,\,\,{\rm{if}}\,\,i \ne j},\\
{1\,\,\,\,{\rm{if}}\,\,i = j},
\end{array}} \right.\,\,\,\,(i,j=1,2,3),
\end{equation}
we obtain a contravariant (reciprocal) basis as
\begin{equation}\label{beam_th_str_reciprocal_basis_init}
{{\boldsymbol{G}}^i} \coloneqq D{{\boldsymbol{x}}_0}^{ - \mathrm{T}}{{\boldsymbol{E}}^i}\,\,\,\,(i=1,2,3).
\end{equation}
For convenience, here we recall the expression of current position vector of any point of the beam at time $t$ from Eq.\,(\ref{intro_beam_th_str_cur_config})
\begin{equation}\label{beam_th_str_cur_config}
{{\boldsymbol{x}}_t} = {{\boldsymbol{x}}_t({\xi^1},{\xi^2},{\xi^3},t)} = {{\boldsymbol{\varphi }}}(s,t) + {{\xi ^\gamma }{{\boldsymbol{d}}_\gamma }(s,t)}.
\end{equation}
A covariant basis, defined as ${{\boldsymbol{g}}_i} \coloneqq \partial {{\boldsymbol{x}}_t}/\partial {\xi ^i}$, is expressed by
\begin{equation}\label{beam_th_str_cur_cov_base}
\left\{ \begin{array}{l}
\begin{aligned}
{{\boldsymbol{g}}_1}({{\xi}^1},{{\xi}^2},{{\xi}^3},t) &= {{\boldsymbol{d}}_1}(s,t),\\
{{\boldsymbol{g}}_2}({{\xi}^1},{{\xi}^2},{{\xi}^3},t) &= {{\boldsymbol{d}}_2}(s,t),\\
{{\boldsymbol{g}}_3}({{\xi}^1},{{\xi}^2},{{\xi}^3},t) &= {{{\boldsymbol{\varphi}}_{\!,s}}}(s,t) + {{\xi ^\gamma }{{\boldsymbol{d}}_{\gamma ,s}}(s,t)}.
\end{aligned}
\end{array} \right.
\end{equation}
The Fr{\'{e}}chet derivative of the mapping $\boldsymbol{x}_t({\xi^1},{\xi^2},{\xi^3},t)$ is written as
\begin{align}\label{beam_th_str_frechet_deriv}
D{{\boldsymbol{x}}_t} \coloneqq {{\boldsymbol{g}}_i} \otimes {{\boldsymbol{E}}^i}.
\end{align}
From the orthogonality condition ${{\boldsymbol{g}}_i} \cdot {{\boldsymbol{g}}^j} = \delta _i^j$ we obtain the contravariant basis as
\begin{equation}\label{beam_th_str_reciprocal_basis}
{{\boldsymbol{g}}^i} \coloneqq D{{\boldsymbol{x}}_t}^{ - {\mathrm{T}}}{{\boldsymbol{E}}^i}\,\,\,\,(i=1,2,3).
\end{equation}
The deformation gradient tensor of the mapping is obtained by
\begin{align} \label{beam_th_str_deform_grad}
{{\boldsymbol{F}}} \coloneqq D{{\boldsymbol{\Phi }}_t}= D{{\boldsymbol{x}}_t}D{{\boldsymbol{x}}_0}^{ - 1}= {\boldsymbol{g}_i}\otimes{\boldsymbol{G}^i}.
\end{align}
The Jacobian of the mapping ${{\boldsymbol{\Phi}}_t}$ is then given by
\begin{equation}\label{beam_th_str_jcb_init_to_cur}
{J_t} \coloneqq \det {{\boldsymbol{F}}} = \frac{{{j_t}}}{{{j_0}}},
\end{equation}
where $\det [\bullet]$ denotes the determinant. Here, $j_0$ and $j_t$ respectively define the Jacobians of the mappings ${\boldsymbol{x}}_0({\xi^1},{\xi^2},{\xi^3})$ and ${\boldsymbol{x}}_t({\xi^1},{\xi^2},{\xi^3},t)$, and can be expressed in terms of the covariant base vectors, as (see Appendix \ref{deriv_jacob} for a derivation)
\begin{equation}\label{beam_th_str_jcb_init}
{j_0} \coloneqq \det D{{\boldsymbol{x}}_0} = \left({{\boldsymbol{G}}_1} \times {{\boldsymbol{G}}_2}\right)\cdot{{\boldsymbol{G}}_3},
\end{equation}
and
\begin{equation}\label{beam_th_str_jcb_cur}
{j_t} \coloneqq \det D{{\boldsymbol{x}}_t} = \left({{\boldsymbol{g}}_1} \times {{\boldsymbol{g}}_2}\right)\cdot{{\boldsymbol{g}}_3}.
\end{equation}
The infinitesimal volume in the reference configuration can be expressed by
\begin{equation}
{\mathrm{d}}{\mathcal{B}} = {\mathrm{d}}{\xi^1}{\mathrm{d}}{\xi^2}{\mathrm{d}}{\xi^3}.
\end{equation}
Then the corresponding infinitesimal volume due to the mappings of Eqs.\,(\ref{beam_th_str_init_config}) and (\ref{beam_th_str_cur_config}) are, respectively, obtained by
\begin{subequations}
\begin{align}\label{beam_inf_vol_jcb}
{\mathrm{d}}{\mathcal{B}_0} &= {j_0}\,{\mathrm{d}}{\mathcal{B}},\\
{\mathrm{d}}{\mathcal{B}_t}&={j_t}\,{\mathrm{d}}{\mathcal{B}}={J_t}\,{\mathrm{d}}{\mathcal{B}_0}.
\end{align}
\end{subequations}
\begin{definition}
\label{remark_lat_bd_surf_new}
\textit{Area change of the lateral boundary surface.} 
Let ${\boldsymbol{\nu }} = {{\nu}_i}{{\boldsymbol{E}}^i}$ denote the outward unit normal vector on the boundary surface $\mathcal{S}\coloneqq{\partial{\mathcal{B}}}$, and ${\mathrm{d}}\mathcal{S}$ represent an infinitesimal area. The surface area vector in the current configuration can be expressed by\footnote{This formula of area change is often called \textit{Nanson's formula}.}
\begin{align} 
{\mathrm{d}}{{\boldsymbol{\mathcal{S}}}_t} &\coloneqq {{\boldsymbol{\nu }}_t}\,{\mathrm{d}}{{\mathcal{S}}_t} = {j_t}\,D{{\boldsymbol{x}}_t}^{ - {\mathrm{T}}}{\boldsymbol{\nu }}\,{\mathrm{d}}{\mathcal{S}}\label{cur_surf_transform_1},
\end{align}
\noindent where $\boldsymbol{\nu}_t$ denotes the outward unit normal vector on the surface $\mathcal{S}_t$, and ${\mathrm{d}}\mathcal{S}_t$ denotes the infinitesimal area. In the same way, the surface area vector in the initial configuration can be expressed by
\begin{align}
{\mathrm{d}}{{\boldsymbol{\mathcal{S}}}_0} \coloneqq {{\boldsymbol{\nu }}_0}\,{\mathrm{d}}{{\mathcal{S}}_0}
= {j_0}\,D{{\boldsymbol{x}}_0}^{ - {\mathrm{T}}}{\boldsymbol{\nu }}\,{\mathrm{d}}{\mathcal{S}},\label{init_surf_transform}
\end{align}
where $\boldsymbol{\nu}_0$ denotes the outward unit normal vector on the surface $\mathcal{S}_0$, and ${\mathrm{d}}\mathcal{S}_0$ denotes the infinitesimal area. Combining Eqs.\,(\ref{cur_surf_transform_1}) and (\ref{init_surf_transform}), we have
\begin{align}\label{cur_surf_transform_2}
{\mathrm{d}}{{\boldsymbol{\mathcal{S}}}_t} = {J_t}\,{{\boldsymbol{F}}^{ - {\mathrm{T}}}}{{\boldsymbol{\nu }}_0}\,{\mathrm{d}}{{\mathcal{S}}_0}.
\end{align}
\end{definition}
\noindent {If the lateral boundary surface $\mathcal{S}^\mathrm{L}_0$ is parameterized by two convective coordinates $\zeta^1$ and $\zeta^2$, i.e., $\mathcal{S}^\mathrm{L}_0: (\zeta^1,\zeta^2)\in\Bbb{R}^2\to\boldsymbol{X}^\mathrm{L}(\zeta^1,\zeta^2)\in\Bbb{R}^3$, the infinitesimal area of lateral boundary surface $\mathcal{S}^{\mathrm{L}}_0$ can be expressed by
\begin{align}
\label{init_lat_aurf_inf_area}
\mathrm{d}{{\mathcal{S}}^\mathrm{L}_0} = \left\| {{{\boldsymbol{A}}_{1}}} \times {{\boldsymbol{A}}_{2}} \right\|\mathrm{d}{\zeta^1}\mathrm{d}{\zeta^2},
\end{align}
where ${{\boldsymbol{A}}_{\alpha}}\coloneqq\partial\boldsymbol{X}^\mathrm{L}/\partial{\zeta^\alpha}\,(\alpha=1,2)$ denotes the surface covariant base vectors. For example, if the lateral boundary surface is parameterized by a NURBS surface, and the convective coordinate $\zeta^1$ represents the coordinate along the central axis, Eq.\,(\ref{init_lat_aurf_inf_area}) can be rewritten, using $\mathrm{d}s={\tilde j}\mathrm{d}\zeta^1$ with ${\tilde j}\coloneqq\left\|\partial\boldsymbol{\varphi}_0/\partial{\zeta^1}\right\|$, as
\begin{align}
\label{inf_area_lat_bd_surf}
\mathrm{d}{{\mathcal{S}}^{\mathrm{L}}_0} = \mathrm{d}{\Gamma _0}\mathrm{d}s\,\,\mathrm{with}\,\,\mathrm{d}{\Gamma _0}\coloneqq \frac{1}{{\tilde j}}\left\| {{{\boldsymbol{A}}_{1}} \times {{\boldsymbol{A}}_{2}}} \right\|\mathrm{d}\zeta^2.
\end{align}
It is clear advantage of using IGA that the beam central axis curve and the lateral boundary surface can be parameterized by the same coordinate in the axial direction, which enables to calculate the exact surface geometrical information like covariant base vectors $\boldsymbol{A}_1$ and ${\boldsymbol{A}_2}$ in Eq.\,(\ref{inf_area_lat_bd_surf}). The significance of geometrical exactness in the calculation of the surface integral might be more significant in laterally loaded beam with varying cross-section. However, in this paper, we deal only with uniform cross-sections along the central axis, and the investigation on the different kinds of parameterization of lateral boundary surface and the significance of geometrical exactness remain future works.
}
\section{Equations of motion}
\label{eq_motion}
\subsection{Three-dimensional elasticity}
We recall the equilibrium equations and boundary conditions of a three-dimensional deformable body, which occupies an open domain $\mathcal{B}_t$ bounded by the boundary surface $\mathcal{S}_t\coloneqq \partial{\mathcal{B}_t}$ in the current configuration. The boundary is composed of a prescribed displacement boundary $\mathcal{S}^{\mathrm{D}}_t$ and a prescribed traction boundary $\mathcal{S}^{\mathrm{N}}_t$, which are mutually disjoint, i.e.\footnote{Strictly speaking, those boundary conditions are defined for each independent component in the global Cartesian frame.}
\begin{align}\label{solid_elas_boundary_disjoint}
\mathcal{S}_t = \mathcal{S}^{\mathrm{D}}_t \cup \mathcal{S}^\mathrm{N}_t,\,\,{\text{and}}\,\,\mathcal{S}_t^{\mathrm{D}} \cap \mathcal{S}_t^\mathrm{N} = \emptyset.
\end{align}
The equations of motion are obtained from the local forms of the balance laws whose derivation can be found in many references on the continuum mechanics, for example, \citet{bonet2010nonlinear}. First, the local conservation of mass is expressed by ${\rho _0} = {\rho _t}{J_t}\,\,\text{in}\,\,{\mathcal{B}}_t$, where $\rho_0$ and $\rho_t$ define the mass densities at the initial and current configurations, respectively. Second, the local balance of linear momentum in a three-dimensional body is expressed as
\begin{align}\label{conserv_lin_mnt_intrinsic}
\text{div}\boldsymbol{\sigma} + {\boldsymbol{b}} = {{\rho}_t}\,{\boldsymbol{x}_{t,tt}}\,\,\text{in}\,\,{\mathcal{B}}_t,
\end{align}
where $\boldsymbol{\sigma}$ denotes the Cauchy stress tensor, and $\text{div}(\bullet)$ represents the divergence operator with respect to the current configuration, and $\boldsymbol{b}$ represents the body force per unit current volume, and $(\bullet)_{,tt}$ represents the second order partial differentiation with respect to time. Third, the local balance of angular momentum in the absence of body moment is expressed by the symmetry of the Cauchy stress tensor, i.e., $\boldsymbol{\sigma} = \boldsymbol{\sigma}^{\mathrm{T}}\,\,\text{in}\,\,{\mathcal{B}}_t$. The non-homogeneous Dirichlet (displacement) boundary condition is given as
\begin{equation}\label{solid_elas_disp_bdc}
\boldsymbol{u}_t = \boldsymbol{\bar u}_0,\,\text{or equivalently}\,\,\boldsymbol{x}_t = {\bar {\boldsymbol{x}}}_0\,\,\text{on}\,\,\mathcal{S}^{\mathrm{D}}_t, 
\end{equation}
where $\boldsymbol{u}_t\coloneqq {\boldsymbol{x}_t}-{\boldsymbol{x}_0}$ denotes the displacement vector, and $\boldsymbol{\bar u}_0$ and $\boldsymbol{\bar x}_0$ are the prescribed values. Taking the first variation of Eq.\,(\ref{solid_elas_disp_bdc}) yields the homogeneous Dirichlet boundary condition
\begin{equation}\label{solid_elas_disp_bdc_homo}
\delta\boldsymbol{u}_t = \boldsymbol{0},\,\,\text{or equivalently}\,\,\delta{\boldsymbol{x}}_t=\boldsymbol{0}\,\,\text{on}\,\,\mathcal{S}^{\mathrm{D}}_t. 
\end{equation}
Further, the natural (traction) boundary condition is given as
\begin{equation}\label{solid_elas_traction_bdc}
\boldsymbol{\sigma}\boldsymbol{\nu}_t = {\bar {\boldsymbol{t}}}_0\,\,\text{on}\,\,\mathcal{S}^{\mathrm{N}}_t,
\end{equation}
where $\boldsymbol{\nu}_t$ defines the unit outward normal vector on $\mathcal{S}^{\mathrm{N}}_t$, and ${\bar {\boldsymbol{t}}}_0$ defines the prescribed surface traction vector in the current configuration. The surface traction can be also defined with respect to the initial configuration, as
\begin{equation}\label{solid_elas_traction_bdc_init}
\boldsymbol{P}\boldsymbol{\nu}_0 = {\bar{\boldsymbol{T}}}_0\,\,\text{on}\,\,\mathcal{S}^{\mathrm{N}}_0,
\end{equation}
where ${\boldsymbol{P}} \coloneqq {J_t}{\boldsymbol{\sigma }}{{\boldsymbol{F}}^{ - \mathrm{T}}}$
denotes the first Piola-Kirchhoff stress tensor, and $\boldsymbol{\nu}_0$ and $\bar{\boldsymbol{T}}_0$ define the unit outward normal vector and the prescribed surface traction vector, respectively, on $\mathcal{S}^{\mathrm{N}}_0$.

\subsection{Resultant linear and director momentum}
The \textit{resultant linear momentum} over the cross-section $\mathcal{A}_t$, with units of linear momentum per unit of initial arc-length, is defined as
\begin{align}\label{lin_momentum_def}
{{\boldsymbol{p}}_t} \coloneqq \int_{\mathcal {A}} {{\rho _t}}\,{\boldsymbol{x}}_{t,t}\,{j_t}\,{\mathrm{d}}{\mathcal{A}}= \int_{\mathcal {A}} {{\rho _0}}\,{\boldsymbol{x}}_{t,t}\,{j_0}\,{\mathrm{d}}{\mathcal{A}},
\end{align}
where $\mathrm{d}{\mathcal{A}} \coloneqq {\mathrm{d}}{\xi ^1}{\mathrm{d}}{\xi ^2}$ denotes the infinitesimal area of the cross-section in the reference domain. $(\bullet)_{,t}$ denotes the partial differentiation with respect to time. As $\boldsymbol{\varphi}(s,t)$ represents the current position of the centroid, the parametric position $({\xi^1},{\xi^2})\in\mathcal{A}$ satisfies
\begin{align}\label{vanish_1st_integ_rho}
\int_{\mathcal{A}} {{\xi^{\gamma}}\,{\rho _0}\,{j_0}\,{\mathrm{d}}{\mathcal{A}}} = 0\,\,\quad(\gamma=1,2).
\end{align}
By substituting Eq.\,(\ref{beam_th_str_cur_config}) into Eq.\,(\ref{lin_momentum_def}) and using Eq.\,(\ref{vanish_1st_integ_rho}), we have
\begin{align}\label{lin_momentum_fin}
{{\boldsymbol{p}}_t} = {{\rho}_A}{\boldsymbol{\varphi}_{\!,t}},
\end{align}
where ${{\rho}_A}$ represents the initial line density (mass per unit of initial arc-length), defined as
\begin{align}\label{area_rho_def}
{{\rho}_A} \coloneqq \int_{\mathcal{A}} {{\rho _0}\,{j_0}\,{\mathrm{d}}{\mathcal{A}}}.
\end{align}
Similarly, we define the \textit{resultant angular momentum} over the cross-section $\mathcal{A}_t$, with units of angular momentum per unit of initial arc-length, as
\begin{align}\label{ang_momentum_def}
{{\boldsymbol{H}}_t} &\coloneqq \int_{\mathcal{A}} {\left\{({{\boldsymbol{x}}_t} - {\boldsymbol{\varphi }}) \times {\rho _t}\,{{\boldsymbol{x}}_{t,t}}\,{j_t}\right\}{\mathrm{d}}{\mathcal{A}}} = {{\boldsymbol{d}}_\gamma } \times {{\boldsymbol{\tilde H}}^{\gamma}_t},
\end{align}
where ${{\boldsymbol{\tilde H}}^{\gamma}_t}$ defines the \textit{resultant director momentum}, given by
\begin{align}\label{dir_momentum_def}
{{\boldsymbol{\tilde H}}^{\gamma}_t} &\coloneqq \int_{\mathcal{A}} {{\xi ^\gamma }{\rho _t}\,{{\boldsymbol{x}}_{t,t}}\,{j_t}\,{\mathrm{d}}{\mathcal{A}}}\,\,\quad(\gamma=1,2).
\end{align}
Substituting Eq. (\ref{beam_th_str_cur_config}) into Eq. (\ref{dir_momentum_def}), we obtain
\begin{align}\label{dir_momentum_1}
{{\boldsymbol{\tilde H}}^{\gamma}_t} = I_\rho ^{\gamma \delta }{{\boldsymbol{d}}_{\delta ,t}}\,\,\quad(\gamma=1,2),
\end{align}
where the components of the second moment of inertia tensor are expressed by
\begin{align}\label{dir_momentum_inertia}
I_\rho ^{\gamma \delta } \coloneqq \int_{\mathcal{A}} {{\rho _t}\,{\xi ^\gamma }\,{\xi ^\delta }{j_t}\,{\mathrm{d}}{\mathcal{A}}}  = \int_{\mathcal{A}} {{\rho _0}\,{\xi ^\gamma }\,{\xi ^\delta }{j_0}\,{\mathrm{d}}{\mathcal{A}}}.
\end{align}
Note that these components of the second moment of inertia tensor do not depend on time.
\subsection{Stress resultants and stress couples}
We formulate the balance equations in terms of stress resultants and director stress couples.
We define the \textit{stress resultant} as the force acting on the cross-section $\mathcal{A}_t$, i.e.
\begin{equation}\label{beam_th_str_def_res_force}
{\boldsymbol{n}} \coloneqq \int_{\mathcal{A}} {{\boldsymbol{\sigma }}{{\boldsymbol{g}}^3}{j_t}\,{\mathrm{d}}{\mathcal{A}}}.
\end{equation}
Similarly, we define the \textit{stress couple} as the moment acting on the cross-section $\mathcal{A}_t$, i.e.
\begin{align}\label{def_strs_couple}
{{\boldsymbol{m}}} &\coloneqq \int_{\mathcal{A}} {({{\boldsymbol{x}}_t} - {\boldsymbol{\varphi }}) \times {\boldsymbol{\sigma }}{{\boldsymbol{g}}^3}{j_t}\,{\mathrm{d}}{\mathcal{A}}} = {{\boldsymbol{d}}_\alpha } \times {{{\boldsymbol{\tilde m}}}^\alpha },
\end{align}
where ${{{\boldsymbol{\tilde m}}}^\alpha}$ defines the \textit{director stress couple}, given by
\begin{align}\label{def_dir_strs_couple}
{{\boldsymbol{\tilde m}}^\alpha } \coloneqq \int_{\mathcal{A}} {{\xi ^\alpha}{\boldsymbol{\sigma }}{{\boldsymbol{g}}^3}{j_t}\,{\mathrm{d}}{\mathcal{A}}}\quad{({\alpha}=1,2).}
\end{align}
We further define the \textit{through-the-thickness stress resultant} as
\begin{equation}\label{def_th_strs_res}
{{\boldsymbol{l}}^\alpha } \coloneqq \int_{\mathcal{A}} {{\boldsymbol{\sigma }}{{\boldsymbol{g}}^\alpha }{j_t}\,{\mathrm{d}}{\mathcal{A}}}\quad{({\alpha}=1,2).}
\end{equation}
\subsection{Momentum balance equations}
Starting from Eq.\,(\ref{conserv_lin_mnt_intrinsic}) the resultant forms of the local linear and director momentum balance equations are respectively derived as (see Appendix \ref{deriv_bal_lin_momentum} for a detailed derivation)
\begin{align}\label{beam_lin_mnt_balance_app_eq}
{{\boldsymbol{n}}_{,s}} + {\boldsymbol{\bar n}} = {{\rho}_A}{\boldsymbol{\varphi}_{\!,tt}},
\end{align}
and
\begin{align}\label{beam_dir_mnt_bal_eq}
{\boldsymbol{\tilde m}}_{,s}^\gamma  - {{\boldsymbol{l}}^\gamma } + {{\boldsymbol{\bar {\tilde m}}}^\gamma } = I_\rho ^{\gamma \delta }{{\boldsymbol{d}}_{\delta ,tt}}\quad(\gamma=1,2).
\end{align}
Here, ${\boldsymbol {\bar n}} = {\boldsymbol {\bar n}}(s,t)$ denotes the \textit{external stress resultant}, with units of external force per unit of initial arc-length, given by
\begin{align}\label{beam_lin_mnt_balance_ext_f}
{\boldsymbol{\bar n}} \coloneqq \int_{\partial {{\mathcal{A}}_0}} {{{{\boldsymbol{\bar T}}}_0}\,{\mathrm{d}}{{\Gamma}_0}}  + \int_{{{\mathcal{A}}}} {{\boldsymbol{b}_0}\,{j_0}\,{\mathrm{d}}{{\mathcal{A}}}},
\end{align}
where $\boldsymbol{b}_0$ denotes the body force per unit initial volume such that ${j_t}\boldsymbol{b}_t={j_0}\boldsymbol{b}_0$. ${{\boldsymbol{\bar {\tilde m}}}^\gamma } = {{\boldsymbol{\bar {\tilde m}}}^\gamma }(s)$ denotes the \textit{external director stress couple}, which is an external moment per unit of initial arc-length due to the surface and body force fields, given by
\begin{align}\label{beam_dir_mnt_balance_ext_m}
{{\boldsymbol{\bar {\tilde m}}}^\gamma } \coloneqq \int_{\partial {{\mathcal{A}}_0}} {{\xi ^\gamma }{{{\boldsymbol{\bar T}}}_0}\,{\mathrm{d}}{{\Gamma}_0}}  + \int_{{{\mathcal{A}}}} {{\xi ^\gamma }{\boldsymbol{b}_0}\,{j_0}\,{\mathrm{d}}{{\mathcal{A}}}}\quad(\gamma=1,2).
\end{align}
We also obtain the resultant form of the balance of angular momentum from the symmetry of the Cauchy stress tensor, as (see Appendix \ref{deriv_bal_ang_momentum} for a detailed derivation)
\begin{align}\label{beam_ang_mnt_balance}
{{\boldsymbol{\varphi }}_{\!,s}} \times {\boldsymbol{n}} + {{\boldsymbol{d}}_{\gamma ,s}} \times {{\boldsymbol{\tilde m}}^\gamma } + {{\boldsymbol{d}}_\gamma } \times {{\boldsymbol{l}}^\gamma } = {\boldsymbol{0}}.
\end{align}
We finally state the static beam problem: Find
${\boldsymbol{y}} \coloneqq {\left[ {{{\boldsymbol{\varphi }}^{\mathrm{T}}},{{\boldsymbol{d}}_1}^{\mathrm{T}},{{\boldsymbol{d}}_2}^{\mathrm{T}}} \right]^{\mathrm{T}}} \in {\left[ {{{\Bbb{R}}^3}} \right]^3}$ that satisfies \citep{naghdi1981finite}
\begin{subequations}
\label{recall_momentum_balance_eq}
\begin{alignat}{2}
{{\boldsymbol{n}}_{,s}} + {\boldsymbol{\bar n}} = {\boldsymbol{0}}&{}\quad\text{(linear momentum balance),}\label{mnt_bal_eqn_lin_mnt}\\
{\boldsymbol{\tilde m}}_{,s}^\gamma  - {{\boldsymbol{l}}^\gamma } + {{{\boldsymbol{\bar {\tilde m}}}}^\gamma } = {\boldsymbol{0}}&{}\quad(\text{director momentum balance),}\label{mnt_bal_eqn_dir_mnt}\\
{{\boldsymbol{\varphi }}_{\!,s}} \times {\boldsymbol{n}} + {{\boldsymbol{d}}_{\gamma ,s}} \times {{{\boldsymbol{\tilde m}}}^\gamma } + {{\boldsymbol{d}}_\gamma } \times {{\boldsymbol{l}}^\gamma } = {\boldsymbol{0}}&{}\quad\text{(angular momentum balance).}\label{final_ang_mnt_bal}
\end{alignat}
\end{subequations}
We define the Dirichlet boundary condition, as 
\begin{align}\label{dirichlet_bdc}
{\boldsymbol{\varphi }} = {{\boldsymbol{\bar \varphi }}_0},\,\,{{\boldsymbol{d}}_1} = {{\boldsymbol{\bar d}}_{01}},\,\,{{\boldsymbol{d}}_2} = {{\boldsymbol{\bar d}}_{02}}\,\,\,\text{on}\,\,{\Gamma_\mathrm{D}},
\end{align}
where the central axis position and director vectors are prescribed at the boundary ${\Gamma_\mathrm{D}}\ni{s}$. The Neumann boundary condition is defined as
\begin{align}\label{natural_bdc}
{\boldsymbol{n}} = {{\boldsymbol{\bar n}}_0},\,\,{{\boldsymbol{{\tilde m}}}^\gamma } = {\boldsymbol{\bar {\tilde m}}}_0^\gamma\,\,\,\text{on}\,\,{\Gamma_\mathrm{N}}\,\,\quad(\gamma=1,2).
\end{align}
It is noted that ${\Gamma_\mathrm{D}} \cap {\Gamma_\mathrm{N}} = \emptyset$, and ${\Gamma_\mathrm{D}} \cup {\Gamma_\mathrm{N}} = \left\{ {0,L} \right\}$. 
\subsection{Effective stress resultant}
The balance of angular momentum given by Eq.\,(\ref{final_ang_mnt_bal}) can be automatically satisfied by representing the balance laws in terms of an effective stress resultant tensor\,\citep{simo1990stress}. We define this effective stress resultant tensor as
\begin{align}\label{beam_th_def_eff_strs_res}
{\boldsymbol{\tilde n}} &\coloneqq {\boldsymbol{n}} \otimes {{\boldsymbol{\varphi}}_{\!,s}} - {{\boldsymbol{d}}_{\gamma ,s}} \otimes {{\tilde {\boldsymbol{m}}}^\gamma } + {{\boldsymbol{l}}^\gamma } \otimes {{\boldsymbol{d}}_\gamma }.
\end{align}
We also recall the identities $\widehat {{\boldsymbol{a}} \times {\boldsymbol{b}}} = {2\,\rm{skew}}[{\boldsymbol{b}} \otimes {\boldsymbol{a}}]$ and ${\rm{skew}}[{\boldsymbol{a}} \otimes {\boldsymbol{b}}] =  - {\rm{skew}}[{\boldsymbol{b}} \otimes {\boldsymbol{a}}]$ for vectors ${\boldsymbol{a}},{\boldsymbol{b}} \in {{\Bbb R}^3}$ where $\widehat {(\bullet)}$ represents the skew-symmetric matrix associated with the vector $(\bullet)\in{{\Bbb R}^3}$, that is, $\widehat {(\bullet)}{\boldsymbol{a}} = (\bullet) \times {\boldsymbol{a}},\,\forall {\boldsymbol{a}} \in {{\Bbb{R}}^3}$, and ${\rm{skew}}[(\bullet)] \coloneqq {\frac{1}{2}}\left\{{(\bullet) - {(\bullet)^{\mathrm{T}}}}\right\}$. Then Eq.\,(\ref{final_ang_mnt_bal}) can be rewritten as the symmetry condition of the effective stress resultant tensor, i.e., ${\tilde{\boldsymbol{n}}}={{\tilde{\boldsymbol{n}}}^{\mathrm{T}}}$. 

Decomposing the stress resultant forces and moment relative to the basis of $\{{{{\boldsymbol{d}}_1},{{\boldsymbol{d}}_2},{{\boldsymbol{\varphi}}_{\!,s}}}\}$ yields
\begin{subequations}
\label{beam_dec_strs_res}
\begin{alignat}{2}
{\boldsymbol{n}} &= {n}{{\boldsymbol{\varphi}}_{\!,s}} + {q^\alpha }{{\boldsymbol{d}}_\alpha },\label{strs_res_f}\\
{{\boldsymbol{\tilde m}}^\alpha } &= {{\tilde m}^{\alpha}}{{\boldsymbol{\varphi}}_{\!,s}} + {{\tilde m}^{\beta \alpha}}{{\boldsymbol{d}}_\beta},\label{strs_res_dir_mnt}\\
{{\boldsymbol{l}}^\alpha} &= {l^{\alpha}}{{\boldsymbol{\varphi}}_{\!,s}} + {l^{\beta\alpha}}{{\boldsymbol{d}}_\beta }.\label{strs_res_dir_f}
\end{alignat}
\end{subequations}
We also decompose ${{\boldsymbol{d}}_{\alpha ,s}}$ in the same basis as
\begin{align}\label{beam_dec_d_s}
{{\boldsymbol{d}}_{\alpha ,s}} = {k} _\alpha{{\boldsymbol{\varphi}}_{\!,s}} + {k} _\alpha ^\beta {{\boldsymbol{d}}_\beta }.
\end{align}
\begin{definition}
\label{remark_curv_k}
\textit{Physical interpretation of current curvatures}. Without loss of generality, we examine the case $\alpha=1$ in Eq.\,(\ref{beam_dec_d_s}). The change of director vector along the central axis has three different components, i.e.
\begin{equation}\label{remark_beam_dec_d_s}
{{\boldsymbol{d}}_{1,s}} = {k}_{1}{{\boldsymbol{\varphi}}_{\!,s}} + {k}_{1}^{1} {{\boldsymbol{d}}_{1}} + {k} _{1}^{2}{{\boldsymbol{d}}_{2}}.
\end{equation}
The components $k_1$, $k_1^2$ represent the \textit{bending} and \textit{torsional} curvatures in the current configuration. However, they are not exactly geometrical curvatures, since the basis $\left\{ {{{\boldsymbol{d}}_1},{{\boldsymbol{d}}_2},{{\boldsymbol{\varphi }}_{\!,s}}} \right\}$ is not orthonormal. $k_1^1$ is associated with a varying cross-section stretch ($\lambda_1$) along the central axis in the current configuration. If the transverse and in-plane cross-section shear deformations are zero (i.e., ${{\boldsymbol{\varphi }}_{,s}} \cdot {{\boldsymbol{d}}_1} = {{\boldsymbol{d}}_1} \cdot {{\boldsymbol{d}}_2} = 0$), we have $k_1^1 = {\lambda _{1,s}}/{\lambda _1}$. In other words, if the cross-section stretch is non-varying along the central axis in the current configuration, we have $k_1^1 = 0$.
\end{definition}
Using the component forms in Eqs.\,(\ref{beam_dec_strs_res}) and (\ref{beam_dec_d_s}), the effective stress resultant tensor of Eq.\,(\ref{beam_th_def_eff_strs_res}) can be rewritten as
\begin{equation}
{\boldsymbol{\tilde n}} = {{\tilde n}}{{\boldsymbol{\varphi}}_{\!,s}} \otimes {{\boldsymbol{\varphi}}_{\!,s}} + {{\tilde q}^\alpha }{{\boldsymbol{d}}_\alpha } \otimes {{\boldsymbol{\varphi}}_{\!,s}}
+ {{\tilde l}^{\alpha }} {{\boldsymbol{\varphi}}_{\!,s}} \otimes {{\boldsymbol{d}}_\alpha } + {{\tilde l}^{\alpha \beta }}{{\boldsymbol{d}}_\alpha } \otimes {{\boldsymbol{d}}_\beta },
\end{equation}
where the following component expressions are defined relative to the basis $\{{{{\boldsymbol{d}}_1},{{\boldsymbol{d}}_2},{{\boldsymbol{\varphi}}_{\!,s}}}\}$
\begin{subequations}
\label{beam_th_strs_res_comp_basis_d123}
\begin{alignat}{2}
{{\tilde n}} &\coloneqq {n} - {{\tilde m}^{\gamma }}k _\gamma\,\,&&({\text{effective axial stress resultant}}),\label{eff_axial_res}\\
{{\tilde q}^\alpha } &\coloneqq {q^\alpha } - {{\tilde m}^{\gamma }}k _\gamma ^\alpha\,\,&&({\text{effective transverse shear stress resultant}}), \label{eff_trans_shear_res}\\
{{\tilde l}^{\alpha }} &\coloneqq {l^{\alpha }} - {{\tilde m}^{\alpha \gamma }}k _\gamma\,\,\,&&({\text{effective longitudinal shear stress resultant}}),\label{eff_sym_shear_res}\\
{{\tilde l}^{\alpha \beta }} &\coloneqq {l^{\beta \alpha }} - {{\tilde m}^{\alpha \gamma }}k _\gamma ^\beta \,\,&&({\text{effective transverse normal and cross-section shear stress resultants).}}\label{eff_trans_nm_cs_shear}
\end{alignat}
\end{subequations}
The symmetry condition ${\tilde{\boldsymbol{n}}}={{\tilde{\boldsymbol{n}}}^{\mathrm{T}}}$ yields the following symmetry conditions on the components
\begin{align}
{\tilde q^\alpha } = {\tilde l^{\alpha }}\,\,{\text{and}}\,\,\,{\tilde l^{\alpha \beta }} = {\tilde l^{\beta \alpha }}.
\end{align}
\section{Variational formulation}
\subsection{Weak form of the governing equation}
\label{var_for_weak_form}
We define a variational space by
\begin{align}\label{var_space}
{\mathcal{V}} \coloneqq \left\{ {\left. {\delta\boldsymbol{y}\coloneqq\left[\delta {\boldsymbol{\varphi }}^{\mathrm{T}},\delta {{\boldsymbol{d}}_1}^{\mathrm{T}},\delta {{\boldsymbol{d}}_2}^{\mathrm{T}}\right]^{\mathrm{T}} \in {\left[{H^1}(0,L)\right]^{d}}} \right|\delta {\boldsymbol{\varphi }} = \delta {{\boldsymbol{d}}_1} = \delta {{\boldsymbol{d}}_2} = \boldsymbol{0}\,\,\text{on}\,\,{\Gamma _\mathrm{D}}} \right\},
\end{align}
where ${H^1}(0,L)$ defines the Sobolev space of order one which is the collection of all continuous functions whose first order derivatives are square integrable in the open domain $(0,L)\ni{s}$. Here the components of $\delta \boldsymbol{y}$ in the global Cartesian frame are considered as independent solution functions, so that the dimension becomes $d=9$. In the following, we restrict our attention to the static case. By multiplying the linear and director momentum balance equations by $\delta\boldsymbol{\varphi}$ and $\delta\boldsymbol{d}_{\gamma}$ ($\gamma=1,2$), respectively, we have
\begin{align}\label{beam_weak_form_static_balance_eq}
\int_0^L {\left\{ {\left( {{{\boldsymbol{n}}_{,s}} + {\boldsymbol{\bar n}}} \right) \cdot \delta {\boldsymbol{\varphi }} + \left( {{\boldsymbol{\tilde m}}_{,s}^\gamma  - {{\boldsymbol{l}}^\gamma } + {{{\boldsymbol{\bar {\tilde m}}}}^\gamma }} \right) \cdot \delta {{\boldsymbol{d}}_\gamma }} \right\}\,{\mathrm{d}}s = 0},
\end{align}
where $\delta (\bullet)$ denotes the first variation. Integration by parts of Eq.\,(\ref{beam_weak_form_static_balance_eq}) leads to the following variational equation\footnote{See Appendix \ref{pdisp_linearize_sec} for the linearization of Eq.\,(\ref{beam_var_eq_balance_eq}) and the configuration update process.}
\begin{align}\label{beam_var_eq_balance_eq}
{G_{{\mathop{\rm int}} }}({\boldsymbol{y}},\delta {\boldsymbol{y}}) = {G_{{\rm{ext}}}}({\boldsymbol{y}},\delta {\boldsymbol{y}}),\,\,{\forall} \delta {\boldsymbol{y}} \in {\mathcal{V}},
\end{align}
where
\begin{align}\label{beam_var_int_vir_work}
{G_{{\mathop{\rm int}} }}({\boldsymbol{y}},\delta {\boldsymbol{y}}) \coloneqq \int_0^L {\left( {{\boldsymbol{n}} \cdot \delta {{\boldsymbol{\varphi }}_{\!,s}} + {{{\boldsymbol{\tilde m}}}^\gamma } \cdot \delta {{\boldsymbol{d}}_{\gamma ,s}} + {{\boldsymbol{l}}^\gamma } \cdot \delta {{\boldsymbol{d}}_\gamma }} \right){\mathrm{d}}s},
\end{align}
and
\begin{align}\label{beam_var_ext_vir_work}
{G_{{\rm{ext}}}}({\boldsymbol{y}},\delta {\boldsymbol{y}}) \coloneqq \left[ {{{{{{\boldsymbol{\bar n}}}_0}}} \cdot \delta {\boldsymbol{\varphi }}} \right]_{\Gamma_\mathrm{N}} + \left[ {{\boldsymbol{\bar {\tilde m}}}^\gamma_0 \cdot \delta {{\boldsymbol{d}}_\gamma }} \right]_{\Gamma_\mathrm{N}} + \int_0^L {\left( {{\boldsymbol{\bar n}} \cdot \delta {\boldsymbol{\varphi }} + {{{\boldsymbol{\bar {\tilde m}}}}^\gamma}\cdot\delta {{\boldsymbol{d}}_\gamma }} \right){\mathrm{d}}s}.
\end{align}
The external virtual work of Eq.\,(\ref{beam_var_ext_vir_work}) depends on the current configuration if a non-conservative load is applied (see for example the distributed follower load in section \ref{ex_beam_end_mnt}, and the external virtual work, expressed by Eq.\,(\ref{pure_bend_vir_work})), and it can be rewritten in compact form by
\begin{equation} \label{ext_vir_work_compact_form}
{G_{{\rm{ext}}}}({\boldsymbol{y}},\delta {\boldsymbol{y}}) = {\left[ {\delta {{{\boldsymbol{y}}}^{\mathrm{T}}}{{{\boldsymbol{\bar R}}}_0}} \right]_{{\Gamma_\mathrm{N}}}} + \int_0^L {\delta {{{\boldsymbol{y}}}^{\mathrm{T}}}{\boldsymbol{\bar R}}\,{\mathrm{d}}s},
\end{equation}
where we define
\begin{equation}
{{\boldsymbol{\bar R}}_0} \coloneqq \left\{ {\begin{array}{l}
\begin{aligned}
{{{{\boldsymbol{\bar n}}}_0}}\\
{{\boldsymbol{\bar {\tilde m}}}_0^1}\\
{{\boldsymbol{\bar {\tilde m}}}_0^2}
\end{aligned}
\end{array}} \right\},\,\,\text{and}\,\,{\boldsymbol{\bar R}} \coloneqq \left\{ {\begin{array}{l}
\begin{aligned}
{{\boldsymbol{\bar n}}}{\,\,\,}\\
{{{{\boldsymbol{\bar {\tilde m}}}}^1}}\\
{{{{\boldsymbol{\bar {\tilde m}}}}^2}}
\end{aligned}
\end{array}} \right\}.
\end{equation}
Using Eqs.\,(\ref{beam_dec_strs_res}) and (\ref{beam_dec_d_s}), the internal virtual work of Eq.\,(\ref{beam_var_int_vir_work}) can be rewritten by the effective stress resultants and director stress couples, as
\begin{align}\label{beam_int_vir_work_effective_strs}
{G_{{\mathop{\rm int}} }}({\boldsymbol{y}},\delta {\boldsymbol{y}}) = \int_0^L {\left( {\tilde n\,\delta \varepsilon + {{\tilde m}^\alpha }\delta {\rho _\alpha } + {{\tilde q}^\alpha }\delta {\delta _\alpha } + {{\tilde m}^{\alpha \beta }}\delta {\gamma _{\alpha \beta }} + {{\tilde l}^{\alpha \beta }}\delta {\chi _{\alpha \beta }}} \right)\,{\mathrm{d}}s},
\end{align}
where the variations of the strain measures (virtual strains) are derived as
\begin{subequations}
\label{beam_var_strains}
\begin{align} 
%
\delta {\varepsilon} &= \delta {{\boldsymbol{\varphi}}_{\!,s}} \cdot {{\boldsymbol{\varphi}}_{\!,s}},
\label{beam_var_strns_eps}\\
\delta {{\rho }_\alpha } &= \delta {{\boldsymbol{\varphi}}_{\!,s}} \cdot {{\boldsymbol{d}}_{\alpha ,s}} + {{\boldsymbol{\varphi}}_{\!,s}} \cdot \delta {{\boldsymbol{d}}_{\alpha ,s}},\label{beam_var_strns_rho}\\
\delta {{\delta }_\alpha } &= \delta {{\boldsymbol{\varphi}}_{\!,s}} \cdot {{\boldsymbol{d}}_\alpha } + {{\boldsymbol{\varphi}}_{\!,s}} \cdot \delta {{\boldsymbol{d}}_\alpha },\label{beam_var_strns_del}\\
\delta {{\gamma }_{\alpha \beta }} &= \delta {{\boldsymbol{d}}_\alpha } \cdot {{\boldsymbol{d}}_{\beta ,s}} + {{\boldsymbol{d}}_\alpha } \cdot \delta {{\boldsymbol{d}}_{\beta ,s}},\label{beam_var_strns_gm}\\
\delta {{\chi }_{\alpha \beta }} &= \frac{1}{2}\left( {\delta {{\boldsymbol{d}}_\alpha } \cdot {{\boldsymbol{d}}_\beta } + {{\boldsymbol{d}}_\alpha } \cdot \delta {{\boldsymbol{d}}_\beta }} \right).\label{beam_var_strns_chi}
\end{align}
\end{subequations}
Using the fact that these strains vanish in the initial beam configuration, we obtain the following strain expressions,
\begin{subequations}
\label{beam_th_strn_comp_basis_d123}
\begin{alignat}{2}
\varepsilon  &\coloneqq \frac{1}{2}({\left\| {{{\boldsymbol{\varphi }}_{\!,s}}} \right\|^2} - 1)\,\,&&({\text{axial stretching strain}}),\\
{{\rho }_\alpha } &\coloneqq {{\boldsymbol{\varphi}}_{\!,s}} \cdot {{\boldsymbol{d}}_{\alpha ,s}} - {{\boldsymbol{\varphi}}_{0,s}} \cdot {{\boldsymbol{D}}_{\alpha ,s}}\,\,&&({\text{bending strain}}),\\
{{\delta }_\alpha } &\coloneqq {{\boldsymbol{\varphi}}_{\!,s}} \cdot {{\boldsymbol{d}}_\alpha }- {{\boldsymbol{\varphi }}_{0,s}} \cdot {{\boldsymbol{D}}_\alpha }\,\,&&({\text{transverse shear strain}}),\\
{{\gamma }_{\alpha \beta }} &\coloneqq {{\boldsymbol{d}}_\alpha } \cdot {{\boldsymbol{d}}_{\beta ,s}} - {{\boldsymbol{D}}_\alpha } \cdot {{\boldsymbol{D}}_{\beta ,s}}\,\,&&({\text{couple shear strain}}),\label{def_b_strn_coup_sh}\\
{{\chi }_{\alpha \beta }} &\coloneqq \frac{1}{2}({{\boldsymbol{d}}_\alpha } \cdot {{\boldsymbol{d}}_\beta } - \boldsymbol{D}_{\alpha}\cdot\boldsymbol{D}_{\beta})\,\,&&({\text{cross-section stretching and shear strains}}).\label{strn_comp_chi}
\end{alignat}
\end{subequations}
\begin{definition} \textit{Physical interpretation of director stress couple components}.
\label{remark_bending_mnt}
Substituting Eq.\,(\ref{strs_res_dir_mnt}) into Eq.\,(\ref{def_strs_couple}) yields
\begin{equation}
{\boldsymbol{m}} = {\tilde m^1}{{\boldsymbol{d}}_1} \times {{\boldsymbol{\varphi }}_{\!,s}} + {\tilde m^2}{{\boldsymbol{d}}_2} \times {{\boldsymbol{\varphi }}_{\!,s}} + \left( {{{\tilde m}^{21}} - {{\tilde m}^{12}}} \right){{\boldsymbol{d}}_1} \times {{\boldsymbol{d}}_2}.
\end{equation}
Here, ${\tilde m}^\alpha\,\left(\alpha=1,2\right)$ represents the bending moment around the axis orthogonal to the current tangent vector to the central axis (i.e., ${\boldsymbol{\varphi}_{\!,s}}$) and director $\boldsymbol{d}_\alpha$, and ${\tilde m}^{12}$ and ${\tilde m}^{21}$ represent torsional moments in the opposite directions around the normal vector of the cross-section. The other components ${\tilde m}^{11}$ and ${\tilde m}^{22}$ are associated with the non-uniform transverse normal stretching in the directions of directors $\boldsymbol{d}_1$ and $\boldsymbol{d}_2$, respectively. Without loss of generality, we examine the component ${\tilde m}^{11}$ and its work conjugate strain $\gamma_{11}$ only. From Eq.\,(\ref{def_b_strn_coup_sh}), we have
\begin{equation}
\label{remark_gamma_11}
\gamma_{11}={\boldsymbol{d}_1}\cdot{\boldsymbol{d}_{1,s}}={\lambda_1}{\lambda_{1,s}},
\end{equation}
where ${{\boldsymbol{D}}_1} \cdot {{\boldsymbol{D}}_{1,s}} = 0$ is used, since we assume $\boldsymbol{D}_1$ is a unit vector. A material fiber aligned in the axial direction rotates, i.e.,~$\gamma_{11}\ne0$ if the transverse normal stretch of the cross-section ($\lambda_1$) is not constant along the central axis, and ${\tilde m}^{11}$ represents the work conjugate moment. If the cross-section deforms uniformly along the central axis, then ${\gamma_{11}}={\tilde m}^{11}=0$.
\end{definition}
\subsection{Hyperelastic constitutive equation}
\label{var_form_constitutive_law}
We can obtain constitutive equations by a reduction of a three-dimensional hyperelastic constitutive model. In what follows, we consider two hyperelastic materials: the St.\,Venant-Kirchhoff material, and the compressible Neo-Hookean material.
\subsubsection{Work conjugate stresses and elasticity tensor}
The Green-Lagrange strain tensor is defined as
\begin{equation} \label{def_GL_strain}
{\boldsymbol{E}} \coloneqq \frac{1}{2}\left( {{{\boldsymbol{F}}^{\mathrm{T}}}{\boldsymbol{F}} - {\boldsymbol{1}}} \right),
\end{equation}
where $\boldsymbol{1}$ represents the identity tensor in $\Bbb{R}^3$. The identity tensor can be expressed in the basis ${\left\{ {{{\boldsymbol{G}}^1},{{\boldsymbol{G}}^2},{{\boldsymbol{G}}^3}} \right\}}$ as
\begin{equation}\label{def_identity_curv}
{\boldsymbol{1}} = {G_{ij}}{{\boldsymbol{G}}^i} \otimes {{\boldsymbol{G}}^j}\,\,\text{where}\,\,G_{ij}\coloneqq{\boldsymbol{G}_i}\cdot{\boldsymbol{G}_j}.
\end{equation}
Using Eq.\,(\ref{beam_th_str_init_cov_base}) the identity tensor can be rewritten as
\begin{align}\label{def_identity_curv_1}
{\boldsymbol{1}} &= {{\boldsymbol{G}}^\alpha } \otimes {{\boldsymbol{G}}^\alpha } + {\xi ^\beta }{{\boldsymbol{D}}_\alpha } \cdot {{\boldsymbol{D}}_{\beta ,s}}\left( {{{\boldsymbol{G}}^\alpha } \otimes {{\boldsymbol{G}}^3} + {{\boldsymbol{G}}^3} \otimes {{\boldsymbol{G}}^\alpha }} \right)\nonumber\\
&+ \left( {1 + 2{\xi ^\alpha }{{\boldsymbol{D}}_{\alpha ,s}} \cdot {{\boldsymbol{D}}_3} + {\xi ^\alpha }{\xi ^\beta }{{\boldsymbol{D}}_{\alpha ,s}} \cdot {{\boldsymbol{D}}_{\beta ,s}}} \right){{\boldsymbol{G}}^3} \otimes {{\boldsymbol{G}}^3}.
\end{align}
Then substituting Eqs.\,(\ref{beam_th_str_deform_grad}) and (\ref{def_identity_curv_1}) into Eq.\,(\ref{def_GL_strain}), the Green-Lagrange strain tensor can be rewritten in terms of the strains in Eq.\,(\ref{beam_th_strn_comp_basis_d123}) as
\begin{align}\label{def_GL_strain_1}
{\boldsymbol{E}} = {E_{\alpha \beta }}{{\boldsymbol{G}}^\alpha } \otimes {{\boldsymbol{G}}^\beta } + {E_{3\gamma }}\left( {{{\boldsymbol{G}}^3} \otimes {{\boldsymbol{G}}^\gamma } + {{\boldsymbol{G}}^\gamma } \otimes {{\boldsymbol{G}}^3}} \right) + {E_{33}}{{\boldsymbol{G}}^3} \otimes {{\boldsymbol{G}}^3},
\end{align}
where the components are
\begin{equation} \label{GL_strn_components}
\left\{ \begin{array}{lcl}
\begin{aligned}
{E_{\alpha \beta }} &= {{\chi }_{\alpha \beta }},\\
{E_{3\alpha }} &= {E_{\alpha 3}} = \frac{1}{2}\left( {{{\delta }_\alpha } + {\xi ^\gamma }{{\gamma }_{\alpha \gamma }}} \right),\\
{E_{33}} &= \varepsilon  + {\xi ^\gamma }{{\rho }_\gamma } + {{\xi ^\gamma }{\xi ^\delta {{\kappa}}_{\gamma \delta} }},
\end{aligned}
\end{array} \right.
\end{equation}
and we define a \textit{high-order bending strain component} as
\begin{align}\label{def_strain_phi}
{{{\kappa}} _{\alpha \beta }} \coloneqq \frac{1}{2}\left( {{{\boldsymbol{d}}_{\alpha ,s}} \cdot {{\boldsymbol{d}}_{\beta ,s}} - {{\boldsymbol{D}}_{\alpha ,s}} \cdot {{\boldsymbol{D}}_{\beta ,s}}} \right).
\end{align}
Taking the first variation of Eq.\,(\ref{def_strain_phi}), we obtain
\begin{equation}\label{def_strain_var_kappa}
\delta {{\kappa }_{\alpha \beta }} = \frac{1}{2}\left( {\delta {{\boldsymbol{d}}_{\alpha ,s}} \cdot {{\boldsymbol{d}}_{\beta ,s}} + {{\boldsymbol{d}}_{\alpha ,s}} \cdot \delta {{\boldsymbol{d}}_{\beta ,s}}} \right).
\end{equation}
For brevity we define the following arrays by exploiting the symmetry of the strains (i.e., $\kappa_{12}=\kappa_{21}$ and $\chi_{12}=\chi_{21}$)
\begin{equation}
{\boldsymbol{\rho }} \coloneqq \left\{ {\begin{array}{*{20}{c}}
{{{\rho }_1}}\\
{{{\rho }_2}}
\end{array}} \right\},\,{\boldsymbol{\kappa }} \coloneqq \left\{ {\begin{array}{*{20}{c}}
{{{\kappa }_{11}}}\\
{{{\kappa }_{22}}}\\
{2{{\kappa }_{12}}}
\end{array}} \right\},\,{\boldsymbol{\delta }} \coloneqq \left\{ {\begin{array}{*{20}{c}}
{{{\delta }_1}}\\
{{{\delta }_2}}
\end{array}} \right\},\,{\boldsymbol{\gamma }} \coloneqq \left\{ {\begin{array}{*{20}{c}}
{{{\gamma }_{11}}}\\
{{{\gamma }_{12}}}\\
{{{\gamma }_{21}}}\\
{{{\gamma }_{22}}}
\end{array}} \right\},\,{\boldsymbol{\chi }} \coloneqq \left\{ {\begin{array}{*{20}{c}}
{{{\chi }_{11}}}\\
{{{\chi }_{22}}}\\
{2{{\chi }_{12}}}
\end{array}} \right\},
\end{equation}
and
\begin{equation}\label{def_eps_hat}
{\boldsymbol{\munderbar \varepsilon}} \coloneqq \left\{ {\begin{array}{*{20}{c}}
{\varepsilon }\\
{{\boldsymbol{\rho }}}\\
{{\boldsymbol{\kappa }}}\\
{{\boldsymbol{\delta }}}\\
{{\boldsymbol{\gamma }}}\\
{{\boldsymbol{\chi }}}
\end{array}} \right\}.
\end{equation}
\begin{definition}{\textit{Incompleteness of the Green-Lagrange strain components in the beam kinematic description of Eq.\,(\ref{intro_beam_th_str_cur_config_2nd}) with ${a_2}={b_1}=0$}. Here it is shown that the kinematic expression of Eq.\,(\ref{intro_beam_th_str_cur_config_2nd}) leads to the Green-Lagrange strain tensor having a complete linear polynomial expression in terms of the coordinates $\xi^1$ and $\xi^2$, but it does not if ${a_2}={b_1}=0$. Based on the kinematic expression of Eq.\,(\ref{intro_beam_th_str_cur_config_2nd}), the covariant base vectors are obtained as
\begin{equation}\label{beam_th_str_cur_cov_base_quad}
\left\{ \begin{array}{l}
\begin{aligned}
{{\boldsymbol{g}}_1} &= {{\boldsymbol{d}}_1} + 2{a_1}{\xi ^1}{{\boldsymbol{d}}_1} + {\xi ^2}\left( {{b_1}{{\boldsymbol{d}}_2} + {a_2}{{\boldsymbol{d}}_1}} \right),\\
{{\boldsymbol{g}}_2} &= {{\boldsymbol{d}}_2} + {\xi ^1}\left( {{a_2}{{\boldsymbol{d}}_1} + {b_1}{{\boldsymbol{d}}_2}} \right) + 2{b_2}{\xi ^2}{{\boldsymbol{d}}_2}.\\
\end{aligned}
\end{array} \right.
\end{equation}
The in-plane components of the Green-Lagrange strain tensor are obtained by substituting Eq.\,(\ref{beam_th_str_cur_cov_base_quad}) into Eq.\,(\ref{def_GL_strain}), as
\begin{equation}
E_{\alpha\beta}={E}^{\mathrm{c}}_{\alpha\beta}+{\tilde E}_{\alpha\beta},
\end{equation}
where the additional parts, after neglecting the quadratic terms of $\xi^1$ and $\xi^2$ \footnote{{The quadratic terms are neglected since the enhanced strain field should satisfy the orthogonality to the constant stress fields \citep{betsch19964}.}}, are
\begin{subequations}
\label{gl_strn_add_part_compat_quad}
\begin{align}
{\tilde E}_{11} &= 2{\xi ^1}{a_1}{{\boldsymbol{d}}_1} \cdot {{\boldsymbol{d}}_1} + {\xi ^2}\left( {{a_2}{{\boldsymbol{d}}_1} \cdot {{\boldsymbol{d}}_1} + {b_1}{{\boldsymbol{d}}_1} \cdot {{\boldsymbol{d}}_2}} \right)+ 2{\xi ^1}{\xi ^2}\left( {{a_1}{a_2}{{\boldsymbol{d}}_1} \cdot {{\boldsymbol{d}}_1} + {a_1}{b_1}{{\boldsymbol{d}}_1} \cdot {{\boldsymbol{d}}_2}} \right),\\
{\tilde E}_{22} &= {\xi ^1}\left( {{a_2}{{\boldsymbol{d}}_1} \cdot {{\boldsymbol{d}}_2} + {b_1}{{\boldsymbol{d}}_2} \cdot {{\boldsymbol{d}}_2}} \right) + 2{b_2}{\xi ^2}{{\boldsymbol{d}}_2} \cdot {{\boldsymbol{d}}_2} + 2{\xi ^1}{\xi ^2}\left( {{a_2}{b_2}{{\boldsymbol{d}}_1} \cdot {{\boldsymbol{d}}_2} + {b_1}{b_2}{{\boldsymbol{d}}_2} \cdot {{\boldsymbol{d}}_2}} \right),\\
2{\tilde E}_{12} &= {\xi ^1}\left( {{a_2}{{\boldsymbol{d}}_1} \cdot {{\boldsymbol{d}}_1} + {b_1}{{\boldsymbol{d}}_1} \cdot {{\boldsymbol{d}}_2} + 2{a_1}{{\boldsymbol{d}}_1} \cdot {{\boldsymbol{d}}_2}} \right) + {\xi ^2}\left( {2{b_2}{{\boldsymbol{d}}_1} \cdot {{\boldsymbol{d}}_2} + {a_2}{{\boldsymbol{d}}_1} \cdot {{\boldsymbol{d}}_2} + {b_1}{{\boldsymbol{d}}_2} \cdot {{\boldsymbol{d}}_2}} \right)\nonumber\\
 &+ {\xi ^1}{\xi ^2}\left( {{a_2}^2{{\boldsymbol{d}}_1} \cdot {{\boldsymbol{d}}_1} + 4{a_1}{b_2}{{\boldsymbol{d}}_1} \cdot {{\boldsymbol{d}}_2} + 2{a_2}{b_1}{{\boldsymbol{d}}_1} \cdot {{\boldsymbol{d}}_2} + {b_1}^2{{\boldsymbol{d}}_2} \cdot {{\boldsymbol{d}}_2}} \right).
\end{align}
\end{subequations}
However, if the bilinear terms in $\xi^1$ and $\xi^2$ are missing in Eq.\,(\ref{intro_beam_th_str_cur_config_2nd}), i.e., $a_2=b_1=0$, those in-plane Green-Lagrange strain components, {after neglecting the quadratic terms}, become
\begin{subequations}
\label{inplane_without_bilinear_GL}
\begin{align}
\tilde E_{11}^{*} &= 2{a_1}{\xi ^1}{{\boldsymbol{d}}_1} \cdot {{\boldsymbol{d}}_1},\label{GL_strn_5p_e11}\\
\tilde E_{22}^{*} &= 2{b_2}{\xi ^2}{{\boldsymbol{d}}_2} \cdot {{\boldsymbol{d}}_2},\label{GL_strn_5p_e22}\\
2\tilde E_{12}^{*} &= 2\left( {{\xi ^1}{a_1} + {\xi ^2}{b_2} + 2{\xi ^1}{\xi ^2}{a_1}{b_2}} \right){{\boldsymbol{d}}_1} \cdot {{\boldsymbol{d}}_2}.\label{GL_strn_5p_e12}
\end{align}
\end{subequations}
\noindent It is noticeable that Eqs.\,(\ref{GL_strn_5p_e11}) and (\ref{GL_strn_5p_e22}) do not have linear terms of the coordinates $\xi^2$ and $\xi^1$, respectively. This means that the kinematic expression of Eq.(\ref{intro_beam_th_str_cur_config_2nd}) without bilinear terms (i.e., $a_2=b_1=0$) is not able to represent trapezoidal deformations of the cross-section, illustrated in Fig.\,\ref{intro_cs_deform_2nd_trapezoid}.}
\end{definition}

We assume that the \textit{strain energy density} (defined as the strain energy per unit undeformed volume) is expressed in terms of the Green-Lagrange strain tensor, as
\begin{equation}
\Psi = \Psi(\boldsymbol{E}).
\end{equation}
The second Piola-Kirchhoff stress tensor, which is \textit{work conjugate} to the Green-Lagrange strain tensor, is obtained by
\begin{equation}\label{2nd_pk_strs_comp}
{\boldsymbol{S}} = {S^{ij}}{{\boldsymbol{G}}_i} \otimes {{\boldsymbol{G}}_j}\,\,\,\text{with}\,\,\,{S^{ij}} = \frac{{\partial \Psi }}{{\partial {E_{ij}}}}.
\end{equation}
The components $S^{11}$,\,$S^{22}$, and $S^{12}$ are typically assumed to be zero in many beam formulations and this zero stress condition has made the application of general nonlinear constitutive laws not straightforward. Exploiting the symmetries, the second order tensors ${\boldsymbol{E}}$ and ${\boldsymbol{S}}$ can be expressed in array form (Voigt notation), as ${\boldsymbol{\munderbar S}} \coloneqq {\left[ {{S^{11}},{S^{22}},{S^{33}},{S^{12}},{S^{13}},{S^{23}}} \right]^{\mathrm{T}}}$, and ${\boldsymbol{\munderbar E}} \coloneqq {\left[ {{E_{11}},{E_{22}},{E_{33}},2{E_{12}},2{E_{13}},2{E_{23}}} \right]^{\mathrm{T}}}$. The total strain energy of the beam can be expressed as
\begin{align}\label{tot_strn_energy_beam}
U = {\int_0^L {\int_{\mathcal{A}} {{{\Psi}}\,{j_0}\,{\mathrm{d}}\mathcal{A}}\,{\mathrm{d}}s}}\,.
\end{align}
The first variation of the strain energy density function can be obtained, by using the chain rule of differentiation, as (see Appendix \ref{1st_var_strn_e_M_mat} for the details)
\begin{align}\label{deriv_energy_general}
\delta \Psi  = \boldsymbol{\munderbar S}^{\mathrm{T}}\delta{\boldsymbol{\munderbar E}} = \boldsymbol{\munderbar S}^{\mathrm{T}}{\boldsymbol{\munderbar D}}\delta {\boldsymbol{\munderbar \varepsilon }}\,\,\,\mathrm{with}\,\,\,\boldsymbol{\munderbar D}\coloneqq{\frac{\partial\boldsymbol{\munderbar E}}{\partial\boldsymbol{\munderbar\varepsilon}}}.
\end{align}
Taking the first variation of the total strain energy of Eq.\,(\ref{tot_strn_energy_beam}) and using Eq.\,(\ref{deriv_energy_general}) we obtain the internal virtual work
\begin{align}\label{tot_strn_energy_beam_time_deriv}
{G_\text{int}}(\boldsymbol{y},\delta\boldsymbol{y})\equiv{\delta U} = \int_0^L {\delta {{{\boldsymbol{\munderbar \varepsilon }}^{\mathrm{T}}}}{{\boldsymbol{R}}}\,{\mathrm{d}}s},
\end{align}
where $\boldsymbol{R}$ defines the array of stress resultants and director stress couples,
\begin{align}
{\boldsymbol{R}} \coloneqq\int_{\mathcal{A}} {{\boldsymbol{\munderbar D}^{\mathrm{T}}}{\boldsymbol{\munderbar S}}\,{j_0}\,{\mathrm{d}}{\mathcal{A}}}= {\left[ {{{\tilde n}},{{\tilde m}^{1}},{{\tilde m}^{2}},{{\tilde h}^{11}},{{\tilde h}^{22}},{{\tilde h}^{12}},{{\tilde q}^1},{{\tilde q}^2},{{\tilde m}^{11}},{{\tilde m}^{12}},{{\tilde m}^{21}},{{\tilde m}^{22}},{{\tilde l}^{11}},{{\tilde l}^{22}},{{\tilde l}^{12}}} \right]^{\mathrm{T}}}.
\end{align}
Here, ${{\tilde h}^{\alpha \beta }}$ defines the component of the \textit{high-order director stress couple}.
For general hyperelastic materials, the constitutive relation between $\boldsymbol{S}$ and $\boldsymbol{E}$ is nonlinear. Thus, we need to linearize the constitutive relation, by taking the directional derivative of $\boldsymbol{S}$,
\begin{equation}\label{linear_rel_dir_deriv_SE}
D{\boldsymbol{S}} \cdot \Delta {\boldsymbol{x}_t} = {\boldsymbol{\mathcal{C}}}:D{\boldsymbol{E}} \cdot \Delta {\boldsymbol{x}_t},
\end{equation}
where $D(\bullet)\cdot(*)$ represents the directional derivative of $(\bullet)$ in direction $(*)$, and $\Delta\boldsymbol{x}_t$ denotes the increment of the material point position at the current configuration. The fourth-order tensor $\boldsymbol{\mathcal{C}}$, called the \textit{Lagrangian} or \textit{material elasticity tensor}, is expressed by
\begin{equation}\label{mat_lag_elasticity_tensor}
{\boldsymbol{\mathcal{C}}} \coloneqq \frac{{\partial {\boldsymbol{S}}}}{{\partial {\boldsymbol{E}}}} = {{\mathcal{C}}^{ijk\ell }}{{\boldsymbol{G}}_i} \otimes {{\boldsymbol{G}}_j} \otimes {{\boldsymbol{G}}_k} \otimes {{\boldsymbol{G}}_\ell }\,\,\,\text{with}\,\,\,{{\mathcal{C}}^{ijk\ell }} = \frac{{{\partial ^2}\Psi }}{{\partial {E_{ij}}\,\partial {E_{k\ell }}}}.
\end{equation}
Note that the elasticity tensor has both major and minor symmetries. For computational purposes we can therefore represent the fourth order tensor $\boldsymbol{\mathcal{C}}$ in matrix form as
\begin{equation}
{\boldsymbol{\munderbar{\munderbar{\mathcal{C}}}}} \coloneqq \left[ {\begin{array}{*{20}{c}}{{{\mathcal{C}}^{1111}}}&{{{\mathcal{C}}^{1122}}}&{{{\mathcal{C}}^{1133}}}&{{{\mathcal{C}}^{1112}}}&{{{\mathcal{C}}^{1113}}}&{{{\mathcal{C}}^{1123}}}\\
{}&{{{\mathcal{C}}^{2222}}}&{{{\mathcal{C}}^{2233}}}&{{{\mathcal{C}}^{2212}}}&{{{\mathcal{C}}^{2213}}}&{{{\mathcal{C}}^{2223}}}\\
{}&{}&{{{\mathcal{C}}^{3333}}}&{{{\mathcal{C}}^{3312}}}&{{{\mathcal{C}}^{3313}}}&{{{\mathcal{C}}^{3323}}}\\
{}&{}&{}&{{{\mathcal{C}}^{1212}}}&{{{\mathcal{C}}^{1213}}}&{{{\mathcal{C}}^{1223}}}\\
{}&{{\rm{sym}}{\rm{.}}}&{}&{}&{{{\mathcal{C}}^{1313}}}&{{{\mathcal{C}}^{1323}}}\\
{}&{}&{}&{}&{}&{{{\mathcal{C}}^{2323}}}
\end{array}} \right].
\end{equation}
In a similar manner to the derivation of Eq.\,(\ref{deriv_energy_general}), the directional derivative of $\boldsymbol{\munderbar S}$ can be derived as
\begin{equation}\label{dir_deriv_2pk}
D{\boldsymbol{\munderbar S}}\cdot \Delta {\boldsymbol{y}}= {\boldsymbol{\munderbar{\munderbar {\mathcal{C}}}}}{\boldsymbol{\munderbar D}}\left(D{\boldsymbol{\munderbar \varepsilon }}\cdot \Delta {\boldsymbol{y}}\right).
\end{equation}
Then, the directional derivative of $\boldsymbol{R}$ is obtained by using Eq.\,(\ref{dir_deriv_2pk}), as
\begin{equation}\label{beam_lin_strs_resultant_R}
D{\boldsymbol{R}} \cdot \Delta {\boldsymbol{y}} = {\Bbb{C}}\left(D{\boldsymbol{\munderbar \varepsilon }} \cdot \Delta {\boldsymbol{y}}\right),
\end{equation}
where $\Delta {\boldsymbol{y}} \coloneqq {\left[ {\Delta {{\boldsymbol{\varphi}}}^{\mathrm{T}},\Delta {{\boldsymbol{d}}_1}^{\mathrm{T}},\Delta {{\boldsymbol{d}}_2}^{\mathrm{T}}} \right]^{\mathrm{T}}}$, and $\Bbb{C}$ represents the symmetric constitutive matrix, defined by
\begin{equation}
{\Bbb{C}} \coloneqq \int_{\mathcal{A}} {\left( {{{\boldsymbol{\munderbar D}}^{\mathrm{T}}}
 {\boldsymbol{\munderbar{\munderbar{\mathcal{C}}}}}{\boldsymbol{\munderbar D}}{j_0}} \right){\mathrm{d}}{\mathcal{A}}}.
\end{equation}
\begin{definition}
\label{num_integ_polar_circ}
\textit{Numerical integration over the circular cross-section}. In this paper, we restrict our discussion to rectangular and circular cross-sections. In the case of circular cross-section of radius $R$, we can simply parametrize the domain by polar coordinates, as
\begin{equation}
{\xi^1}=r\,{\mathrm{cos}}\,\theta\,\,\text{and}\,\,{\xi^2}=r\,{\mathrm{sin}}\,\theta\,\,\text{with}\,\,0 \le r \le R,\,\,\text{and}\,\,0 \le \theta  < 2\pi.
\end{equation}
Then, the infinitesimal area simply becomes 
\begin{equation}
{\mathrm{d}}{\mathcal{A}} = r\,{\mathrm{d}}r\,{\mathrm{d}}\theta,\,\,r = \sqrt {{{\left( {{\xi ^1}} \right)}^2} + {{\left( {{\xi ^2}} \right)}^2}}.
\end{equation}
\end{definition}
\subsubsection{St.\,Venant-Kirchhoff material}
In the St.\,Venant-Kirchhoff material model, the strain energy density is expressed by
\begin{equation}\label{mat_stvk_def_energy_func}
\Psi = \frac{1}{2}\lambda {\left( {{\mathrm{tr}}{\boldsymbol{E}}} \right)^2} + \mu {\boldsymbol{E}}:{\boldsymbol{E}},
\end{equation}
where $\lambda$ and $\mu$ are the Lam{\'{e}} constants, which are related to Young's modulus $E$ and Poisson's ratio $\nu$ by
\begin{equation}\label{mat_lame_cnst_emod_shear_mod}
\lambda  = \frac{{E\nu }}{{(1 + \nu )(1 - 2\nu )}}\,\,\text{and}\,\,\mu  = \frac{E}{{2(1 + \nu )}}.
\end{equation}
The second Piola-Kirchhoff stress tensor is then obtained by
\begin{equation}\label{def_2nd_pk_stress_tilde}
{\boldsymbol{S}} = \frac{{\partial \Psi}}{{\partial {\boldsymbol{E}}}} = \lambda \left( {\mathrm{tr}{\boldsymbol{E}}} \right){\boldsymbol{1}} + 2\mu {\boldsymbol{E}}.
\end{equation}
Note the linearity in the constitutive relation of Eq.\,(\ref{def_2nd_pk_stress_tilde}), which restricts the applicability of this material law to moderate strains. The contravariant component of $\boldsymbol{S}$ follows as
\begin{equation}
{S^{ij}} = {\boldsymbol{S}}:{{\boldsymbol{G}}^i} \otimes {{\boldsymbol{G}}^j} = {C^{ijk\ell }}{E_{k\ell }},
\end{equation}
where
\begin{align}\label{c_comp_st_venant}
{C^{ijk\ell }} =\lambda {G^{ij}}{G^{k\ell }} + \mu \left( {{G^{ik}}{G^{j\ell }} + {G^{i\ell }}{G^{jk}}} \right).
\end{align}
\subsubsection{Compressible Neo-Hookean material}
The stored energy function of the three-dimensional compressible Neo-Hookean material is defined as
\begin{align}\label{nh_mat_stored_efunc}
\Psi  = \frac{\mu }{2}({\mathrm{tr}}{\boldsymbol{C}} - 3) - \mu \ln J + \frac{\lambda }{2}{(\ln J)^2},
\end{align}
where $\boldsymbol{C}\coloneqq{{\boldsymbol{F}}^{\mathrm{T}}}{\boldsymbol{F}}$ is the right Cauchy-Green deformation tensor. The second Piola-Kirchhoff stress tensor follows as \citep{bonet2010nonlinear}
\begin{align}\label{nh_mat_2nd_pk}
{\boldsymbol{S}} = \frac{{\partial \Psi }}{{\partial {\boldsymbol{E}}}} = \mu ({\boldsymbol{1}} - {{\boldsymbol{C}}^{ - 1}}) + \lambda (\ln J){{\boldsymbol{C}}^{ - 1}}.
\end{align}
The contravariant components of $\boldsymbol{S}$ can then be derived as
\begin{align}
{S^{ij}} = {\boldsymbol{S}}:{{\boldsymbol{G}}^i} \otimes {{\boldsymbol{G}}^j} = \mu \left\{ {{G^{ij}} - {{\left( {{{\boldsymbol{C}}^{ - 1}}} \right)}^{ij}}} \right\} + \lambda (\ln J){\left( {{{\boldsymbol{C}}^{ - 1}}} \right)^{ij}}.
\end{align}
The corresponding Lagrangian elasticity tensor follows as \citep{bonet2010nonlinear}
\begin{align}\label{nh_mat_lag_elas_tensor}
{\boldsymbol{\mathcal{C}}} = \lambda {{\boldsymbol{C}}^{ - 1}} \otimes {{\boldsymbol{C}}^{ - 1}} + 2(\mu  - \lambda \ln J){\boldsymbol{\mathcal{I}}},
\end{align}
where 
\begin{align}\label{nh_mat_inv_c_tensor}
{{\boldsymbol{C}}^{ - 1}} \otimes {{\boldsymbol{C}}^{ - 1}} = {\left( {{{\boldsymbol{C}}^{ - 1}}} \right)^{ij}}{\left( {{{\boldsymbol{C}}^{ - 1}}} \right)^{k\ell }}{{\boldsymbol{G}}_i} \otimes {{\boldsymbol{G}}_j} \otimes {{\boldsymbol{G}}_k} \otimes {{\boldsymbol{G}}_\ell},
\end{align}
and the fourth order tensor $\boldsymbol{\mathcal{I}}$ can be expressed in terms of the covariant basis, as (see Appendix \ref{app_constitutive_nh} for the derivation)
\begin{align}\label{nh_mat_lag_elas_tensor}
{\boldsymbol{\mathcal{I}}} \coloneqq - \frac{{\partial {{\boldsymbol{C}}^{ - 1}}}}{{\partial {\boldsymbol{C}}}} = \frac{1}{2}\left\{ {{{\left( {{{\boldsymbol{C}}^{ - 1}}} \right)}^{ik}}{{\left( {{{\boldsymbol{C}}^{ - 1}}} \right)}^{j\ell}} + {{\left( {{{\boldsymbol{C}}^{ - 1}}} \right)}^{i\ell}}{{\left( {{{\boldsymbol{C}}^{ - 1}}} \right)}^{jk}}} \right\}{{\boldsymbol{G}}_i} \otimes {{\boldsymbol{G}}_j} \otimes {{\boldsymbol{G}}_k} \otimes {{\boldsymbol{G}}_\ell}.
\end{align}
Then the contravariant components of $\boldsymbol{\mathcal{C}}$ are obtained as
\begin{align}\label{nh_mat_lag_c_cont_comp}
{C^{ijk\ell }} = \lambda {\left( {{{\boldsymbol{C}}^{ - 1}}} \right)^{ij}}{\left( {{{\boldsymbol{C}}^{ - 1}}} \right)^{k\ell }} + (\mu  - \lambda \ln J)\left\{ {{{\left( {{{\boldsymbol{C}}^{ - 1}}} \right)}^{ik}}{{\left( {{{\boldsymbol{C}}^{ - 1}}} \right)}^{j\ell }} + {{\left( {{{\boldsymbol{C}}^{ - 1}}} \right)}^{i\ell }}{{\left( {{{\boldsymbol{C}}^{ - 1}}} \right)}^{jk}}} \right\}.
\end{align}
\subsection{Isogeometric discretization}
\subsubsection{NURBS curve}
The geometry of the beam's central axis can be represented by a NURBS curve. Here we summarize the construction of a NURBS curve. More detailed explanation on the properties of NURBS and geometric algorithms like knot insertion and degree elevation can be found in \citet{piegl2012nurbs}. Further discussions on the important properties of NURBS in the analysis can be found in \citet{hughes2005isogeometric}. For a given knot vector ${\tilde \varXi}={\left\{{\xi_1},{\xi_2},...,{\xi_{{n_{\mathrm{cp}}}+p+1}}\right\}}$, where ${\xi_i}\in{\Bbb R}$ is the $i$th knot, $p$ is the degree of basis function, and ${n_{\mathrm{cp}}}$ is the number of basis functions (or control points), B-spline basis functions are recursively defined \citep{piegl2012nurbs}. For $p=0$, they are defined by
\begin{equation} \label{Bspline_basis_0}
 B_I^0(\xi ) = 
  \begin{cases} 
   1&{{\rm{if~~ }}{\xi _I} \le \xi  < {\xi _{I + 1}}},\\
   0&{{\text{otherwise,  }}}
  \end{cases}
\end{equation}
and for $p=1,2,3,...,$ they are defined by
\begin{equation} \label{Bspline_basis_p}
B_I^p(\xi ) = \frac{{\xi  - {\xi _I}}}{{{\xi _{I + p}} - {\xi _I}}}B_I^{p - 1}(\xi ) + \frac{{{\xi _{I + p + 1}} - \xi }}{{{\xi _{I + p + 1}} - {\xi _{I + 1}}}}B_{I + 1}^{p - 1}(\xi ),
\end{equation}
where $\xi\in\varXi\subset{\Bbb R}$ denotes the parametric coordinate, and $\varXi\coloneqq\left[\xi_{1},\xi_{{n_{\mathrm{cp}}}+p+1}\right]$ represents the parametric space. From the B-spline basis functions the NURBS basis functions are defined by 
\begin{equation}\label{nurbs_basis_1d_def}
{N_I}(\xi ) = \frac{{B_I^p(\xi )\,{w_I}}}{{\sum\limits_{J = 1}^{n_{\mathrm{cp}}} {B_J^p(\xi )\,{w_J}} }},
\end{equation}
where ${w_I}$ denotes the given weight of the $I$th control point. If weights are equal, NURBS becomes B-spline. The geometry of the initial beam central axis can be represented by a NURBS curve, as
\begin{equation}\label{beam_curve_pos_nurbs}
{\boldsymbol{X}}(\xi ) = \sum\limits_{I = 1}^{n_{\mathrm{cp}}} {{N_I}(\xi )\,{{\boldsymbol{X}}_{\!I}}},
\end{equation}
where $\boldsymbol{X}_{\!I}$ are the control point positions. The arc-length parameter along the initial central axis can be expressed by the mapping $s(\xi):{\varXi}\to{\left[0,L\right]}$, defined by
\begin{equation}\label{beam_curve_pos_nurbs_alen_map}
s(\xi )\coloneqq \int_{{\xi _1}}^{\eta  = \xi } {\left\| {{{\boldsymbol{X}}_{\!,\eta }}(\eta )} \right\|{\mathrm{d}}\eta }.
\end{equation}
Then the Jacobian of the mapping is derived as
\begin{align}\label{beam_curve_pos_nurbs_alen_map_jcb}
\tilde j\coloneqq \frac{{{\mathrm{d}}s}}{{{\mathrm{d}}\xi }}= \left\| {{{\boldsymbol{X}}_{\!,\xi }}(\xi )} \right\|.
\end{align}
In the discretization of the variational form, we often use the notation ${N_{I,s}}$ for brevity, which is defined by
\begin{align}\label{beam_curve_pos_nurbs_alen_map_jcb}
{N_{I,s}} \coloneqq {N_{I,\xi }}\frac{{{\mathrm{d}}\xi }}{{{\mathrm{d}}s}} = \frac{1}{{\tilde j}}{N_{I,\xi }},
\end{align}
where ${N_{I,\xi }}$ denotes the differentiation of the basis function ${N_{I}(\xi)}$ with respect to $\xi$.\\
\subsubsection{Discretization of the variational form}
In the discretization of the variational form using NURBS basis functions, an \textit{element} in one-dimension is defined as the nonzero \textit{knot span}, which means the span between two distinct knot values. Let $\varXi_{e}$ denote the $e$th nonzero knot span (element), then the entire parametric domain is the sum of the whole knot spans, i.e., $\varXi  = {\varXi _1} \cup {\varXi _2} \cup  \cdots  \cup {\varXi_{n_{\mathrm{el}}}}$, where $n_{\mathrm{el}}$ denotes the total number of nonzero knot spans. Using the NURBS basis of Eq.\,(\ref{nurbs_basis_1d_def}), the variations of the central axis position and the two director vectors at $\xi\in{\varXi}_e$ are discretized as
\begin{equation}\label{beam_disp_director_disc_nurbs}
\delta {{\boldsymbol{y}}^h}(s(\xi )) = \left[ {\begin{array}{*{20}{c}}
{{N_1}(\xi )\,{{\bf{1}}_{9 \times 9}}}& \cdots &{{N_{{n_{{e}}}}}(\xi )\,{{\bf{1}}_{9 \times 9}}}
\end{array}} \right]\left\{ {\begin{array}{*{20}{c}}
{\delta {{\bf{y}}_1}}\\
 \vdots \\
{\delta {{\bf{y}}_{{n_{{e}}}}}}
\end{array}} \right\} \eqqcolon {{\Bbb{N}}_e}\delta {{\bf{y}}^e},\,\,\text{with}\,\,\delta {{\bf{y}}_I} \coloneqq \left\{ {\begin{array}{*{20}{c}}
{\delta {{\boldsymbol{\varphi }}_I}}\\
{\delta {{\boldsymbol{d}}_{1I}}}\\
{\delta {{\boldsymbol{d}}_{2I}}}
\end{array}} \right\},
\end{equation}
where ${{\delta {\boldsymbol{\varphi}}}_I} \in {{\Bbb R}^3}$ and ${{\delta {\boldsymbol{d}}}_{\alpha I}} \in {{\Bbb R}^3}$ denote the displacement and director coefficient vectors, and ${\bf{1}}_{m\times{m}}$ denotes the identity matrix of dimension $m\times{m}$. $n_e$ denotes the number of basis functions having local support in the knot span ${\varXi}_e$.
\begin{definition}
{It is noted that the spatial discretization is applied to the increment (variation) of the director vectors, not to the total director vectors. This is because the initial directors are assumed to be orthonormal, and the spatial discretization by NURBS basis functions does not preserve the orthonormality. The initial orthonormal director vectors at an arbitrary position on the central axis may be calculated in many different ways. For example, for a given $C^1$ continuous curve, the \textit{smallest rotation method} gives a smooth parameterization of initial orthonormal directors. More details on this method can be found in \citet{meier2014objective} and \citet{choi2019isogeometric}.}
\end{definition}
Using Eq.\,(\ref{beam_disp_director_disc_nurbs}) and the standard element assembly operator ${\bf{A}}$, we obtain
\begin{equation}
\label{new_disc_int_force_vec}
{G_{{\mathop{\rm int}} }}({{\boldsymbol{y}}^h},\delta {{\boldsymbol{y}}^h}) = \delta {{\bf{y}}}^{\mathrm{T}}{\bf{F}}_{{\mathop{\rm int}} },\,\,\text{with}\,\,{\bf{F}}_{{\mathop{\rm int}} } \coloneqq \mathop {\bf{A}}\limits_{e = 1}^{{n_{{\rm{el}}}}} {\bf{F}}_{{\mathop{\rm int}}}^e\,\,{\rm{and}}\,\,\delta {{\bf{y}}} \coloneqq \mathop {\bf{A}}\limits_{e = 1}^{{n_{{\rm{el}}}}} \delta {{\bf{y}}^e},
\end{equation}
where the element internal force vector is obtained, from Eq.\,(\ref{beam_int_vir_work_compact_Form}), by
\begin{equation}
{\bf{F}}^e_{{\mathop{\rm int}} } \coloneqq \int_{{\varXi _e}} {{{{{\Bbb{B}}_{\rm{total}}^{e\,{\mathrm{T}}}}}}{\boldsymbol{R}}\,\tilde j\,{\mathrm{d}}\xi },
\end{equation}
and the matrix ${{{\Bbb{B}}_{\rm{total}}^e}}$ is defined in Eq.\,(\ref{disc_grad_operator_B_tot}). The external virtual work of Eq.\,(\ref{ext_vir_work_compact_form}) is also discretized as
\begin{equation}
\label{new_disc_ext_vir_work_compact_form}
{G_{{\rm{ext}}}}({{\boldsymbol{y}}^h},\delta {{\boldsymbol{y}}^h}) = \delta {{\bf{y}}}^{\mathrm{T}}{\bf{F}}_{{\rm{ext}}},\,\,\text{with}\,\,{\bf{F}}_{{\rm{ext}}} \coloneqq \mathop {\bf{A}}\limits_{e = 1}^{{n_{{\rm{el}}}}} {\bf{F}}^{e}_{\text{ext}} + {\bf{A}}{\left[ {{{{\boldsymbol{\bar R}}}_0}} \right]_{{\Gamma _{\text{N}}}}},
\end{equation}
where the second term on the right-hand side represents the assembly of load vector at the boundary $\Gamma_{\text{N}}$, and the element external load vector is obtained by
\begin{equation}
{\bf{F}}^e_{{\rm{ext}}} \coloneqq \int_{{\varXi _e}} {{{\Bbb{N}}_e^\mathrm{T}}{\boldsymbol{\bar R}}\,\tilde j\,{\mathrm{d}}\xi }.
\end{equation}
Similarly, the linearized internal virtual work of Eq.\,(\ref{beam_tangent_stiff_cont_form}) is discretized as
\begin{equation}
\label{new_disc_lin_int_force}
\Delta G_{{\mathop{\rm int}} }({{\boldsymbol{y}}^h};\delta {{\boldsymbol{y}}^h},\Delta {{\boldsymbol{y}}^h}) = \delta {{\bf{y}}}^{\mathrm{T}}{{\bf{K}}_\mathrm{int}}\,\Delta {{\bf{y}}}\,\,\,\text{with}\,\,{{\bf{K}}_\mathrm{int}} \coloneqq \mathop {\bf{A}}\limits_{e = 1}^{{n_{{\rm{el}}}}} {{\bf{K}}^e_\mathrm{int}}.
\end{equation}
The element tangent stiffness matrix is obtained by
\begin{equation}\label{elem_tan_mat_fin}
{{\bf{K}}^e_\mathrm{int}} = \int_{{\varXi _e}} {\left( {{{{{\Bbb{B}}_{{\rm{total}}}^{e\,{\mathrm{T}}}}}}{\Bbb{C}}{\Bbb{B}}_{{\rm{total}}}^e + {{\Bbb{Y}}_e}^{\mathrm{T}}{{\boldsymbol{k}}_\mathrm{G}}{{\Bbb{Y}}_e}} \right)\tilde j\,{\mathrm{d}}\xi },
\end{equation}
where ${{\Bbb{Y}}_e}$ is defined in Eq.\,(\ref{def_Y_e_g_tan}). It is noted that the global tangent stiffness matrix $\bf{K}_\mathrm{int}$ is symmetric, since ${\Bbb{C}}$ and ${{\boldsymbol{k}}_\mathrm{G}}$ are symmetric. Substituting Eqs.\,(\ref{new_disc_int_force_vec}), (\ref{new_disc_ext_vir_work_compact_form}), and Eq.\,(\ref{new_disc_lin_int_force}) into Eq.\,(\ref{new_config_update_lin_eq}) leads to
%
\begin{equation}
\label{disc_var_eq_global}
\delta {{\bf{y}}}^{\mathrm{T}}\,\leftidx{^{n + 1}}{{{\bf{K}}}}^{(i - 1)}\,\Delta {{\bf{y}}} = \delta {{\bf{y}}}^{\mathrm{T}}\,\leftidx{^{n + 1}}{{{\bf{R}}}}^{(i - 1)},
\end{equation}
where $\bf{K}\coloneqq{\bf{K}_\mathrm{int}}-{\bf{K}_\mathrm{ext}}$, and the global load stiffness matrix ${\bf{K}_\mathrm{ext}}$ appears, e.g., due to non-conservative follower loads, and it is generally unsymmetric (see for example Eq.\,(\ref{pure_bend_lstiff_op_disc_new})). The global residual vector is
\begin{equation}
{\bf{R}}\coloneqq \mathop {\bf{A}}\limits_{e = 1}^{{n_{{\rm{el}}}}} \left( {{\bf{F}}^e_{\rm{ext}} - {\bf{F}}^e_{{\mathop{\rm int}} }} \right).
\end{equation}
After applying the kinematic boundary conditions to Eq.\,(\ref{disc_var_eq_global}), we obtain
\begin{equation}\label{enhanced_global_eq_reduced}
\leftidx{^{n + 1}}{{{\bf{K}}}}^{(i - 1)}_\mathrm{r}\,\Delta {{{\bf{y}}}_{\text{r}}} = \leftidx{^{n + 1}}{{{{\bf{R}}}_{\text{r}}^{(i - 1)}}},
\end{equation}
where $(\bullet)_{\text{r}}$ denotes the \textit{reduced} vector or matrix after applying the kinematic boundary conditions.
\begin{definition} The symmetry of the global tangent stiffness matrix $\bf{K}$ depends solely on whether the external loading is conservative. If a non-conservative load is applied, the load stiffness leads to unsymmetric tangent stiffness matrix.
\end{definition}
\section{{Alleviation of Poisson locking by the EAS method}}
\label{eas_formulation}
{In order to alleviate Poisson locking, the in-plane strain field in the cross-section should be at least linear. We employ the EAS method, and we modify the Green-Lagrange strain tensor as
\begin{equation}\label{modify_green_lag_strn_enhanced}
{\boldsymbol{E}} = \underbrace {{{\boldsymbol{E}^{\text{c}}}}}_{{\rm{compatible}}} + \underbrace {{\boldsymbol{\tilde E}}}_{{\rm{enhanced}}},
\end{equation}
where the compatible strain part is the same as in Eq.\,(\ref{def_GL_strain_1}), and the additional strain part ${\boldsymbol{\tilde E}}$, which is incompatible, is intended to enhance the in-plane strain components of the cross-section, expressed by
\begin{equation}
{{\boldsymbol{\tilde E}}} = {\tilde E_{\alpha \beta }}\,{{\boldsymbol{G}}^\alpha } \otimes {{\boldsymbol{G}}^\beta }.
\end{equation}
The enhanced strain components are assumed as the linear and the bi-linear terms of the coordinates $\xi^1$ and $\xi^2$ in the cross-section, i.e.,
\begin{equation}\label{enhanced_strn}
\left\{ {\begin{array}{*{20}{c}}
{{{\tilde E}_{11}}}\\
{{{\tilde E}_{22}}}\\
{2{{\tilde E}_{12}}}
\end{array}} \right\} = \left[ {\begin{array}{*{20}{c}}
{{\xi ^1}}&{{\xi ^2}}&{{\xi ^1}{\xi ^2}}&0&0&0&0&0&0\\
0&0&0&{{\xi ^1}}&{{\xi ^2}}&{{\xi ^1}{\xi ^2}}&0&0&0\\
0&0&0&0&0&0&{{\xi ^1}}&{{\xi ^2}}&{{\xi ^1}{\xi ^2}}
\end{array}} \right]\left\{ {\begin{array}{*{20}{c}}
{{\alpha _1}}\\
{{\alpha _2}}\\
 \vdots \\ 
{{\alpha _8}}\\
{{\alpha _9}}
\end{array}} \right\} \eqqcolon {\boldsymbol{\Gamma \alpha }},
\end{equation}
where nine independent enhanced strain parameters $\alpha_i\in{L_2}(0,L)\,\,(i=1\sim9)$ are introduced. ${L_2}(0,L)$ defines the collection of all the functions, which are square integrable in the domain $(0,L)\ni{s}$. Even though the additional Green-Lagrange strain parts may include quadratic or higher order terms, we enrich the linear strain field only, since the enhanced strain is required to be orthogonal to constant stress fields in order to satisfy the stability condition \citep{betsch19964}. 
\begin{definition} 
\label{remark_5param_form_eas}
In this paper, it is shown that it may lead to erroneous results if the expression of Eq.\,(\ref{inplane_without_bilinear_GL}) is applied. For example, following Eq.\,(\ref{inplane_without_bilinear_GL}), one could define the enhanced strain part, as 
\begin{align}
\label{eas_strn_5param_form}
\left\{ {\begin{array}{*{20}{c}}
{\tilde E_{11}^ * }\\
{\tilde E_{22}^ * }\\
{2\tilde E_{12}^ * }
\end{array}} \right\} = \left[ {\begin{array}{*{20}{c}}
{{\xi ^1}}&0&0&0&0\\
0&{{\xi ^2}}&0&0&0\\
0&0&{{\xi ^1}}&{{\xi ^2}}&{{\xi ^1}{\xi ^2}}
\end{array}} \right]\left\{ {\begin{array}{*{20}{c}}
{{\alpha^{*}_1}}\\
{{\alpha^{*}_2}}\\
{{\alpha^{*}_3}}\\
{{\alpha^{*}_4}}\\
{{\alpha^{*}_5}}
\end{array}} \right\},
\end{align}
where five enhanced strain parameters $\alpha^{*}_i\in{L_2}(0,L)\,\,(i=1\sim5)$ are introduced. In numerical examples of section \ref{ex_cant_b_end_f}, it is shown that the EAS method based on Eq.\,(\ref{eas_strn_5param_form}) still suffers from significant Poisson locking.
\end{definition}
Applying the modified Green-Lagrange strain tensor to the three-field Hu-Washizu variational principle, the total strain energy is written in terms of the modified Green-Lagrange strain tensor as \citep{bischoff1997shear}
\begin{equation}
\tilde U\left(\boldsymbol{y},\boldsymbol{\tilde {E}},\boldsymbol{\tilde {S}}\right) = \int_0^L {\int_{\mathcal{A}} {\left\{ {\Psi ({{\boldsymbol{E}}^{\rm{c}}} + {\boldsymbol{\tilde E}}) - {\boldsymbol{\tilde S}}:{\boldsymbol{\tilde E}}} \right\}{j_0}\,{\mathrm{d}}{\mathcal{A}}\,{\mathrm{d}}s} }.
\end{equation}
The following condition that the stress field is $L_2$-orthogonal to the enhanced strain field enables to eliminate the stress field from the formulation, which leads to a \textit{two-field variational formulation}.
\begin{equation}
\label{ortho_condition}
\int_0^L {\int_{\mathcal{A}} {\left({\boldsymbol{\tilde S}}:{\boldsymbol{\tilde E}}\,{j_0}\right)\,{\mathrm{d}}{\mathcal{A}}\,{\mathrm{d}}s} }=0.
\end{equation}
The independent stress field $\boldsymbol{\tilde S}$, which satisfy the orthogonality condition of Eq.\,(\ref{ortho_condition}), does not explicitly appear in the subsequent formulation, and is generally different from the stress field $\boldsymbol{S}$, which is calculated by the constitutive law\footnote{See page 2,557 of \cite{buchter1994three}.}.
The first variation of the total strain energy is obtained by 
\begin{align}\label{int_vir_work_modified_enhanced}
\delta {\tilde U} &= \int_0^L {\int_{\mathcal{A}} {\left( {\frac{{\partial \Psi }}{{\partial {\boldsymbol{E}}}}:\delta {{\boldsymbol{E}}^{\text{c}}}\,{j_0}} \right){\mathrm{d}}{\mathcal{A}}}\,{\mathrm{d}}s} + \int_0^L {\int_{\mathcal{A}} {\left( {\frac{{\partial \Psi }}{{\partial {\boldsymbol{E}}}}} :\delta {\boldsymbol{\tilde E}}\,{j_0}\right)\,{\mathrm{d}}{\mathcal{A}}}\,{\mathrm{d}}s}.
\end{align}
We rewrite Eq.\,(\ref{int_vir_work_modified_enhanced}), using Eqs.\,(\ref{tot_strn_energy_beam_time_deriv}), (\ref{del_eps_hat_compact}), and (\ref{enhanced_strn}), as
\begin{align}\label{re_mod_int_virt_work_enhance}
{G_{{\mathop{\rm int}} }}(\boldsymbol{\eta},{\delta \boldsymbol{\eta}}) \equiv{\delta{\tilde U}}= \int_0^L {\left( {\delta {{\boldsymbol{y}}^{\mathrm{T}}}{{{{{\Bbb{B}}_{\text{total}}^{\mathrm{T}}}}}}{\boldsymbol{R}}} \right){\mathrm{d}}s}  + \int_0^L {{\delta {{\boldsymbol{\alpha }}^{\mathrm{T}}}\boldsymbol{s}}\,{\mathrm{d}}s},
\end{align}
where $\delta\boldsymbol{\eta}\coloneqq\left[\delta {\boldsymbol{y}^\mathrm{T}}, \delta {\boldsymbol{\alpha}}^\mathrm{T}\right]^\mathrm{T}$, and 
\begin{align}
{\boldsymbol{s}} \coloneqq \int_{\mathcal{A}} {\left( {{\boldsymbol{\Gamma }}^\mathrm{T}}{\munderbar{\boldsymbol{\hat S}}}\,{j_0} \right){\mathrm{d}}{\mathcal{A}}}\,\,\mathrm{with}\,\,{\munderbar{\boldsymbol{\hat S}}} \coloneqq {\left[S^{11}, S^{22}, S^{12}\right]^\mathrm{T}} = {\left[ {\frac{{\partial \Psi }}{{\partial {E_{11}}}},\frac{{\partial \Psi }}{{\partial {E_{22}}}},\frac{{\partial \Psi }}{{\partial {E_{12}}}}} \right]^\mathrm{T}}.
\end{align}
\subsection{Linearization}
The directional derivative of the internal virtual work of Eq.\,(\ref{re_mod_int_virt_work_enhance}) in the direction of $\Delta\boldsymbol{\eta}\coloneqq\left[\Delta {\boldsymbol{y}^\mathrm{T}}, \Delta {\boldsymbol{\alpha}}^\mathrm{T}\right]^\mathrm{T}$ is given by
\begin{alignat}{5}
\Delta G_{{\mathop{\rm int}} } ({\boldsymbol{\eta }};\delta {\boldsymbol{\eta }},\Delta {\boldsymbol{\eta }}) &&&\coloneqq D{G_{{\mathop{\rm int}} }} \cdot {\Delta \boldsymbol{\eta }}\nonumber\\
&&&= \int_0^L {\delta {{\boldsymbol{\eta }}^{\mathrm{T}}}\left[ {\begin{array}{*{20}{c}}
{{{\Bbb{B}}^{{\rm{total}}}}^{\mathrm{T}}{\Bbb{C}}{{\Bbb{B}}^{{\rm{total}}}} + {{\boldsymbol{Y}}^{\mathrm{T}}}{{\boldsymbol{k}}_{\rm{G}}}{\boldsymbol{Y}}}&{{{\Bbb{B}}^{{\rm{total}}}}^{\mathrm{T}}{{\Bbb{C}}^{{\rm{ay}}}}^{\mathrm{T}}}\\
{\mathrm{sym.}}&{{{\Bbb{C}}^{{\rm{aa}}}}}
\end{array}} \right]\Delta {\boldsymbol{\eta }}\,{\mathrm{d}}s},
\end{alignat}
where we use the following matrices
\begin{equation}
{{\Bbb{C}}^{\mathrm{ay}}} \coloneqq \int_{\mathcal{A}} {\left( {{{\boldsymbol{\Gamma }}^{\mathrm{T}}}{\Bbb{\bar C}}^{\mathrm{ay}}{{\munderbar {\boldsymbol{D}}}}\,{j_0}} \right){\mathrm{d}}{\mathcal{A}}}\,\,\,\mathrm{with}\,\,\,{\Bbb{\bar C}}^{\mathrm{ay}}\coloneqq{\left[ {\begin{array}{*{20}{c}}
{{C^{1111}}}&{{C^{1122}}}&{{C^{1133}}}&{{C^{1112}}}&{{C^{1113}}}&{{C^{1123}}}\\
{{C^{2211}}}&{{C^{2222}}}&{{C^{2233}}}&{{C^{2212}}}&{{C^{2213}}}&{{C^{2223}}}\\
{{C^{1211}}}&{{C^{1222}}}&{{C^{1233}}}&{{C^{1212}}}&{{C^{1213}}}&{{C^{1223}}}
\end{array}} \right]},
\end{equation}
and
\begin{equation}
{{\Bbb{C}}^{\text{aa}}} \coloneqq \int_{\mathcal{A}} {\left( {{{\boldsymbol{\Gamma }}^{\mathrm{T}}}{\bar {\Bbb{C}}}^{\text{aa}}\,{\boldsymbol{\Gamma }}\,{j_0}} \right){\mathrm{d}}{\mathcal{A}}}\,\,\,\mathrm{with}\,\,\,{{\Bbb{\bar C}}^{\text{aa}}} \coloneqq \left[ {\begin{array}{*{20}{c}}
{{C^{1111}}}&{{C^{1122}}}&{{C^{1112}}}\\
{}&{{C^{2222}}}&{{C^{2212}}}\\
{{\rm{sym}}{\rm{.}}}&{}&{{C^{1212}}}
\end{array}} \right].
\end{equation}
%
\subsection{Solution update procedure}
The iterative process to find solution ${}^{n + 1}{\boldsymbol{\eta}} \coloneqq {\left[ {{}^{n + 1}{{\boldsymbol{y}}^{\mathrm{T}}},{}^{n + 1}{{\boldsymbol{\alpha}}^{\mathrm{T}}}} \right]^{\mathrm{T}}}$ at the $(n+1)$th load step is stated as: For a given solution ${}^{n + 1}{\boldsymbol{\eta }}^{(i-1)}$ at the $(i-1)$th iteration of the $(n+1)$th load step, find the solution increment $\Delta {\boldsymbol{\eta }}$,
where $\Delta {\boldsymbol{y}} \in \mathcal{V}$ and $\Delta {\boldsymbol{\alpha }} \in \left[{L_2}(0,L)\right]^{d}$, such that 
\begin{equation}\label{enhance_strn_lin_var_eq_newton}
{\Delta G}_{{\mathop{\rm int}} }({}^{n + 1}{\boldsymbol{\eta }}^{(i-1)};\delta {\boldsymbol{\eta }},\Delta {\boldsymbol{\eta }}) = {G_{{\rm{ext}}}}(\delta {\boldsymbol{y}}) - {G_{{\rm{int}}}}({}^{n + 1}{\boldsymbol{\eta }}^{(i-1)},\delta {\boldsymbol{\eta }}),\,\,{\forall}{\delta {\boldsymbol{y}}} \in \mathcal{V},\,\,{\rm{and}}\,\,{\forall}\delta {\boldsymbol{\alpha }}\in \left[{L_2}(0,L)\right]^{d},
\end{equation}
where the dimension of the solution space of enhanced strain parameters can be $d=9$ or $d=5$ (see Remark \ref{remark_5param_form_eas}). Since the enhanced strain parameters are chosen to belong to the space ${L_2}(0,L)$, no inter-element continuity is required. Thus, it is possible to condense out those additional degrees-of-freedom at element level \citep{bischoff1997shear}. The solution is updated by
\begin{alignat}{2}
\left.\begin{array}{c}
\begin{aligned}
{}^{n + 1}{\boldsymbol{y}}^{(i)} 			&= {}^{n + 1}{\boldsymbol{y}}^{(i-1)} + \Delta {\boldsymbol{y}},&{{}^{n + 1}{\boldsymbol{y}}^{(0)}}&= {}^n{\boldsymbol{y}},\\
{}^{n + 1}{\boldsymbol{\alpha }}^{(i)}  &= {}^{n + 1}{\boldsymbol{\alpha }}^{(i-1)} + \Delta {\boldsymbol{\alpha }},&{{}^{n + 1}{\boldsymbol{\alpha }}^{(0)}}&= {}^n{\boldsymbol{\alpha}}.\\
\end{aligned}
\end{array} \right\}
\end{alignat}}
\subsection{{Discretization of the enhanced strain parameters and static condensation}}
{We reparameterize each of the $n_\text{el}$ elements of the central axis by a parametric coordinate ${\tilde \xi} \in \left[ {-1,1}\right]$. We define a linear mapping between the parametric domain of the $e$th element ${\varXi _e} = \left[ {\xi _e^1,\xi _e^2} \right]\ni \xi$ and $\left[ { - 1,1} \right] \ni \tilde \xi$, as
\begin{equation}
\tilde \xi  = 1 - 2\left( {\frac{{\xi _e^2 - \xi }}{{\xi _e^2 - \xi _e^1}}} \right).
\end{equation}
Then, within each element the vector of virtual enhanced strain parameters ${\delta\boldsymbol{\alpha }}={\delta\boldsymbol{\alpha }}(\tilde \xi )$ is linearly interpolated as
\begin{equation}\label{interp_var_alpha_beta}
{\delta\boldsymbol{\alpha }^h}(\tilde \xi ) = \left[ {\begin{array}{*{20}{c}}
{{{\tilde N}_1}(\tilde \xi )\,{{\bf{1}}_{9 \times 9}}}&{{{\tilde N}_2}(\tilde \xi )\,{{\bf{1}}_{9 \times 9}}}
\end{array}} \right]\left\{ {\begin{array}{*{20}{c}}
{\delta{\boldsymbol{\alpha }}_1}\\
{\delta{\boldsymbol{\alpha }}_2}
\end{array}} \right\} \eqqcolon \,{{\tilde{\Bbb{N}}}_{e}(\tilde \xi)}{\delta{\boldsymbol{\alpha }}^e},
\end{equation}
with nodal vectors of enhanced strain parameters ${\delta\boldsymbol{\alpha}_i}\,(i=1,2)$. In this paper, we use linear basis functions, given by
\begin{equation}
\left. {\begin{array}{lcl}\begin{aligned}
{{\tilde N}_1}(\tilde \xi ) &= (1 - \tilde \xi )/2\\
{{\tilde N}_2}(\tilde \xi ) &= (1 + \tilde \xi )/2
\end{aligned}\end{array}} \right\},\,\,\tilde \xi  \in \left[ { - 1,1} \right].
\end{equation}
Similarly, the vector of incremental enhanced strain parameters is interpolated within each element, as
\begin{equation}\label{interp_del_alpha}
{\Delta\boldsymbol{\alpha }^h}(\tilde \xi ) = {{{\tilde{\Bbb{N}}}}_{e}(\tilde \xi)}\,{\Delta{\boldsymbol{\alpha }}^e}.
\end{equation}
Substituting Eq.\,(\ref{interp_var_alpha_beta}) into the internal virtual work of Eq.\,(\ref{re_mod_int_virt_work_enhance}), and using the standard element assembly process, we have
\begin{equation}\label{disc_re_mod_int_vir_work_enhance}
{G_{{\mathop{\rm int}} }}({{\boldsymbol{\eta}}^h},\delta {{\boldsymbol{\eta}}^h}) = \mathop {\bf{A}}\limits_{e = 1}^{{n_{{\rm{el}}}}} {\left\{ {\begin{array}{*{20}{c}}
{\delta {{\bf{y}}^e}}\\
{\delta {{\boldsymbol{\alpha }}^e}}\\
\end{array}} \right\}^{\mathrm{T}}}\left\{ {\begin{array}{*{20}{c}}
{{\bf{F}}^e_{{\mathop{\rm int}} }}\\
{{{\bf{s}}^e}}
\end{array}} \right\},
\end{equation}
where we use
\begin{equation}
{{\bf{s}}^e} \coloneqq \int_{{\varXi _e}} {\left\{ {{\tilde j}\,{{{\tilde{\Bbb{N}}}_e}^{\mathrm{T}}}\int_{\mathcal{A}} {\left( {{{\boldsymbol{\Gamma }}^{\mathrm{T}}}{\munderbar{\boldsymbol{\hat S}}}\,{j_0}} \right){\mathrm{d}}{\mathcal{A}}} } \right\}{\mathrm{d}}\xi }.
\end{equation}
The linearized variational equation (Eq.\,(\ref{enhance_strn_lin_var_eq_newton})) is discretized as follows. For the given solution at the $(i-1)$th iteration of the $(n+1)$th load step, we find the solution increment such that
\begin{align}\label{disc_lin_var_eq_newton}
&\mathop {\bf{A}}\limits_{e = 1}^{{n_{{\rm{el}}}}} {\left({\left\{ {\begin{array}{*{20}{c}}
{\delta {{\bf{y}}^e}}\\
{\delta {{\boldsymbol{\alpha }}^e}}
\end{array}} \right\}^{\mathrm{T}}}\,\leftidx{^{n+1}}{\left[ {\begin{array}{*{20}{c}}
{{{\bf{K}}_\mathrm{int}^e}}&{{\bf{K}}{{^e_{\text{ay}}}^{\mathrm{T}}}}\\
{\mathrm{sym.}}&{{\bf{K}}^e_{\text{aa}}} 
\end{array}} \right]}^{(i-1)}\left\{ {\begin{array}{*{20}{c}}
{\Delta {{\bf{y}}^e}}\\
{\Delta {{\boldsymbol{\alpha }}^e}}
\end{array}} \right\}\right)} \nonumber\\
&= \mathop {\bf{A}}\limits_{e = 1}^{{n_{{\rm{el}}}}} {\left({\left\{ {\begin{array}{*{20}{c}}
{\delta {{\bf{y}}^e}}\\
{\delta {{\boldsymbol{\alpha }}^e}}
\end{array}} \right\}^{\mathrm{T}}}\,\leftidx{^{n+1}}{\left\{ {\begin{array}{*{20}{c}}
{{\bf{F}}^e_{\rm{ext}} - {\bf{F}}^e_{{\mathop{\rm int}} }}\\
{-{{\bf{s}}^e}}
\end{array}} \right\}}^{(i-1)}\right)},
\end{align}
where we use
\begin{equation}
{\bf{K}}^e_{\text{ay}} \coloneqq \int_{{\varXi _e}} {\left( {{\Bbb{\tilde N}}{{_e}^{\mathrm{T}}}{{\Bbb{C}}^{\text{ay}}}{\Bbb{B}}_e^{{\rm{total}}}\,\tilde j} \right){\mathrm{d}}\xi },
\end{equation}
and
\begin{equation}
{\bf{K}}^e_{\text{aa}} \coloneqq \int_{{\varXi _e}} {\left( {{\Bbb{\tilde N}}{{_e}^{\mathrm{T}}}{{\Bbb{C}}^{\text{aa}}}{{\Bbb{\tilde N}}}_e\,\tilde j} \right){\mathrm{d}}\xi }.
\end{equation}
Since we allow for a discontinuity of the enhanced strain field between adjacent elements, Eq.\,(\ref{disc_lin_var_eq_newton}) can be rewritten as
\begin{subequations}
\begin{alignat}{3}
\mathop {\bf{A}}\limits_{e = 1}^{{n_{{\rm{el}}}}} \left\{ {\delta {{\bf{y}}^e}^{\mathrm{T}}\left( {{{{\bf{K}}_\mathrm{int}^e}}\Delta {{\bf{y}}^e} + {{\bf{K}}{{^e_{\text{ay}}}^{\mathrm{T}}}}\Delta {{\boldsymbol{\alpha }}^e}} \right)} \right\} &= \mathop {\bf{A}}\limits_{e = 1}^{{n_{{\rm{el}}}}} \left\{ {\delta {{\bf{y}}^e}^{\mathrm{T}}\,\left( {{\bf{F}}^e_{\text{ext}} - {\bf{F}}^e_{{\mathop{\rm int}} }} \right)}\right\}\label{enhance_elem_y_eq},\\
\delta {{\boldsymbol{\alpha }}^e}^{\mathrm{T}}\left( {{{ {{\bf{K}}^e_{\text{ay}}}}}\Delta {{\bf{y}}^e} + {{{{\bf{K}}^e_{\text{aa}}}}}\Delta {{\boldsymbol{\alpha }}^e}}\right) &= -\delta {{\boldsymbol{\alpha }}^e}^{\mathrm{T}}{{\bf{s}}^e},\,e =1,2,...,{n_{{\rm{el}}}}.\label{enhance_elem_alpha_eq}
\end{alignat}
\end{subequations}
From Eq.\,(\ref{enhance_elem_alpha_eq}), we obtain
\begin{equation}\label{static_cond_del_alpha_eq}
\Delta {{\boldsymbol{\alpha }}_e} =  - \leftidx{^{n + 1}}{\left[ {{\bf{K}}{{^e_{\text{aa}}}^{ - 1}}} \right]}^{(i - 1)}\left( {{}^{n + 1}\left[{{\bf{s}}^e}\right]^{(i - 1)} + {}^{n + 1}{{\left[ {{\bf{K}}^e_{\text{ay}}} \right]}^{(i - 1)}}\Delta {{\bf{y}}^e}} \right),\,e = 1,2,...,{n_{{\rm{el}}}}.
\end{equation}
Substituting Eq.\,(\ref{static_cond_del_alpha_eq}) into Eq.\,(\ref{enhance_elem_y_eq}) leads to
\begin{equation}
\label{eas_var_eq_1}
\delta {{\bf{y}}}^{\mathrm{T}}\leftidx{^{n + 1}}{{{\bf{\tilde K}}}}^{(i - 1)}\Delta {{\bf{y}}} = \delta {{\bf{y}}}^{\mathrm{T}}\leftidx{^{n + 1}}{{{\bf{\tilde R}}}}^{(i - 1)},
\end{equation}
where the global tangent stiffness matrix is defined as
\begin{equation}\label{enhanced_global_eq}
{\bf{\tilde K}} \coloneqq \mathop {\bf{A}}\limits_{e = 1}^{{n_{{\rm{el}}}}} \left( {{{\bf{K}}_\mathrm{int}^e} - {\bf{K}}{{^e_{\text{ay}}}^{\mathrm{T}}}{\bf{K}}{{^e_{\text{aa}}}^{ - 1}}{\bf{K}}^e_{\text{ay}}} \right),
\end{equation}
and the global residual vector is
\begin{equation}
{\bf{\tilde R}}\coloneqq \mathop {\bf{A}}\limits_{e = 1}^{{n_{{\rm{el}}}}} \left( {{\bf{F}}^e_{\rm{ext}} - {\bf{F}}^e_{{\mathop{\rm int}} } + {\bf{K}}{{^{e\,\mathrm{T}}_{\text{ay}}}}{\bf{K}}{{^{e\,{-1}}_{\text{aa}}}}{{\bf{s}}^e}} \right).
\end{equation}
After applying the kinematic boundary conditions to Eq.\,(\ref{eas_var_eq_1}), we obtain
\begin{equation}\label{enhanced_global_eq_reduced}
{{\leftidx{^{n + 1}}{{\bf{\tilde K}}}_{\text{r}}^{(i - 1)}}}\Delta {{{\bf{y}}}_{\text{r}}} = {{\leftidx{^{n + 1}}{{\bf{\tilde R}}}_{\text{r}}^{(i - 1)}}}.
\end{equation}}
\section{Numerical examples}
\label{num_ex}
We verify the presented beam formulation by comparison with reference solutions from the isogeometric analysis of three-dimensional hyperelasticity using brick elements. The brick elements use different degrees of basis functions in each parametric coordinate direction. We denote this by `$\mathrm{deg.}=({p_{\mathrm{L}}},{p_{\mathrm{W}}},{p_{\mathrm{H}}})$', where $p_{\mathrm{L}}$, $p_{\mathrm{W}}$, and $p_{\mathrm{H}}$ denote the degrees of basis functions along the length (L), width (W), and height (H), respectively. Further, we indicate the number of elements in each of those directions by $n_{\mathrm{el}}={n_{\mathrm{el}}^{\mathrm{L}}}\times{n_{\mathrm{el}}^{\mathrm{W}}}\times{n_{\mathrm{el}}^{\mathrm{H}}}$.
We employed two different hyperelastic material models: St.\,Venant-Kirchhoff and compressible Neo-Hookean types, which are abbreviated by `SVK' and `NH', respectively. We also use the following abbreviations to indicate our three beam formulations.
\begin{itemize}
\item{Ext.-dir.-std.: The standard extensible director beam formulation.}
\item{Ext.-dir.-EAS: The EAS method with nine enhanced strain parameters, i.e., Eq.\,(\ref{enhanced_strn}).}
\item{Ext.-dir.-EAS-5p.: The EAS method with five enhanced strain parameters, i.e., Eq.\,(\ref{eas_strn_5param_form}).}
\end{itemize}
In the beam formulation, the integration over the cross-section is evaluated numerically. We use standard Gauss integration for the central axis and cross-section, where $(p+1)$ integration points are used for the central axis, and the number of integration points for the cross-section is mentioned in each numerical example. Here $p$ denotes the order of basis functions approximating the central axis displacement and director fields.
\subsection{Uniaxial tension of a straight beam}
\begin{figure}[htp]
\centering
\includegraphics[width=0.375\linewidth]{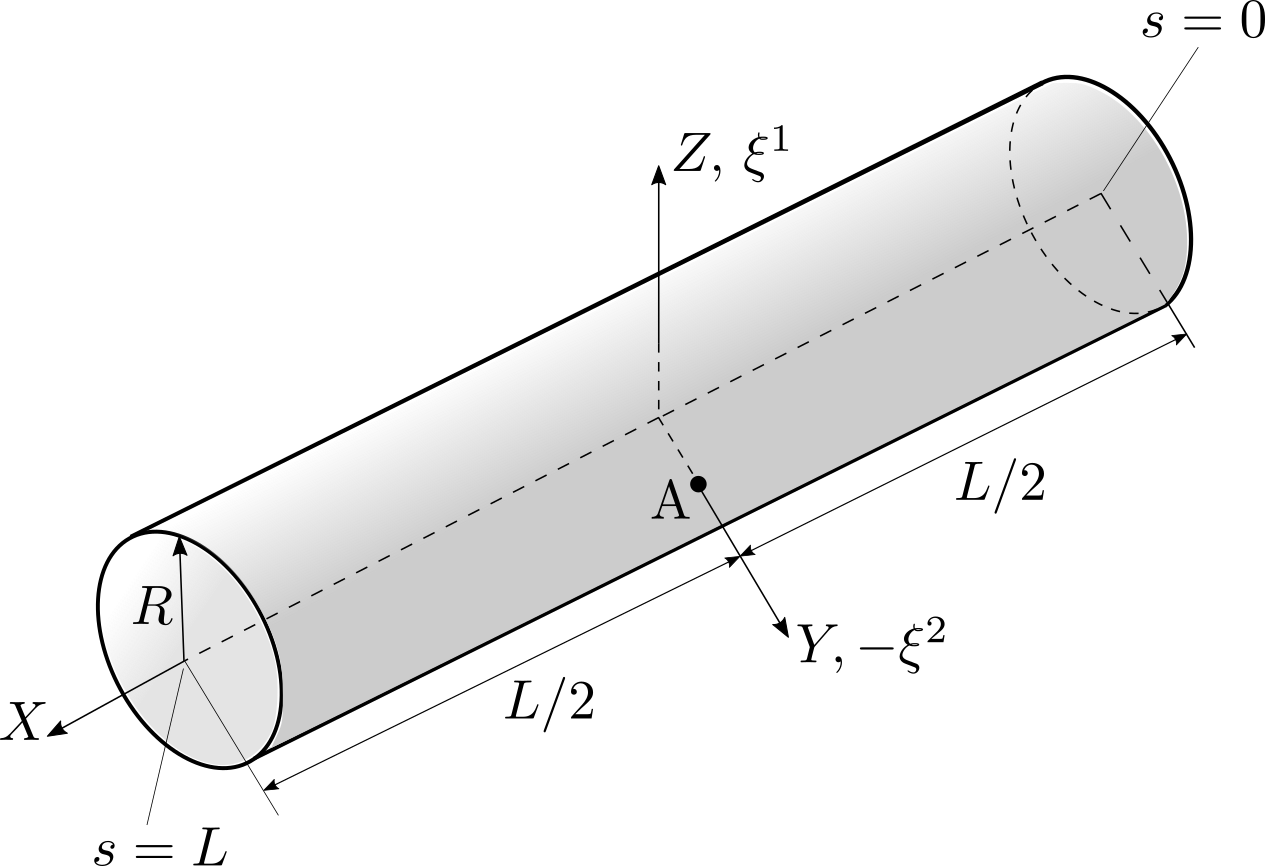}
\caption{Uniaxial tension of a straight beam: Undeformed configuration. The directions of ${\xi^1}$ and $\xi^2$ represent the chosen principal directions of the circular cross-section.}
\label{uniaxial_tens_undeformed}	
\end{figure}

\begin{figure}[htp]
	\centering
	\begin{subfigure}[b] {0.375\textwidth} \centering
		\includegraphics[width=\linewidth]{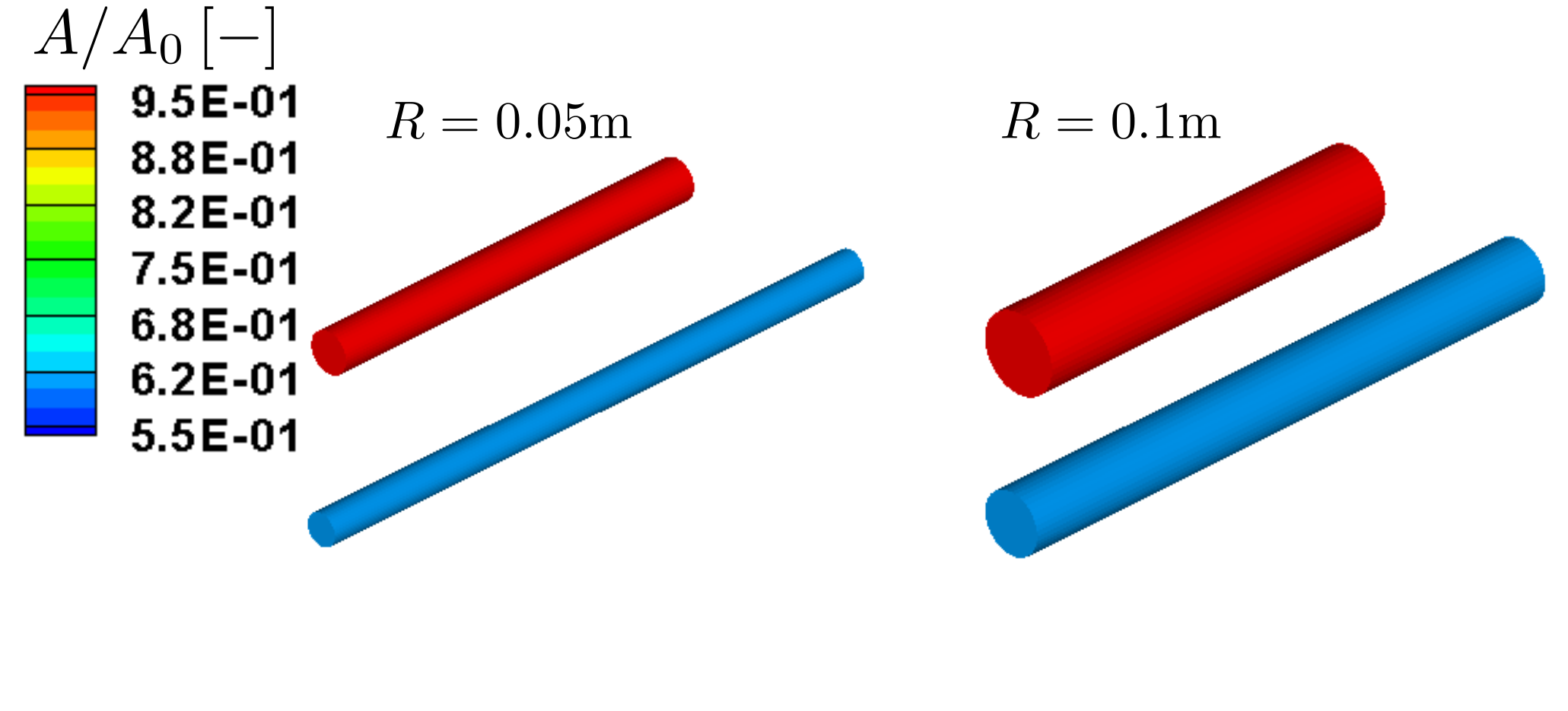}
		\caption{End directors free (SVK).}
		\label{uniaxial_tens_deformed_d_free_svk}
	\end{subfigure}	
	\begin{subfigure}[b] {0.525\textwidth} \centering
		\includegraphics[width=\linewidth]{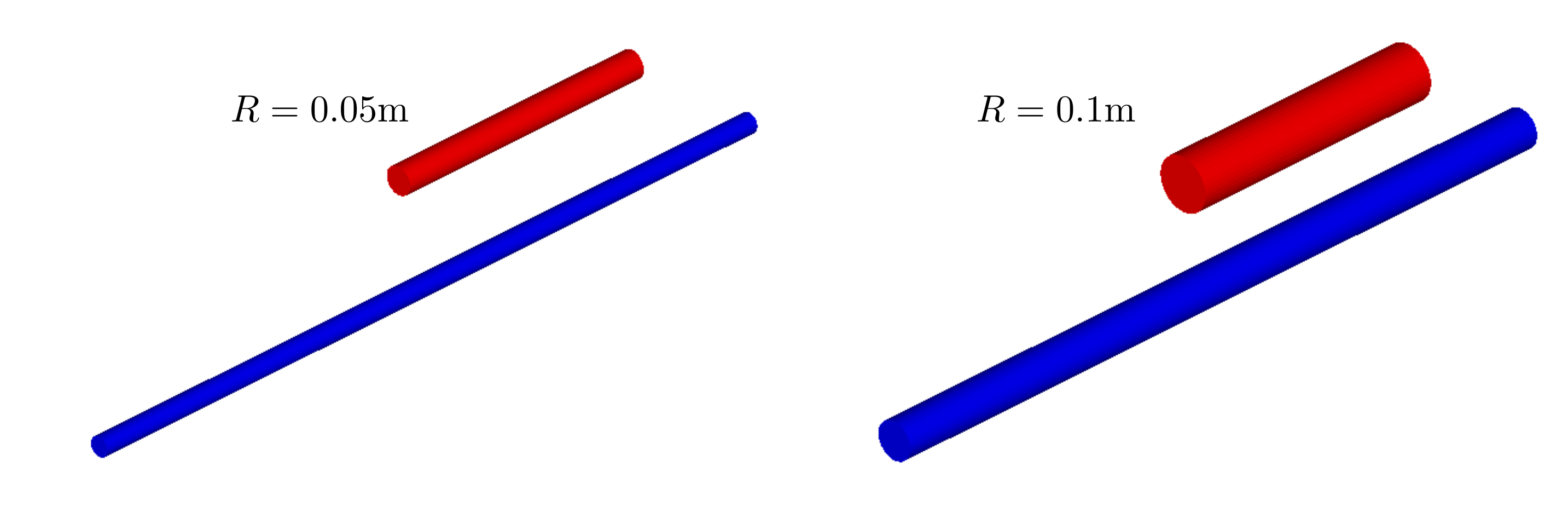}
		\caption{End directors free (NH).}
		\label{uniaxial_tens_deformed_d_free_nh}
	\end{subfigure}
	\begin{subfigure}[b] {0.325\textwidth} \centering
		\includegraphics[width=\linewidth]{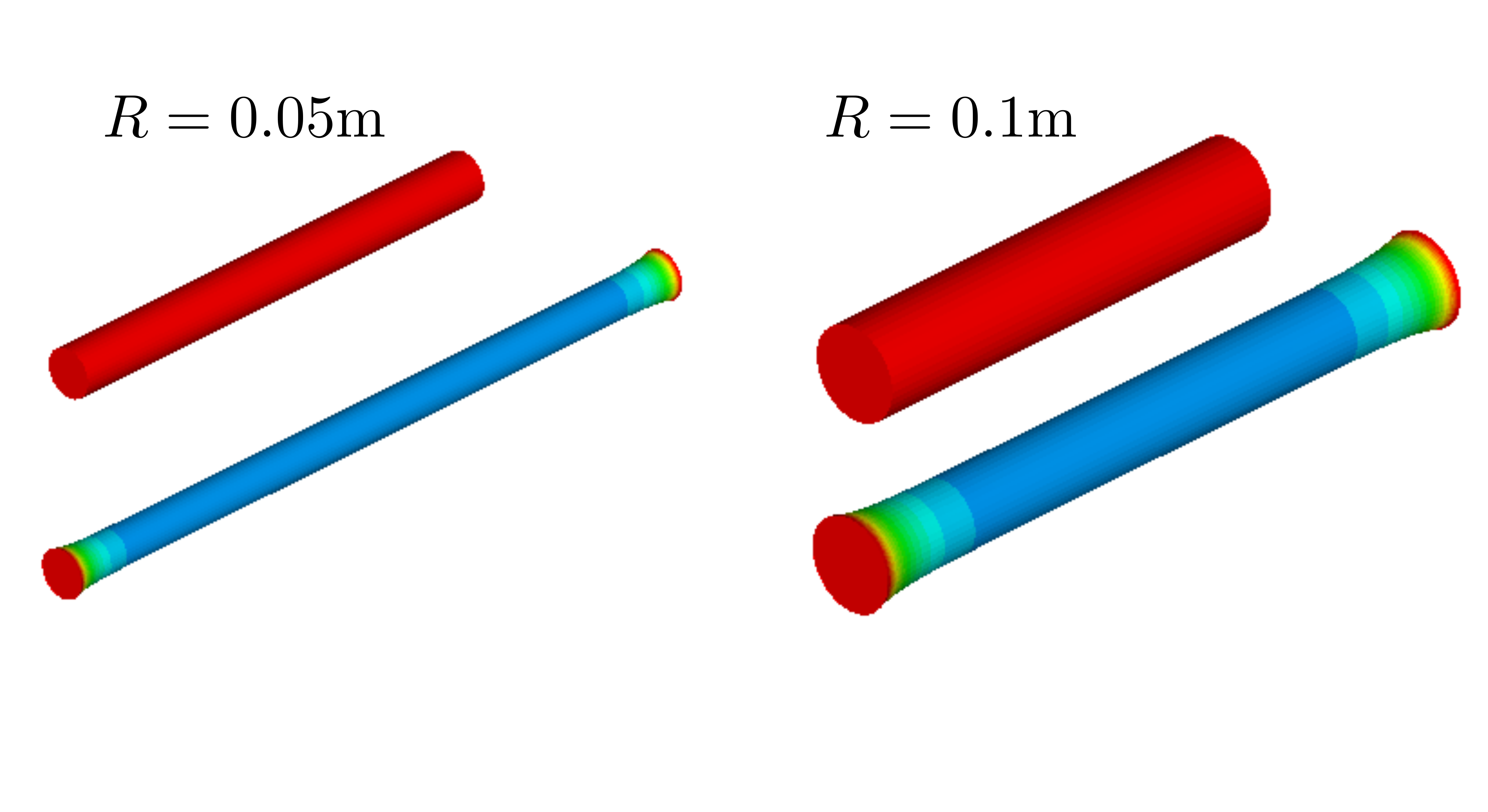}
		\caption{End directors fixed (SVK).}
		\label{uniaxial_tens_deformed_d_free_svk}
	\end{subfigure}	
	\begin{subfigure}[b] {0.525\textwidth} \centering
		\includegraphics[width=\linewidth]{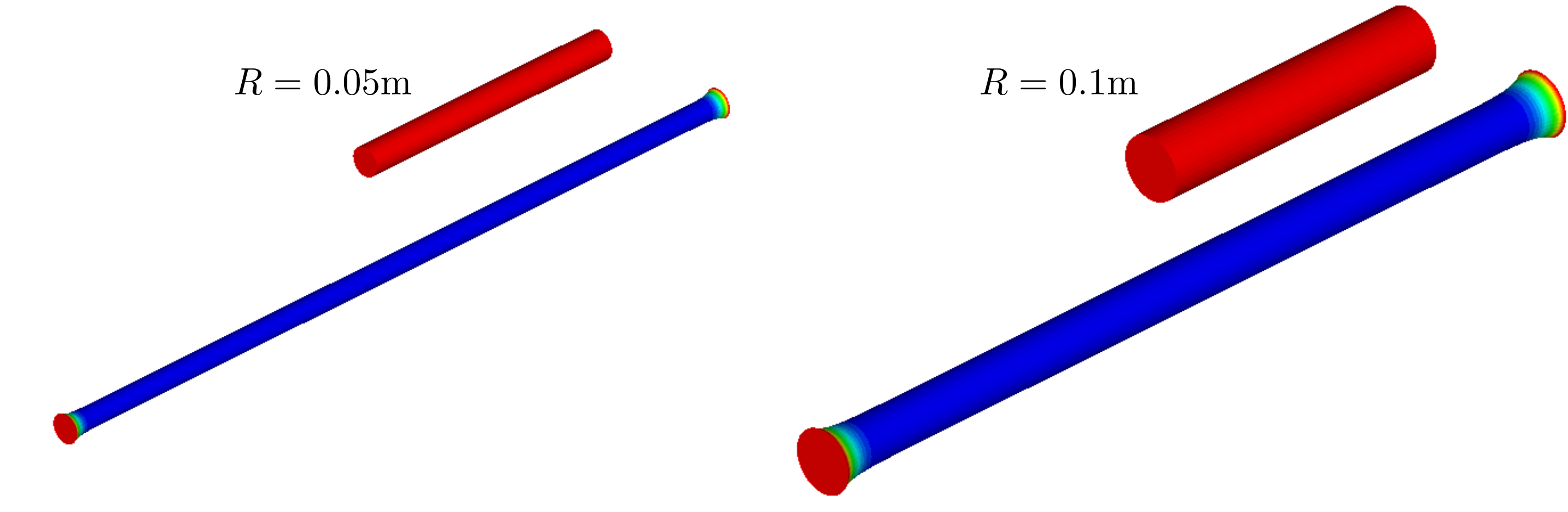}
		\caption{End directors fixed (NH).}
		\label{uniaxial_tens_deformed_d_free_nh}
	\end{subfigure}\hspace{2.5mm}			
\caption{Uniaxial tension of a straight beam: Undeformed and deformed configurations. The color represents the ratio of the current cross-sectional area ($A$) to the initial one ($A_0$). 40 cubic B-spline elements have been used for the analysis.}
\label{deform_str_uni_axial_tens}	
\end{figure}

\begin{figure}[htp]	
\centering
	\begin{subfigure}[b] {0.495\textwidth} \centering
		\includegraphics[width=\linewidth]{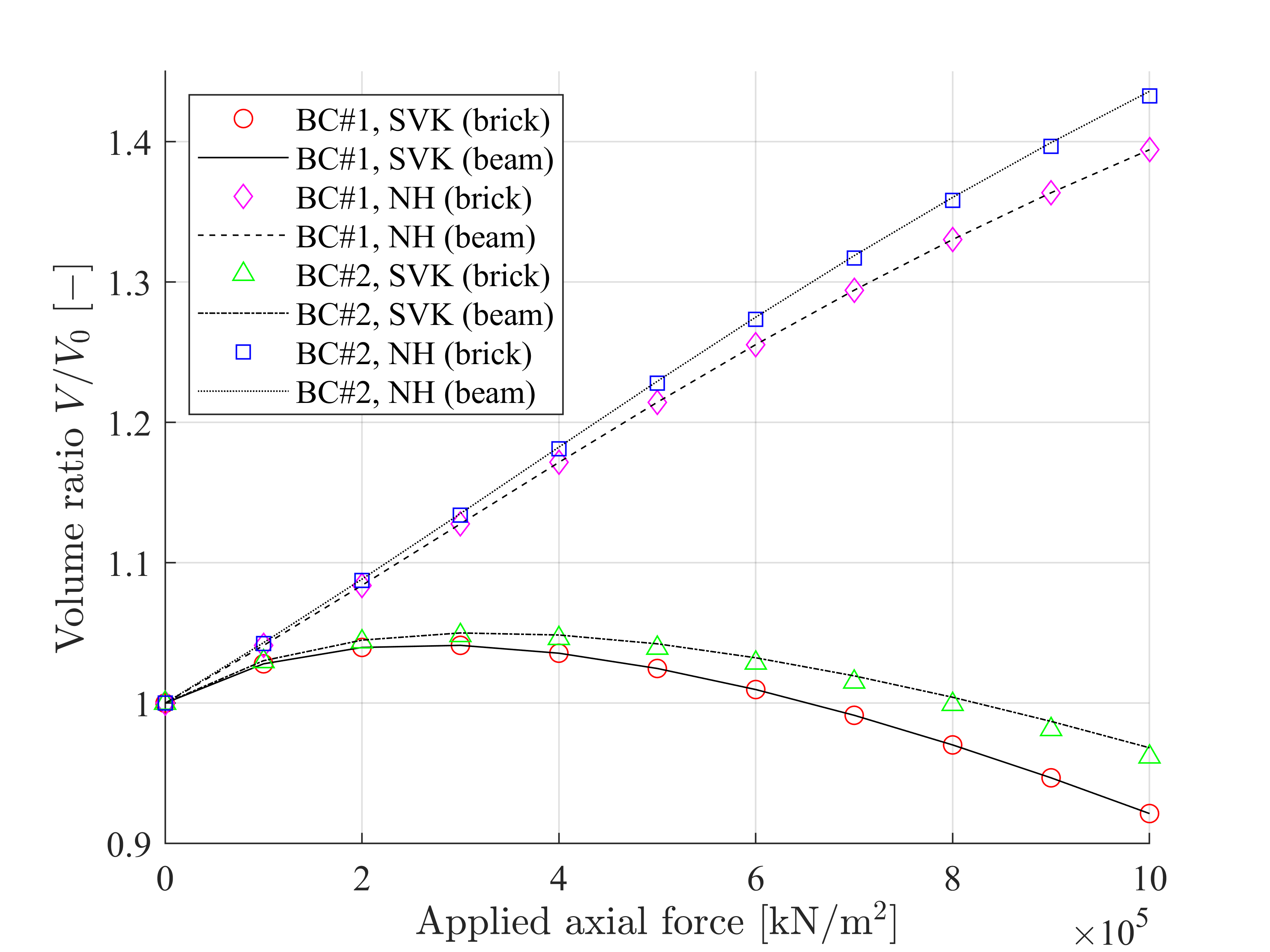}
		\caption{$R=0.05\mathrm{m}$}
		\label{uniaxial_tens_vol_change_r005}	
	\end{subfigure}			
	\begin{subfigure}[b] {0.495\textwidth} \centering
		\includegraphics[width=\linewidth]{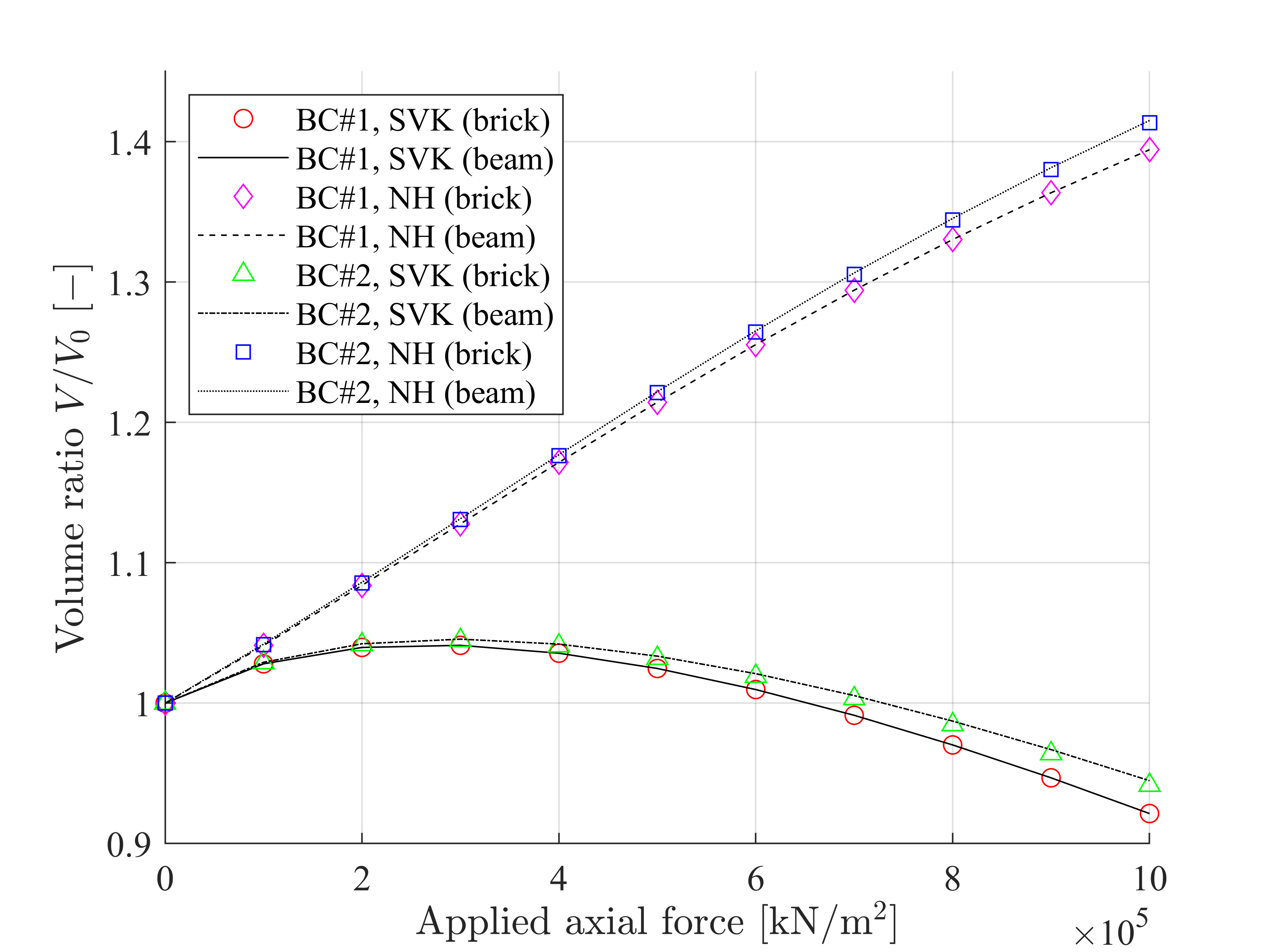}
		\caption{$R=0.1\mathrm{m}$}
		\label{uniaxial_tens_vol_change_r010}	
	\end{subfigure}
\caption{Uniaxial tension of a straight beam: Comparison of volume change in uniaxial tension with brick elements and beam elements for the two different material models and cross-section radii with two cases of kinematic boundary conditions.}
\label{uniaxial_tens_vol_ratio}		
\end{figure}

In order to verify the capability of the presented beam formulation in representing finite axial and transverse normal strain, we consider uniaxial tension of a straight beam having nonzero Poisson's ratio. The beam has length $L=1\text{m}$ and a circular cross-section with two cases for its radius, $R=0.05\text{m}$ and $R=0.1\text{m}$, while Young's modulus and Poisson's ratio are $E=1\text{GPa}$ and $\nu=0.3$, respectively. Two different kinematic boundary conditions at the two ends of beam (i.e., $s\in\left\{0,L\right\}$) are considered. First, the cross-section is allowed to deform at the both ends (BC{\#}1), and second, this is not allowed (BC{\#}2). A traction of ${{\boldsymbol{\bar T}}_0} = {\left[ {{{\bar T}_0},0,0} \right]^{\mathrm{T}}}$ where ${{\bar T}_0} = {10^6}\mathrm{kN}/{\mathrm{m}^2}$ is applied on the undeformed cross-section at $s=L$. In the beam model, these two boundary conditions are implemented as follows. 
\begin{itemize}
\item BC{\#}1: Central axis displacement is constrained at one end, but the end directors are free, i.e.,
\begin{equation}
\Delta {\boldsymbol{\varphi }} = {\bf{0}}\,\,\text{at}\,\,s = 0,\,\,\text{and}\,\,{\boldsymbol{d}_1}\,\,\text{and}\,\,{\boldsymbol{d}_2}\,\,\text{are free at}\,\,s \in \left\{ {0,L} \right\}.\nonumber
\end{equation}
\item BC{\#}2: All degrees-of-freedom are constrained at one end, and the directors are fixed at the other end, that is, 
\begin{equation}
\Delta {\boldsymbol{\varphi }} = \Delta {{\boldsymbol{d}}_1} = \Delta {{\boldsymbol{d}}_2} = {\bf{0}}\,\,\text{at}\,\,s = 0,\,\,\text{and}\,\,\Delta {{\boldsymbol{d}}_1} = \Delta {{\boldsymbol{d}}_2} = {\bf{0}}\,\,\text{at}\,\,s = L.\nonumber
\end{equation} 
\end{itemize}
{In the numerical integration over the circular cross-section of the beam, we employ polar coordinates $r$ and $\theta$ (see Remark \ref{num_integ_polar_circ}), and four Gauss integration points are used for each of the variables $r$ and $\theta$ in each quarter of the domain $0\leq{\theta}<{2\pi}$.} Fig.\,\ref{uniaxial_tens_undeformed} shows the undeformed configuration, and Fig.\,\ref{deform_str_uni_axial_tens} shows the deformed configurations for the different boundary conditions and material models, where the decrease of cross-sectional area is noticeable. We compare the lateral displacement at surface point\,A, indicated in Fig.\,\ref{uniaxial_tens_undeformed}, with the reference solutions obtained from IGA using brick elements (convergence results for the lateral displacement at point A and the volume change can be found in Tables \ref{app_utens_conv_test_svk_010_fixed} and \ref{app_utens_conv_test_nh_010_fixed}). Tables \ref{str_utens_verif_surf_disp_svk} and \ref{str_utens_verif_surf_disp_nh} compare the lateral ($Y$-directional) displacements. The results from the developed beam model are in excellent agreement with the reference solution. In Fig.\,\ref{uniaxial_tens_vol_ratio}, we can also verify that the volume change of the beam agrees with the reference solutions in all cases of the selected materials and cross-section radii. As expected those two material models show similar behavior within the small strain range; however, the behavior become different for large strains. Note that the SVK material shows unphysical volume decrease beyond certain strains, which shows the unsuitability of this material model for large strains.

\begin{table}[]
\small
\centering
\caption{Uniaxial tension of a straight beam: Verification of the lateral displacement at surface point\,$\mathrm{A}$ (St.\,Venant-Kirchhoff material). All results are obtained by IGA.}
\label{str_utens_verif_surf_disp_svk}
\begin{tabular}{lcclcclcr}
\Xhline{3\arrayrulewidth}
&\multicolumn{2}{c}{End directors free}& &\multicolumn{2}{c}{End directors fixed}& &\multicolumn{2}{c}{Ratio}                                                                                  
\\ \cline{2-3} \cline{5-6} \cline{8-9} 
\multicolumn{1}{c}{\begin{tabular}[c]{@{}c@{}}$R$\\{[}m{]}\end{tabular}} &\multicolumn{1}{c}{\begin{tabular}[c]{@{}c@{}}Brick, deg.=(2,2,2),\\${n_\mathrm{el}}=320\times20\times20$,\\{[}m{]} (a)\end{tabular}} & \multicolumn{1}{c}{\begin{tabular}[c]{@{}c@{}}Beam, $p=3$,\\${n_\mathrm{el}}=40$\\{[}m{]} (b)\end{tabular}} &  & \multicolumn{1}{c}{\begin{tabular}[c]{@{}c@{}}Brick, deg.=(3,3,3),\\${n_\mathrm{el}}=320\times15\times15$\\{[}m{]} (c)\end{tabular}} & \multicolumn{1}{c}{\begin{tabular}[c]{@{}c@{}}Beam, $p=3$,\\${n_\mathrm{el}}=40$\\{[}m{]} (d)\end{tabular}} &  & \begin{tabular}[c]{@{}c@{}}(b)/(a)\\ {[}{\%}{]}\end{tabular} & \begin{tabular}[c]{@{}c@{}}(d)/(c)\\ {[}{\%}{]}\end{tabular} \\
\Xhline{3\arrayrulewidth} 
0.05& -1.1089E-02  & -1.1089E-02 & &	-1.1089E-02    & -1.1089E-02  &   & 100.00      & 100.00\\
0.1 & -2.2178E-02  & -2.2178E-02 & &	-2.2181E-02    & -2.2177E-02  &   & 100.00      & 99.98\\
\Xhline{3\arrayrulewidth}
\end{tabular}
\end{table}

\begin{table}[]
\small
\centering
\caption{Uniaxial tension of a straight beam: Verification of the lateral displacement at surface point\,$\mathrm{A}$ (compressible Neo-Hookean material). All results are obtained by IGA.}
\label{str_utens_verif_surf_disp_nh}
\begin{tabular}{lcclcclcr}
\Xhline{3\arrayrulewidth}
&\multicolumn{2}{c}{End directors free}& &\multicolumn{2}{c}{End directors fixed}& &\multicolumn{2}{c}{Ratio}                                                                                  
\\ \cline{2-3} \cline{5-6} \cline{8-9} 
\multicolumn{1}{c}{\begin{tabular}[c]{@{}c@{}}$R$\\{[}m{]}\end{tabular}} &\multicolumn{1}{c}{\begin{tabular}[c]{@{}c@{}}Brick, deg.=(2,2,2),\\${{n_\mathrm{el}}}=320\times20\times20$,\\{[}m{]} (a)\end{tabular}} & \multicolumn{1}{c}{\begin{tabular}[c]{@{}c@{}}Beam, $p=3$,\\${{n_\mathrm{el}}}=40$,\\{[}m{]} (b)\end{tabular}} &  & \multicolumn{1}{c}{\begin{tabular}[c]{@{}c@{}}Brick, deg.=(2,2,2),\\${{n_\mathrm{el}}}=320\times20\times20$,\\{[}m{]} (c)\end{tabular}} & \multicolumn{1}{c}{\begin{tabular}[c]{@{}c@{}}Beam, $p=3$,\\${{n_\mathrm{el}}}=40$,\\{[}m{]} (d)\end{tabular}} &  & \begin{tabular}[c]{@{}c@{}}(b)/(a)\\ {[}{\%}{]}\end{tabular} & \begin{tabular}[c]{@{}c@{}}(d)/(c)\\ {[}{\%}{]}\end{tabular} \\
\Xhline{3\arrayrulewidth} 
0.05& -1.4593E-02  & -1.4593E-02 & &	-1.4593E-02  &-1.4593E-02   &   & 100.00      & 100.00\\
0.1 & -2.9186E-02  & -2.9186E-02 & &	-2.9186E-02  & -2.9186E-02  &   & 100.00      & 100.00\\
\Xhline{3\arrayrulewidth}
\end{tabular}
\end{table}

\subsection{Cantilever beam under end moment}
\label{ex_beam_end_mnt}
\begin{figure}[htp]
\centering
\includegraphics[width=0.6\linewidth]{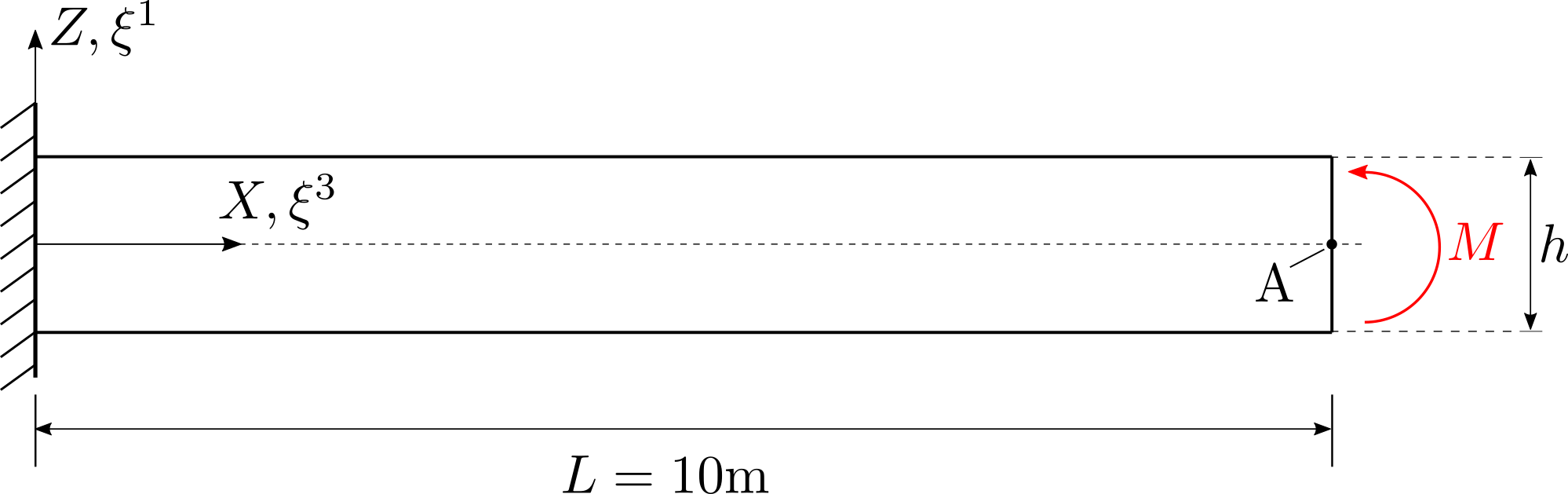}
\caption{Cantilever beam under end moment: Undeformed configuration and boundary conditions.}
\label{cant_beam_end_moment}	
\end{figure}

An initially straight beam of length $L=10\text{m}$ with rectangular cross-section of width ${w}=1\text{m}$ and height $h$ is clamped at one end and subject to bending moment $M$ on the other end (see Fig.\,\ref{cant_beam_end_moment}). The material properties are Young's modulus $E=1.2\times{10^7}\text{Pa}$, and Poisson's ratio $\nu=0$. Under the assumption of pure bending, an applied moment $M$ deforms the beam central axis into a circle with radius $R=EI/M$, where the $X$- and $Z$-displacements at the tip of the central axis (point $\mathrm{A}$ in Fig.\,\ref{cant_beam_end_moment}) can be derived, respectively, as
\begin{subequations}
\label{beam_end_mnt_exact_sol}
\begin{alignat}{2}
{u_{\mathrm{A}}} &= R\sin \frac{L}{R} - L,\label{beam_end_mnt_exact_sol_x}\\
{w_{\mathrm{A}}} &= R\left( {1 - \cos \frac{L}{{R}}} \right).\label{beam_end_mnt_exact_sol_z}
\end{alignat}
\end{subequations}
Since the presented extensible director beam formulation contains no rotational degrees of freedom, we cannot directly apply the bending moment. There are several ways to implement the moment load: A coupling element was introduced in \citet{frischkorn2013solid}, and the virtual work contribution of the boundary moment was directly discretized in the rotation-free thin shell formulation of \citet{duong2017new}. We adopt another way presented in \citet{betsch1995assumed} to use a distributed follower load acting on the end face. At the loaded end face, the following linear distribution of the first Piola-Kirchhoff stress is prescribed,
\begin{equation}\label{end_moment_first_pk}
{\boldsymbol{P}} = p\,{{\boldsymbol{\nu }}_t} \otimes {{\boldsymbol{\nu }}_0}\,\,\text{with}\,\,p\coloneqq{-\frac{M}{I}{\xi ^1}}\,\,\mathrm{and}\,\,{I=\frac{wh^3}{12}}\,\,\mathrm{at}\,\,s\in{\Gamma_\mathrm{N}}\,\left(s=L\right),
\end{equation}
where the outward unit normal vector on the initial end face is $\boldsymbol{\nu}_0={\boldsymbol{e}_1}$ since the beam central axis is aligned with the $X$-axis, and the outward unit normal vector on the current end face is 
\begin{equation}
{{\boldsymbol{\nu }}_t} = \boldsymbol{d}_3\,\,\text{with}\,\,{{\boldsymbol{d}}_3} = \frac{{{{\boldsymbol{d}}_1} \times {{\boldsymbol{d}}_2}}}{{\left\| {{{\boldsymbol{d}}_1} \times {{\boldsymbol{d}}_2}} \right\|}},\,\,\mathrm{and}\,\,{\boldsymbol{d}_2}=-{\boldsymbol{e}_2}.
\end{equation}
From Eq.\,(\ref{end_moment_first_pk}), we can simply obtain the prescribed traction vector ${\boldsymbol{\bar T}}_0$, as
\begin{equation}
\label{num_ex_end_mnt_prescribed_traction}
{{\boldsymbol{\bar T}}_0} = {\boldsymbol{P}}{{\boldsymbol{\nu }}_0} = p\,{{\boldsymbol{d}}_3}\,\,\mathrm{at}\,\,s\in{\Gamma_\mathrm{N}}.
\end{equation}
Substituting the traction vector of Eq.\,(\ref{num_ex_end_mnt_prescribed_traction}) into Eqs.\,(\ref{app_nbdc_strs_res}) and (\ref{app_nbdc_strs_coup}), we obtain
\begin{subequations}
\begin{gather}
\label{ex_end_mnt_n0_m_condition}
{{\boldsymbol{\bar n}}_0} = \int_{{\mathcal{A}}_0} {{\boldsymbol{\bar T}_0}\,{\mathrm{d}}{{\mathcal{A}}_0}} = {\boldsymbol{0}},\\
{{\boldsymbol{\bar {\tilde m}}}_0^1} = \int_{{\mathcal{A}}_0} {\xi^1}{{\boldsymbol{\bar T}_0}\,{\mathrm{d}}{{\mathcal{A}}_0}} = - M{\boldsymbol{d}_3},\,\,\,\mathrm{and}\,\,{{\boldsymbol{\bar {\tilde m}}}_0^2} = {\boldsymbol{0}}.
\end{gather}
\end{subequations}
That is, the Neumann boundary condition at $s\in{\Gamma_\mathrm{N}}$ is given by
\begin{subequations}
\label{ex_end_mnt_neumann_bdc}
\begin{gather}
\boldsymbol{n}=\boldsymbol{0},\label{nbdc_n_end_mnt}\\
{{\boldsymbol{\tilde m}}^1}=-M{\boldsymbol{d}_3},\,\,\mathrm{and}\,\,{{\boldsymbol{\tilde m}}^2}={\boldsymbol{0}}.\label{nbdc_m_td_end_mnt}
\end{gather}
\end{subequations}
A detailed expression of the external virtual work and the load stiffness operator can be found in Appendix \ref{mnt_load_follower_load_exp}.
\begin{figure}[!htpt]	
	\centering
	\begin{subfigure}[b] {0.435\textwidth} \centering
		\includegraphics[width=\linewidth]{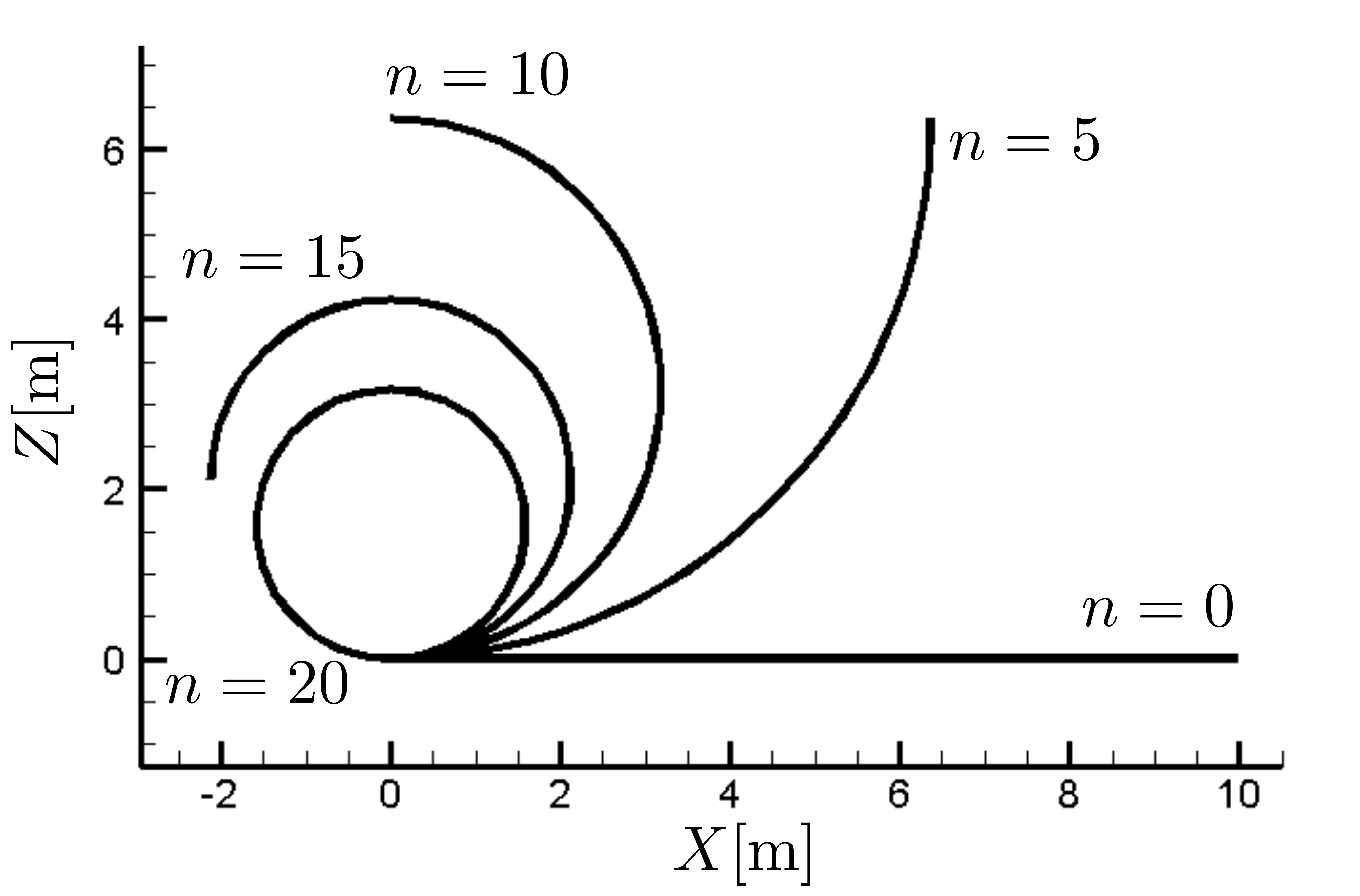}
		\caption{Initial cross-section height ${h}=0.1\rm{m}$}
		\label{pure_bend_deform_h010}
	\end{subfigure}\hspace{2.5mm}			
	\begin{subfigure}[b] {0.435\textwidth} \centering
		\includegraphics[width=\linewidth]{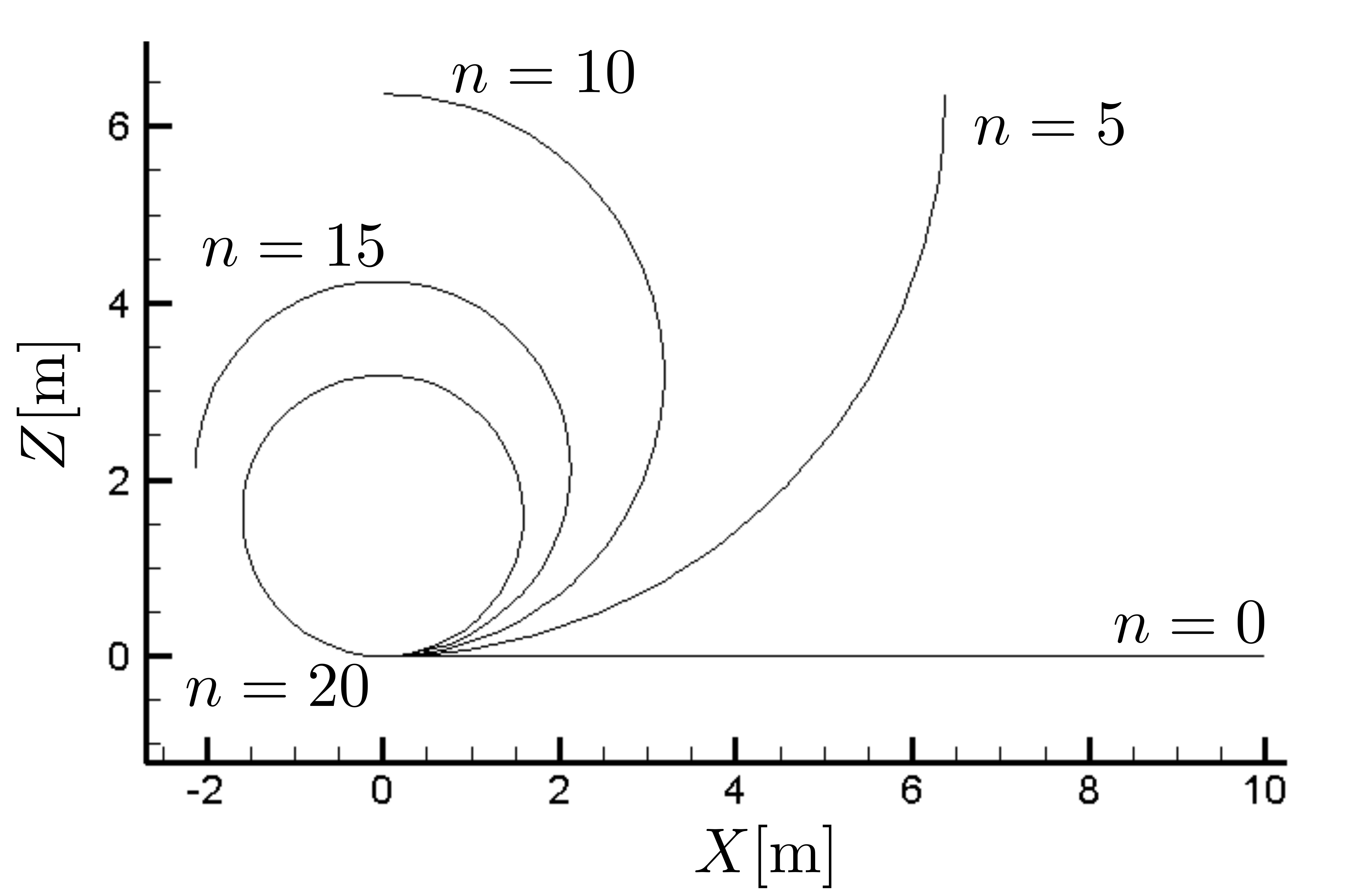}
		\caption{Initial cross-section height ${h}=0.01\rm{m}$}
		\label{pure_bend_deform_h001}
	\end{subfigure}\hspace{2.5mm}			
\caption{Cantilever beam under end moment: Deformed configurations for two different cross-section heights. $n$ denotes the load step number, where the applied end moment is $M=0.1n\pi EI/L$. Figure (b) shows the central axis only, because the cross-section is too thin to clearly visualize. The beam solutions are calculated by IGA with $p=4$ and $n_{\mathrm{el}}=160$.}
\label{cant_beam_end_moment_deformed}	
\end{figure}
\begin{figure}[!htpt]	
	\centering
	\begin{subfigure}[b] {0.4875\textwidth} \centering
		\includegraphics[width=\linewidth]{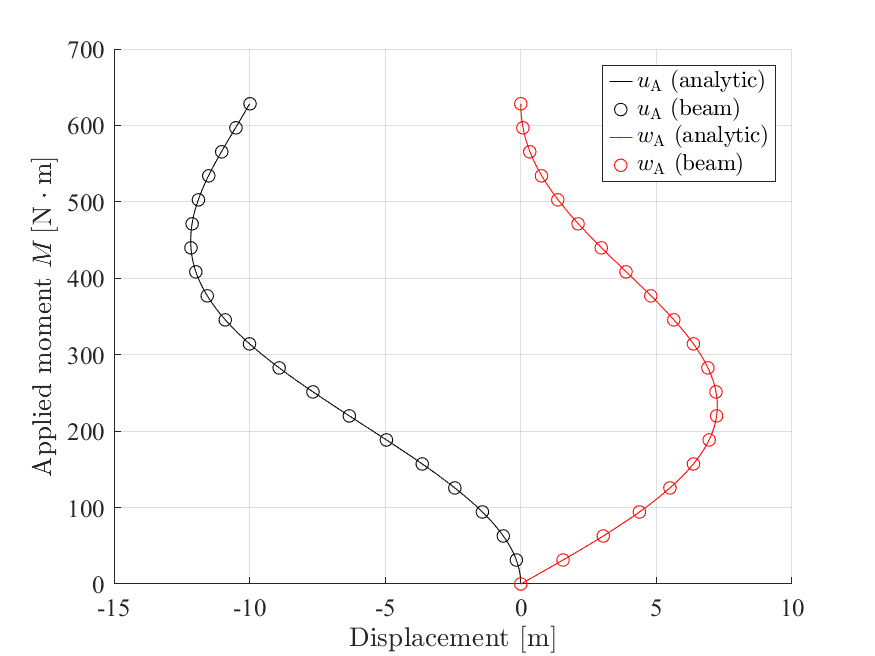}
		\caption{Initial cross-section height ${h}=0.1\rm{m}$}
		\label{pure_bend_compare_exact_h010}
	\end{subfigure}		
	\begin{subfigure}[b] {0.4875\textwidth} \centering
		\includegraphics[width=\linewidth]{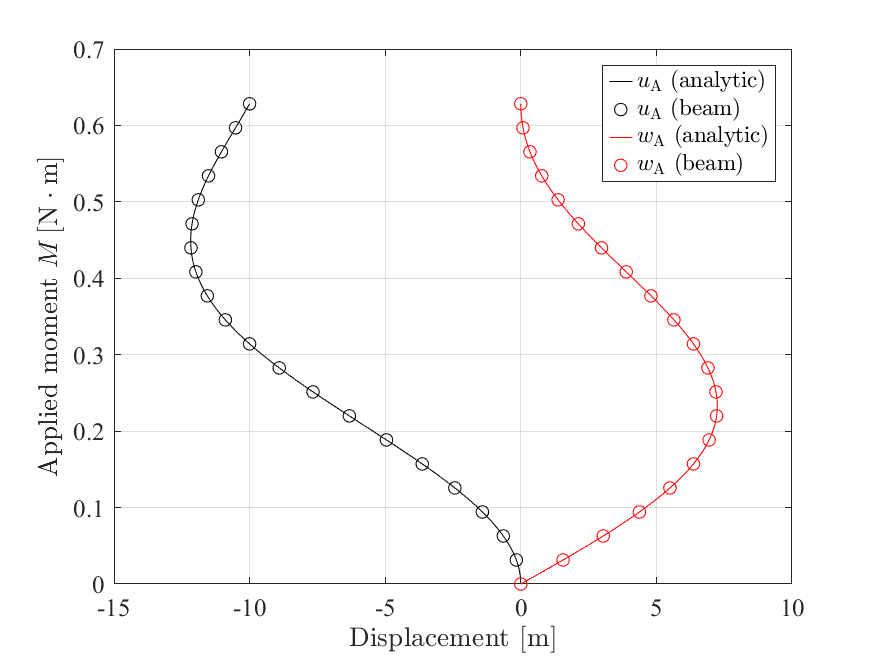}
		\caption{Initial cross-section height ${h}=0.01\rm{m}$}
		\label{pure_bend_compare_exact_h001}
	\end{subfigure}		
\caption{Cantilever beam under end moment: Comparison of the $X$- and $Z$-displacements at the tip of the central axis with the analytical solutions for the different initial cross-section heights. The beam solutions are calculated by IGA with $p=4$ and $n_{\mathrm{el}}=160$.}
\label{cant_beam_end_moment_compare_exact_sol}	
\end{figure}

Figs.\,\ref{pure_bend_deform_h010} and \ref{pure_bend_deform_h001} show the deformed configurations of the cantilever for initial heights ${h}=0.1\text{m}$ and ${h}=0.01\text{m}$, respectively. The external load is incrementally applied in 20 uniform steps. The final deformed configurations are very close to circles, but are not \textit{exact} circles. As Fig.\,\ref{cant_beam_end_moment_compare_exact_sol} shows, the $X$- and $Z$-displacements at the end are in very good agreement with the analytic solution of Eq.\,(\ref{beam_end_mnt_exact_sol}). However, it turns out that a slight difference persists even in the converged solutions. This difference in the converged solution is attributed to the fact that axial strain in the central axis and the transverse normal strain in the cross-section are induced by the bending deformation, which is not considered in the analytical solution under the pure bending assumption. 
\subsubsection{Coupling between bending and axial strains}
\label{ex_end_mnt_subsub_axial}
\begin{figure}[htp]
\centering
\includegraphics[width=0.65\linewidth]{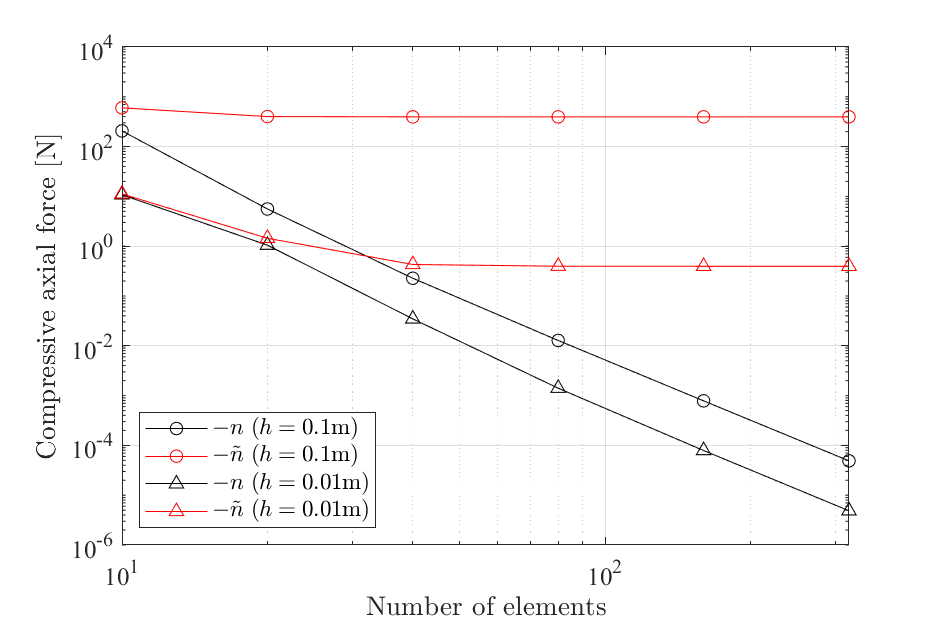}
\caption{Cantilever beam under end moment: Convergence of axial stress resultant $n$ and effective axial stress resultant $\tilde n$ for the two different cross-section heights $h$. {IGA with $p=4$ is used}. As expected, $n$ vanishes, while $\tilde n$ approaches a constant. The applied bending moment is $M=2{\pi}EI/L$. {The converged values of $-\tilde n$ at $n_\mathrm{el}=320$ are $395.7\mathrm{N}$ and $0.3948\mathrm{N}$ for the cases of $h=0.1\mathrm{m}$ and $0.01\mathrm{m}$, respectively, which is consistent with the theoretical estimation of convergence rate of ${\tilde n}\sim{h^3}$ discussed below Eq.\,(\ref{theo_est_axial_strs_res}).}}
\label{pure_bend_plot_n_ntilde}	
\end{figure}

\begin{figure}[htp]
	\centering
	\begin{subfigure}[b] {0.465\textwidth} \centering
		\includegraphics[width=\linewidth]{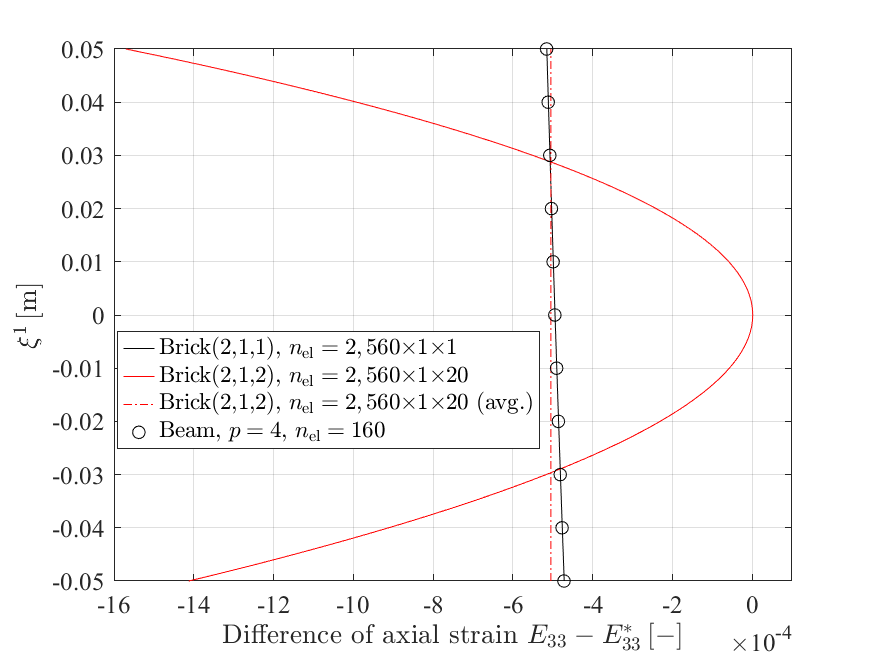}
		\caption{Initial cross-section height ${h}=0.1\rm{m}$}
		\label{dist_GL_strn_axial_diff_h010}
	\end{subfigure}		
	\begin{subfigure}[b] {0.465\textwidth} \centering
		\includegraphics[width=\linewidth]{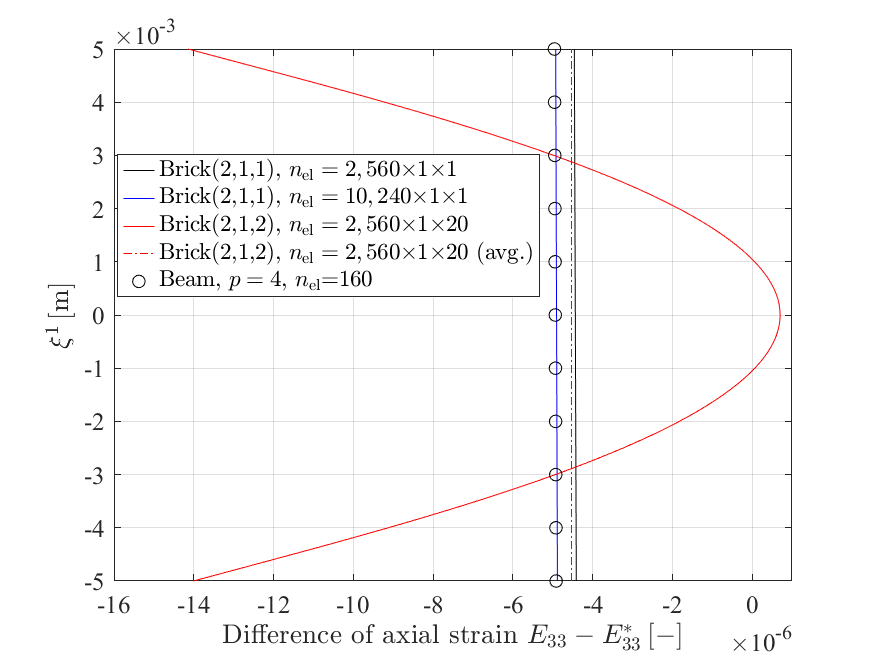}
		\caption{Initial cross-section height ${h}=0.01\rm{m}$}
		\label{dist_GL_strn_axial_diff_h001}
	\end{subfigure}		
\caption{Cantilever beam under end moment: Difference of the axial strain component along the cross-section height at the loaded end ($s=L$), and the applied moment $M=2{\pi}EI/L$. `avg.' represents the average. Note that, in the solid red line of (a), $E_{33}=1.5580\times{10^{-6}}$ at ${\xi^1}=0$, i.e., the central axis is slightly stretched.}
\label{dist_GL_strn_axial_diff}	
\end{figure}

\begin{figure}[htp]
	\centering
	\begin{subfigure}[b] {0.4875\textwidth} \centering
		\includegraphics[width=\linewidth]{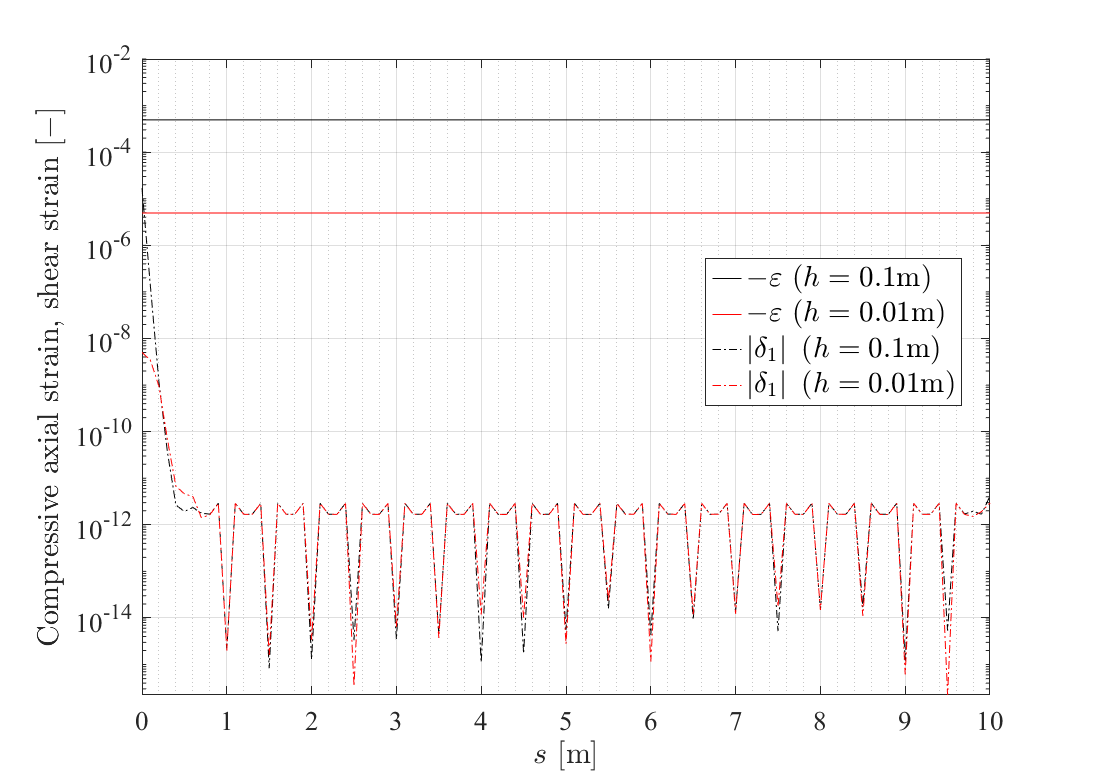}
		\caption{Strain components $\varepsilon$ and $\delta_1$}
		\label{ex_end_mnt_e_field}
	\end{subfigure}
	\begin{subfigure}[b] {0.4875\textwidth} \centering
		\includegraphics[width=\linewidth]{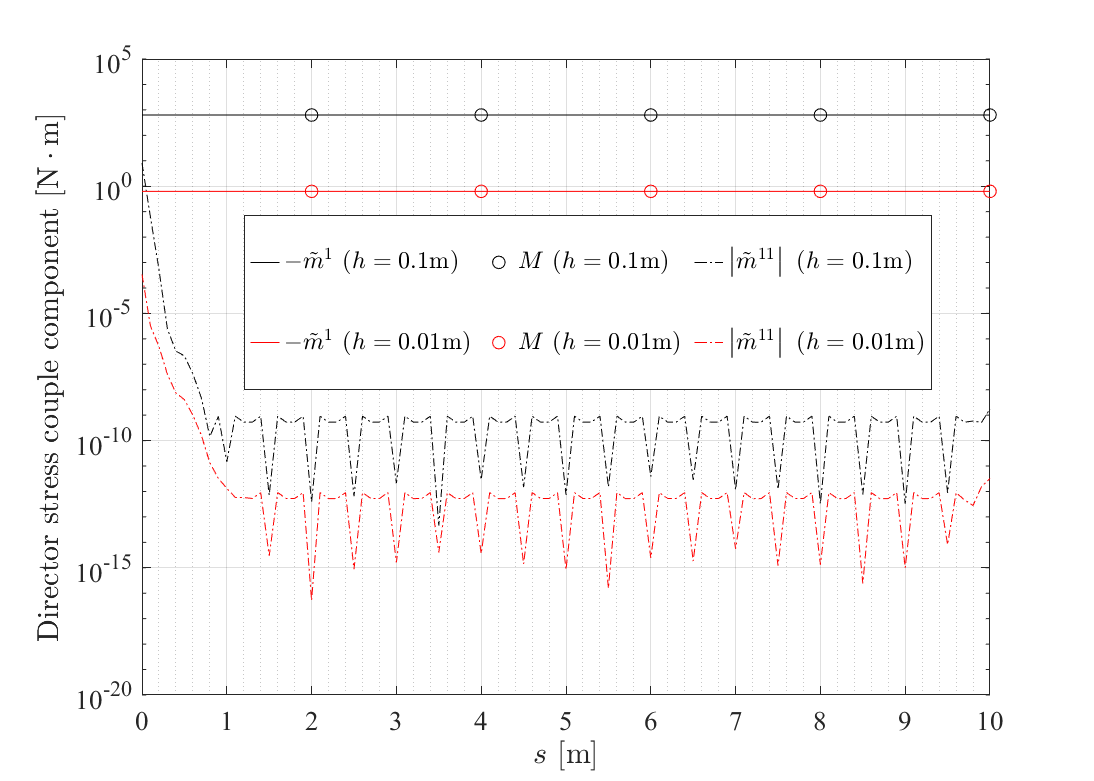}
		\caption{Director stress couples ${\tilde m}^1$ and ${\tilde m}^{11}$}
		\label{ex_end_mnt_m_field}
	\end{subfigure}			
\caption{{Cantilever beam under end moment: Distribution of the axial strain ($\varepsilon$), transverse shear strain ($\delta_1$) and director stress couple components (${\tilde m}^{11}$ and ${\tilde m}^{1}$) of the beam along the central axis. The results are from IGA with $p=4$, and $n_{\mathrm{el}}=320$.}}
\label{dist_e_m_field}
\end{figure}

\begin{figure}[htp]	
	\centering
	\begin{subfigure}[b] {0.475\textwidth} \centering
		\includegraphics[width=\linewidth]{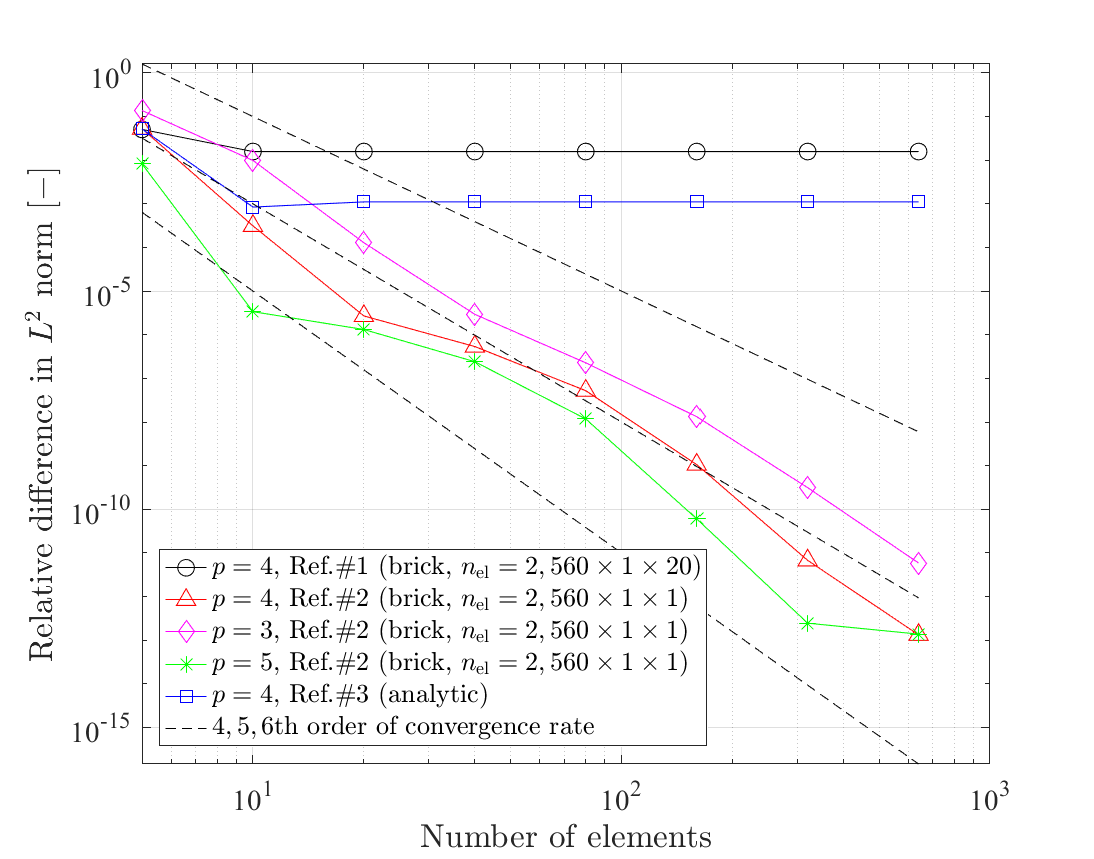}
		\caption{Initial cross-section height ${h}=0.1\rm{m}$}
		\label{pure_bend_conv_test_ux}
	\end{subfigure}\hspace{2.5mm}			
	\begin{subfigure}[b] {0.475\textwidth} \centering
		\includegraphics[width=\linewidth]{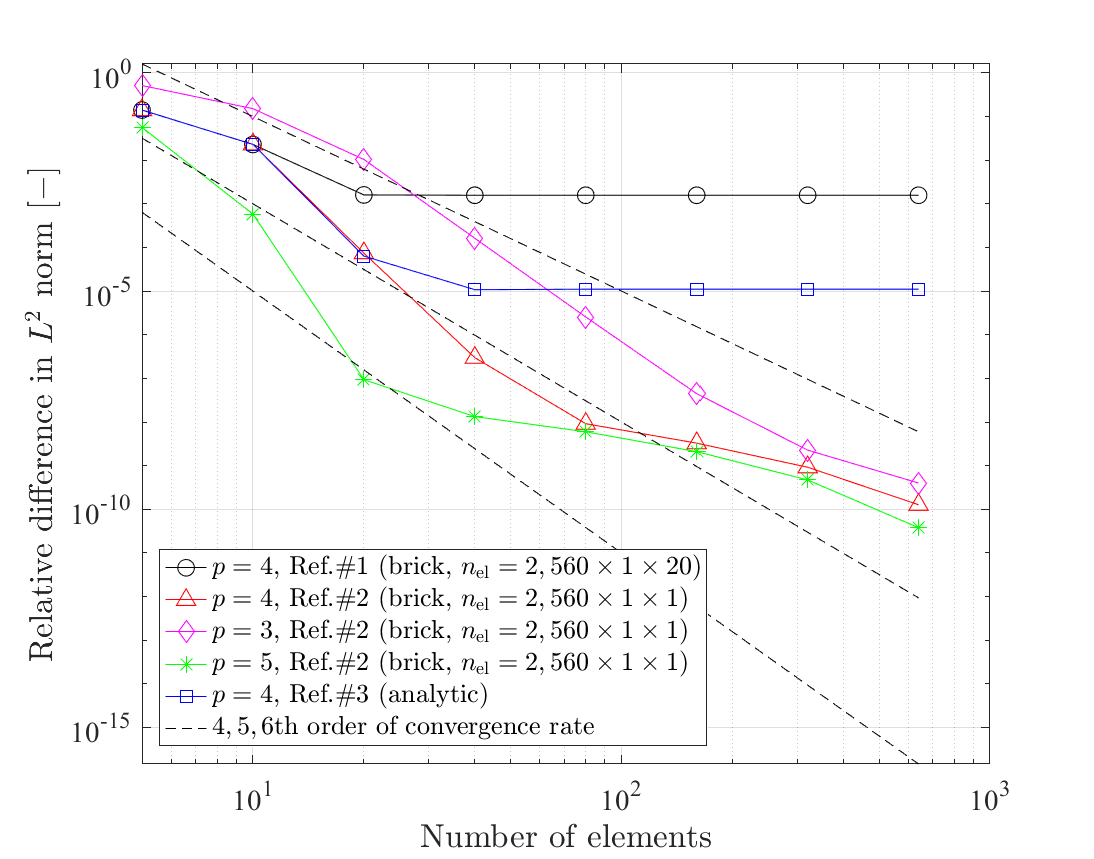}
		\caption{Initial cross-section height ${h}=0.01\rm{m}$}
		\label{pure_bend_conv_test_ux}
	\end{subfigure}			
\caption{{Cantilever beam under end moment: Convergence of the relative difference of the $X$-displacement in the central axis. The applied bending moment is $M=2{\pi}EI/L$. The dashed lines represent the theoretical convergence rate of ${\bar h}^{p+1}$, where $\bar h$ denotes the element size.}}
\label{cant_beam_end_moment_conv_rate}	
\end{figure}

The axial strain ($\varepsilon$) is not zero, but decreases with $h$. To verify this, we show that the effective stress resultant $\tilde n$, which is work conjugate to the axial strain $\varepsilon$ (see Eq.\,(\ref{tot_strn_energy_beam_time_deriv})), is not zero. From Eq.\,(\ref{nbdc_m_td_end_mnt}), we obtain ${{\tilde m}}^{1}=-M/\left({\boldsymbol{\varphi}_{\!,s}\cdot{\boldsymbol{d}_3}}\right)$ by using Eq.\,(\ref{strs_res_dir_mnt}) and the relations ${\boldsymbol{d}_3}\cdot{\boldsymbol{d}_{\alpha}}=0$ and ${\boldsymbol{\varphi}_{\!,s}}\cdot{\boldsymbol{d}_3}>0$ (postulation of Eq.\,(\ref{beam_th_str_calc_d3_vec})). From Eq.\,(\ref{nbdc_n_end_mnt}), it follows that $n={\boldsymbol{n}}\cdot{\boldsymbol{d}_3}/\left({{\boldsymbol{\varphi}_{\!,s}}\cdot{\boldsymbol{d}_3}}\right)$, obtained by using Eq.\,\ref{strs_res_f}, vanishes $\mathrm{at}\,\,s\in{\Gamma_\mathrm{N}}$. Therefore, using Eq.\,(\ref{eff_axial_res}), we obtain the effective axial stress resultant
\begin{equation}
\label{theo_est_axial_strs_res}
\tilde n =  - {\tilde m^1}{k_1},
\end{equation}
where the current bending curvature is $k_1\approx{1/R}$. Thus, $\tilde n$ does not vanish $\mathrm{at}\,\,s\in{\Gamma_\mathrm{N}}$. This is a high order effect of beam theory that disappears quickly for {decreasing $h$: $\tilde n$ decreases with the initial cross-section height $h$ due to ${\tilde m^1}\sim{M}\sim{h^3}$, i.e., ${\tilde n}\sim{h^3}$. Therefore, since the cross-sectional area is proportional to $h$, the work conjugate axial strain is $\varepsilon \sim {h^2}$. Then, the membrane strain energy is
\begin{equation}
\label{theo_memb_strn_e}
{\Pi _\varepsilon } \coloneqq \int_0^L {\tilde n\varepsilon {\rm{d}}s}\sim{h^5}.
\end{equation}
Further, for the given end moment $M\sim{h^3}$, the bending strain $\rho_1$ is nearly constant with respect to $h$, then the bending strain energy is
\begin{equation}
\label{theo_bend_strn_e}
{\Pi _\rho} \coloneqq \int_0^L {{{\tilde m}^1}{\rho_1}{\rm{d}}s}\sim{h^3}.
\end{equation}
}Fig.\,\ref{pure_bend_plot_n_ntilde} shows the convergence of axial stress resultant $n$ and the effective stress resultant $\tilde n$ with the mesh refinement in the beam. We calculate $\boldsymbol{n}$ using Eq.\,(\ref{beam_th_str_def_res_force}), from which we can extract $n$. It is observed that the condition of vanishing $n$ is weakly satisfied. 
We compare the axial strain field on the loaded end face in the presented beam formulation with the following three different reference solutions.
\begin{itemize}
\item {Ref.\#1}: IGA with ${n_\mathrm{el}}=2,560\times1\times20$ brick elements and $\mathrm{deg.}=(2,1,2)$. One element along the beam width is sufficient since $\nu=0$.
\item {Ref.\#2}: IGA with ${{n_\mathrm{el}}}=2,560\times1\times1$ brick elements and $\mathrm{deg.}=(2,1,1)$. {In the calculation of the relative difference of the displacement in the $L^2$ norm in Fig.\,\ref{cant_beam_end_moment_conv_rate}, we use IGA with ${{n_\mathrm{el}}}=2,560\times1\times1$ brick elements and $\mathrm{deg.}=(4,1,1)$ in order to obtain the convergence of the difference to machine precision. {It is noted that three Gauss integration points are used in the direction of cross-section height for brick and beam element solutions.}}
\item {Ref.\#3}: The analytic solution under the pure bending assumption.
\end{itemize}
In the reference solution using brick elements, we apply the end moment in the same way as in the beam formulation, that is, we apply the distributed follower load on the end face. In the following, we derive the analytical solution of the axial strain under the pure bending assumption (Ref.\,\#3). In pure bending, every material fiber deforms into a circle and is being stretched in the axial direction, where the amount of stretch linearly varies through the cross-section height. If the central axis deforms into a full circle with radius $R=L/{2\pi}$, we have the following expression for the axial stretch
\begin{align}
\label{ex_end_mnt_analytic_axial_stretch}
U_{33}^* = \frac{\ell}{L} = \frac{2\pi\left(R-{{\xi}^1}\right)}{L} = 1 - \frac{{2\pi }}{L}{\xi ^1},\,\,{\xi ^1} \in \left[ { - h/2,h/2} \right],
\end{align}
where $\ell$ denotes the current length of each material fiber. Then, the axial component of the Green-Lagrange strain is obtained by
\begin{align}
\label{ex_end_mnt_axial_comp_GL}
E_{33}^* = \frac{1}{2}\left\{ {{{\left( {1 - \frac{{2\pi }}{L}{\xi ^1}} \right)}^2} - 1} \right\} =  - 2\pi \frac{{{\xi ^1}}}{L} + 2{\pi ^2}{\left( {\frac{{{\xi ^1}}}{L}} \right)^2},\,\,{\xi ^1} \in \left[ { - h/2,h/2} \right].
\end{align}
In this analytical expression, it should be noted that the axial strain is zero at the central axis ${\xi^1}=0$. Since the cross-section height $h$ is much smaller than the beam length $L$, the quadratic order term in Eq.\,(\ref{ex_end_mnt_axial_comp_GL}) almost vanishes, so that the axial strain has nearly linear distribution along the coordinate $\xi^1$ (see Fig.\,\ref{pure_bend_analytic_graph}). Fig.\,\ref{dist_GL_strn_axial_diff} shows the differences between $E^*_{33}$ and the axial strains of reference solutions Ref.\#1, Ref.\#2, and the presented beam formulation. It is noticeable that the axial strain is nonzero in the results using brick elements as well. The beam solution agrees very well with that of Ref.\#2, since the Ref.\#2 also assumes a linear displacement field along the cross-section height. Especially, in case of ${h}=0.01\mathrm{m}$, it is observed that as we increase the number of elements along the central axis, the reference solution (${{n_\mathrm{el}}}=10,240\times1\times1$) approaches the beam solution. The solution of Ref.\#1 shows that the cross-section does not remain plane but undergoes \textit{warping}. Therefore there are large differences in the axial strain between Ref.\#1 and the beam solution; however, it is remarkable that the average of the solution in Ref.\#1 still agrees very well with the beam solution. Fig.\,\ref{ex_end_mnt_e_field} shows that the axial strain of the beam is nearly constant along the central axis, and decreases with the initial cross-section height $h$. Also, the shear strain is negligible, which is consistent with $-{\tilde m}^1\approx{M}$, shown in Fig.\,\ref{ex_end_mnt_m_field}. The slight shear strain near the clamped boundary is associated with the drastic change of current cross-section height there. At the clamped boundary, the cross-section does not deform. Thus, the gradient $\boldsymbol{d}_{1,s}$ does not vanish, i.e., $k^1_1\ne0$ (see Remark \ref{remark_curv_k}), which generates the effective shear stress ${\tilde q}^1$of Eq.\,(\ref{eff_trans_shear_res}). Similarly, the gradient $\boldsymbol{d}_{1,s}$ at the clamped boundary generates the nonvanishing strain $\gamma_{11}$, and its work conjugate $\tilde{m}^{11}$ (see Fig.\,\ref{ex_end_mnt_m_field}). However, $\tilde{m}^{11}$ is almost zero elsewhere in the domain, and this means that the current cross-section height is almost uniform (see Remark \ref{remark_bending_mnt} for the relavant explanation). 
\subsubsection{Coupling between bending and through-the-thickness stretch}
\label{ex_end_mnt_subsub_th}
The through-the-thickness stretch ${\chi}^{11}$ is also coupled with the bending deformation, and decreases quickly with the initial cross-section height $h$. In the absence of an external director stress couple, $\bar{\tilde {\boldsymbol{m}}}^{\gamma}=\boldsymbol{0}$, substituting Eq.\,(\ref{strs_res_dir_mnt}) into Eq.\,(\ref{mnt_bal_eqn_dir_mnt}), and using the fact that torsional deformation is absent, i.e., ${\tilde m}^{21}=0$, we obtain
\begin{align}
\label{ex_end_mnt_l1_vector}
{{\boldsymbol{l}}^1} = {\tilde m^1_{,s}}{{\boldsymbol{\varphi }}_{\!,s}} + {\tilde m^1}{{\boldsymbol{\varphi }}_{\!,ss}} + {\tilde m^{11}_{,s}}{{\boldsymbol{d}}_1} + {\tilde m^{11}}{{\boldsymbol{d}}_{1,s}}\approx{\tilde m^1}{{\boldsymbol{\varphi }}_{\!,ss}},
\end{align}
since ${\tilde m}^1$ is nearly constant, and ${\tilde m}^{11}$ is negligible in the domain $s\in(0,L)$. Let $\tilde s$ be the arc-length coordinate along the current central axis. Then, $\boldsymbol{\varphi}_{\!,{\tilde s}{\tilde s}}$ represents the curvature vector such that $\kappa  \coloneqq \left\| {{{\boldsymbol{\varphi }}_{\!,\tilde s\tilde s}}} \right\|$ denotes the curvature of the deformed central axis, which is given by $\kappa\approx{1/R}$ in the example. Using the relation $\mathrm{d}{\tilde s}=\sqrt{1+{2\varepsilon}}\mathrm{d}{s}$ and the chain rule of differentiation, we find
\begin{align}
\label{ex_end_mnt_2nd_mnt_caxis}
{{\boldsymbol{\varphi }}_{\!,ss}} = \frac{\varepsilon _{,s}}{\sqrt{1+2\varepsilon}}{{\boldsymbol{\varphi }}_{\!,\tilde s}} + {(1 + 2\varepsilon )}{{\boldsymbol{\varphi }}_{\!,\tilde s\tilde s}} \approx {\frac{1}{{\lambda_1}R}}{(1 + 2\varepsilon )}{{\boldsymbol{d}}_1},
\end{align}
since $\varepsilon$ is nearly constant, and the shear deformation is negligible such that the unit normal vector of the central axis is approximated by $\boldsymbol{\varphi}_{\!,{\tilde s}{\tilde s}}/\kappa={\boldsymbol{d}_1}/{\lambda_1}$. Substituting Eq.\,(\ref{ex_end_mnt_2nd_mnt_caxis}) into Eq.\,(\ref{ex_end_mnt_l1_vector}) and using the decomposition of Eq.\,(\ref{strs_res_dir_f}), we obtain
\begin{equation}
{{\tilde l}^{11}} \approx \frac{1}{{\lambda_1}{R}}{(1 + 2\varepsilon )}{\tilde m^1}.
\end{equation}
This means that the transverse normal stress ${\tilde l}^{11}$ does not vanish, but decreases with the initial cross-section height $h$ through the relation ${\tilde m}^1\sim{h^3}$, i.e., ${\tilde l}^{11}\sim{h^3}$. {Therefore, since the cross-sectional area is proportional to $h$, the work conjugate strain is ${\chi}_{11}\sim{h^2}$, and the in-plane strain energy of the cross-section is
\begin{align}
\label{the_strn_e_inp_cs}
{\Pi _\chi } \coloneqq \int_0^L {{{\tilde l}^{11}}{\chi _{11}}{\rm{d}}s}\sim{h^5}.
\end{align} 
}

\begin{figure}[htp]	
	\centering
	\begin{subfigure}[b] {0.44\textwidth} \centering
		\includegraphics[width=\linewidth]{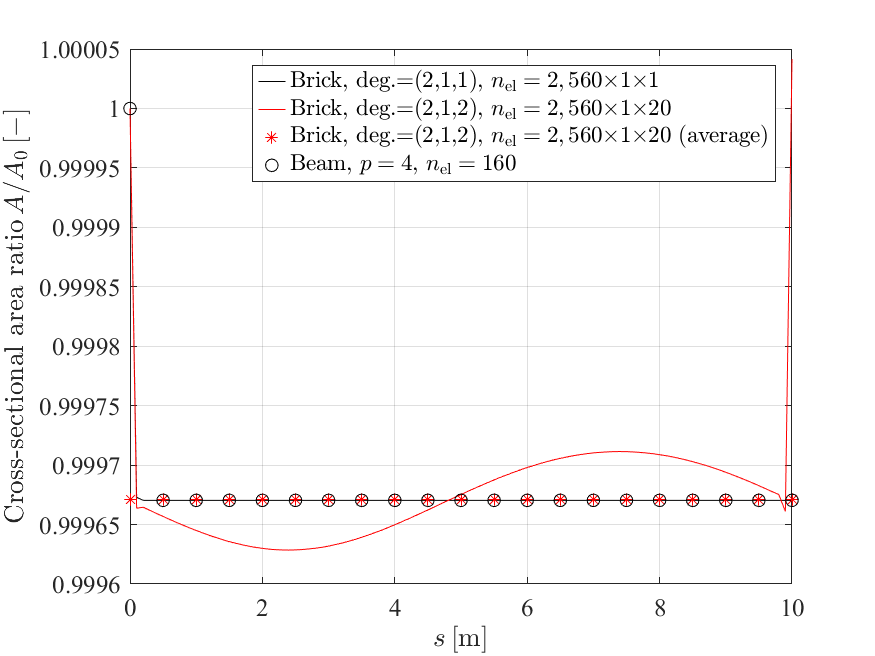}
		\caption{Initial cross-section height ${h}=0.1\rm{m}$}
		\label{pure_bend_carea_h010}
	\end{subfigure}		
	\begin{subfigure}[b] {0.44\textwidth} \centering
		\includegraphics[width=\linewidth]{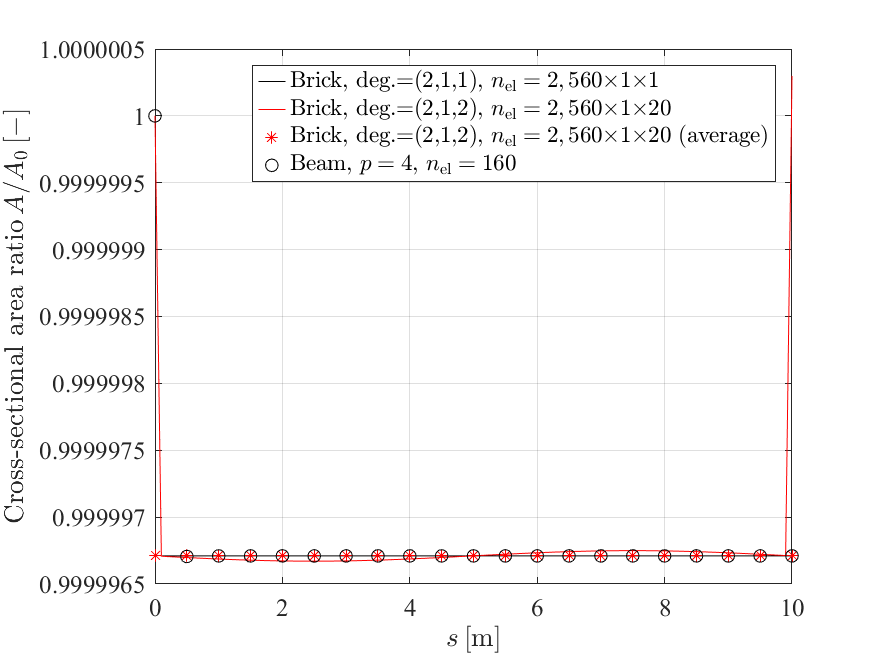}
		\caption{Initial cross-section height ${h}=0.01\rm{m}$}
		\label{pure_bend_carea_h001}
	\end{subfigure}			
	\caption{Cantilever beam under end moment: Distribution of the current cross-sectional area along the central axis. `average' denotes the average in the whole domain of the central axis where the two boundary points are excluded. The applied bending moment is $M=2{\pi}EI/L$.}
	\label{dist_GL_strn_th}
\end{figure}

\begin{figure}[htp]	
	\centering
	\begin{subfigure}[b] {0.44\textwidth} \centering
		\includegraphics[width=\linewidth]{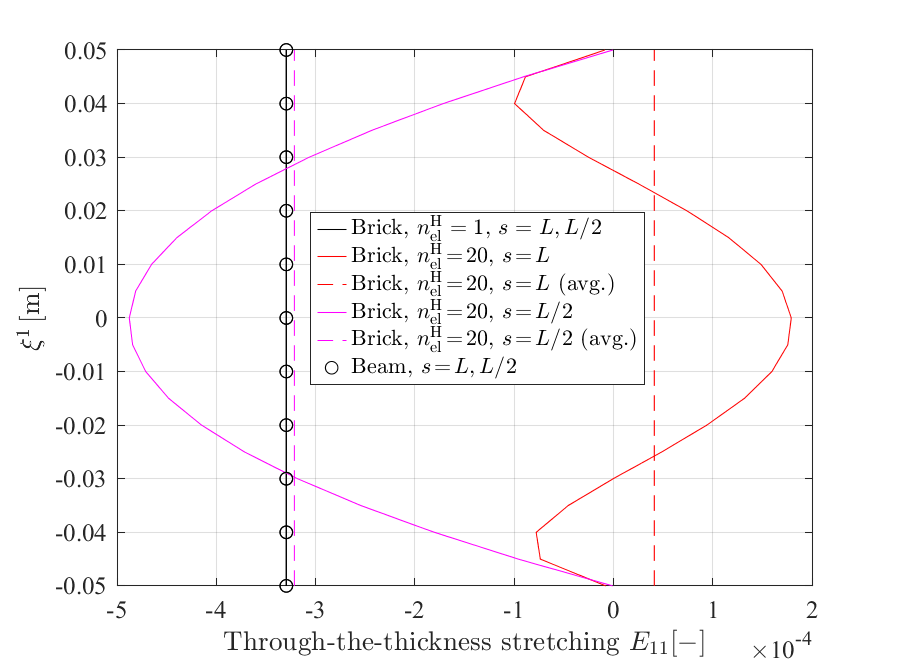}
		\caption{Initial cross-section height ${h}=0.1\rm{m}$}
		\label{pure_bend_E11_dist_thick_h010}
	\end{subfigure}		
	\begin{subfigure}[b] {0.44\textwidth} \centering
		\includegraphics[width=\linewidth]{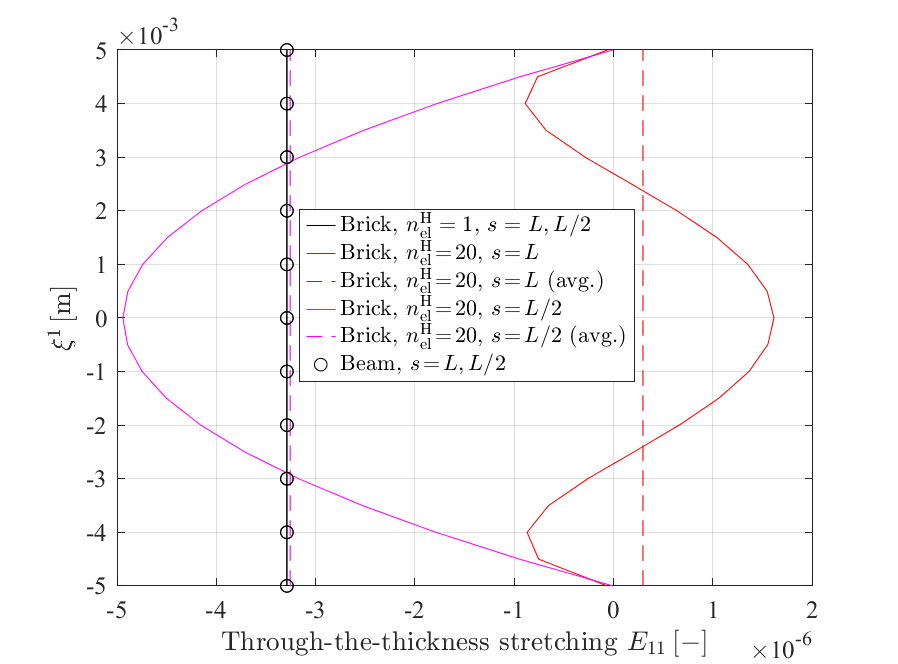}
		\caption{Initial cross-section height ${h}=0.01\rm{m}$}
		\label{pure_bend_E11_dist_thick_h001}
	\end{subfigure}			
\caption{Cantilever beam under end moment: Distribution of the transverse normal (through-the-thickness) component of the Green-Lagrange strain along the cross-section height at the loaded end face. `avg.' denotes the average, and $n_\mathrm{el}^{\mathrm{H}}$ denotes the number of brick elements in the direction of cross-section height. For brick element, $\mathrm{deg.}=(2,1,1)$ for $n_{\mathrm{el}}^{\mathrm{H}}=1$, and $\mathrm{deg.}=(2,1,2)$ in the other cases. For beam element, $p=4$, and $n_{\mathrm{el}}=160$. The applied bending moment is $M=2{\pi}EI/L$.}
\label{dist_GL_through_height_strn_th}
\end{figure}

Fig.\,\ref{dist_GL_strn_th} compares the change of cross-sectional area along the axis for the beam and the reference solutions. It is noticeable that the cross-sectional area also decreases when using brick elements. The cross-sectional area in the beam solution agrees very well with that of Ref.\,\#2, since both assume constant transverse normal (through-the-thickness) strain of the cross-section (see also Fig.\,\ref{dist_GL_through_height_strn_th}). Also, Fig.\,\ref{dist_GL_strn_th} shows that the amount of change in cross-sectional area decreases with $h$. The deformation of the cross-section in Ref.\,\#1 is more complicated than for the other cases, since it allows for warping, i.e., the cross-section does not remain plane after deformation. Especially, at the loaded end face, the cross-sectional area slightly increases, since the central axis is stretched. It is shown in Fig.\,\ref{dist_GL_through_height_strn_th} (red curves) that the cross-section is stretched along the transverse direction at the center (${\xi^1}=0$, $s=L$), so that the average of the transverse normal strain is positive, i.e., the cross-section is stretched in average. On the other hand, apart from the boundary, the through-the-thickness compressive force coupled with the bending deformation is dominant, so that the cross-sectional area decreases. In Fig.\,\ref{dist_GL_strn_th}, it is remarkable that the average cross-sectional area of Ref.\,\#1 in the domain $s\in\left(0,L\right)$ coincides with that of the beam and Ref.\,\#2. Further, in Fig.\,\ref{dist_GL_through_height_strn_th}, the average of the transverse normal strain at the middle of the central axis ($s=L/2$) agrees very well with that of the beam and Ref.\,\#2.

\subsubsection{Verification of displacements}
Fig.\,\ref{cant_beam_end_moment_conv_rate} compares the relative difference of the $X$-displacement of the beam from the three different reference solutions, where the relative $L^2$ norm of the difference in the $X$-displacement $u$ in the domain of the central axis $(0,L)$ is calculated by
\begin{equation}\label{def_rel_l2_err}
{\left\| {{e_u}} \right\|_{{L^2}}} = \sqrt {\frac{{\int_0^L {{{\left( {u - {u_{{\rm{ref}}}}} \right)}^2}\,{\mathrm{d}}s } }}{{\int_0^L {{u_{{\rm{ref}}}}^2\,{\mathrm{d}}s } }}},
\end{equation}
where $u_\mathrm{ref}$ denotes the reference solution of the displacement component. The convergence test results of the reference solutions are given in Tables \ref{app_conv_test_xdisp_tip_h010} and \ref{app_conv_test_xdisp_tip_h001}. In Fig.\,\ref{cant_beam_end_moment_conv_rate}, Ref.\,\#2 shows the smallest differences from the beam solution. {The difference is even smaller than the analytical solution and vanishes to machine precision, since Ref.\,\#2 is kinematically the same as the beam formulation with Poisson's ratio $\nu=0$. Ref.\,\#1 shows the largest differences, but they are only around $1\%$ and $0.1\%$ in the cases of $h=0.1\mathrm{m}$ and $h=0.01\mathrm{m}$, respectively. We also compare the convergence rate in several different orders of basis function ($p=3,4,5$) with the asymptotic and optimal convergence rate of ${\bar h}^{p+1}$, where $\bar h$ denotes the element size. The beam solution shows comparable or even better rate of convergence than the optimal one, especially in the coarser level of mesh discretization.} 
\subsubsection{Instability in thin beam limit}
\label{instab_thin_b_lim_end_mnt}
Tables\,\ref{str_end_mnt_iter_history_h010} and \ref{str_end_mnt_iter_history_h001} compare the total number of load steps and iterations in the cases of $h=0.1\mathrm{m}$ and $h=0.01\mathrm{m}$, respectively. Ref.\,\#1 requires larger number of iterations than Ref.\#2 and the beam solution. This is mainly attributed to more complicated deformations of the cross-section. It is also shown that more iterations are required for the thinner cross-section case. It has been investigated in the shell formulation with extensible director \citep{simo1990stress} that the instability in the thin limit ($h\to0$) is associated with the coupling of bending and through-the-thickness stretching. A couple of methods to alleviate this instability has been presented, for example based on a multiplicative decomposition of the extensible director into an inextensible direction vector and a scalar stretch \citep{simo1990stress}, and based on the mass scaling in dynamic problems \citep{hokkanen2019isogeometric}. In this paper, we restrict our application of the developed beam formulation to low to moderate slenderness ratios, and further investigation on the alleviation of the instability remains future work.
\begin{table}[!htpt]
\caption{Cantilever beam under end moment: History of the Newton-Raphson iteration for $M=0.1nEI\pi L$ at step $n=20$, and the total number of load steps and iterations (initial cross-section height ${h}=0.1\mathrm{m}$).}
\label{str_end_mnt_iter_history_h010}
\centering
\scriptsize
\begin{tabular}{clcclcclcc}
\Xhline{3\arrayrulewidth}
\multirow{3}{*}{\begin{tabular}[c]{@{}c@{}}Iteration\\ number\\ at step\\ $n=20$\end{tabular}} &  & \multicolumn{5}{c}{Brick}                                                                                                                                                                                                                                                                                              & \multicolumn{1}{c}{} & \multicolumn{2}{c}{Beam}                                                                                                                                           \\ \cline{3-7} \cline{9-10} 
                                                                                                 &  & \multicolumn{2}{c}{\begin{tabular}[c]{@{}c@{}}IGA, deg.=(2,1,2),\\${{n_\mathrm{el}}}=2,560\times1\times20$\end{tabular}}                                           & \multicolumn{1}{c}{} & \multicolumn{2}{c}{\begin{tabular}[c]{@{}c@{}}IGA, deg.=(2,1,1),\\ ${{n_\mathrm{el}}}=2,560\times1\times1$\end{tabular}}                                                                                   & \multicolumn{1}{c}{} & \multicolumn{2}{c}{\begin{tabular}[c]{@{}c@{}}IGA, $p=4$,\\${{n_\mathrm{el}}}=160$\end{tabular}}                                                                                                                                       \\ \cline{3-4} \cline{6-7} \cline{9-10} 
                                                                                                 &  & \begin{tabular}[c]{@{}c@{}}Euclidean\\      norm of residual\end{tabular} & \begin{tabular}[c]{@{}c@{}}Energy\\      norm\end{tabular} & \multicolumn{1}{c}{} & \multicolumn{1}{c}{\begin{tabular}[c]{@{}c@{}}Euclidean\\      norm of residual\end{tabular}} & \multicolumn{1}{c}{\begin{tabular}[c]{@{}c@{}}Energy\\      norm\end{tabular}} & \multicolumn{1}{c}{} & \multicolumn{1}{c}{\begin{tabular}[c]{@{}c@{}}Euclidean\\      norm of residual\end{tabular}} & \multicolumn{1}{c}{\begin{tabular}[c]{@{}c@{}}Energy\\      norm\end{tabular}} \\
\Xhline{3\arrayrulewidth}
1   &  & 1.6E+02       & 1.1E+01         &     & 3.1E+02     & 9.8E+00       &        & 3.1E+01  & 9.8E+00                                                    \\
2   &  & 6.4E+04       & 1.5E+04         &     & 6.2E+04     & 1.1E+04       &        & 6.7E+04  & 1.1E+04                                                    \\
3   &  & 4.0E+03       & 6.6E+01         &     & 3.1E+03     & 2.7E+01       &        & 3.9E+03  & 2.7E+01                                                    \\
4   &  & 1.3E+03       & 8.0E+00         &     & 1.2E+01     & 1.4E-02       &        & 1.8E+01  & 1.4E-02                                                    \\
5   &  & 1.1E+03       & 5.8E+00         &     & 3.5E+01     & 3.6E-03       &        & 4.5E+01  & 3.6E-03                                                    \\
6   &  & 5.5E+02       & 1.3E+00         &     & 8.7E-01     & 2.0E-04       &        & 8.9E-01  & 2.0E-04                                                    \\
7   &  & 1.4E+01       & 5.6E-02         &     & 1.3E+00     & 5.8E-06       &        & 1.2E+00  & 5.8E-06                                                    \\
8   &  & 1.6E+02       & 1.1E-01         &     & 1.7E-03     & 8.7E-10       &        & 1.6E-03  & 8.7E-10                                                    \\
9   &  & 8.5E-01       & 5.9E-03         &     & 5.8E-06     & 1.2E-16       &        & 5.2E-06  & 1.2E-16                                                    \\
10  &  & 3.2E+01       & 4.3E-03         &     & 1.4E-06     & 4.3E-20       &        & 4.5E-08  & 1.3E-22                                                    \\
11  &  & 9.3E-03       & 3.2E-06         &     &             &               &        &          &                                                            \\
12  &  & 1.8E-02       & 1.4E-09         &     &             &               &        &          &                                                            \\
13  &  & 5.8E-07       & 4.9E-19         &     &             &               &        &          &                                                            \\ \cline{1-1} \cline{3-4} \cline{6-7} \cline{9-10} 
\#load steps                                                                                     &  & \multicolumn{2}{c}{20}                                                                                                     &                      & \multicolumn{2}{c}{20}                                                                                                     &                      & \multicolumn{2}{c}{20}                                                                                                     \\ \cline{1-1} \cline{3-4} \cline{6-7} \cline{9-10} 
\#iterations                                                                                     &  & \multicolumn{2}{c}{445}                                                                                                    &                      & \multicolumn{2}{c}{200}                                                                                                    &                      & \multicolumn{2}{c}{200}                                                                                                    \\
\Xhline{3\arrayrulewidth}
\end{tabular}
\end{table}

\begin{table}[!htpt]
\caption{Cantilever beam under end moment: History of the Newton-Raphson iteration for $M=0.1nEI\pi L$ at step $n=20$, and the total number of load steps and iterations (initial cross-section height ${h}=0.01\mathrm{m}$).}
\label{str_end_mnt_iter_history_h001}
\centering
\scriptsize
\begin{tabular}{clcclcclcc}
\Xhline{3\arrayrulewidth}
\multirow{3}{*}{\begin{tabular}[c]{@{}c@{}}Iteration\\ number\\ at step\\ $n=20$\end{tabular}} &  & \multicolumn{5}{c}{Brick}                                                                                                                                                                                                                                                                                              & \multicolumn{1}{c}{} & \multicolumn{2}{c}{Beam}                                                                                                                                           \\ \cline{3-7} \cline{9-10} 
                                                                                                 &  & \multicolumn{2}{c}{\begin{tabular}[c]{@{}c@{}}IGA, deg.=(2,1,2),\\${{n_\mathrm{el}}}=2,560\times1\times20$\end{tabular}}                                           & \multicolumn{1}{c}{} & \multicolumn{2}{c}{\begin{tabular}[c]{@{}c@{}}IGA, deg.=(2,1,1),\\${{n_\mathrm{el}}}=2,560\times1\times1$\end{tabular}}                                                                                   & \multicolumn{1}{c}{} & \multicolumn{2}{c}{\begin{tabular}[c]{@{}c@{}}IGA, $p=4$,\\${{n_\mathrm{el}}}=160$\end{tabular}}                                                                                                                                       \\ \cline{3-4} \cline{6-7} \cline{9-10} 
                                                                                                 &  & \begin{tabular}[c]{@{}c@{}}Euclidean\\      norm of residual\end{tabular} & \begin{tabular}[c]{@{}c@{}}Energy\\      norm\end{tabular} & \multicolumn{1}{c}{} & \multicolumn{1}{c}{\begin{tabular}[c]{@{}c@{}}Euclidean\\      norm of residual\end{tabular}} & \multicolumn{1}{c}{\begin{tabular}[c]{@{}c@{}}Energy\\      norm\end{tabular}} & \multicolumn{1}{c}{} & \multicolumn{1}{c}{\begin{tabular}[c]{@{}c@{}}Euclidean\\      norm of residual\end{tabular}} & \multicolumn{1}{c}{\begin{tabular}[c]{@{}c@{}}Energy\\      norm\end{tabular}} \\
\Xhline{3\arrayrulewidth}
1    &  & 1.6E+00    & 1.0E-02    &     & \multicolumn{1}{c}{3.1E+00}   & \multicolumn{1}{c}{9.9E-03}    &       & \multicolumn{1}{c}{3.1E-02}    & \multicolumn{1}{c}{9.9E-03}   \\
2    &  & 4.7E+04    & 8.5E+02    &     & \multicolumn{1}{c}{5.4E+04}   & \multicolumn{1}{c}{1.1E+03}    &       & \multicolumn{1}{c}{6.7E+03}    & \multicolumn{1}{c}{1.1E+03}   \\
3    &  & 2.4E+03    & 2.4E+00    &     & \multicolumn{1}{c}{2.6E+03}   & \multicolumn{1}{c}{2.7E+00}    &       & \multicolumn{1}{c}{4.0E+02}    & \multicolumn{1}{c}{2.8E+00}   \\
4    &  & 1.6E+02    & 1.2E-02    &     & \multicolumn{1}{c}{8.7E+00}   & \multicolumn{1}{c}{3.8E-05}    &       & \multicolumn{1}{c}{1.8E+00}    & \multicolumn{1}{c}{3.8E-05}   \\
     &  & {\rvdots}  & 			     &     & {\rvdots}								&     										 &       & {\rvdots}								  &    											\\
10   &  & 7.0E+01    & 2.1E-03    &     & \multicolumn{1}{c}{1.8E+00}   & \multicolumn{1}{c}{1.4E-06}    &       & \multicolumn{1}{c}{1.6E-01}    & \multicolumn{1}{c}{1.2E-06}                                                    \\
11   &  & 3.7E+01    & 6.8E-04    &     & \multicolumn{1}{c}{7.3E-05}   & \multicolumn{1}{c}{5.5E-10}    &       & \multicolumn{1}{c}{3.6E-06}    & \multicolumn{1}{c}{4.8E-10}                                                    \\
12   &  & 8.8E+01    & 3.3E-03    &     & \multicolumn{1}{c}{3.2E-03}   & \multicolumn{1}{c}{4.4E-12}    &       & \multicolumn{1}{c}{2.6E-04}    & \multicolumn{1}{c}{3.4E-12}                                                    \\
13   &  & 1.0E+01    & 7.0E-05    &     & \multicolumn{1}{c}{1.7E-07}   & \multicolumn{1}{c}{1.6E-20}    &       & \multicolumn{1}{c}{6.5E-09}    & \multicolumn{1}{c}{5.9E-21}                                                    \\
     &  & {\rvdots}  &     		  &     &                               &                                &       &                                &                                                                                \\
29   &  & 1.6E+01    & 1.1E-04    &     &                               &                                &       &                                &                                                                                \\
30   &  & 2.1E-04    & 2.2E-09    &     &                               &                                &       &                                &                                                                                \\
31   &  & 1.1E-02    & 5.5E-11    &     &                               &                                &       &                                &                                                                                \\
32   &  & 3.9E-06    & 6.2E-18    &     &                               &                                &       &                                &                                                                                \\ \cline{1-1} \cline{3-4} \cline{6-7} \cline{9-10} 
\multicolumn{1}{l}{\#load steps}                                                                 &  & \multicolumn{2}{c}{20}                                                                                                     &                      & \multicolumn{2}{c}{20}                                                                                                                                             &                      & \multicolumn{2}{c}{20}                                                                                                                                             \\
\multicolumn{1}{l}{\#iterations}                                                                 &  & \multicolumn{2}{c}{787}                                                                                                    &                      & \multicolumn{2}{c}{260}                                                                                                                                            &                      & \multicolumn{2}{c}{260}                                                                                                                                            \\ 
\Xhline{3\arrayrulewidth}
\end{tabular}
\end{table}
\subsubsection{{Alleviation of membrane, transverse shear, and curvature-thickness locking}}
\label{ex_beam_end_mnt_allev_lock}
\begin{figure}[htp]
	\centering
	\includegraphics[width=0.5\linewidth]{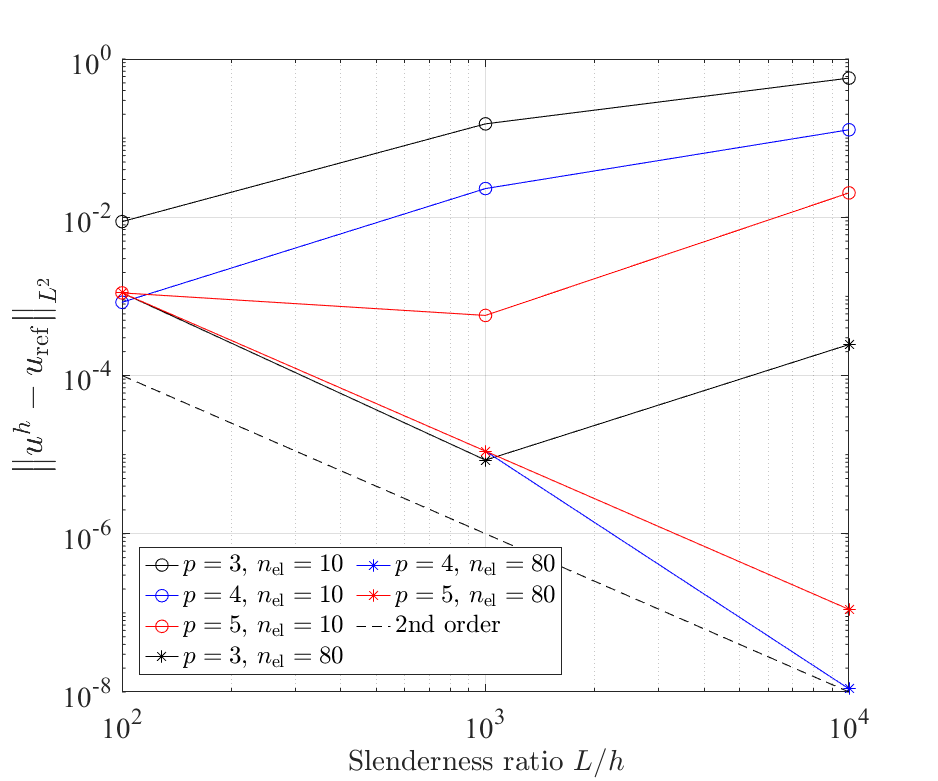}
\caption{{Cantilever beam under end moment: Change of the relative difference of the $X-$displacement in the $L^2$ norm (w.r.t. Ref.\#3) with increasing slenderness ratio. The dashed line represents the theoretically estimated convergence rate of order 2, which agrees very well with the solution of using $p=5$ and $n_\mathrm{el}=80$.}}
\label{cant_beam_end_moment_sl_ratio_l2_error_xdisp}	
\end{figure}

\begin{figure*}[!htbp]	
	\centering
	\begin{subfigure}[b] {0.5\textwidth} \centering
		\includegraphics[width=\linewidth]{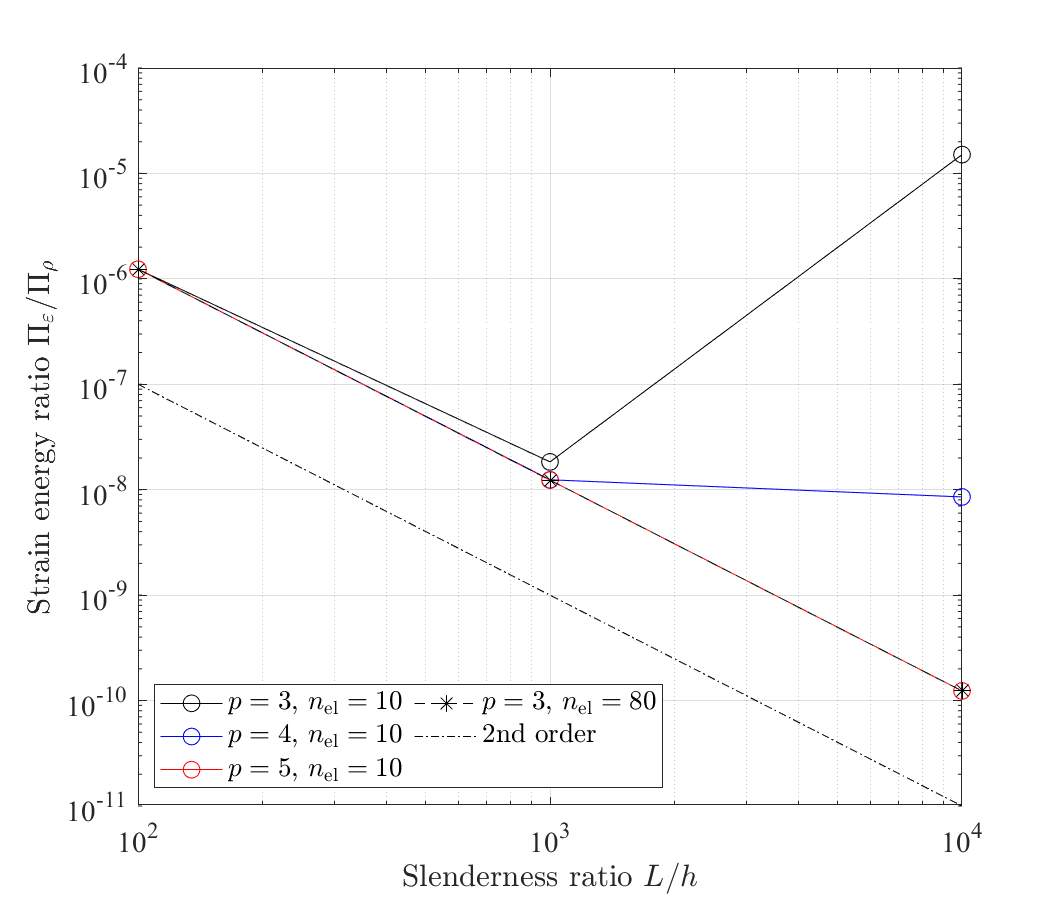}
		\caption{Membrane (axial) strain energy}
		\label{end_mnt_strn_e_memb}
	\end{subfigure}		
	\begin{subfigure}[b] {0.5\textwidth} \centering
		\includegraphics[width=\linewidth]{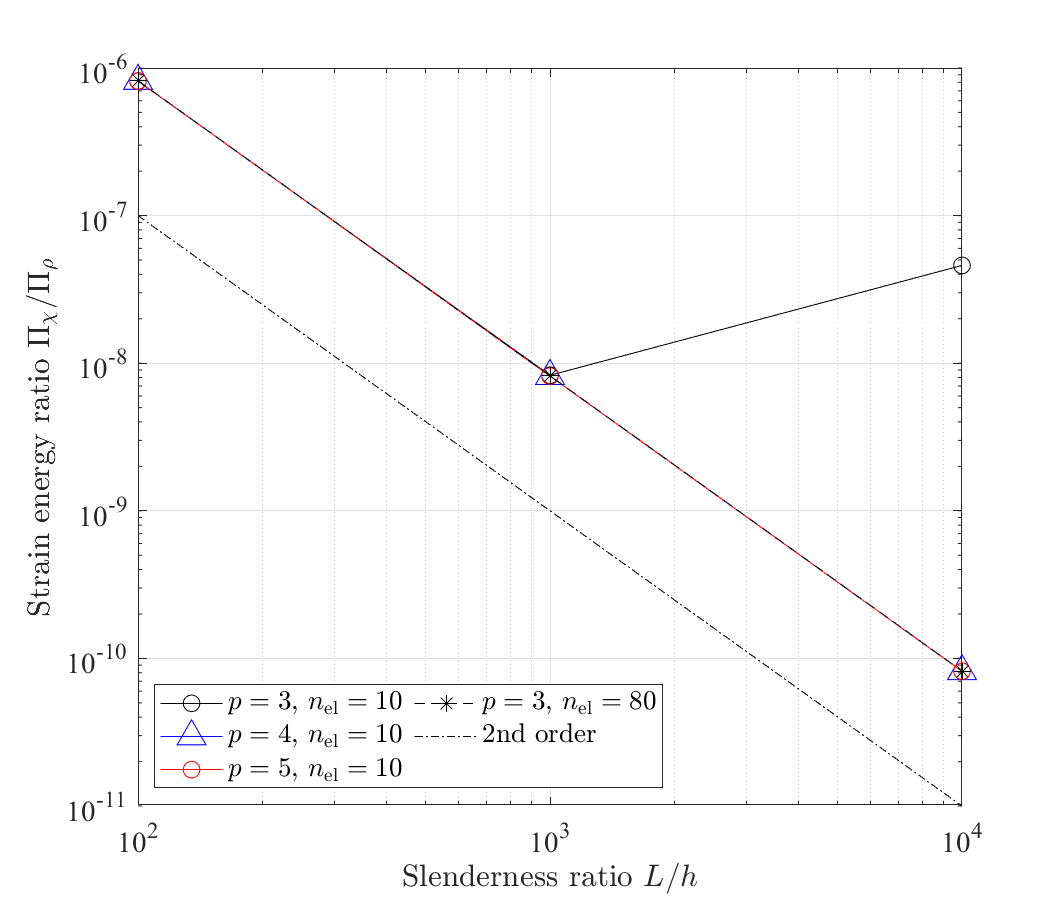}
		\caption{In-plane cross-section strain energy}
		\label{end_mnt_strn_e_inp}
	\end{subfigure}		
	\begin{subfigure}[b] {0.5\textwidth} \centering
		\includegraphics[width=\linewidth]{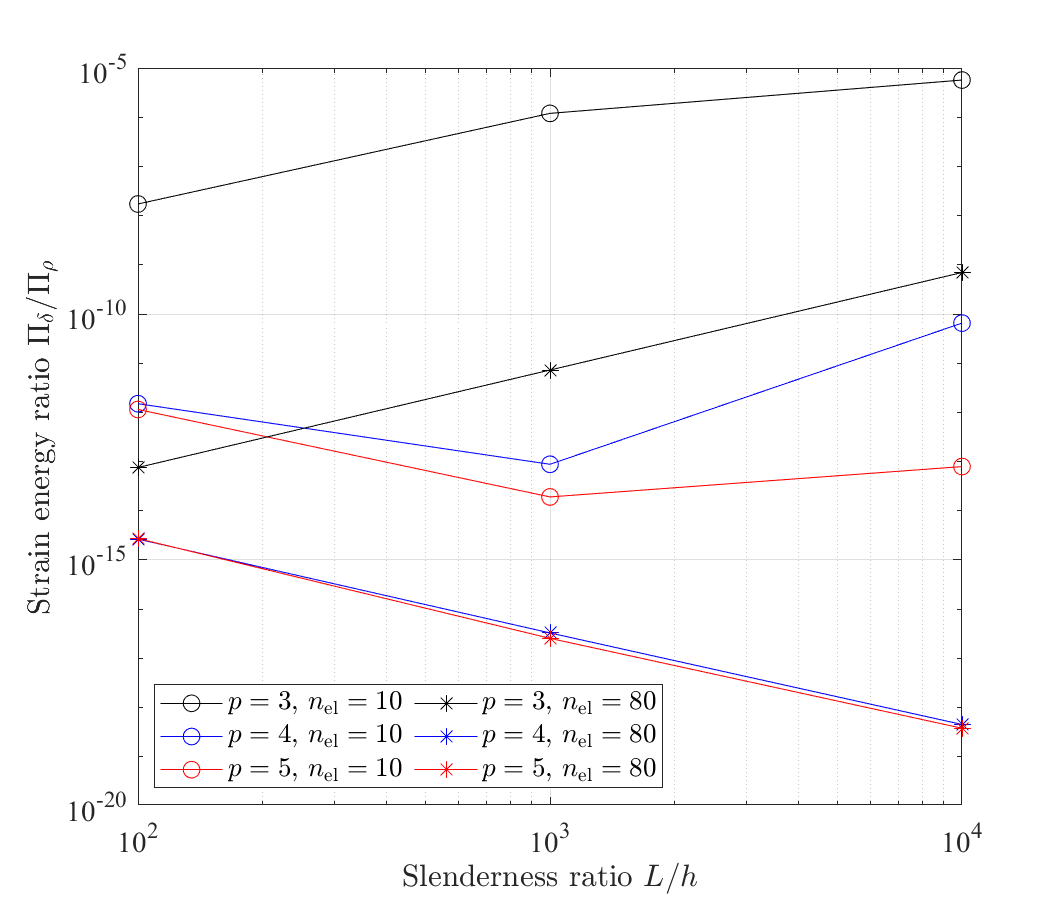}
		\caption{Transverse shear strain energy}
		\label{end_mnt_strn_e_shear}
	\end{subfigure}		
\caption{{Cantilever beam under end moment: Comparison of the ratio of the membrane, in-plane cross-section, and transverse shear strain energy to the bending strain energy. It is noticeable that the solution of $p=3$ and $n_\mathrm{el}=80$ recovers the analytically estimated convergence rate of order 2 in (a) and (b). It is noted that, in (a) and (b), the case $p=3, n_\mathrm{el}=80$ shows the same result as the case of $p=5, n_\mathrm{el}=10$.}}
\label{end_mnt_strn_e_conv}	
\end{figure*}

{We investigate the effect of mesh refinement and higher-order of basis function on the alleviation of membrane, transverse shear, and curvature-thickness locking. We compare, in Fig.\,\ref{cant_beam_end_moment_sl_ratio_l2_error_xdisp}, the relative difference of $X$-displacement from the analytical solution (Ref.\#3) in the $L^2$ norm with increasing slenderness ratio. The difference of the displacement between our beam formulation and Ref.\#3 is attributed to the aforementioned coupling between bending strain and axial/through-the-thickness stretching strains. However, it is shown that both axial ($\varepsilon$) and through-the-thickness stretching ($\chi^{11}$) strains diminish with the rate of $h^2$. Therefore, it is expected that the resulting displacement difference from Ref.\#3 should also decrease with the rate of $h^2$. In Fig.\,\ref{cant_beam_end_moment_sl_ratio_l2_error_xdisp}, it is seen that mesh refinement improves the convergence rate, and it is noticeable that the solution of using $p=5$ with $n_\mathrm{el}=80$ shows the estimated convergence rate of order 2. Further, Fig.\,\ref{end_mnt_strn_e_conv} shows the ratio of membrane ($\Pi_\varepsilon$), through-the-thickness stretching ($\Pi_\chi$), and transverse shear ($\Pi_\delta$) strain energy to bending strain energy ($\Pi_\rho$). In Figs.\,\ref{end_mnt_strn_e_memb} and \ref{end_mnt_strn_e_inp}, it is seen that mesh refinement or higher-order basis functions lead to the expected convergence rate for the strain energy ratio of order 2, i.e., ${\Pi_\varepsilon}/{\Pi_\rho}\sim{h^2}$ and ${\Pi_\chi}/{\Pi_\rho}\sim{h^2}$. This means that the membrane-bending and curvature-thickness locking are alleviated. Further, we investigate the transverse shear strain energy, defined by
\begin{equation}
\label{theo_bend_strn_e}
{\Pi _\delta} \coloneqq \int_0^L {{{\tilde q}^1}{\delta_1}{\rm{d}}s}.
\end{equation}
In Fig.\,\ref{end_mnt_strn_e_shear}, by using higher-order basis functions and mesh refinement ($p=4,5$ with $n_\mathrm{el}=80$), the spurious transverse shear strain energy (transverse shear-bending locking) is alleviated. It should be noted that this result does not mean those locking issues are completely resolved. For example, as discussed in \citet{adam2014improved}, if higher-order basis function is used, the membrane and transverse shear locking are less significant but still existing, due to the field-inconsistency paradigm, which is more pronounced in higher slenderness ratio. However, in this paper, we focus on low to moderate slenderness ratios, and further investigation on the reduced integration method and mixed-variational formulation remains future work.}
\subsection{Cantilever beam under end force}
\label{ex_cant_b_end_f}
{The third example illustrates Poisson locking in the standard extensible director beam formulation, and its alleviation by the EAS method. We further show that the EAS formulation based on Eq.\,(\ref{eas_strn_5param_form}) (i.e., ``ext.-dir.-EAS-5p.'') still suffers from significant Poisson locking due to its incomplete enrichment of the cross-section strains. A beam of length $L=10\mathrm{m}$ and cross-section dimension $h=w=1\mathrm{m}$ is clamped at one end, and subjected to a $Z$-directional force of magnitude $F={10^5}\mathrm{N}/\mathrm{m}^2$ acting on the other end (see Fig.\,\ref{thin_bend_undeformed}). The compressible Neo-Hookean material is selected, and Young's modulus is chosen as $E=10^7\mathrm{Pa}$, and two different Poisson's ratios are considered: $\nu=0$ and $\nu=0.3$.
\begin{figure*}[htp!]	
	\centering
	\begin{subfigure}[b] {0.7\textwidth} \centering
		\includegraphics[width=0.625\linewidth]{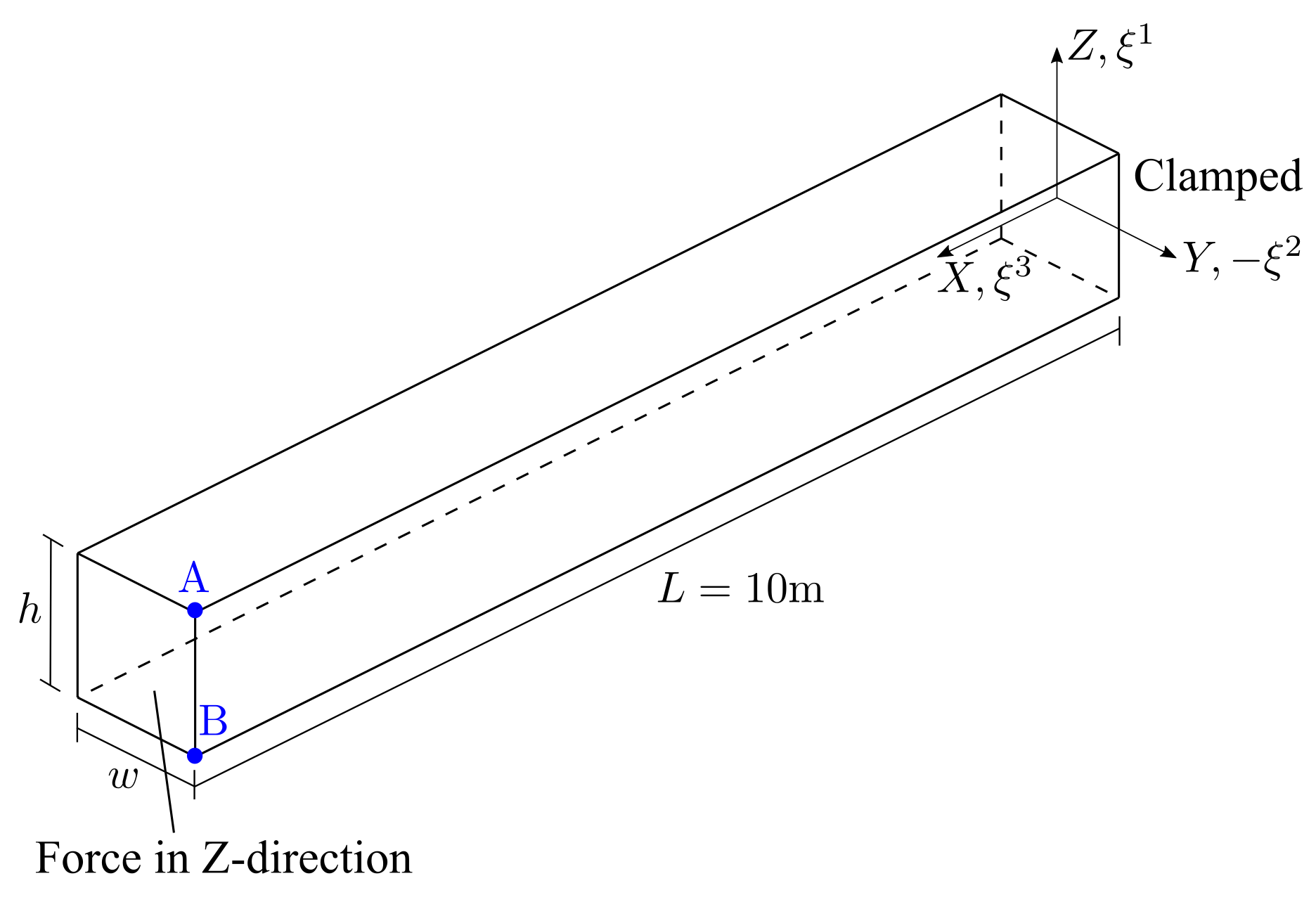}
	\end{subfigure}			
\caption{{Cantilever beam under end force: Undeformed configuration and boundary conditions.}}
\label{thin_bend_undeformed}	
\end{figure*}
{We determine reference solutions by using IGA brick elements of $\mathrm{deg.}=(2,1,2)$ with ${{n_\mathrm{el}}}=200\times1\times15$ and $\mathrm{deg.}=(3,3,3)$ with ${{n_\mathrm{el}}}=200\times20\times20$ for those cases $\nu=0$ and $\nu=0.3$, respectively (the convergence test result can be found in Table \ref{app_conv_test_xdisp_tip}).} {For beams, we use $4\times4$ Gauss integration points for the integration over the cross-section.} Fig.\,\ref{convergence_test_thin_bend} shows the convergence of beam solutions based on the presented extensible director kinematics for those two different cases of Poisson's ratios. If zero Poisson's ratio is considered ($\nu=0$), the results of the standard method are very close to the reference solution, and the EAS method gives the same results as the standard method. However, in the case of nonzero Poisson's ratio, since the standard method only allows for constant transverse normal strains, the coupled bending stiffness increases. This leads to a much smaller deflection than the reference solution. This Poisson locking is alleviated by the EAS method as it enhances the in-plane strain field of the cross-section. The EAS solution gives larger displacements that are much closer to the reference solutions (see also Table \ref{conv_test_beam_xyz_disp_atA}). It is also noticeable that the EAS solution `ext.-dir.-EAS-5p.' gives smaller deflections than the results of `ext.-dir.-EAS', since its enriched linear strain field is incomplete, so that Poisson locking is not effectively alleviated. In case of nonzero Poisson's ratio, we have a lateral ($Y$-directional) displacement. Fig.\,\ref{thin_bend_lat_disp_edge_ab_dist} compares the lateral displacement along the edge $\overline {\mathrm{BA}}$, indicated in Fig.\,\ref{thin_bend_undeformed}. In the EAS solution, the magnitude of lateral displacement increases and becomes closer to the average displacement of the reference one, compared with the solution by the standard method. Although the lateral displacement at the point A (${\xi^1}=0.5\,\mathrm{m}$) in the standard method is closer to the reference solution than that of the EAS solution (see also Table \ref{conv_test_beam_xyz_disp_atA}), it is shown that the accuracy of lateral displacement improves substantially by the EAS method in an average sense (see also the difference in $L^2$ norm in Fig.\,\ref{thin_bend_lat_disp_edge_ab_l2}). Further, it is seen that the 5-parameter EAS formulation (ext.-dir.-EAS-5p.) shows smaller magnitude of lateral displacement in Fig.\,\ref{thin_bend_lat_disp_edge_ab_dist}, and larger $L^2$ norm of difference in Fig.\,\ref{thin_bend_lat_disp_edge_ab_l2} due to the incomplete enrichment of in-plane strain field, compared with the 9-parameter formulation (ext.-dir.-EAS). Fig.\,\ref{thin_bend_deformed_config_compare} shows that the standard beam formulation shows much smaller deflection than the other formulations in the final deformed configuration due to Poisson locking. Table\,\ref{convergence_test_thin_bend_instab} shows that the beam solutions use less number of load steps and iterations than the brick element solution.
\clearpage
\begin{figure*}[!htbp]	
	\centering
	\begin{subfigure}[b] {0.4875\textwidth} \centering
		\includegraphics[width=\linewidth]{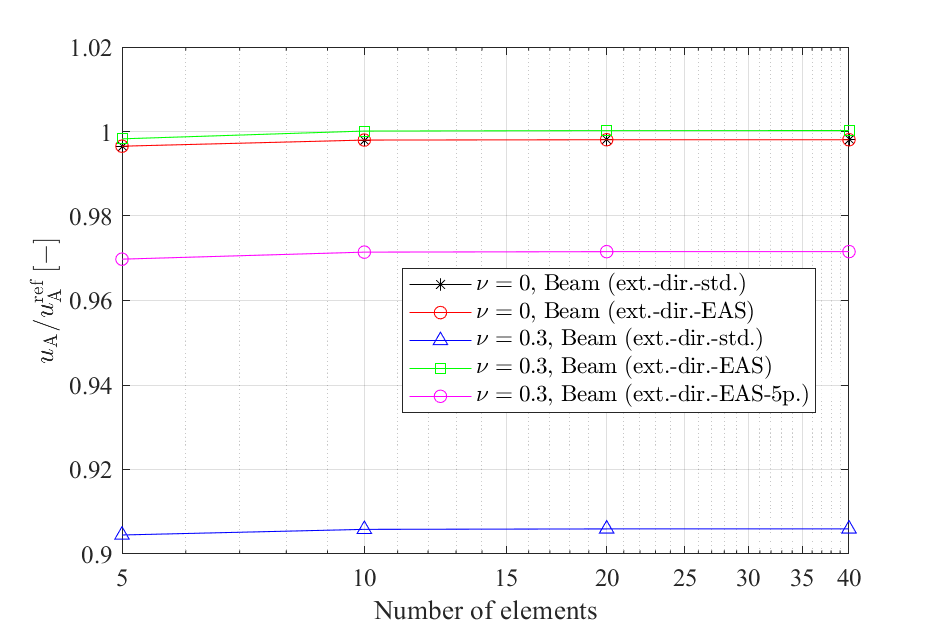}
		\vskip -2pt		
		\caption{$X$-displacement}
		\label{thin_bend_pr0}
	\end{subfigure}		
	\begin{subfigure}[b] {0.4875\textwidth} \centering
		\includegraphics[width=\linewidth]{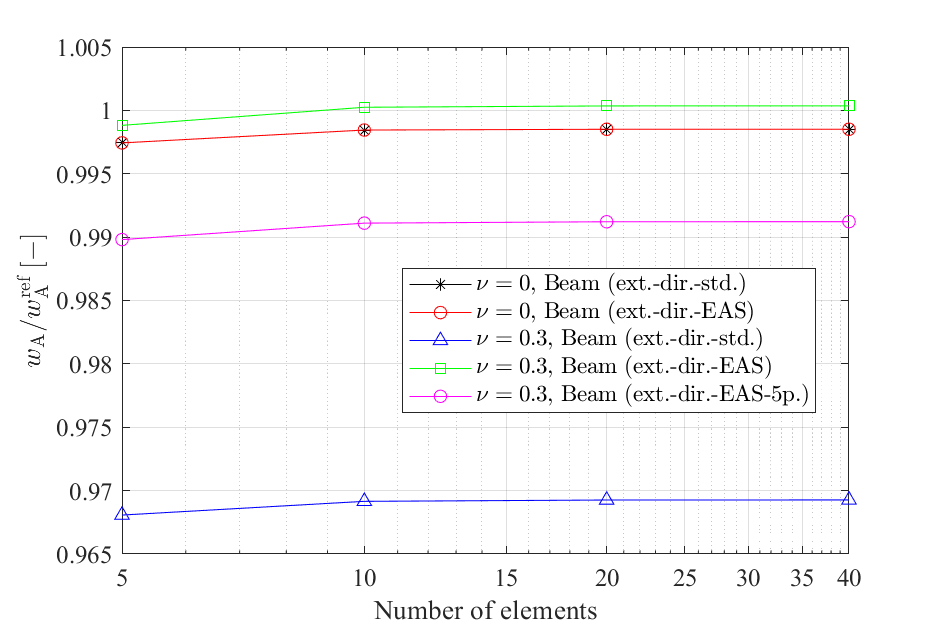}
		\vskip -2pt		
		\caption{$Z$-displacement}
		\label{thin_bend_pr03}
	\end{subfigure}		
		\vskip -2pt
\caption{{Cantilever beam under end force: Convergence of the normalized displacements at point $\mathrm{A}$ for two different cases of Poisson's ratio. The displacement is normalized by the reference solution using brick elements, where $\mathrm{deg.}=(2,1,2)$ and ${{n_\mathrm{el}}}=200\times1\times15$ for $\nu=0$, and $\mathrm{deg.}=(3,3,3)$ and ${{n_\mathrm{el}}}=200\times20\times20$ for $\nu=0.3$.} {The beam solutions are obtained by IGA with $p=3$.}}
\label{convergence_test_thin_bend}	
\end{figure*}
\begin{figure}[htp!]	
	\centering
	\begin{subfigure}[b] {0.52\textwidth} \centering
		\includegraphics[width=\linewidth]{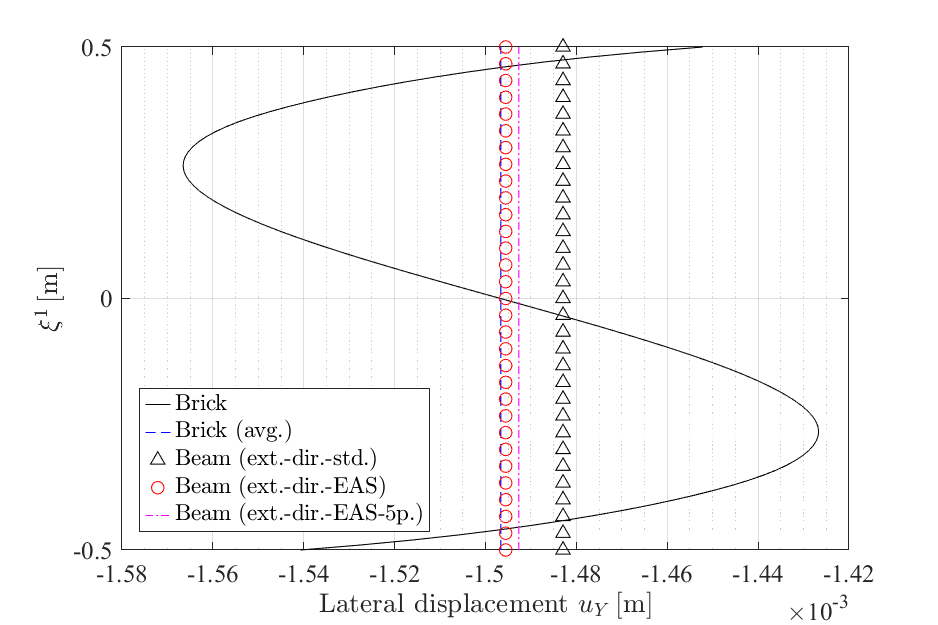}
		\caption{Lateral displacement ($u_Y$)}
		\label{thin_bend_lat_disp_edge_ab_dist}
	\end{subfigure}		
	\begin{subfigure}[b] {0.4675\textwidth} \centering
		\includegraphics[width=\linewidth]{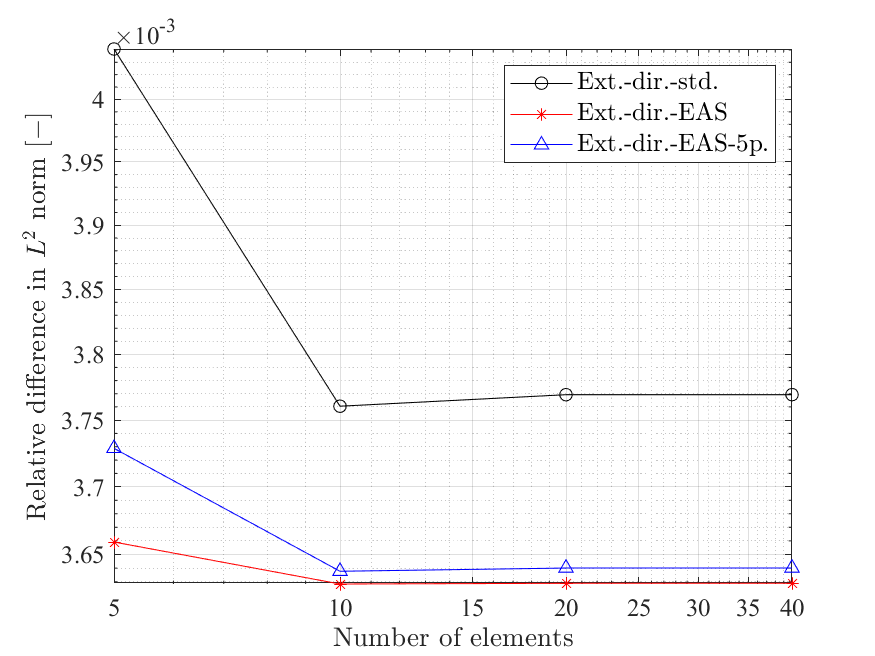}
		\caption{Relative $L^2$ error of $u_Y$}
		\label{thin_bend_lat_disp_edge_ab_l2}
	\end{subfigure}			
\caption{Cantilever beam under end force: Comparison of the lateral displacement along the edge $\overline {\mathrm{BA}}$ in the case of $\nu=0.3$. {The beam solutions are obtained by IGA with $p=3$. Also, we use $n_\mathrm{el}=40$ for the beam solutions in Figure (a).} }
\label{thin_bend_lateral_disp_edge}	
\end{figure}
\begin{figure*}[!htbp]	
	\centering
	\begin{subfigure}[b] {0.32\textwidth} \centering
		\includegraphics[width=\linewidth]{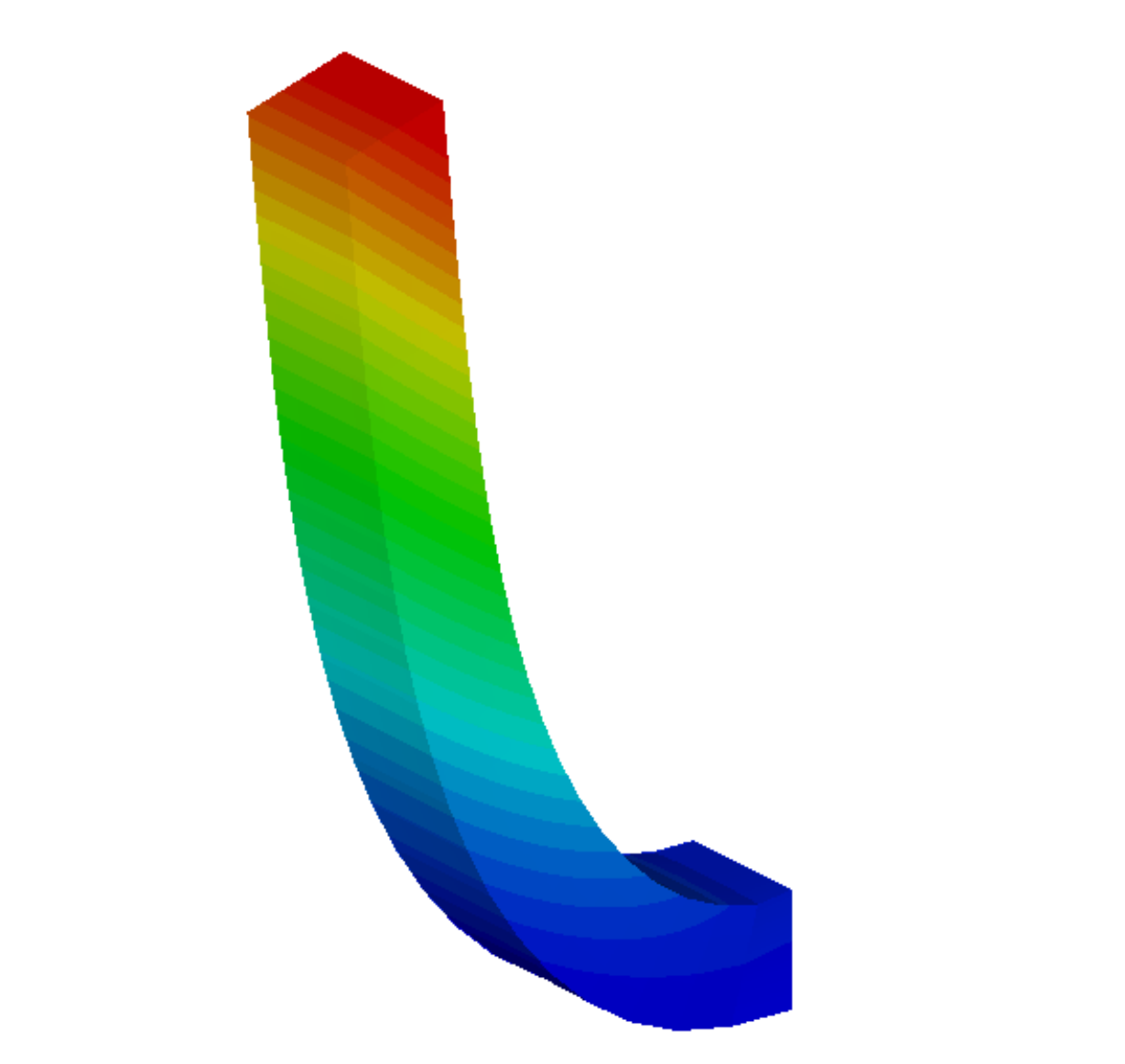}
		\caption{Brick element}
		\label{thin_bend_instab_beam}
	\end{subfigure}		
	\begin{subfigure}[b] {0.32\textwidth} \centering
		\includegraphics[width=\linewidth]{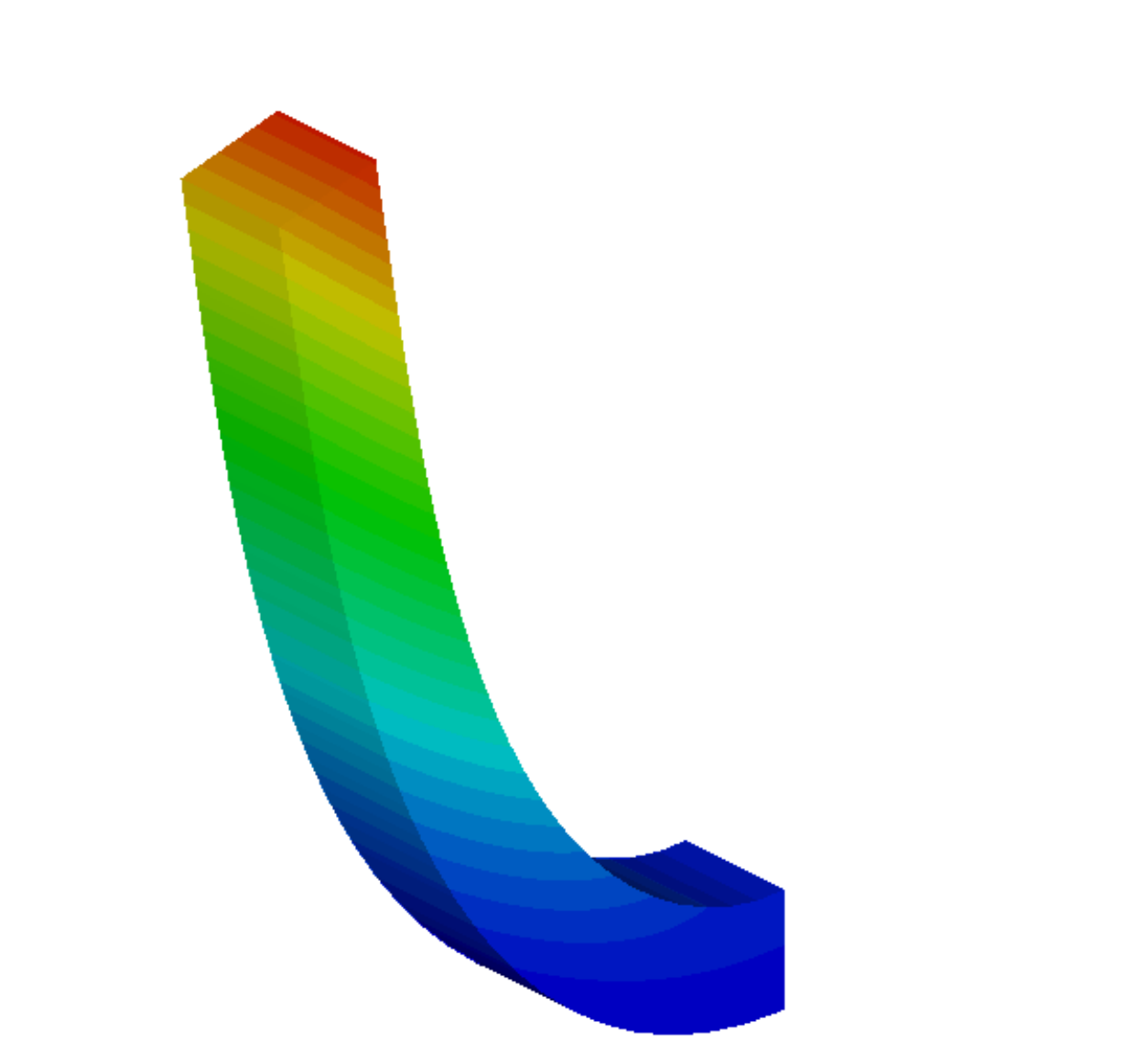}
		\caption{Beam (ext.-dir.-std.)}
		\label{thin_bend_instab_solid}
	\end{subfigure}			
	\begin{subfigure}[b] {0.32\textwidth} \centering
		\includegraphics[width=\linewidth]{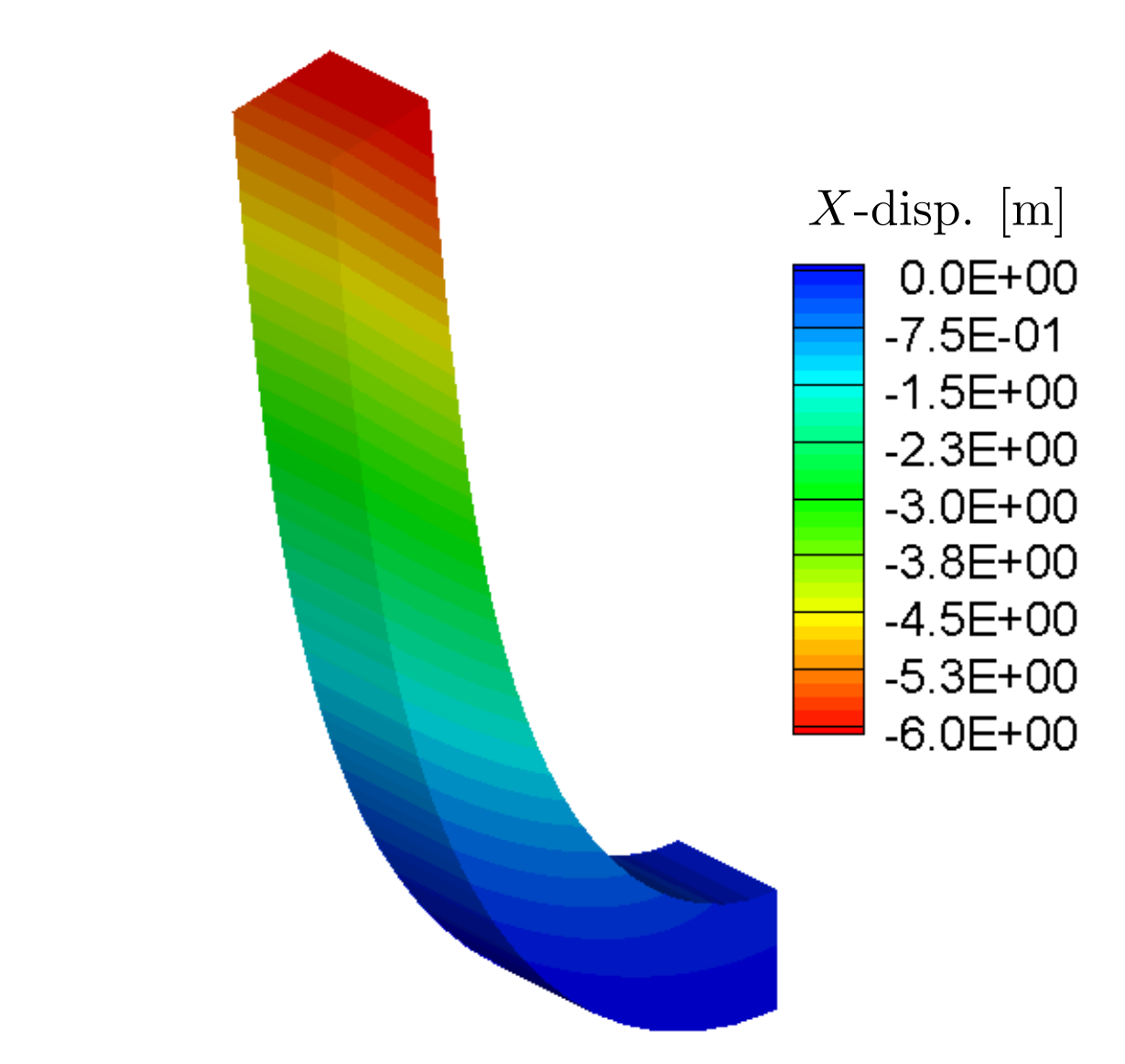}
		\caption{Beam (ext.-dir.-EAS)}
		\label{thin_bend_instab_solid}
	\end{subfigure}		
\caption{{Cantilever beam under end force: Comparison of deformed configurations. Deformations by (a) brick elements with deg.=$(3,3,3)$, ${{n_\mathrm{el}}}=200\times10\times10$, (b) beam elements (ext.-dir.-std.) with $p=3$ and ${{n_\mathrm{el}}}=40$, and (c) beam elements (ext.-dir.-EAS) with the same discretization with (b). The color represents the $X$-displacement.}}
\label{thin_bend_deformed_config_compare}	
\end{figure*}
\begin{table}[!htbp]
\small
\begin{center}
\caption{{Cantilever beam under end force: Convergence test of normalized displacements at the point A for $\nu=0.3$. $u_\mathrm{A}$, $v_\mathrm{A}$, and $w_\mathrm{A}$ denote the $X$-, $Y$-, and $Z$-displacements at the point A, respectively. $(\bullet)^\mathrm{ref}$ denotes the reference solution.} All results are obtained by IGA with $p=3$.}
\label{conv_test_beam_xyz_disp_atA}
\begin{tabular}{cccccccc}
\Xhline{3\arrayrulewidth} 
\multicolumn{4}{c}{Beam (ext.-dir.-std.)}    &  & \multicolumn{3}{c}{Beam (ext.-dir.-EAS)} \\ \cline{2-4} \cline{6-8} 
${n_\mathrm{el}}$ & ${u_\mathrm{A}}/{u_\mathrm{A}^\mathrm{ref}}$   & ${v_\mathrm{A}}/{v_\mathrm{A}^\mathrm{ref}}$     & ${w_\mathrm{A}}/{w_\mathrm{A}^\mathrm{ref}}$         &  & ${u_\mathrm{A}}/{u_\mathrm{A}^\mathrm{ref}}$    & ${v_\mathrm{A}}/{v_\mathrm{A}^\mathrm{ref}}$  & ${w_\mathrm{A}}/{w_\mathrm{A}^\mathrm{ref}}$\\ 
\Xhline{3\arrayrulewidth} 
5  & 9.0455E-01 & 1.0141E+00 & 9.6811E-01 &  & 9.9831E-01  & 1.0261E+00  & 9.9886E-01 \\
10 & 9.0588E-01 & 1.0214E+00 & 9.6919E-01 &  & 1.0002E+00  & 1.0300E+00  & 1.0003E+00 \\
20 & 9.0598E-01 & 1.0211E+00 & 9.6930E-01 &  & 1.0003E+00  & 1.0298E+00  & 1.0004E+00 \\
40 & 9.0599E-01 & 1.0211E+00 & 9.6930E-01 &  & 1.0003E+00  & 1.0298E+00  & 1.0004E+00 \\ 
\Xhline{3\arrayrulewidth} 
\end{tabular}
\end{center}
\end{table}
\begin{table}[]
\centering
\small
\caption{{Cantilever beam under end force: Comparison of Newton-Raphson iteration history for $\nu=0.3$. A uniform load increment is used.}}
\label{convergence_test_thin_bend_instab}	
\begin{tabular}{ccclcclcc}
\Xhline{3\arrayrulewidth}
\multirow{2}{*}{\begin{tabular}[c]{@{}c@{}}Iter.\#\\(last load\\step)\end{tabular}} & \multicolumn{2}{c}{\begin{tabular}[c]{@{}c@{}}Brick, deg.=(3,3,3),\\${n_\mathrm{el}}$=$200\times10\times10$\end{tabular}}                         &  & \multicolumn{2}{c}{\begin{tabular}[c]{@{}c@{}}Beam (ext.-dir.-std.),\\$p=3$, ${n_\mathrm{el}}$=40\end{tabular}}                                    &  & \multicolumn{2}{c}{\begin{tabular}[c]{@{}c@{}}Beam (ext.-dir.-EAS),\\$p=3$, ${n_\mathrm{el}}$=40\end{tabular}}                                    \\ \cline{2-3} \cline{5-6} \cline{8-9} 
                                                                                               & \begin{tabular}[c]{@{}c@{}}Euclidean\\      norm of residual\end{tabular} & \begin{tabular}[c]{@{}c@{}}Energy\\      norm\end{tabular} &  & \begin{tabular}[c]{@{}c@{}}Euclidean\\      norm of residual\end{tabular} & \begin{tabular}[c]{@{}c@{}}Energy\\      norm\end{tabular} &  & \begin{tabular}[c]{@{}c@{}}Euclidean\\      norm of residual\end{tabular} & \begin{tabular}[c]{@{}c@{}}Energy\\      norm\end{tabular} \\ 
\Xhline{3\arrayrulewidth}
1                                                                                              & 4.4E+02                                                       & 2.9E+02                                                    &  & 1.0E+04                                                       & 1.5E+03                                                    &  & 1.0E+04                                                       & 1.2E+03                                                    \\
2                                                                                              & 6.4E+02                                                       & 3.7E+00                                                    &  & 2.2E+04                                                       & 1.2E+02                                                    &  & 1.7E+04                                                       & 7.7E+01                                                    \\
3                                                                                              & 8.6E-01                                                       & 2.3E-04                                                    &  & 1.0E+02                                                       & 1.7E-01                                                    &  & 8.2E+01                                                       & 7.7E-02                                                    \\
4                                                                                              & 6.9E-04                                                       & 4.3E-12                                                    &  & 3.4E+00                                                       & 2.4E-06                                                    &  & 1.4E+00                                                       & 4.8E-07                                                    \\
5                                                                                              & 2.8E-08                                                       & 1.1E-21                                                    &  & 2.5E-06                                                       & 9.5E-17                                                    &  & 7.6E-07                                                       & 5.4E-18                                                    \\
6                                                                                              &                                                               &                                                            &  & 7.7E-08                                                       & 1.8E-22                                                    &  & 7.2E-08                                                       & 1.3E-22                                                    \\ \cline{1-3} \cline{5-6} \cline{8-9} 
\#load steps                                                                                   & \multicolumn{2}{c}{20}                                                                                                     &  & \multicolumn{2}{c}{10}                                                                                                     &  & \multicolumn{2}{c}{10}                                                                                                     \\
\#iterations                                                                                   & \multicolumn{2}{c}{124}                                                                                                    &  & \multicolumn{2}{c}{73}                                                                                                     &  & \multicolumn{2}{c}{78}                                                                                                     \\ 
\Xhline{3\arrayrulewidth}
\end{tabular}
\end{table}
}
\subsection{Laterally loaded beam}
\label{ex_lat_load_b}
\begin{figure*}[!htbp] 
\centering	
\includegraphics[width=0.6\linewidth]{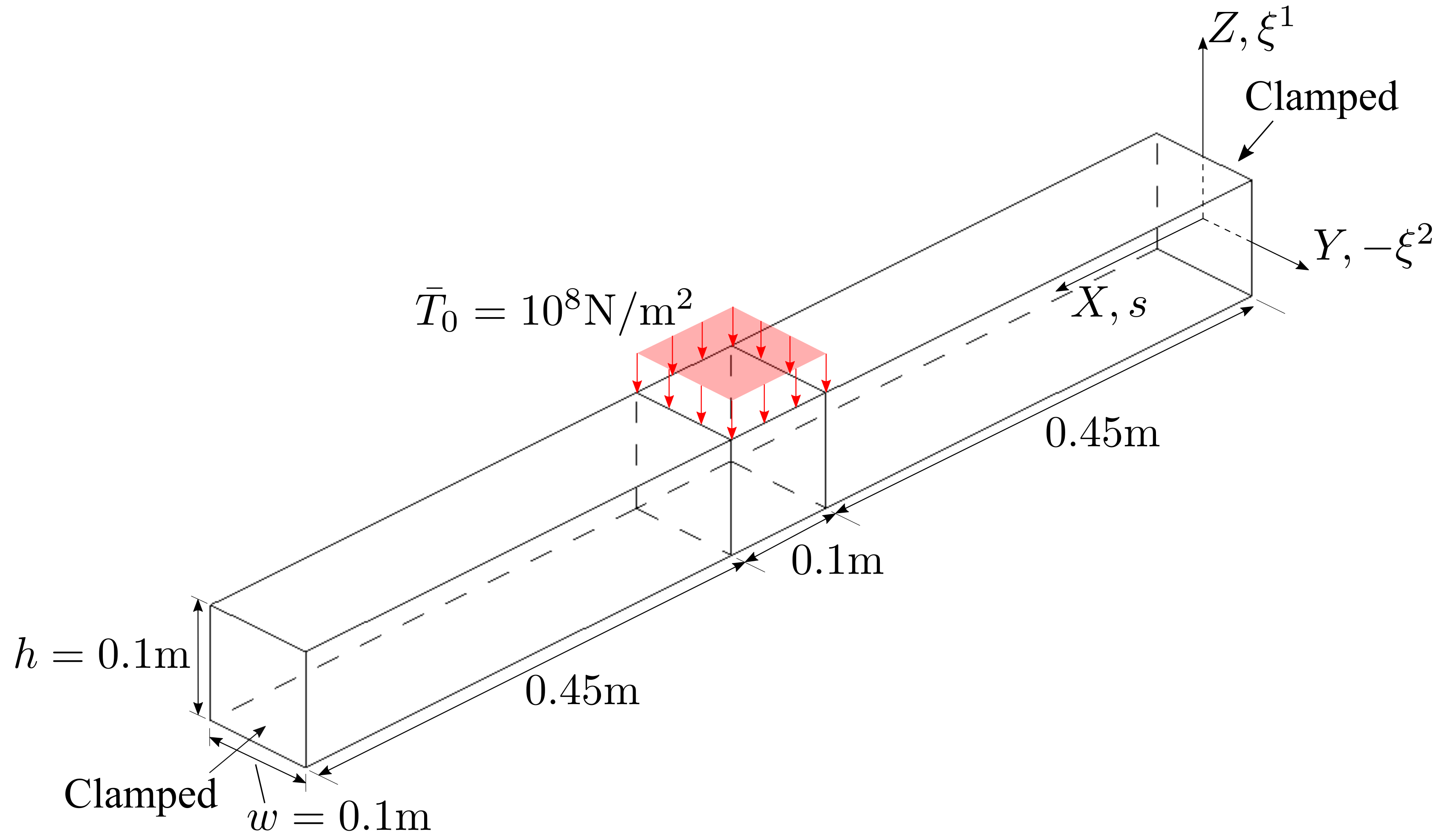}
\caption{Laterally loaded beam: Undeformed configuration and boundary conditions.}
\label{model_des_str_b_con_f}
\end{figure*}

\begin{figure}[H] \centering	
\includegraphics[width=0.6125\linewidth]{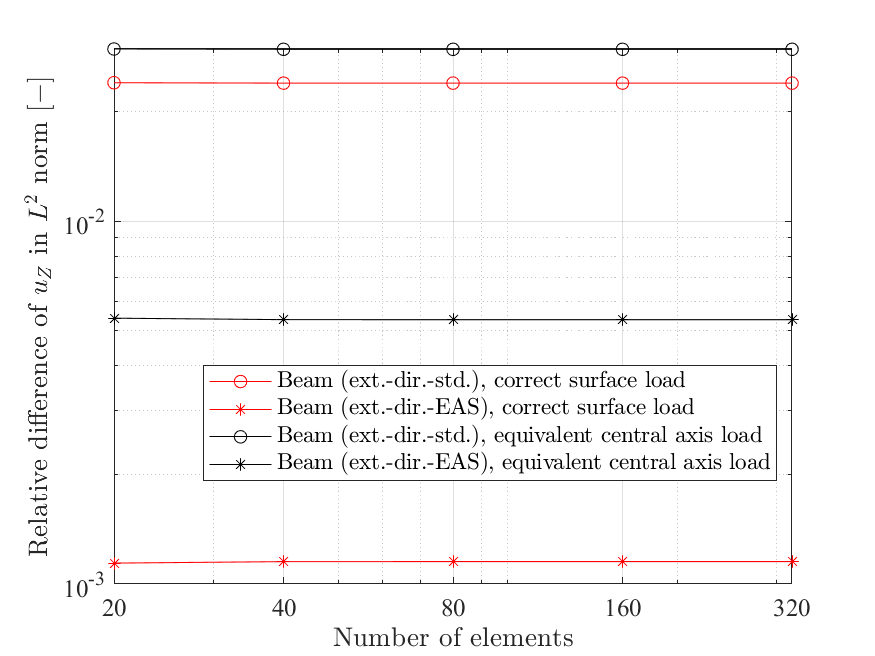}
\caption{{Laterally loaded beam: Comparison of relative difference in the $Z$-displacement on the central axis.} {The results are obtained by IGA with $p=3$.}}
\label{disp_diff_comp_str_b_con_f}
\end{figure}

\begin{figure*}[!htbp]	
	\centering
	\begin{subfigure}[b] {0.3125\textwidth} \centering
		\includegraphics[width=\linewidth]{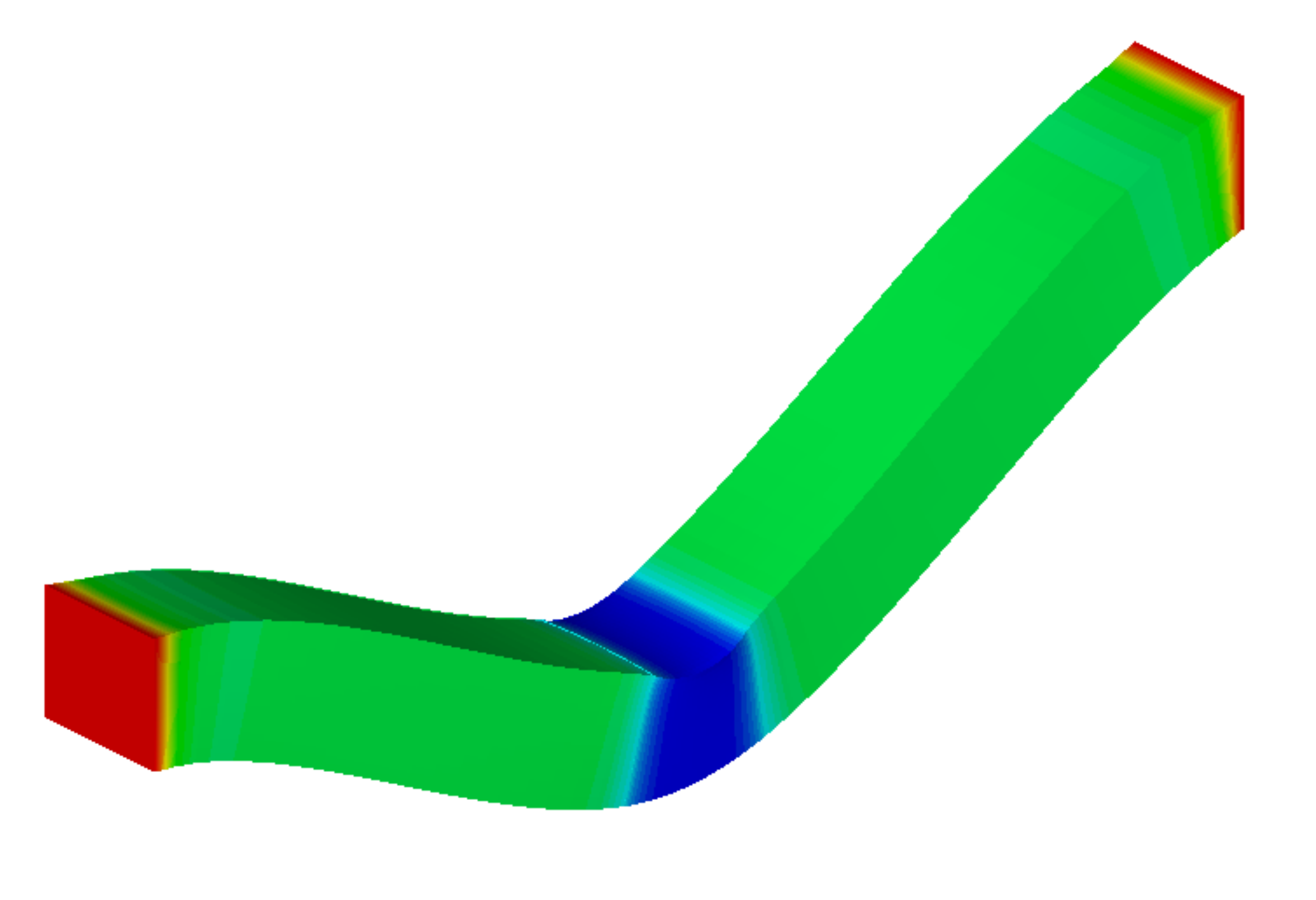}
		\caption{Brick element (surface load)}
		\label{deform_str_cent_f_deformed_solid}
	\end{subfigure}\hspace{2.5mm}		
	\begin{subfigure}[b] {0.3125\textwidth} \centering
		\includegraphics[width=\linewidth]{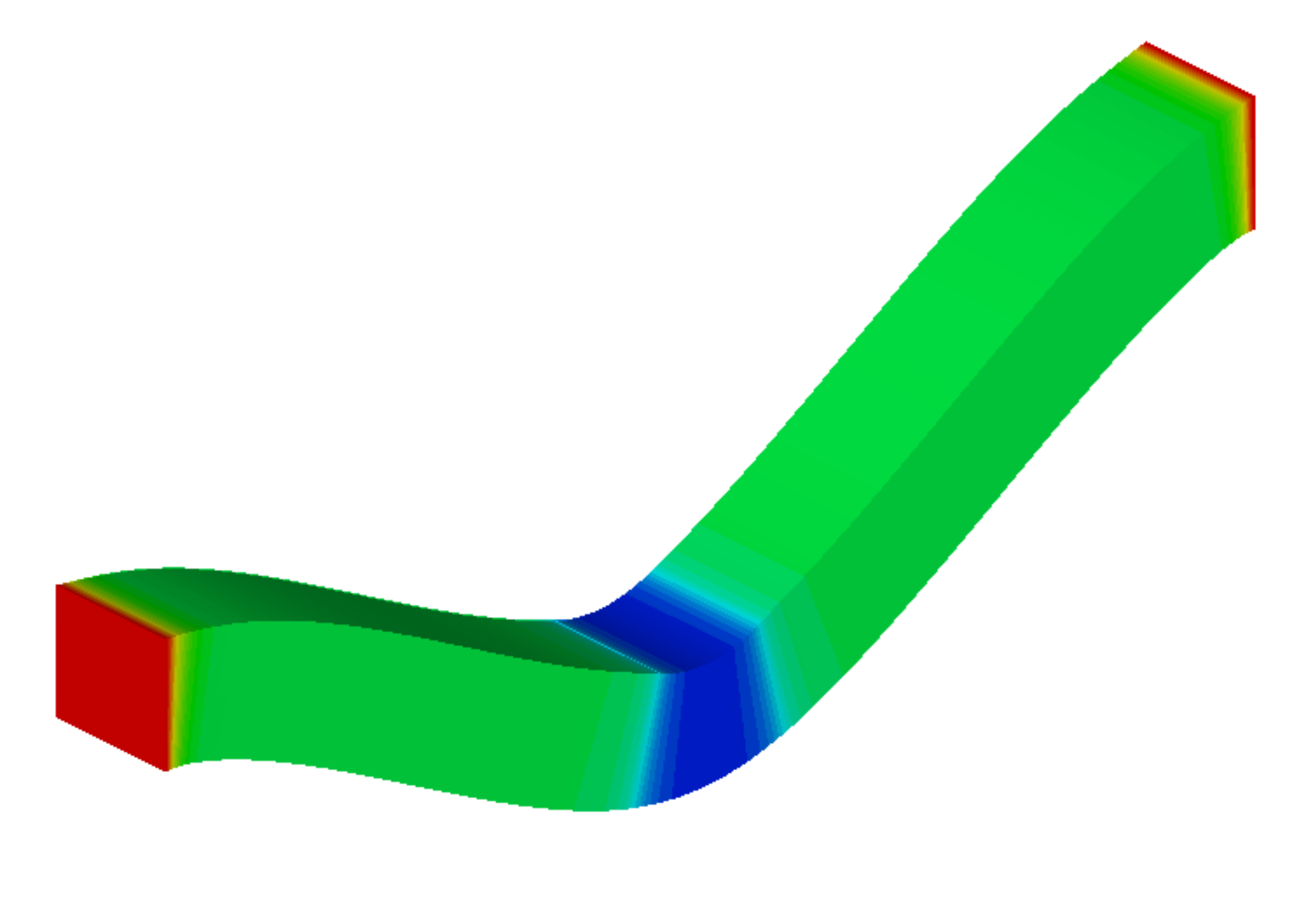}
		\caption{Beam element (surface load)}
		\label{deform_str_cent_f_deformed_beam}
	\end{subfigure}\hspace{2.5mm}			
	\begin{subfigure}[b] {0.3125\textwidth} \centering
		\includegraphics[width=\linewidth]{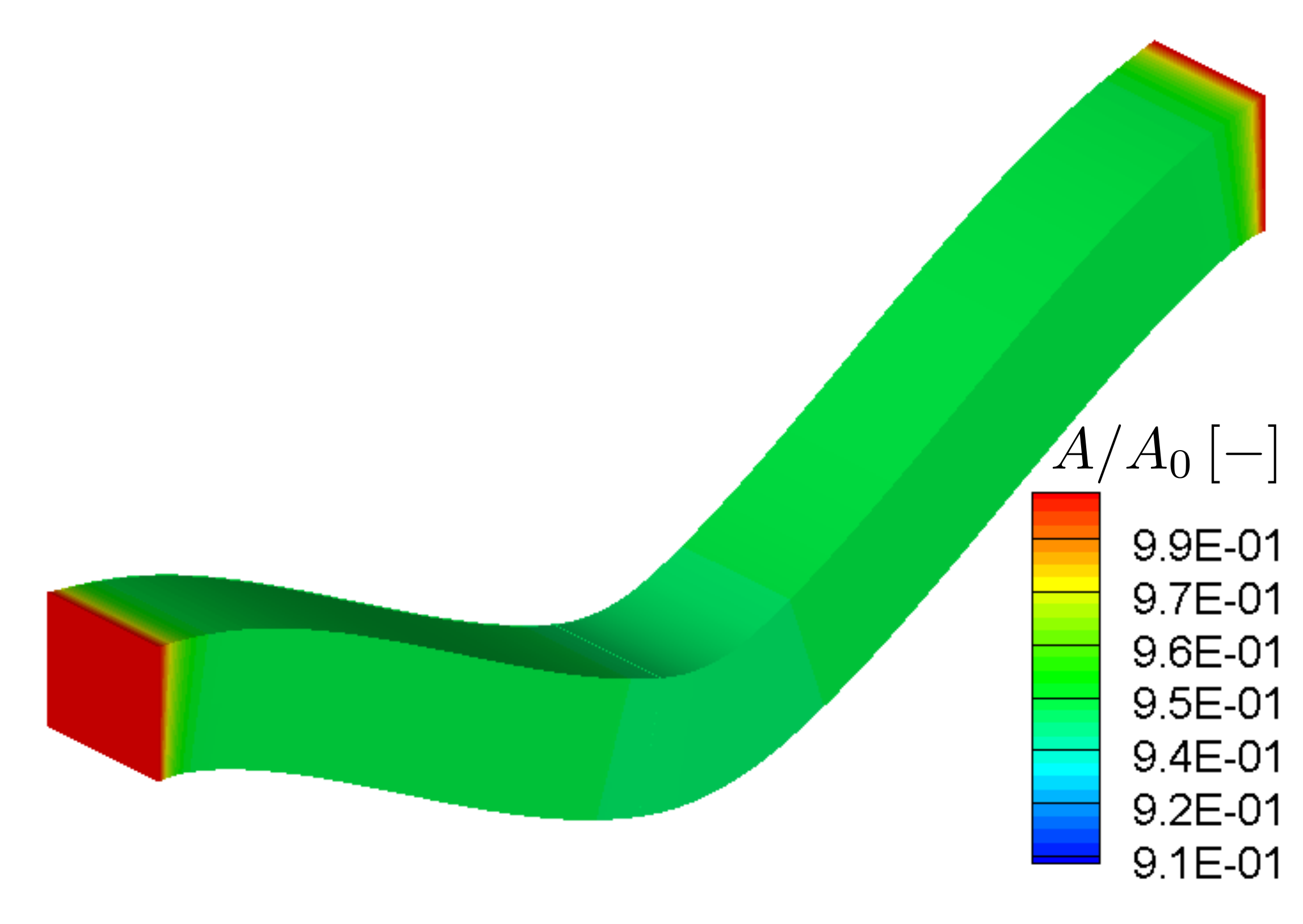}
		\caption{Beam element (central axis load)}
		\label{deform_str_cent_f_deformed_beam_equiv_f}
	\end{subfigure}\hspace{2.5mm}			
\caption{{Laterally loaded beam: Comparison of deformed configurations. The results are obtained by IGA using (a) brick elements with deg.\,=(3,3,3) and ${{n_\mathrm{el}}}=320\times15\times15$, (b) beam elements (ext.-dir.-EAS) with $p=3$ and ${{n_\mathrm{el}}}=40$ with correct surface load, and (c) the same spatial discretization with (b) but with equivalent load directly applied to the central axis. The color represents the ratio of current cross-sectional area ($A$) to the initial one ($A_0$).}}
\label{deform_str_b_clamped_cf}	
\end{figure*}

{The fourth example investigates the significance of considering the correct surface load rather than applying an equivalent load directly to the central axis, which is typically assumed in the analysis of thin beams. The significance was also discussed in the shell formulation based on an extensible director in \cite{simo1990stress}. We consider a clamped-clamped straight beam, and a distributed force of magnitude ${{\bar T}_0}={10^8}\mathrm{N}/\mathrm{m}^2$ in the negative $Z$-direction is applied over $0.1\mathrm{m}$ along the middle of the beam, as illustrated in Fig.\,\ref{model_des_str_b_con_f}. The beam has initial length $L=1\mathrm{m}$ and a square cross-section of dimension $h=w=0.1\mathrm{m}$, and the compressible Neo-Hookean material with Young's modulus $E=1\mathrm{GPa}$ and Poisson's ratio $\nu=0.3$. We model the geometry using three NURBS patches such that the basis functions have ${C^0}$-continuity at the boundaries of the loaded area ($s=0.45\mathrm{m}\,\,\mathrm{and}\,\,0.55\mathrm{m}$) in order to satisfy the discontinuity of the distributed load. {For the beam, $4\times4$ Gauss integration points are used for the integration over the cross-section.} Fig.\,\ref{disp_diff_comp_str_b_con_f} compares the relative difference of the $Z$-displacement at the central axis between the beam formulation and the reference solution obtained by IGA using brick elements with deg.=(3,3,3) and ${{n_\mathrm{el}}}=320\times15\times15$ (Table\,\ref{app_lat_load_conv_test_ref_sol} shows the convergence result of the brick element solution). As expected, it is seen in Fig.\,\ref{disp_diff_comp_str_b_con_f} that the EAS formulation gives much smaller differences than the standard formulation due to the alleviation of Poisson locking. Fig.\,\ref{disp_diff_comp_str_b_con_f} also illustrates the difference between two ways of applying the surface load: One follows the common practice in the analysis of thin beams that applies an equivalent load directly to the central axis, and is termed as \textit{equivalent central axis load}. The second, termed \textit{the correct surface load}, calculates the external stress resultant ${\bar{\boldsymbol{n}}}_0$ and external director stress couple ${\bar{\tilde {\boldsymbol{m}}}}_0^1$ by substituting ${\bar{\boldsymbol{T}}}_0=-{{\bar T}_0}{{\boldsymbol{e}}_3}$ into Eqs.\,(\ref{beam_lin_mnt_balance_ext_f}) and (\ref{beam_dir_mnt_balance_ext_m}), respectively. On the other hand, in the \textit{equivalent central axis load}, the force per unit arc-length is calculated by ${\bar{\boldsymbol{n}}}_0=-{{\bar T}_0}w{{\boldsymbol{e}}_3}$, and the effect of the director stress couple is neglected, i.e., ${\bar{\tilde {\boldsymbol{m}}}}_0^1=\boldsymbol{0}$, since the load is assumed to be directly applied to the central axis. In Fig.\,\ref{disp_diff_comp_str_b_con_f}, the beam solutions using the correct surface load show much smaller difference in both of standard and EAS formulations, compared with the results using the equivalent central axis load. Further, Fig.\,\ref{deform_str_b_clamped_cf} compares the deformed configurations and the change of cross-sectional area in three different formulations; the brick element solution with correct surface load, the beam element solution with EAS method and correct surface load, and the beam element solution with EAS method and equivalent central axis load. It is noticeable that the beam solution using the equivalent load shows much smaller change of cross-sectional area at the loaded part, since it neglects the effect of external director stress couple. Table\,\ref{newton_iter_lateral_load_compare} shows that the brick element formulation requires larger number of load steps to achieve the convergence.}

\begin{table}[]
\small
\begin{center}
\caption{{Laterally loaded beam: Comparison of the total number of load steps and iterations in the Newton-Raphson method. A uniform load increment is used. Brick elements are deg.=(3,3,3) and $n_{\mathrm{el}}=320\times15\times15$, and all the beam elements are $p=3$ and $n_{\mathrm{el}}=320$.}}
\label{newton_iter_lateral_load_compare}
\begin{tabular}{cccclcc}
\Xhline{3\arrayrulewidth}
                                 & \multicolumn{3}{c}{Correct surface load}                                                              &  & \multicolumn{2}{l}{Equivalent central axis load}             \\ \cline{2-4} \cline{6-7} 
                                 & \multicolumn{1}{l}{Brick} & \multicolumn{1}{l}{ext.-dir.-std} & \multicolumn{1}{l}{ext.-dir.-EAS} &  & ext.-dir.-std        & \multicolumn{1}{l}{ext.-dir.-EAS} \\ \Xhline{3\arrayrulewidth}
\multicolumn{1}{c}{\#load steps} & 5                         & 1                                   & 2                                   &  & \multicolumn{1}{c}{2}  & 2                                   \\
\multicolumn{1}{c}{\#iterations} & 29                        & 15                                  & 16                                  &  & \multicolumn{1}{c}{15} & 16                                  \\ 
\Xhline{3\arrayrulewidth}
\end{tabular}
\end{center}
\end{table}
\subsection{{45$^{\circ}$-arch cantilever beam under end force}}
\label{ex_45_cant_beam_end_f}
\begin{figure}[H] \centering	
\includegraphics[width=0.6\linewidth]{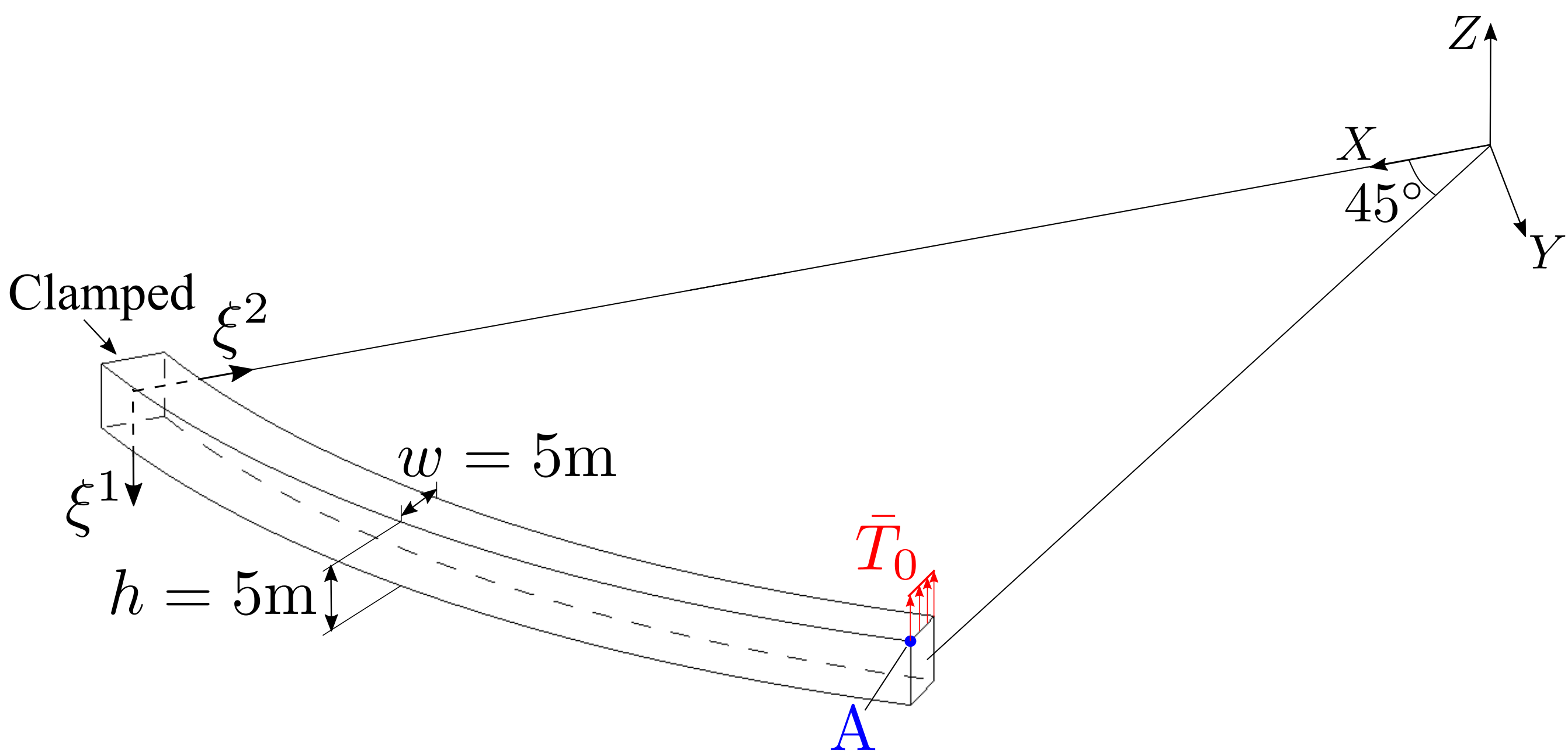}
\caption{{45$^{\circ}$-arch cantilever beam: Undeformed configuration and boundary conditions. The axes of $\xi^1$ and $\xi^2$ represent two principal directions of the cross-section.}}
\label{model_des_cant45}
\end{figure}

\begin{figure}[htp]
	\centering	
	\begin{subfigure}[b] {0.49\textwidth} \centering
		\includegraphics[width=\linewidth]{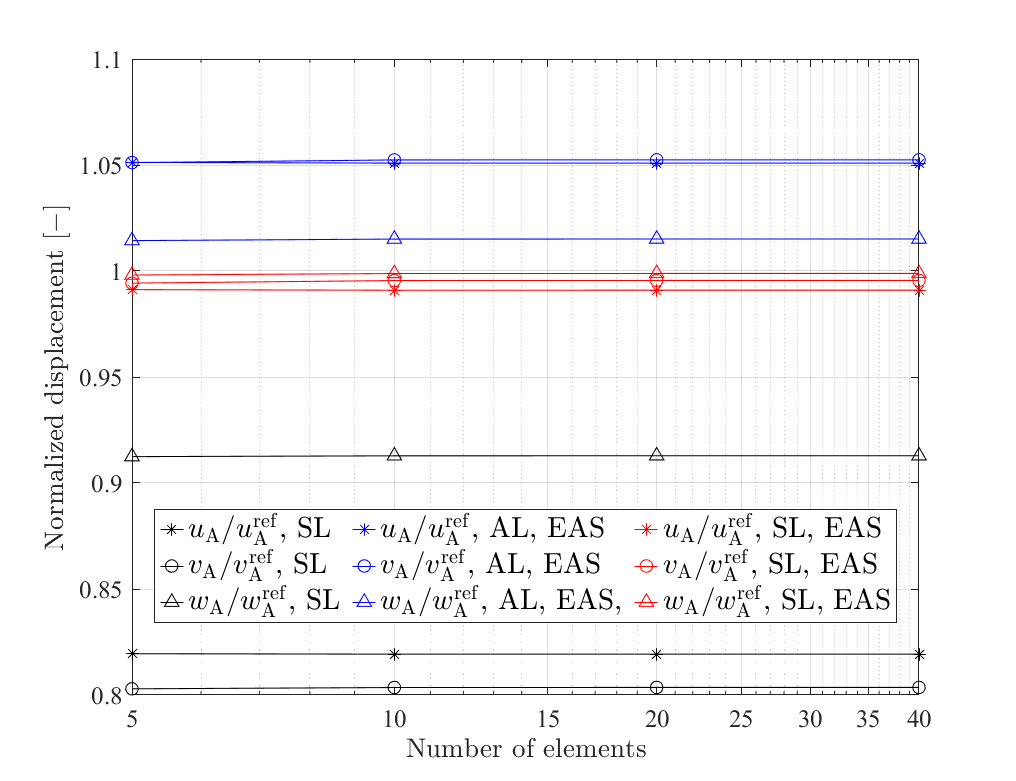}
 		\vskip -4pt
		\caption{Normalized displacements}
		\label{cant45deg_conv_norm_disp_graph}		
	\end{subfigure}		
	\begin{subfigure}[b] {0.49\textwidth} \centering
		\includegraphics[width=\linewidth]{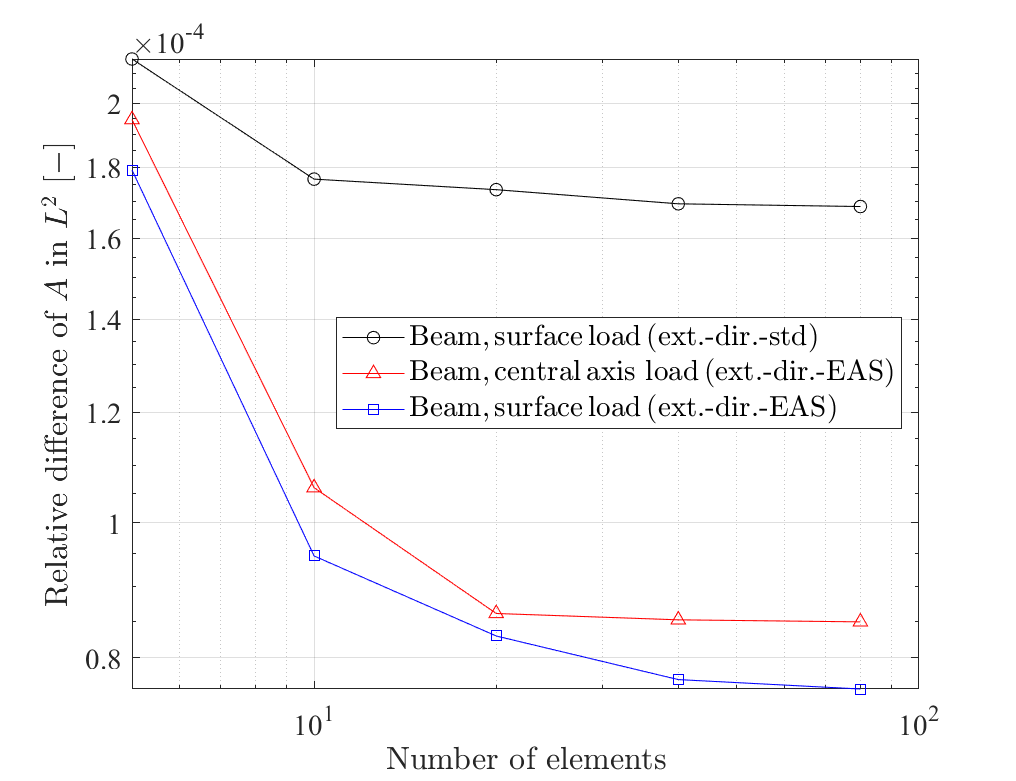}
 		\vskip -4pt
		\caption{Cross-sectional area ($A$)}
		\label{cant45deg_conv_area_diff_graph}			
	\end{subfigure}			
\caption{{45$^{\circ}$-arch cantilever beam: (a) Convergence of normalized displacements at the point A, and (b) the relative difference of the cross-sectional area $A$ from the brick element result in $L^2$ norm. `AL' and `SL' denote the equivalent central axis load and the correct surface load, respectively. $u_\mathrm{A}$, $v_\mathrm{A}$, and $w_\mathrm{A}$ denote the $X$-, $Y$-, and $Z$-displacements at the point A, respectively. $(\bullet)^\mathrm{ref}$ denotes the reference solution. Also, `EAS' represents the 9-parameter EAS formulation, i.e., `ext.-dir.-EAS'.} All results are obtained by IGA with $p=3$.}
\label{deform_45deg_conv_disp}	
\end{figure}

\begin{figure*}[!htbp]	
	\centering	
	\begin{subfigure}[b] {0.3\textwidth} \centering
		\includegraphics[width=\linewidth]{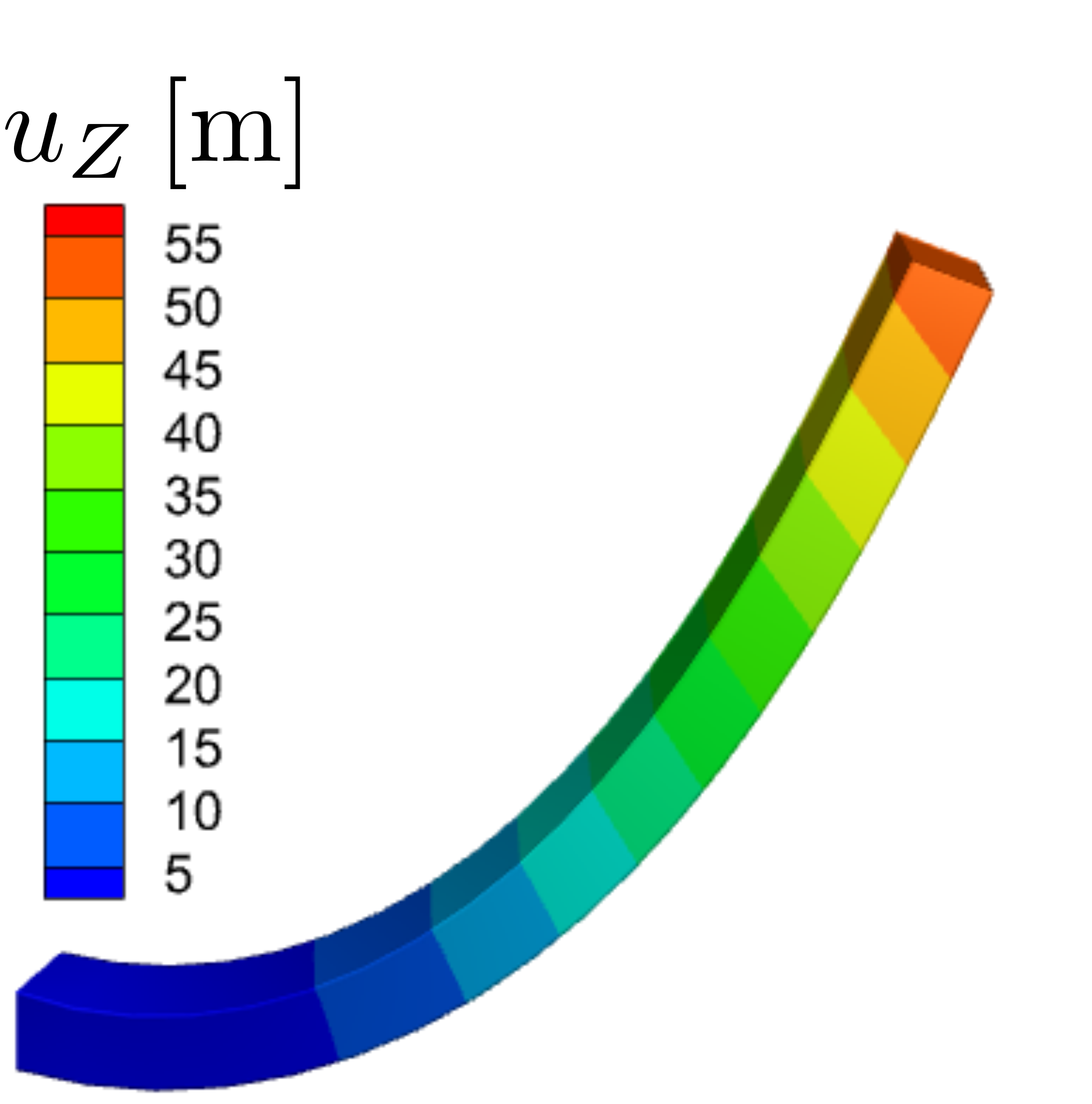}
		\caption{Brick (correct surface load)}
		\label{deform_cant45_h5_beam_solid}
	\end{subfigure}
	\begin{subfigure}[b] {0.3\textwidth} \centering
		\includegraphics[width=\linewidth]{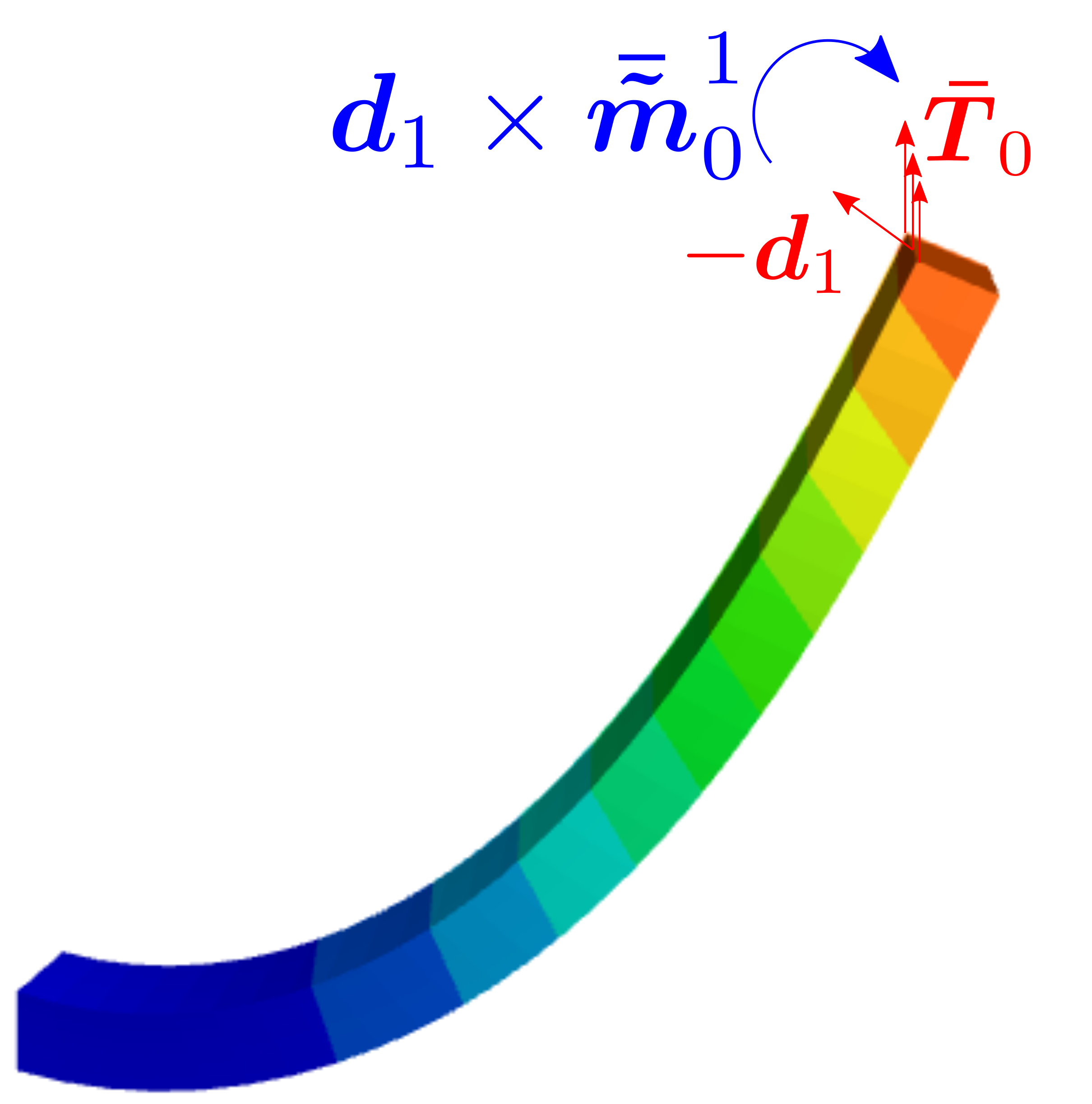}
		\caption{Beam (correct surface load)}
		\label{deform_cant45_h5_beam_cor}
	\end{subfigure}	
	\begin{subfigure}[b] {0.3\textwidth} \centering
		\includegraphics[width=\linewidth]{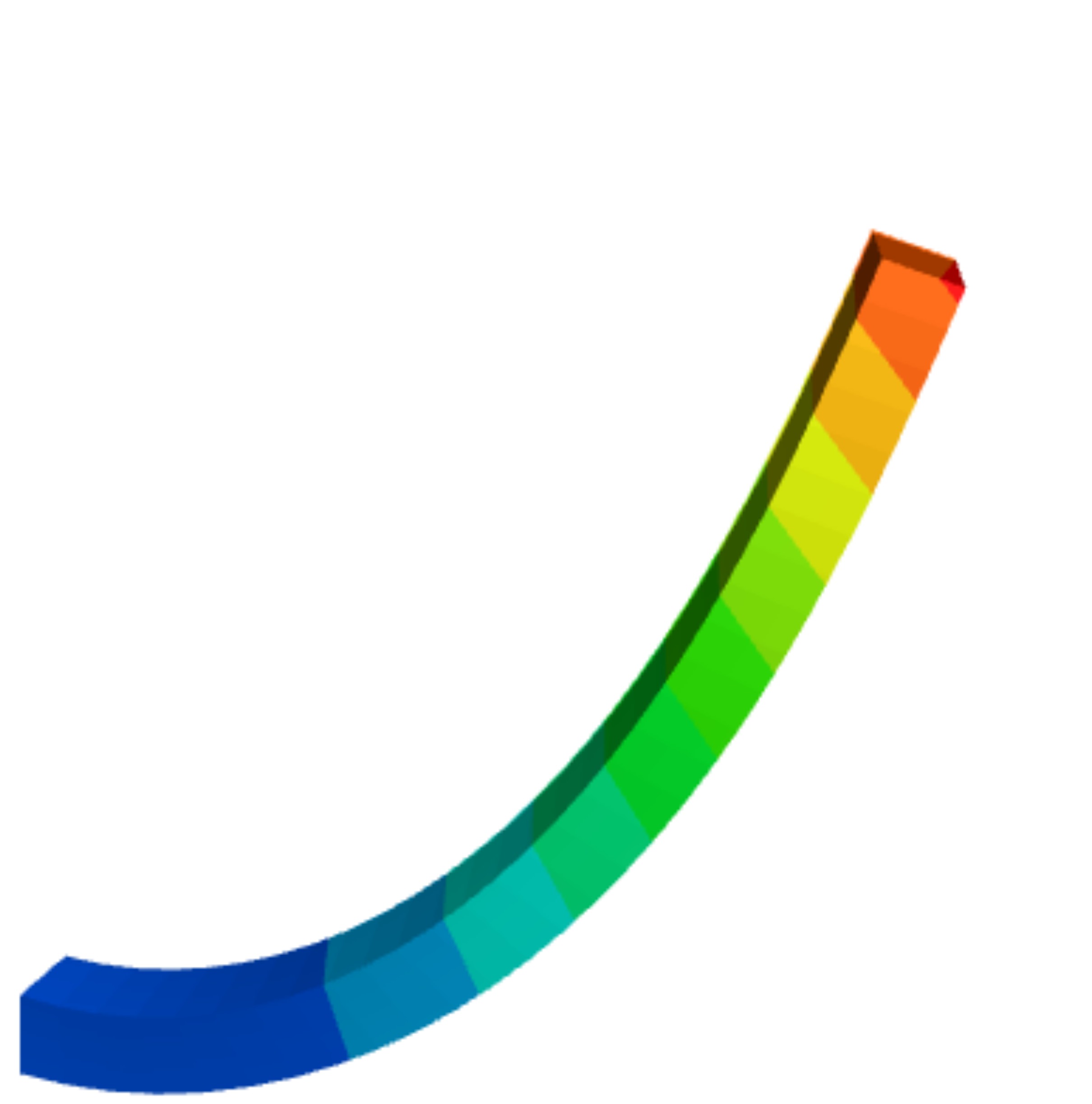}
		\caption{Beam (central axis load)}
		\label{deform_cant45_h5_beam_equiv_f}
	\end{subfigure}	
\caption{{45$^{\circ}$-arch cantilever beam: Comparison of deformed configurations. The color represents the displacement in $Z$-direction.} The brick element solution is obtained by IGA with $\mathrm{deg.}=(3,3,3)$ and $n_\mathrm{el}=240\times15\times15$, and the beam element solutions are obtained by 9 parameter EAS formulation and IGA with $p=3$ and $n_\mathrm{el}=40$.}
\label{deform_45deg_bend_deformed}	
\end{figure*}

{We verify the alleviation of Poisson locking by the EAS method, and the significance of correct surface load in a curved beam example as well. The initial beam central axis lies on the $XY$-plane and describes an $1/8$ of a full circle with radius $100\mathrm{m}$, and the cross-section has a square shape of dimension $h=w=5\mathrm{m}$. A $Z$-directional force of magnitude ${{\bar T}_0}=7.5\times{10^4}\mathrm{N/m}$ is applied on the upper edge of the end face, and the other end is clamped (see Fig.\,\ref{model_des_cant45}). We select the compressible Neo-Hookean material with Young's modulus $E=10^7\mathrm{Pa}$ and Poisson's ratio $\nu=0.3$. {For beams, we use $3\times3$ Gauss integration points for the integration over the cross-section.} The surface load ${{\boldsymbol{\bar T}}_0}=\left[0,0,{\bar T}_0\right]^\mathrm{T}$ leads to the external director stress couple ${\boldsymbol{\bar {\tilde m}}}_0^1 = {\left[ {0,0,-{{\bar T}_0}wh/2} \right]^\mathrm{T}}$, since the loaded edge is located at ${\xi^1}=-h/2$. Consequently, the following external stress couple is applied at the loaded end (see Fig.\,\ref{deform_cant45_h5_beam_cor}).
\begin{equation}
{{\bf{\bar m}}_0} \coloneqq {{\boldsymbol{d}}_\gamma } \times {\boldsymbol{\bar {\tilde m}}}_0^\gamma  = {{\boldsymbol{d}}_1} \times {\boldsymbol{\bar {\tilde m}}}_0^1\ne\boldsymbol{0}.
\end{equation}
Fig.\,\ref{cant45deg_conv_norm_disp_graph} shows the beam displacement at point A normalized by the reference solution based on brick elements with deg.=(3,3,3) and $n_{\mathrm{el}}=240\times15\times15$ (see Tables \ref{app_45deg_conv_test_ref_sol} and \ref{cant45_conv_normalized_disp} for the convergence results of the brick and beam element solutions, respectively). As expected, the results of the standard formulation (black curves), which is combined with the correct surface load condition, exhibit significantly smaller displacements due to Poisson locking, and this is improved by employing the EAS method. Since the \textit{equivalent central axis load} neglects the external director stress couple (i.e., ${\boldsymbol{\bar {\tilde m}}^1_0}=\boldsymbol{0}$), it significantly overestimates the displacement at the point A, while the beam solution based on the \textit{correct surface load} is in a very good agreement with the reference solution (see also the comparison of $Z$-displacement contours in Fig.\,\ref{deform_45deg_bend_deformed}). Further, Fig.\,\ref{cant45deg_conv_area_diff_graph} compares the relative difference of the cross-sectional area from the brick element result. It is seen that the EAS formulation more accurately captures the change of cross-sectional area, compared with the standard formulation, and the \textit{equivalent central axis load} leads to a larger difference of the change of cross-sectional area from the brick element result, compared with the result of \textit{correct surface load}. Table \ref{num_iter_45_deg} compares the total number of load steps and iterations in the iterative solution process.}

\begin{table}[]
\small
\begin{center}
\caption{{{45$^{\circ}$-arch cantilever beam: Convergence of the normalized displacements at the point A. $u_\mathrm{A}$, $v_\mathrm{A}$, and $w_\mathrm{A}$ denote the $X$-, $Y$-, and $Z$-displacements at the point A, respectively. $(\bullet)^\mathrm{ref}$ denotes the reference solution.}} All results are obtained by IGA with $p=3$.}
\label{cant45_conv_normalized_disp}
\begin{tabular}{cccclccc}
\Xhline{3\arrayrulewidth}
\multirow{2}{*}{${n_\mathrm{el}}$} & \multicolumn{3}{c}{Beam (ext.-dir.-stand.)} &  & \multicolumn{3}{c}{Beam (ext.-dir.-EAS)} \\ \cline{2-4} \cline{6-8} 
                    & ${u_\mathrm{A}}/{u_\mathrm{A}^\mathrm{ref}}$   & ${v_\mathrm{A}}/{v_\mathrm{A}^\mathrm{ref}}$  & ${w_\mathrm{A}}/{w_\mathrm{A}^\mathrm{ref}}$   &  & ${u_\mathrm{A}}/{u_\mathrm{A}^\mathrm{ref}}$    & ${v_\mathrm{A}}/{v_\mathrm{A}^\mathrm{ref}}$   & ${w_\mathrm{A}}/{w_\mathrm{A}^\mathrm{ref}}$         \\ 
\Xhline{3\arrayrulewidth}
5                   & 9.6802E-01  & 9.3753E-01  & 9.7859E-01 &  & 9.9660E-01  & 9.9514E-01  & 9.9583E-01 \\
10                  & 9.6783E-01  & 9.4193E-01  & 9.8154E-01 &  & 9.9633E-01  & 1.0018E+00  & 1.0002E+00 \\
20                  & 9.6785E-01  & 9.4222E-01  & 9.8178E-01 &  & 9.9635E-01  & 1.0021E+00  & 1.0005E+00 \\
40                  & 9.6785E-01  & 9.4224E-01  & 9.8181E-01 &  & 9.9635E-01  & 1.0022E+00  & 1.0005E+00 \\ 
\Xhline{3\arrayrulewidth}
\end{tabular}
\end{center}
\end{table}
\begin{table}[]
\caption{{45$^{\circ}$-arch cantilever beam: Comparison of the total number of load steps and iterations. In each formulation, the load is uniformly incremented.}}
\label{num_iter_45_deg}
\begin{center}
\begin{tabular}{cccc}
\Xhline{3\arrayrulewidth}
                                 & Brick element                                    & Beam (ext.-dir.-std.)            & Beam (ext.-dir.-EAS)             \\
                                 & ${\mathrm{deg.}=(3,3,3), n_\mathrm{el}=200\times15\times15}$ & ${p=3, n_\mathrm{el}=40}$ & $p=3$, $n_\mathrm{el}=40$\\ 
\Xhline{3\arrayrulewidth}
\multicolumn{1}{c}{\#load steps} & 10                                               & 10                               & 10                               \\
\multicolumn{1}{c}{\#iterations} & 80                                              & 74 & 80                              \\ 
\Xhline{3\arrayrulewidth}
\end{tabular}
\end{center}
\end{table}

\section{Conclusions}
In this paper, we present an isogeometric finite element formulation of geometrically exact Timoshenko beams with extensible directors. The presented beam formulation has the following advantages.
\begin{itemize}
\item The extensible director vectors allow for the accurate and efficient description of in-plane cross-sectional deformations.
\item They belong to the space $\Bbb R^3$, so that the configuration can be additively updated. 
\item {In order to alleviate Poisson locking, the complete in-plane strain field has been added in the form of incompatible modes by the EAS method.}
\item The formulation does not require the zero stress assumption in the constitutive law, and offers a straightforward interface to employ general three-dimensional constitutive law like hyperelasticity. 
\item In the analysis of beams, the external load is often assumed to be directly applied to the central axis. It is shown that this equivalent central axis load leads to significant error. 
\item We verify the accuracy and efficiency of the developed beam formulation by comparison with the results of brick elements.
\end{itemize}
The following areas could be interesting future research directions.
\begin{itemize}
\item {Incorporation of out-of-plane deformation of cross-sections: In this paper, cross-section warping has not been considered, which restricts the range of application to compact convex cross-sections, where the out-of-plane cross-sectional deformation is less pronounced. In order to consider open and thin-walled cross-sections, one can incorporate cross-section warping by employing an additional degree-of-freedom as in \citet{simo1991geometrically}, \citet{gruttmann1998geometrical}, and \citet{coda2009solid}.
One can also refer to the works of \citet{wackerfuss2009mixed, wackerfuss2011nonlinear}, where additional strain and stress parameters are eliminated at the element level, so that the finite element formulation finally has three translation and three rotational degrees-of-freedom per node.}
\item Incorporation of \textit{exact} geometry of initial boundary surface of beams with non-uniform cross-section: {Eq.\,(\ref{inf_area_lat_bd_surf}) can be applied to non-uniform cross-sections along the central axis.} IGA has an advantage over conventional finite element formulation in perspective of straightforwardly utilizing \textit{exact} geometrical information of the initial boundary NURBS surface.  
\item {Numerical instability in high slenderness ratio: There are several factors that limit the slenderness ratio within which the presented beam formulation can be properly utilized. First, the coupling between bending and through-the-thickness stretching can lead to an ill-conditioned stiffness matrix in the thin beam limit (see section \ref{instab_thin_b_lim_end_mnt}). These issues can be alleviated by existing techniques such as a multiplicative decomposition of directors in \citet{simo1990stress} and mass scaling techniques in \citet{hokkanen2019isogeometric}. Second, a mixed variational formulation or an optimal quadrature rule of reduced integration needs to be developed to alleviate membrane, shear, and curvature-thickness locking. Since the developed beam formulation has additional degrees-of-freedom for the in-plane cross-sectional deformation and the numerical integration over the cross-section, it may not be straightforward to directly employ existing reduced integration methods like the recent development in \citet{zou2021galerkin}.}
\item Enforcement of rotation continuity at intersections with slope discontinuity: As the developed beam formulation does not rely on rotational degrees-of-freedom, describing rigid joint connections between multiple beams becomes a challenge. A selective introduction of rotational degrees-of-freedom associated with the variation (increment) of director vectors can be utilized.
\item Beam contact problems: One can investigate the advantages of incorporating the transverse normal strain in beam contact problems. For example, the coupling between transverse normal stretching and bending deformations was illustrated in \citet{naghdi1989significance}.
\item Incompressible and nearly incompressible hyperelastic materials: One can extend the presented formulation to incorporate incompressibility constraint.  
\item Strain objectivity and energy-momentum conservation: It has been shown in several works including \citet{romero2002objective}, \citet{betsch2002frame}, and \citet{eugster2014director}
that the direct interpolation of director fields satisfies the objectivity of strain measures. Furthermore, this can facilitate the straightforward application of time integration schemes with energy-momentum conservation\footnote{Refer to the comments in \citet{eugster2014director} and references therein.}. An in-depth discussion on the objectivity and energy-momentum conservation property in the developed beam formulation is planned for subsequent work.
Although a relevant numerical study on the objectivity, path-independence, and energy-momentum conservation was performed in \citet{coda2009solid} and \citet{coda2011fem} based on beam kinematics with extensible directors, further investigation including an analytical verification seems still necessary.
\end{itemize}
\nopagebreak[4] 
\appendix
\gdef\thesection{\Alph{section}} 
\makeatletter
\counterwithin{figure}{section}
\counterwithin{table}{section}
\renewcommand\@seccntformat[1]{\csname the#1\endcsname.\hspace{0.5em}}
\makeatother
\counterwithin*{equation}{subsection}
\renewcommand{\theequation}{\thesubsection.\arabic{equation}}
\section{Appendix to the beam formulation}	
\label{app_theory}
\subsection{Jacobians of the mappings $\boldsymbol{x}_0$ and $\boldsymbol{x}_t$}
\label{deriv_jacob}
We recall the following Piola identity for a linear transformation ${\bf{A}}\in {\Bbb{R}}^{m\times n}$ and vectors ${\bf{a}},\,{\bf{b}}\in {\Bbb{R}}^{n}$
\begin{align}\label{piola_identiy_recall}
{\bf{Aa}} \times {\bf{Ab}} = \left( {\det {\bf{A}}} \right){{\bf{A}}^{ - {\mathrm{T}}}}\left( {{\bf{a}} \times {\bf{b}}} \right).
\end{align}
Then, using Eq.\,(\ref{piola_identiy_recall}), the triple product of the covariant base vectors in the current configuration can be expressed by
\begin{align}\label{triple_product_cov_bases_cur_config}
\left({{\boldsymbol{g}}_1} \times {{\boldsymbol{g}}_2}\right) \cdot{{\boldsymbol{g}}_3} = \det {D{{\boldsymbol{x}}_t}}\left( {D{{\boldsymbol{x}}_t}{{\boldsymbol{E}}_3}} \right) \cdot \left({{D{{\boldsymbol{x}}_t}}^{ - {\mathrm{T}}}}{{\boldsymbol{E}}_3}\right)= \det {D{{\boldsymbol{x}}_t}} \eqqcolon{j_t}.
\end{align}
In the same way, it can be easily shown that 
\begin{align}\label{triple_product_cov_bases_cur_config}
\left({{\boldsymbol{G}}_1} \times {{\boldsymbol{G}}_2}\right)\cdot{{\boldsymbol{G}}_3}= \det {D{{\boldsymbol{x}}_0}}\eqqcolon{j_0}.
\end{align}

\subsection{Derivation of momentum balance equations for the beam}
\label{deriv_bal_eq}
\subsubsection{Linear and director momentum balance}
\label{deriv_bal_lin_momentum}
The divergence of the Cauchy stress tensor can be expressed in terms of the basis $\left\{{{\boldsymbol{g}^1},{\boldsymbol{g}^2},{\boldsymbol{g}^3}}\right\}$, as
\begin{align}\label{strs_div_curv_basis}
{\text{div}}{\boldsymbol{\sigma }} = \frac{1}{{{j_t}}}{\left( {{j_t}{\boldsymbol{\sigma }}{{\boldsymbol{g}}^i}} \right)_{\!,i}},
\end{align}
where $(\bullet)_{,i}$ denotes the partial differentiation with respect to $\xi_i$. For a detailed derivation one can refer to chapter 10 of \,\citet{zienkiewicz2014finite}. Then, the local form of the balance of linear momentum given in Eq.\,(\ref{conserv_lin_mnt_intrinsic}) can be rewritten as
\begin{align}\label{conserv_lin_mnt_curvilinear}
{\left( {{j_t}{\boldsymbol{\sigma }}{{\boldsymbol{g}}^i}} \right)_{\!,i}} + {{{j_t}}}{\boldsymbol{b}} = {{{j_t}}}\,{{\rho_t}\,{\boldsymbol{x}}_{t,tt}}.
\end{align}
From Eq.\,(\ref{conserv_lin_mnt_curvilinear}), using the principle of virtual work, the variational identity follows
\begin{align}\label{conserv_lin_mnt_mult_del_x}
\int_{\mathcal{B}} {\delta {{\boldsymbol{x}}_t} \cdot {{\left({j_t}{\boldsymbol{\sigma }}{{\boldsymbol{g}}^i}\right)}_{\!,i}}\,{\mathrm{d}}{\mathcal{B}}}  = \int_{\mathcal{B}} {{j_t}\,{\rho _t}\,\delta {{\boldsymbol{x}}_t} \cdot {{\boldsymbol{x}}_{t,tt}}\,{\mathrm{d}}{\mathcal{B}}}  - \int_{\mathcal{B}} {{j_t}\,\delta {{\boldsymbol{x}}_t} \cdot {\boldsymbol{b}}\,{\mathrm{d}}{\mathcal{B}}}.
\end{align}
Then, using the divergence theorem and Eq.\,(\ref{init_surf_transform}), we have
\begin{align}\label{transformation_area_vec}
\int_{\mathcal{B}} {{{\left( {\delta {{\boldsymbol{x}}_t} \cdot {j_t}{\boldsymbol{\sigma }}{{\boldsymbol{g}}^i}} \right)}_{\!,i}}{\mathrm{d}}{\mathcal{B}}} = \int_{\mathcal{S}} {\left( {\delta {{\boldsymbol{x}}_t} \cdot {j_t}{\boldsymbol{\sigma }}D{{\boldsymbol{x}}_t}^{ - {\mathrm{T}}}{\boldsymbol{\nu}}} \right){\mathrm{d}}{\mathcal{S}}} = \int_{{{\mathcal{S}}_0}} {\left( {\delta {{\boldsymbol{x}}_t} \cdot {{\boldsymbol{P}}}{{\boldsymbol{\nu }}_0}} \right){\mathrm{d}}{{\mathcal{S}}_0}}.
\end{align}
After applying the boundary conditions of Eqs.\,(\ref{solid_elas_disp_bdc_homo}) and (\ref{solid_elas_traction_bdc_init}), we have
\begin{align}\label{transformation_area_vec}
\int_{\mathcal{B}} {{{\left( {\delta {{\boldsymbol{x}}_t} \cdot {j_t}{\boldsymbol{\sigma }}{{\boldsymbol{g}}^i}} \right)}_{\!,i}}{\mathrm{d}}{\mathcal{B}}} = \int_{{{\mathcal{S}}^{\mathrm{N}}_0}} {\left( {\delta {{\boldsymbol{x}}_t} \cdot {\bar{\boldsymbol{T}}_0}} \right){\mathrm{d}}{{\mathcal{S}}^{\mathrm{N}}_0}}.
\end{align}
Then, the left-hand side of Eq.\,(\ref{conserv_lin_mnt_mult_del_x}) can be rewritten as
\begin{align}\label{LHS_rewrite}
\int_{\mathcal{B}} {\delta {{\boldsymbol{x}}_t} \cdot {{\left({j_t}{\boldsymbol{\sigma }}{{\boldsymbol{g}}^i}\right)}_{\!,i}}{\mathrm{d}}{\mathcal{B}}}  = \int_{{{\mathcal{S}}^{\mathrm{N}}_0}} {\left( {\delta {{\boldsymbol{x}}_t} \cdot {\bar{\boldsymbol{T}}_0}} \right){\mathrm{d}}{{\mathcal{S}}^{\mathrm{N}}_0}}- \int_{\mathcal{B}} {\left( {\delta {{\boldsymbol{x}}_{t,s}} \cdot {j_t}{\boldsymbol{\sigma }}{{\boldsymbol{g}}^3}} \right){\mathrm{d}}{\mathcal{B}}}  - \int_{\mathcal{B}} {\left( {\delta {{\boldsymbol{x}}_{t,\alpha }} \cdot {j_t}{\boldsymbol{\sigma }}{{\boldsymbol{g}}^\alpha }} \right){\mathrm{d}}{\mathcal{B}}}.
\end{align}
From the beam kinematics in Eq.\,(\ref{beam_th_str_cur_config}), we obtain the following relations
\begin{alignat}{2}
\label{del_x_t_deriv}
\left.\begin{array}{c}
\begin{aligned}
\delta {{\boldsymbol{x}}_t} &= \delta {{\boldsymbol{\varphi }}} + {\xi ^\gamma }\delta {{\boldsymbol{d}}_\gamma },\\
\delta {{\boldsymbol{x}}_{t,s}} &= \delta {{\boldsymbol{\varphi }}_{,s}} + {\xi ^\gamma }\delta {{\boldsymbol{d}}_{\gamma ,s}},\\
\delta {{\boldsymbol{x}}_{t,\alpha }} &= \delta {{\boldsymbol{d}}_\alpha }.
\end{aligned}
\end{array}\right\}
\end{alignat}
Rearranging terms after substituting Eq.\,(\ref{del_x_t_deriv}) into Eq.\,(\ref{LHS_rewrite}) and using Eq.\,(\ref{inf_area_lat_bd_surf}), we have
\begin{align}\label{lin_momentum_bal_LHS}
\int_{\mathcal{B}} {\delta {{\boldsymbol{x}}_t} \cdot {{({j_t}{\boldsymbol{\sigma }}{{\boldsymbol{g}}^i})}_{\!,i}}{\mathrm{d}}{\mathcal{B}}} &=  - \int_0^L {\left( {\delta {{\boldsymbol{\varphi }}_{,s}} \cdot {\boldsymbol{n}} + \delta {{\boldsymbol{d}}_{\gamma ,s}} \cdot {{{\boldsymbol{\tilde m}}}^\gamma } + \delta {{\boldsymbol{d}}_\alpha } \cdot {{\boldsymbol{l}}^\alpha }} \right){\mathrm{d}}s} \nonumber\\
&+ \int_0^L {\left( {\delta {{\boldsymbol{\varphi }}} \cdot {\int_{\partial {{\mathcal{A}}_0}} {{{\bar{\boldsymbol{T}}_0}}{\mathrm{d}}{{\Gamma}_0}}}} + {\delta {{\boldsymbol{d}}_\gamma } \cdot {\int_{\partial {{\mathcal{A}}_0}} {{\xi ^\gamma }{{\bar{\boldsymbol{T}}_0}}{\mathrm{d}}{{\Gamma}_0}}}} \right){\mathrm{d}}s} \nonumber\\
&+ {\left[ {\delta {{\boldsymbol{\varphi }}} \cdot {{\int_{{{\mathcal{A}}_0}} {{\bar{\boldsymbol{T}}_0}\,{\mathrm{d}}{{\mathcal{A}}_0}} }} + \delta {{\boldsymbol{d}}_\gamma } \cdot {\int_{{\mathcal{A}}_0} {{{\xi ^\gamma }{{\bar{\boldsymbol{T}}_0}}}\,{\mathrm{d}}{{\mathcal{A}}_0}} } } \right]_{s\in{\Gamma_{\mathrm{N}}}}}.
\end{align}
Furthermore, substituting Eq.\,(\ref{del_x_t_deriv}) into the first and second terms of the right-hand side of Eq.\,(\ref{conserv_lin_mnt_mult_del_x}), respectively, we have
\begin{align}\label{RHS_inertia_res}
\int_{\mathcal{B}} {{\rho _t}{j_t}\delta {{\boldsymbol{x}}_t} \cdot {{\boldsymbol{x}}_{t,tt}}\,{\mathrm{d}}{\mathcal{B}}} = \int_0^L {\left( {\delta {\boldsymbol{\varphi }} \cdot {\rho_A}{{\boldsymbol{\varphi }}_{\!,tt}} + \delta {{\boldsymbol{d}}_\gamma } \cdot I_\rho ^{\gamma \delta }{{\boldsymbol{d}}_{\delta ,tt}}} \right){\mathrm{d}}s},
\end{align}
and
\begin{align}\label{RHS_body_force_res}
\int_{\mathcal{B}} {{j_t}\delta {{\boldsymbol{x}}_t} \cdot {\boldsymbol{b}}\,{\mathrm{d}}{\mathcal{B}}}  = \int_0^L {\left( {\delta {\boldsymbol{\varphi }} \cdot \int_{\mathcal{A}} {{j_t}{\boldsymbol{b}}\,{\mathrm{d}}{\mathcal{A}}} } \right){\mathrm{d}}s}  + \int_0^L {\left({\delta {{\boldsymbol{d}}_\gamma } \cdot \int_{\mathcal{A}} {{j_t}{\xi ^\gamma }{\boldsymbol{b}}\,{\mathrm{d}}{\mathcal{A}}} } \right){\mathrm{d}}s}.
\end{align}
Finally, substituting Eqs.\,(\ref{lin_momentum_bal_LHS})-(\ref{RHS_body_force_res}) into Eq.\,(\ref{conserv_lin_mnt_mult_del_x}) and rearranging terms yields
\begin{align}\label{re_var_identity_1}
&\int_0^L {\left( {\delta {{\boldsymbol{\varphi }}_{\!,s}} \cdot {\boldsymbol{n}} + \delta {{\boldsymbol{d}}_{\gamma ,s}} \cdot {{{\boldsymbol{\tilde m}}}^\gamma } + \delta {{\boldsymbol{d}}_\alpha } \cdot {{\boldsymbol{l}}^\alpha }} \right){\mathrm{d}}s}  + \int_0^L {\left( {\delta {\boldsymbol{\varphi }} \cdot {\rho_A}{{\boldsymbol{\varphi }}_{\!,tt}} + \delta {{\boldsymbol{d}}_\gamma } \cdot I_\rho ^{\gamma \delta }{{\boldsymbol{d}}_{\delta ,tt}}} \right){\mathrm{d}}s} \nonumber\\
&= \int_0^L {\left( {\delta {{\boldsymbol{\varphi }}} \cdot {\boldsymbol{\bar n}} + \delta {{\boldsymbol{d}}_\gamma } \cdot {{{\boldsymbol{\bar {\tilde m}}}}^\gamma }} \right){\mathrm{d}}s} + {\left[ {\delta {{\boldsymbol{\varphi }}} \cdot {{{\boldsymbol{\bar n}}}_0} + \delta {{\boldsymbol{d}}_\gamma } \cdot {\boldsymbol{\bar {\tilde m}}}_0^\gamma } \right]_{s\in{\Gamma _{\mathrm{N}}}}},
\end{align}
where the prescribed stress resultant and stress couple are defined as
\begin{equation}
\label{app_nbdc_strs_res}
{{\boldsymbol{\bar n}}_0} \coloneqq {\left[ {\int_{{\mathcal{A}}_0} {{\boldsymbol{\bar T}_0}\,{\mathrm{d}}{{\mathcal{A}}_0}} } \right]_{s \in {\Gamma_{\mathrm{N}}}}},
\end{equation}
and
\begin{equation}
\label{app_nbdc_strs_coup}
{\boldsymbol{\bar {\tilde m}}}_0^\gamma  \coloneqq {\left[ {\int_{{\mathcal{A}}_0} {{{\xi ^\gamma }\boldsymbol{\bar T}_0}\,{\mathrm{d}}{{\mathcal{A}}_0}} } \right]_{s \in {\Gamma _{\mathrm{N}}}}},
\end{equation}
and the distributed external resultant stress and stress couple are defined as
\begin{equation}
{\boldsymbol{\bar n}} \coloneqq \int_{\partial {{\mathcal{A}}_0}} {{{{\boldsymbol{\bar T}}}_0}{\mathrm{d}}{{\Gamma}_0}}  + \int_{{{\mathcal{A}}}} {{\boldsymbol{b}_0}{j_0}{\mathrm{d}}{{\mathcal{A}}}},
\end{equation}
and
\begin{equation}
{{\boldsymbol{\bar {\tilde m}}}^\gamma } \coloneqq \int_{\partial {{\mathcal{A}}_0}} {{\xi ^\gamma }{{{\boldsymbol{\bar T}}}_0}{\mathrm{d}}{{\Gamma}_0}}  + \int_{{{\mathcal{A}}}} {{\xi ^\gamma }{\boldsymbol{b}_0}{j_0}{\mathrm{d}}{{\mathcal{A}}}}.
\end{equation}
Applying integration by parts and the homogeneous displacement boundary condition to the first term on the left-hand side of Eq.\,(\ref{re_var_identity_1}) yields
\begin{align}
&\int_0^L {\left\{ {\delta {{\boldsymbol{\varphi }}} \cdot \left( {{{\boldsymbol{n}}_{,s}} + {\boldsymbol{\bar n}} - {\rho_A}{{\boldsymbol{\varphi }}_{\!,tt}}} \right) + \delta {{\boldsymbol{d}}_\gamma } \cdot \left( {{\boldsymbol{\tilde m}}_{,s}^\gamma  + {{{\boldsymbol{\bar {\tilde m}}}}^\gamma } - {{\boldsymbol{l}}^\gamma} - I_\rho ^{\gamma \delta }{{\boldsymbol{d}}_{\delta ,tt}}} \right)} \right\}{\mathrm{d}}s} \nonumber\\
&+ {\left[ {\delta {{\boldsymbol{\varphi}}} \cdot \left( {{\boldsymbol{n}} - {{{\boldsymbol{\bar n}}}_0}} \right) + \delta {{\boldsymbol{d}}_\gamma } \cdot \left( {{{{\boldsymbol{\tilde m}}}^\gamma } - {\boldsymbol{\bar {\tilde m}}}_0^\gamma } \right)} \right]_{s\in{\Gamma _{\mathrm{N}}}}} = 0.
\end{align}
Then, applying the standard localization theorem, we finally obtain the local momentum balance equations under the boundary conditions: We find $\left( {{\boldsymbol{\varphi }},{{\boldsymbol{d}}_1},{{\boldsymbol{d}}_2}} \right) \in {{\Bbb{R}}^3} \times {{\Bbb{R}}^3} \times {{\Bbb{R}}^3}$ such that
\begin{subequations}
\begin{align}
{{\boldsymbol{n}}_{,s}} + {\boldsymbol{\bar n}} &= {\rho_A}{{\boldsymbol{\varphi }}_{\!,tt}},\\
{\boldsymbol{\tilde m}}_{,s}^\gamma  - {{\boldsymbol{l}}^\gamma } + {{{\boldsymbol{\bar {\tilde m}}}}^\gamma } &= I_\rho ^{\gamma \delta }{{\boldsymbol{d}}_{\delta ,tt}}\quad(\gamma=1,2),
\end{align}
\end{subequations}
with the Neumann boundary conditions
\begin{equation}
{\boldsymbol{n}} = {{\boldsymbol{\bar n}}_0}\,\,\text{and}\,\,{{\boldsymbol{\tilde m}}^\gamma } = {\boldsymbol{\bar {\tilde m}}}_0^\gamma\,\,\text{at}\,\,s \in {\Gamma _{\mathrm{N}}},
\end{equation}
and the Dirichlet boundary conditions
%
\begin{align}
{\boldsymbol{\varphi }} = {{{\boldsymbol{\bar \varphi }}}_0},\,\,{{\boldsymbol{d}}^\gamma } = {\boldsymbol{\bar d}}_0^\gamma\,\,\mathrm{at}\,\,s \in {\Gamma _{\mathrm{D}}}.	
\end{align}
\subsubsection{Angular momentum balance}
\label{deriv_bal_ang_momentum}
The symmetry of the Cauchy stress tensor implies\,\citep{simo1989stress}
\begin{equation}\label{sym_cauchy_strs_1}
{{\boldsymbol{g}}_i} \times {\boldsymbol{\sigma }}{{\boldsymbol{g}}^i} = {\boldsymbol{0}}.
\end{equation}
Integration of Eq.\,(\ref{sym_cauchy_strs_1}) over $\mathcal{A}$ after multiplying the Jacobian $j_t$ and using Eqs.\,(\ref{beam_th_str_cur_config}) and (\ref{beam_th_str_cur_cov_base}) yields
\begin{align}
\int_{\mathcal A} {{{\boldsymbol{g}}_i} \times {\boldsymbol{\sigma }}{{\boldsymbol{g}}^i}{j_t}\,{\mathrm{d}}{\mathcal{A}}} = {{\boldsymbol{\varphi }}_{\!,s}} \times \int_{\mathcal{A}} {{\boldsymbol{\sigma }}{{\boldsymbol{g}}^3}{j_t}\,{\mathrm{d}}{\mathcal{A}}}  + {{\boldsymbol{d}}_{\gamma ,s}} \times \int_{\mathcal{A}} {{\xi ^\gamma }{\boldsymbol{\sigma }}{{\boldsymbol{g}}^3}{j_t}\,{\mathrm{d}}{\mathcal{A}}}  + {{\boldsymbol{d}}_\gamma } \times\int_{\mathcal{A}} {{\boldsymbol{\sigma }}{{\boldsymbol{g}}^\gamma }{j_t}\,{\mathrm{d}}{\mathcal{A}}}.
\end{align}
Thus, the local angular momentum balance equation is expressed as
\begin{align}
{{\boldsymbol{\varphi }}_{\!,s}} \times {\boldsymbol{n}} + {{\boldsymbol{d}}_{\gamma ,s}} \times {{\boldsymbol{\tilde m}}^\gamma } + {{\boldsymbol{d}}_\gamma } \times {{\boldsymbol{l}}^\gamma } = {\bf{0}}.
\end{align}
\subsection{Constitutive equation}
\subsubsection{The first variation of the strain energy density function}
\label{1st_var_strn_e_M_mat}
The first variation of $\boldsymbol{\munderbar E}$ can be expressed as
\begin{align}
\delta \boldsymbol{\munderbar E} ={\boldsymbol{\munderbar D}}\delta {\boldsymbol{\munderbar \varepsilon }},
\end{align}
where
\begin{equation}
{\boldsymbol{\munderbar D}}\coloneqq\frac{{\partial {\boldsymbol{\munderbar E}}}}{{\partial {\boldsymbol{\munderbar \varepsilon }}}}=\left[ {\begin{array}{*{20}{c}}
0&0&0&0&0&0&0&0&0&0&0&0&1&0&0\\
0&0&0&0&0&0&0&0&0&0&0&0&0&1&0\\
1&{{\xi ^1}}&{{\xi ^2}}&{{\xi ^1}{\xi ^1}}&{{\xi ^2}{\xi ^2}}&{{\xi ^1}{\xi ^2}}&0&0&0&0&0&0&0&0&0\\
0&0&0&0&0&0&0&0&0&0&0&0&0&0&1\\
0&0&0&0&0&0&1&0&{{\xi ^1}}&{{\xi ^2}}&0&0&0&0&0\\
0&0&0&0&0&0&0&1&0&0&{{\xi ^1}}&{{\xi ^2}}&0&0&0
\end{array}} \right].
\end{equation}
\subsubsection{Compressible Neo-Hookean material}
\label{app_constitutive_nh}
Here we obtain the contravariant components of the fourth order tensor ${\boldsymbol{\mathcal{I}}}$. In the standard Cartesian basis $\left\{ {{{\boldsymbol{e}}_1},{{\boldsymbol{e}}_2},{{\boldsymbol{e}}_3}} \right\}$, we have \citep{bonet2010nonlinear}
\begin{equation}
{\boldsymbol{\mathcal{I}}} \coloneqq  - \frac{{\partial {{\boldsymbol{C}}^{ - 1}}}}{{\partial {\boldsymbol{C}}}} = \frac{1}{2}\left\{ {{{\left( {{{\boldsymbol{C}}^{ - 1}}} \right)}_{\!AC}}{{\left( {{{\boldsymbol{C}}^{ - 1}}} \right)}_{\!BD}} + {{\left( {{{\boldsymbol{C}}^{ - 1}}} \right)}_{\!AD}}{{\left( {{{\boldsymbol{C}}^{ - 1}}} \right)}_{\!BC}}} \right\}{{\boldsymbol{e}}_A} \otimes {{\boldsymbol{e}}_B} \otimes {{\boldsymbol{e}}_C} \otimes {{\boldsymbol{e}}_D}.
\end{equation}
The contravariant components are thus obtained by 
\begin{align}
{I^{ijk\ell }} &= {\boldsymbol{\mathcal{I}}}::{{\boldsymbol{G}}^i} \otimes {{\boldsymbol{G}}^j} \otimes {{\boldsymbol{G}}^k} \otimes {{\boldsymbol{G}}^\ell }\nonumber\\
 &= \frac{1}{2}\left\{ {{{\left( {{{\boldsymbol{C}}^{ - 1}}} \right)}_{\!AC}}{{\left( {{{\boldsymbol{C}}^{ - 1}}} \right)}_{\!BD}} + {{\left( {{{\boldsymbol{C}}^{ - 1}}} \right)}_{\!AD}}{{\left( {{{\boldsymbol{C}}^{ - 1}}} \right)}_{\!BC}}} \right\}( {{{\boldsymbol{e}}_A} \cdot {{\boldsymbol{G}}^i}})( {{{\boldsymbol{e}}_B} \cdot {{\boldsymbol{G}}^j}})( {{{\boldsymbol{e}}_C} \cdot {{\boldsymbol{G}}^k}} )( {{{\boldsymbol{e}}_D} \cdot {{\boldsymbol{G}}^\ell }})\nonumber\\
 &= \frac{1}{2}\left\{\left( {{{\boldsymbol{G}}^i} \cdot {{\boldsymbol{C}}^{ - 1}}{{\boldsymbol{G}}^k}} \right)\left( {{{\boldsymbol{G}}^j} \cdot {{\boldsymbol{C}}^{ - 1}}{{\boldsymbol{G}}^\ell }} \right) + \left( {{{\boldsymbol{G}}^i} \cdot {{\boldsymbol{C}}^{ - 1}}{{\boldsymbol{G}}^\ell }} \right)\left( {{{\boldsymbol{G}}^j} \cdot {{\boldsymbol{C}}^{ - 1}}{{\boldsymbol{G}}^k}} \right)\right\}\nonumber\\
 &= \frac{1}{2}\left\{{\left( {{{\boldsymbol{C}}^{ - 1}}} \right)^{ik}}{\left( {{{\boldsymbol{C}}^{ - 1}}} \right)^{j\ell }} + {\left( {{{\boldsymbol{C}}^{ - 1}}} \right)^{i\ell }}{\left( {{{\boldsymbol{C}}^{ - 1}}} \right)^{jk}}\right\}.
\end{align}
\subsection{Linearization of the weak form}
\label{lin_weak_form}
\label{pdisp_linearize_sec}
\subsubsection{Configuration update}
We employ the Newton-Raphson method to solve the nonlinear equation of Eq.\,(\ref{beam_var_eq_balance_eq}). An external load is incrementally applied, and the solution at the $(n+1)$th load step is found based on the equilibrium at the previous $n$th load step. The following steps are repeated until a given convergence criterion is satisfied. The iterative scheme to find solution ${}^{n + 1}{\boldsymbol{y}} \coloneqq \left[ {{}^{n + 1}{\boldsymbol{\varphi }}^{\mathrm{T}},{}^{n + 1}{{\boldsymbol{d}}_1}^{\mathrm{T}},{}^{n + 1}{{\boldsymbol{d}}_2}^{\mathrm{T}}} \right]^{\mathrm{T}}$ is stated as: For a given solution ${}^{n + 1}{{\boldsymbol{y}}^{(i - 1)}}$ at the $(i-1)$th iteration of the $(n+1)$th load step, find the solution increment $\Delta {{\boldsymbol{y}}} \coloneqq \left[ \Delta {{\boldsymbol{\varphi }}}^{\mathrm{T}},\Delta {{{\boldsymbol{d}}_1}}^{\mathrm{T}},\Delta {{{\boldsymbol{d}}_2}}^{\mathrm{T}} \right]^{\mathrm{T}} \in \mathcal{V}$ such that 
\begin{align}
\label{new_config_update_lin_eq}
{{\Delta}G}\left( {{}^{n + 1}{{\boldsymbol{y}}^{(i - 1)}};\delta {{\boldsymbol{y}}},\Delta {\boldsymbol{y}}} \right)= {G_\mathrm{ext}}\left({}^{n + 1}{{\boldsymbol{y}}^{(i - 1)}},{\delta {\boldsymbol{y}}} \right) - {G_{{\mathop{\rm int}} }}\left( {{}^{n + 1}{{\boldsymbol{y}}^{(i - 1)}},\delta {\boldsymbol{y}}} \right),\,{\,\forall }\delta {\boldsymbol{y}} \in \mathcal{V},
\end{align}
and we update the configuration by
\begin{alignat}{3}
\left.\begin{array}{c}
\begin{aligned}
{}^{n + 1}{{\boldsymbol{\varphi }}^{(i)}} &= {}^{n + 1}{{\boldsymbol{\varphi }}^{(i - 1)}} + \Delta {{\boldsymbol{\varphi }}},&{}^{n + 1}{{\boldsymbol{\varphi }}^{(0)}} &= {}^n{\boldsymbol{\varphi }},\\
{}^{n + 1}{{\boldsymbol{d}}_\gamma }^{(i)} &= {}^{n + 1}{{\boldsymbol{d}}_\gamma }^{(i - 1)} + \Delta {{\boldsymbol{d}}_\gamma },&{}^{n + 1}{{\boldsymbol{d}}_\gamma }^{(0)}&= {}^n{{\boldsymbol{d}}_\gamma }.
\end{aligned}
\end{array}\right\}
\end{alignat}
${\Delta}G\coloneqq{\Delta}G_{{\mathop{\rm int}}}-{\Delta}G_{{\mathop{\rm ext}}}$ represents the \textit{tangent stiffness}, and the first part ${\Delta}G_{{\mathop{\rm int}}}$ is given by the linearization of the internal virtual work in the following. The second part ${\Delta}G_{{\mathop{\rm ext}}}$ is the \textit{load stiffness} which appears, e.g., due to a non-conservative load (see for example Eq.\,(\ref{pure_bend_vir_work_lin})).
\subsubsection{Operator expressions}
We define the following strain operators in order to have compact forms for the virtual strains of Eqs.\,(\ref{beam_var_strains}) and (\ref{def_strain_var_kappa}).
\begingroup
\allowdisplaybreaks
\begin{subequations}
\label{beam_virtual_strain_compact_form}
\begin{alignat}{3}
\delta \varepsilon  &= {\left[ {\begin{array}{*{20}{c}}
{{{\boldsymbol{\varphi }}_{\!,s}}^{\mathrm{T}}{{(\bullet)}_{,s}}}&{{{\bf{0}}_{1 \times 6}}}
\end{array}} \right]_{1 \times 9}}\delta {\boldsymbol{y}}&&\eqqcolon {{\Bbb{B}}_{\varepsilon}}\delta {\boldsymbol{y}},\\
\delta {\boldsymbol{\rho }} &= {\left[ {\begin{array}{*{20}{c}}
{{{\boldsymbol{d}}_{1,s}}^{\mathrm{T}}{{(\bullet)}_{,s}}}&{{{\boldsymbol{\varphi }}_{\!,s}}^{\mathrm{T}}{{(\bullet)}_{,s}}}&{{{\bf{0}}^{\mathrm{T}}}}\\
{{{\boldsymbol{d}}_{2,s}}^{\mathrm{T}}{{(\bullet)}_{,s}}}&{{{\bf{0}}^{\mathrm{T}}}}&{{{\boldsymbol{\varphi }}_{\!,s}}^{\mathrm{T}}{{(\bullet)}_{,s}}}
\end{array}} \right]_{2 \times 9}}{\delta\boldsymbol{y}}&&\eqqcolon {{\Bbb{B}}_{{\rho}}}\delta {\boldsymbol{y}},\\
\delta {{\boldsymbol{\kappa }}} &= {\left[ {\begin{array}{*{20}{c}}
{{{\bf{0}}^{\mathrm{T}}}}&{{{\boldsymbol{d}}_{1,s}}^{\mathrm{T}}{{(\bullet)}_{,s}}}&{{{\bf{0}}^{\mathrm{T}}}}\\
{{{\bf{0}}^{\mathrm{T}}}}&{{{\bf{0}}^{\mathrm{T}}}}&{{{\boldsymbol{d}}_{2,s}}^{\mathrm{T}}{{(\bullet)}_{,s}}}\\
{{{\bf{0}}^{\mathrm{T}}}}&{{{\boldsymbol{d}}_{2,s}}^{\mathrm{T}}{{(\bullet)}_{,s}}}&{{{\boldsymbol{d}}_{1,s}}^{\mathrm{T}}{{(\bullet)}_{,s}}}
\end{array}} \right]_{3 \times 9}}{\delta\boldsymbol{y}}&&\eqqcolon {{\Bbb{B}}_{\kappa}}\delta {\boldsymbol{y}},\\
\delta {\boldsymbol{\delta }} &= {\left[ {\begin{array}{*{20}{c}}
{{{\boldsymbol{d}}_1}^{\mathrm{T}}{{(\bullet)}_{,s}}}&{{{\boldsymbol{\varphi }}_{\!,s}}^{\mathrm{T}}}&{{{\bf{0}}^{\mathrm{T}}}}\\
{{{\boldsymbol{d}}_2}^{\mathrm{T}}{{(\bullet)}_{,s}}}&{{{\bf{0}}^{\mathrm{T}}}}&{{{\boldsymbol{\varphi }}_{\!,s}}^{\mathrm{T}}}
\end{array}} \right]_{2 \times 9}}{\delta\boldsymbol{y}}&&\eqqcolon {{\Bbb{B}}_{{\delta}}}\delta {\boldsymbol{y}},\\
\delta {\boldsymbol{\gamma }} &={\left[ {\begin{array}{*{20}{c}}
{{{\bf{0}}^{\mathrm{T}}}}&{{{\boldsymbol{d}}_1}^{\mathrm{T}}{{(\bullet)}_{,s}} + {{\boldsymbol{d}}_{1,s}}^{\mathrm{T}}}&{{{\bf{0}}^{\mathrm{T}}}}\\
{{{\bf{0}}^{\mathrm{T}}}}&{{{\boldsymbol{d}}_{2,s}}^{\mathrm{T}}}&{{{\boldsymbol{d}}_1}^{\mathrm{T}}{{(\bullet)}_{,s}}}\\
{{{\bf{0}}^{\mathrm{T}}}}&{{{\boldsymbol{d}}_2}^{\mathrm{T}}{{(\bullet)}_{,s}}}&{{{\boldsymbol{d}}_{1,s}}^{\mathrm{T}}}\\
{{{\bf{0}}^{\mathrm{T}}}}&{{{\boldsymbol{0}}^{\mathrm{T}}}}&{{{\boldsymbol{d}}_2}^{\mathrm{T}}{{(\bullet)}_{,s}} + {{\boldsymbol{d}}_{2,s}}^{\mathrm{T}}}
\end{array}} \right]_{4 \times 9}}{\delta\boldsymbol{y}}&&\eqqcolon {{\Bbb{B}}_{\gamma}}{\delta\boldsymbol{y}},\\
\delta {\boldsymbol{\chi }} &= {\left[ {\begin{array}{*{20}{c}}
{{{\bf{0}}^{\mathrm{T}}}}&{{{\boldsymbol{d}}_1}^{\mathrm{T}}}&{{{\boldsymbol{0}}^{\mathrm{T}}}}\\
{{{\bf{0}}^{\mathrm{T}}}}&{{{\boldsymbol{0}}^{\mathrm{T}}}}&{{{\boldsymbol{d}}_2}^{\mathrm{T}}}\\
{{{\bf{0}}^{\mathrm{T}}}}&{{{\boldsymbol{d}}_2}^{\mathrm{T}}}&{{{\boldsymbol{d}}_1}^{\mathrm{T}}}
\end{array}} \right]_{3 \times 9}}{\delta\boldsymbol{y}} &&\eqqcolon {{\Bbb{B}}_{\chi}}{\delta\boldsymbol{y}},\label{operator_strain_chi}
\end{alignat}
\end{subequations}
\endgroup
where ${\left[\bullet \right]_{m \times n}}$ indicates that the matrix $\left[\bullet \right]$ has dimension $m\times n$. Combining Eqs.\,(\ref{def_eps_hat}) and (\ref{beam_virtual_strain_compact_form}) leads to 
\begin{equation}\label{del_eps_hat_compact}
\delta {\boldsymbol{\munderbar \varepsilon }} = {{\Bbb{B}}_{{\rm{total}}}}\,\delta {\boldsymbol{y}},
\end{equation}
where we use
\begin{equation}
\Bbb{B}_{\text{total}}\coloneqq{\left[ {\begin{array}{*{20}{c}}
{{{\Bbb{B}}_\varepsilon }}\\
{{{\Bbb{B}}_\rho }}\\
{{{\Bbb{B}}_\kappa }}\\
{{{\Bbb{B}}_\delta }}\\
{{{\Bbb{B}}_\gamma }}\\
{{{\Bbb{B}}_\chi }}
\end{array}} \right]_{15 \times 9}}.
\end{equation}
Then Eq.\,(\ref{tot_strn_energy_beam_time_deriv}) can be rewritten as
\begin{align} \label{beam_int_vir_work_compact_Form}
{G_{\text{int}}}(\boldsymbol{y},\delta\boldsymbol{y}) = \int_0^L {{{\delta{\boldsymbol{y}}}^{\mathrm{T}}}{{\Bbb{B}}_{\text{total}}^{\mathrm{T}}}{\boldsymbol{R}}\,\mathrm{d}s}.
\end{align}
\subsubsection{Material part}
Taking the directional derivative of the internal virtual work of Eq.\,(\ref{beam_int_vir_work_compact_Form}) with the virtual strain part held constant and using Eq.\,(\ref{beam_lin_strs_resultant_R}) yields
\begin{align}\label{beam_mat_tan_operator}
{D_{\mathrm{M}}}{{\mathrm{G}}_{\text{int}}} \cdot {\Delta{\boldsymbol{y}}}\coloneqq \int_0^L {{{\delta{\boldsymbol{y}}}^{\mathrm{T}}}{{\Bbb{B}}_{\text{total}}^{\mathrm{T}}}D{\boldsymbol{R}}\cdot{\Delta\boldsymbol{y}}\,\mathrm{d}s}= \int_0^L {{{{\delta{\boldsymbol{y}}}}^{\mathrm{T}}}{{\Bbb{B}}_{\text{total}}^{\mathrm{T}}}{\Bbb{C}}{{\Bbb{B}}_{\text{total}}}{\Delta{\boldsymbol{y}}}\,\mathrm{d}s}.
\end{align}

\subsubsection{Geometric part}
Taking the directional derivative of the internal virtual work of Eq.\,(\ref{beam_int_vir_work_compact_Form}) with $\boldsymbol{R}$ held constant yields
\begin{equation}\label{beam_geom_tan_oper}
{D_{\mathrm{G}}}{G_{\text{int}}} \cdot {\Delta{\boldsymbol{y}}} \coloneqq \int_0^L {{{\boldsymbol{R}}^{\mathrm{T}}}D\delta {\boldsymbol{\munderbar \varepsilon}} \cdot {\Delta{\boldsymbol{y}}}\,\mathrm{d}s}  = \int_0^L {{{{\delta{\boldsymbol{y}}}}^{\mathrm{T}}}{{\boldsymbol{Y}}^{\mathrm{T}}}{{{\boldsymbol{k}}}_{\text{G}}}{\boldsymbol{Y}}{\Delta{\boldsymbol{y}}}\,\mathrm{d}s},
\end{equation}
where we define
%
\begin{equation}\label{geom_stiff_mat_k_g}
{{\boldsymbol{k}}_{\mathrm{G}}} \coloneqq \left[ {\begin{array}{*{20}{c}}
\begin{aligned}
{{\boldsymbol{k}}_\varepsilon }\,\,\,\,\,\,\,{{\boldsymbol{k}}_\rho }\,\,\,\,\,{{\boldsymbol{k}}_\delta }\\
{\,\,}\,\,\,\,\,\,\,{{\boldsymbol{k}}_\kappa }\,\,\,\,\,{{\boldsymbol{k}}_\gamma }\\
{{\rm{sym}}{\rm{.}}}\,\,\,\,\,{\,\,\,\,\,}\,\,\,\,\,{{\boldsymbol{k}}_\chi }
\end{aligned}
\end{array}} \right]_{15\times15}
\end{equation}
with the submatrices
\begingroup
\allowdisplaybreaks
\begin{subequations}
\begin{alignat}{3}
{\boldsymbol{k}}_\varepsilon  &\coloneqq \tilde n{{\boldsymbol{1}}_{3\times3}},\\
{\boldsymbol{k}}_\rho  &\coloneqq \left[ {\begin{array}{*{20}{c}}
{{{\tilde m}^1}{{\boldsymbol{1}}_{3\times3}}}&{{{\tilde m}^2}{{\boldsymbol{1}}_{3\times3}}}
\end{array}} \right],\\
{\boldsymbol{k}}_\delta  &\coloneqq \left[ {\begin{array}{*{20}{c}}
{{{\tilde q}^1}{{\boldsymbol{1}}_{3\times3}}}&{{{\tilde q}^2}{{\boldsymbol{1}}_{3\times3}}}
\end{array}} \right],\\
{\boldsymbol{k}}_\kappa  &\coloneqq \left[ {\begin{array}{*{20}{c}}
{{{\tilde h}^{11}}{{\boldsymbol{1}}_{3\times3}}}&{{{\tilde h}^{12}}{{\boldsymbol{1}}_{3\times3}}}\\
{{\rm{sym}}{\rm{.}}}&{{{\tilde h}^{22}}{{\boldsymbol{1}}_{3\times3}}}
\end{array}} \right],\\
{\boldsymbol{k}}_\gamma  &\coloneqq \left[ {\begin{array}{*{20}{c}}
{{{\tilde m}^{11}}{{\boldsymbol{1}}_{3\times3}}}&{{{\tilde m}^{21}}{{\boldsymbol{1}}_{3\times3}}}\\
{{{\tilde m}^{12}}{{\boldsymbol{1}}_{3\times3}}}&{{{\tilde m}^{22}}{{\boldsymbol{1}}_{3\times3}}}
\end{array}} \right],\\
{\boldsymbol{k}}_\chi  &\coloneqq \left[ {\begin{array}{*{20}{c}}
{{{\tilde l}^{11}}{{\boldsymbol{1}}_{3\times3}}}&{{{\tilde l}^{21}}{{\boldsymbol{1}}_{3\times3}}}\\
{{\rm{sym}}{\rm{.}}}&{{{\tilde l}^{22}}{{\boldsymbol{1}}_{3\times3}}}
\end{array}} \right],
\end{alignat}
\end{subequations}
\endgroup
and the operator
\begin{equation}\label{def_matY_oper}
{\boldsymbol{Y}} \coloneqq {\left[ {\begin{array}{*{20}{c}}
{{{(\bullet)}_{,s}}{{\boldsymbol{1}}_{3\times3}}}&{{{\bf{0}}_{3 \times 3}}}&{{{\bf{0}}_{3 \times 3}}}\\
{{{\bf{0}}_{3 \times 3}}}&{{{(\bullet)}_{,s}}{{\boldsymbol{1}}_{3\times3}}}&{{{\bf{0}}_{3 \times 3}}}\\
{{{\bf{0}}_{3 \times 3}}}&{{{\bf{0}}_{3 \times 3}}}&{{{(\bullet)}_{,s}}{{\boldsymbol{1}}_{3\times3}}}\\
{{{\bf{0}}_{3 \times 3}}}&{{\boldsymbol{1}}_{3\times3}}&{{{\bf{0}}_{3 \times 3}}}\\
{{{\bf{0}}_{3 \times 3}}}&{{{\bf{0}}_{3 \times 3}}}&{{\boldsymbol{1}}_{3\times3}}
\end{array}} \right]_{15 \times 9}}.
\end{equation}
Here ${\bf{0}}_{m \times n}$ denotes the null matrix of dimension $m \times n$. Combining Eqs.\,(\ref{beam_mat_tan_operator}) and (\ref{beam_geom_tan_oper}) we finally obtain the following linearization of the internal virtual work\footnote{Strictly speaking, this is the increment in the linearization of the internal virtual work.}
\begin{equation}\label{beam_tangent_stiff_cont_form}
{\Delta}G_{\text{int}}({\boldsymbol{y}};{\delta \boldsymbol{y}},{\Delta \boldsymbol{y}})\coloneqq D{G_{\text{int}}} \cdot {\Delta{\boldsymbol{y}}} = \int_0^L {{{{\delta{\boldsymbol{y}}}}^{\mathrm{T}}}\left( {{{\Bbb{B}}_{\text{total}}^{\mathrm{T}}}{\Bbb{C}}{{\Bbb{B}}_{\text{total}}} + {{\boldsymbol{Y}}^{\mathrm{T}}}{{{\boldsymbol{k}}}_{\text{G}}}{{\boldsymbol{Y}}}} \right){\Delta{\boldsymbol{y}}}\,\mathrm{d}s}.
\end{equation}
\subsection{Discretization of the variational form}
\label{def_operators_disc_weak_form}
\subsubsection{Discretization of the internal virtual work}
The discretization of virtual strains at $\xi\in{\varXi_e}$ is expressed in compact form as follows.
\begingroup
\allowdisplaybreaks
\begin{subequations}\label{beam_vir_strn_disc_compact}
\begin{alignat}{3}
\delta {\varepsilon ^h} &= {\left[ {\begin{array}{*{20}{c}}
{{{\boldsymbol{\varphi }}_{\!,s}}^{\mathrm{T}}{N_{I,s}}}&{{{\bf{0}}_{1 \times 6}}}
\end{array}} \right]_{1 \times 9}}\delta {{\bf{y}}_I}&&\eqqcolon {\Bbb{B}}_{\varepsilon}^I\delta {{\bf{y}}_I},\\
\delta {{\boldsymbol{\rho }}^h} &= {\left[ {\begin{array}{*{20}{c}}
{{{\boldsymbol{d}}_{1,s}}^{\mathrm{T}}{N_{I,s}}}&{{{\boldsymbol{\varphi}}_{\!,s}}^{\mathrm{T}}{N_{I,s}}}&{{{\bf{0}}^{\mathrm{T}}}}\\
{{{\boldsymbol{d}}_{2,s}}^{\mathrm{T}}{N_{I,s}}}&{{{\bf{0}}^{\mathrm{T}}}}&{{{\boldsymbol{\varphi}}_{\!,s}}^{\mathrm{T}}{N_{I,s}}}
\end{array}} \right]_{2 \times 9}}\delta {{\bf{y}}_I} &&\eqqcolon {\Bbb{B}}_{{\rho}}^I\delta {{\bf{y}}_I},\\
\delta {{\boldsymbol{\kappa }}^h} & = {\left[ {\begin{array}{*{20}{c}}
{{{\bf{0}}^{\mathrm{T}}}}&{{{\boldsymbol{d}}_{1,s}}^{\mathrm{T}}{N_{I,s}}}&{{{\bf{0}}^{\mathrm{T}}}}\\
{{{\bf{0}}^{\mathrm{T}}}}&{{{\bf{0}}^{\mathrm{T}}}}&{{{\boldsymbol{d}}_{2,s}}^{\mathrm{T}}{N_{I,s}}}\\
{{{\bf{0}}^{\mathrm{T}}}}&{{{\boldsymbol{d}}_{2,s}}^{\mathrm{T}}{N_{I,s}}}&{{{\boldsymbol{d}}_{1,s}}^{\mathrm{T}}{N_{I,s}}}
\end{array}} \right]_{3 \times 9}}\delta {{\bf{y}}_I} &&\eqqcolon {\Bbb{B}}_\kappa ^I\delta {{\bf{y}}_I},\\
\delta {{\boldsymbol{\delta }}^h} &= {\left[ {\begin{array}{*{20}{c}}
{{{\boldsymbol{d}}_1}^{\mathrm{T}}{N_{I,s}}}&{{{\boldsymbol{\varphi}}_{\!,s}}^{\mathrm{T}}{N_I}}&{{{\bf{0}}^{\mathrm{T}}}}\\
{{{\boldsymbol{d}}_2}^{\mathrm{T}}{N_{I,s}}}&{{{\bf{0}}^{\mathrm{T}}}}&{{{\boldsymbol{\varphi}}_{\!,s}}^{\mathrm{T}}{N_I}}
\end{array}} \right]_{2 \times 9}}\delta {{\bf{y}}_I} &&\eqqcolon {\Bbb{B}}_{\delta}^I\delta {{\bf{y}}_I},\\
\delta {{\boldsymbol{\gamma }}^h} &= {\left[ {\begin{array}{*{20}{c}}
{{{\bf{0}}^{\mathrm{T}}}}&{{{\boldsymbol{d}}_1}^{\mathrm{T}}{N_{I,s}} + {{\boldsymbol{d}}_{1,s}}^{\mathrm{T}}{N_I}}&{{{\bf{0}}^{\mathrm{T}}}}\\
{{{\bf{0}}^{\mathrm{T}}}}&{{{\boldsymbol{d}}_{2,s}}^{\mathrm{T}}{N_I}}&{{{\boldsymbol{d}}_1}^{\mathrm{T}}{N_{I,s}}}\\
{{{\bf{0}}^{\mathrm{T}}}}&{{{\boldsymbol{d}}_2}^{\mathrm{T}}{N_{I,s}}}&{{{\boldsymbol{d}}_{1,s}}^{\mathrm{T}}{N_I}}\\
{{{\bf{0}}^{\mathrm{T}}}}&{{{\boldsymbol{0}}^{\mathrm{T}}}}&{{{\boldsymbol{d}}_2}^{\mathrm{T}}{N_{I,s}} + {{\boldsymbol{d}}_{2,s}}^{\mathrm{T}}{N_I}}
\end{array}} \right]_{4 \times 9}}\delta {{\bf{y}}_I} &&\eqqcolon {\Bbb{B}}_\gamma ^I\delta {{\bf{y}}_I},\\
\delta {{\boldsymbol{\chi }}^h} &= {\left[ {\begin{array}{*{20}{c}}
{{{\bf{0}}^{\mathrm{T}}}}&{{{\boldsymbol{d}}_1}^{\mathrm{T}}{N_I}}&{{{\bf{0}}^{\mathrm{T}}}}\\
{{{\bf{0}}^{\mathrm{T}}}}&{{{\bf{0}}^{\mathrm{T}}}}&{{{\boldsymbol{d}}_2}^{\mathrm{T}}{N_I}}\\
{{{\bf{0}}^{\mathrm{T}}}}&{{{\boldsymbol{d}}_2}^{\mathrm{T}}{N_I}}&{{{\boldsymbol{d}}_1}^{\mathrm{T}}{N_I}}
\end{array}} \right]_{3 \times 9}}\delta {{\bf{y}}_I} &&\eqqcolon {\Bbb{B}}_\chi ^I\delta {{\bf{y}}_I},\label{beam_vir_strn_disc_compact_chi}
\end{alignat}
\end{subequations}
\endgroup
where the repeated index $I$ implies summation over values from $1$ to $n_e$. $n_e$ denotes the number of basis functions having local supports in the knot span ${\varXi _e}$. Having Eq. (\ref{beam_vir_strn_disc_compact}) we obtain
\begin{equation}\label{disc_grad_operator_B_tot}
\delta {{\boldsymbol{\munderbar \varepsilon}}^h} = {\left[ {\begin{array}{*{20}{c}}\begin{aligned}
{{\Bbb{B}}_\varepsilon ^e}\\
{{\Bbb{B}}_\rho ^e}\\
{{\Bbb{B}}_\kappa ^e}\\
{{\Bbb{B}}_\delta ^e}\\
{{\Bbb{B}}_\gamma ^e}\\
{{\Bbb{B}}_\chi ^e}\end{aligned}
\end{array}} \right]_{15 \times 9}}\left\{ {\begin{array}{*{20}{c}}
{\delta {{\bf{y}}_1}}\\
 \vdots \\
{\delta {{\bf{y}}_{{n_e}}}}
\end{array}} \right\}\eqqcolon{{\Bbb{B}}_\mathrm{total}^e}\delta {{\bf{y}}^e},
\end{equation}
where we define
\begin{equation}
\left\{ \begin{array}{c} \begin{aligned}
{\Bbb{B}}_\varepsilon ^e &\coloneqq {\left[ {\begin{array}{*{20}{c}}
{{\Bbb{B}}_\varepsilon ^1}&{{\Bbb{B}}_\varepsilon ^2}& \cdots &{{\Bbb{B}}_\varepsilon^{{n_e}}}
\end{array}} \right]_{1 \times 9{n_e}}},\\
{\Bbb{B}}_\rho ^e &\coloneqq {\left[ {\begin{array}{*{20}{c}}
{{\Bbb{B}}_\rho ^1}&{{\Bbb{B}}_\rho ^2}& \cdots &{{\Bbb{B}}_\rho ^{{n_e}}}
\end{array}} \right]_{2 \times 9{n_e}}},\\
{\Bbb{B}}_\kappa ^e &\coloneqq {\left[ {\begin{array}{*{20}{c}}
{{\Bbb{B}}_\kappa ^1}&{{\Bbb{B}}_\kappa ^2}& \cdots &{{\Bbb{B}}_\kappa ^{{n_e}}}
\end{array}} \right]_{3 \times 9{n_e}}},\\
{\Bbb{B}}_\delta ^e &\coloneqq {\left[ {\begin{array}{*{20}{c}}
{{\Bbb{B}}_\delta ^1}&{{\Bbb{B}}_\delta ^2}& \cdots &{{\Bbb{B}}_\delta ^{{n_e}}}
\end{array}} \right]_{2 \times 9{n_e}}},\\
{\Bbb{B}}_\gamma ^e &\coloneqq {\left[ {\begin{array}{*{20}{c}}
{{\Bbb{B}}_\gamma ^1}&{{\Bbb{B}}_\gamma ^2}& \cdots &{{\Bbb{B}}_\gamma ^{{n_e}}}
\end{array}} \right]_{4 \times 9{n_e}}},\\
{\Bbb{B}}_\chi ^e &\coloneqq {\left[ {\begin{array}{*{20}{c}}
{{\Bbb{B}}_\chi ^1}&{{\Bbb{B}}_\chi ^2}& \cdots &{{\Bbb{B}}_\chi ^{{n_e}}}
\end{array}} \right]_{3 \times 9{n_e}}}. \end{aligned}
\end{array} \right.
\end{equation}
Also, in the discretization of the geometric part of the tangent stiffness, we use the following operator, obtained from Eq.\,(\ref{def_matY_oper}).
\begin{equation}
\label{def_Y_e_g_tan}
{{\Bbb{Y}}_e} \coloneqq {\left[ {\begin{array}{*{20}{c}}
{{{\boldsymbol{Y}}_1}}&{{{\boldsymbol{Y}}_2}}& \cdots &{{{\boldsymbol{Y}}_{{{n_{e}}}}}}
\end{array}} \right]_{15 \times 9{{n_{e}}}}},
\end{equation}
where
\begin{equation} \label{def_matY_oper_disc}
{\boldsymbol{Y}}_I \coloneqq {\left[ {\begin{array}{*{20}{c}}
{{N_{I,s}}{{\boldsymbol{1}}_{3\times3}}}&{{{\bf{0}}_{3 \times 3}}}&{{{\bf{0}}_{3 \times 3}}}\\
{{{\bf{0}}_{3 \times 3}}}&{{N_{I,s}}{{\boldsymbol{1}}_{3\times3}}}&{{{\bf{0}}_{3 \times 3}}}\\
{{{\bf{0}}_{3 \times 3}}}&{{{\bf{0}}_{3 \times 3}}}&{{N_{I,s}}{{\boldsymbol{1}}_{3\times3}}}\\
{{{\bf{0}}_{3 \times 3}}}&{{N_I}{{\boldsymbol{1}}_{3\times3}}}&{{{\bf{0}}_{3 \times 3}}}\\
{{{\bf{0}}_{3 \times 3}}}&{{{\bf{0}}_{3 \times 3}}}&{{N_I}{{\boldsymbol{1}}_{3\times3}}}
\end{array}} \right]_{15 \times 9}}.
\end{equation}
\subsection{Implementation of a moment load by a distributed follower load}
\label{mnt_load_follower_load_exp}
The external virtual work due to the Neumann boundary condition of Eq.\,(\ref{ex_end_mnt_neumann_bdc}) is expressed by
\begin{equation}\label{pure_bend_vir_work}
{G_{{\rm{ext}}}}(\boldsymbol{y},\delta\boldsymbol{y}) =\left[
\delta {\boldsymbol{y}}^{\mathrm{T}}{{\boldsymbol{\bar R}}_0}\right]_{s\in\Gamma_\mathrm{N}}\,\,\,\mathrm{with}\,\,\,{{\boldsymbol{\bar R}}_0} = {\left[ {{{\bf{0}}^\mathrm{T}}, - M{{\boldsymbol{d}}_3}^\mathrm{T},{{\bf{0}}^\mathrm{T}}} \right]^\mathrm{T}}.
\end{equation}
The directional derivative of the external virtual work can be derived as
\begin{align}\label{pure_bend_vir_work_lin}
{\Delta}G_\mathrm{ext}\left(\boldsymbol{y};\delta\boldsymbol{y},\Delta{\boldsymbol{y}}\right)&\coloneqq{DG_{{\rm{ext}}} \cdot \Delta {\boldsymbol{y}}}= \left[\delta {{\boldsymbol{y}}^{\mathrm{T}}}{{\boldsymbol{S}}_{{\rm{nc}}}}\Delta {\boldsymbol{y}}\right]_{s\in\Gamma_\mathrm{N}}.
\end{align}
${{\boldsymbol{S}}_{{\rm{nc}}}}$ is the \textit{load stiffness operator} given by
\begin{align}
{{\boldsymbol{S}}_{{\rm{nc}}}}\coloneqq {\frac{M}{{\left\| {{{\boldsymbol{d}}_1} \times {{\boldsymbol{d}}_2}} \right\|}}\left[ {\begin{array}{*{20}{c}}
{{{\bf{0}}_{3 \times 3}}}&{{{\bf{0}}_{3 \times 3}}}&{{{\bf{0}}_{3 \times 3}}}\\
{{{\bf{0}}_{3 \times 3}}}&{\left( {{\boldsymbol{1}} - {{\boldsymbol{d}}_3} \otimes {{\boldsymbol{d}}_3}} \right){{{\boldsymbol{\hat d}}}_2}}&{\left( {{{\boldsymbol{d}}_3} \otimes {{\boldsymbol{d}}_3} - {\boldsymbol{1}}} \right){{{\boldsymbol{\hat d}}}_1}}\\
{{{\bf{0}}_{3 \times 3}}}&{{{\bf{0}}_{3 \times 3}}}&{{{\bf{0}}_{3 \times 3}}}
\end{array}} \right]},
\end{align}
where ${\boldsymbol{\hat d}}_1$ and ${\boldsymbol{\hat d}}_2$ represent the skew-symmetric matrices associated with the dual vectors ${\boldsymbol{d}}_1$ and ${\boldsymbol{d}}_2$. Note that the load stiffness operator $\boldsymbol{S}_{\mathrm{nc}}$ is non-symmetric. The global force vector due to the external virtual work and the global load stiffness matrix can be simply obtained by a finite element assembly, respectively, as
\begin{align}\label{pure_bend_ext_v_work_disc_new}
{\bf{F}}_\mathrm{ext}={\bf{A}}\left[\bar{\boldsymbol{R}}_0\right]_{s\in\Gamma_\mathrm{N}},
\end{align}
and
\begin{align}\label{pure_bend_lstiff_op_disc_new}
{\bf{K}}_\mathrm{ext}={\bf{A}}\left[{\boldsymbol{S}}_\mathrm{nc}\right]_{s\in\Gamma_\mathrm{N}}.
\end{align}
%
\section{Appendix to numerical examples}
\label{app_hypelas_conv_test}
\subsection{Uniaxial tension of a straight beam}
\label{app_conv_test_uni_tens}
\begin{figure}[htp]	
	\centering
	\begin{subfigure}[b] {0.3\textwidth} \centering
		\includegraphics[width=\linewidth]{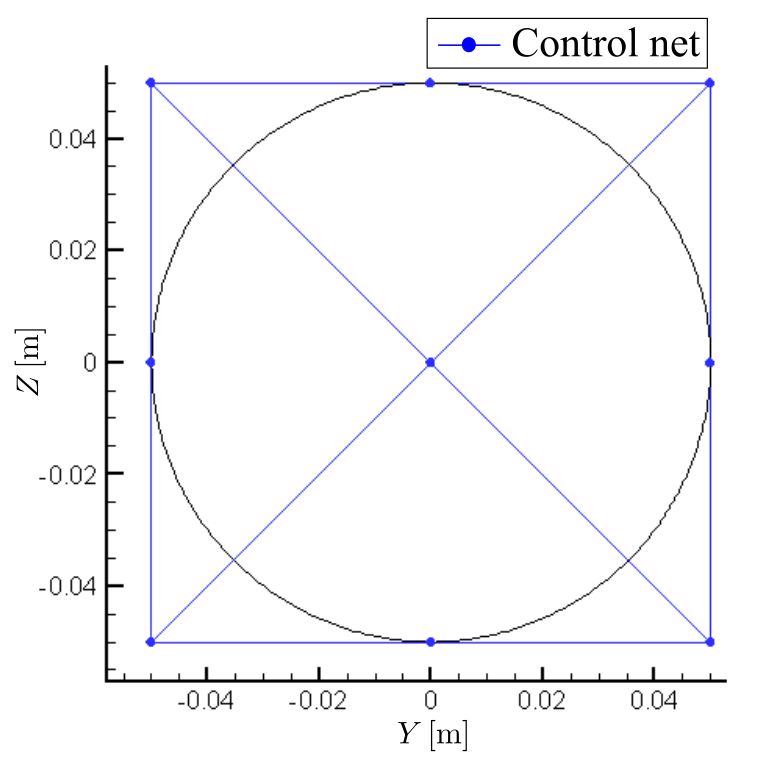}
		\vskip -2pt
		\caption{}
		\label{circle_single_patch}
	\end{subfigure} 
	\begin{subfigure}[b] {0.3\textwidth} \centering
		\includegraphics[width=\linewidth]{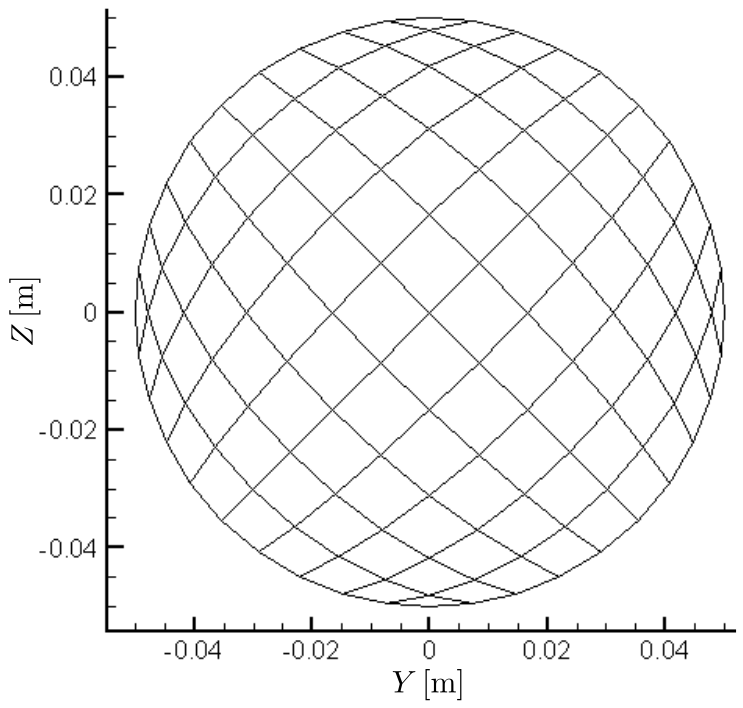}
		\vskip -2pt
		\caption{}
		\label{mesh_10by10}
	\end{subfigure} 
	\begin{subfigure}[b] {0.3\textwidth} \centering
		\includegraphics[width=\linewidth]{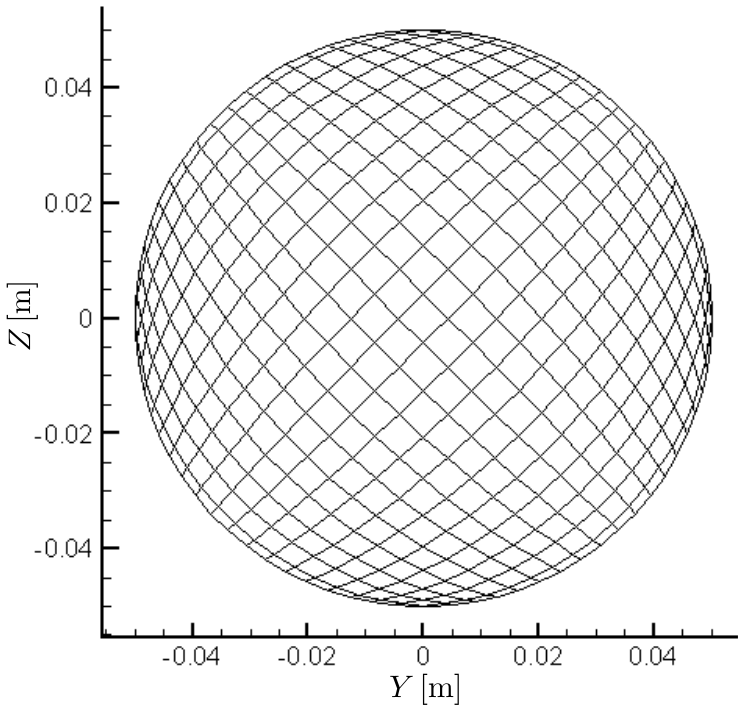}
		\vskip -2pt
		\caption{}
		\label{mesh_20by20}
	\end{subfigure} 
\caption{Uniaxial tension of a straight beam: Modeling of circular plane (cross-section) having radius $R=0.05\mathrm{m}$ and $h$-refinement, used in brick element results. (a) Modeling of circular domain by a single quadratic NURBS patch. (b) $10{\times}10$, (c) $20{\times}20$ elements.}
\label{uni_axial_tens_mesh_profile}	
\end{figure}

The exact geometry of circular section of cylindrical structure can be modeled by a single quadratic NURBS patch (see Fig.\,\ref{circle_single_patch}). Figs.\,\ref{mesh_10by10} and \ref{mesh_20by20} show knot spans in two different levels of $h$-refinement.
Tables \ref{app_utens_conv_test_svk_010_fixed} and \ref{app_utens_conv_test_nh_010_fixed} show the convergence of the reference solution using brick elements and the beam solution. For the conciseness of paper, we present only one case of cross-section radius $R=0.1\mathrm{m}$, and BC{\#}2. The other cases show similar characteristics of convergence.
\begin{table}[]
\small
\centering
\caption{Uniaxial tension of a straight beam: Convergence result of the lateral displacement at A and the volume ($R=0.1\mathrm{m}$, BC{\#}2, St.\,Venant-Kirchhoff material).}
\label{app_utens_conv_test_svk_010_fixed}
\begin{tabular}{rcclccc}
\Xhline{3\arrayrulewidth} 
\multicolumn{3}{c}{Brick, IGA, deg.=(3,3,3)} &  & \multicolumn{3}{c}{Beam, IGA, $p=3$} \\ \cline{1-3} \cline{5-7} 
\multicolumn{1}{c}{${n_\mathrm{el}}$} & $v_{\mathrm{A}}\,[\mathrm{m}]$   	& $V$ [$\mathrm{m}^3$]       	&  & \multicolumn{1}{c}{${{n_\mathrm{el}}}$}  	& $v_{\mathrm{A}}\,[\mathrm{m}]$      	& $V\,[\mathrm{m}^3]$\\ 
\Xhline{3\arrayrulewidth}  
$80\times10\times10$       & -2.2181E-02  		& 3.0215E-02  					&  & 10    				& -2.2179E-02    		& 3.0420E-02    \\
$160\times10\times10$      & -2.2181E-02  		& 3.0214E-02  					&  & 20    				& -2.2177E-02    		& 3.0418E-02    \\
$320\times10\times10$      & -2.2181E-02  		& 3.0214E-02  					&  & 40    				& -2.2177E-02    		& 3.0418E-02    \\
$320\times15\times15$      & -2.2181E-02  		& 3.0214E-02  					&  & 80    				& -2.2177E-02    		& 3.0418E-02    \\ 
\Xhline{3\arrayrulewidth} 
\end{tabular}
\end{table}

\begin{table}[]
\small
\centering
\caption{Uniaxial tension of a straight beam: Convergence test of the lateral displacement at A and the volume ($R=0.1\mathrm{m}$, BC{\#}2, compressible Neo-Hookean material)}
\label{app_utens_conv_test_nh_010_fixed}
\begin{tabular}{rcclccc}
\Xhline{3\arrayrulewidth} 
\multicolumn{3}{c}{Brick, IGA, deg.=(2,2,2)} &  & \multicolumn{3}{c}{Beam, IGA, $p=3$} \\ \cline{1-3} \cline{5-7} 
\multicolumn{1}{c}{${n_\mathrm{el}}$} & $v_{\mathrm{A}}\,[\mathrm{m}]$   	& $V\,[\mathrm{m}^3]$       	&  & \multicolumn{1}{c}{${{n_\mathrm{el}}}$}  	& $v_{\mathrm{A}}\,[\mathrm{m}]$      	& $V\,[\mathrm{m}^3]$\\ 
\Xhline{3\arrayrulewidth}  
$80\times10\times10$       & -2.9186E-02  		& 4.5010E-02					&  & 10    				& -2.9137E-02    		& 4.5185E-02    \\
$160\times10\times10$      & -2.9186E-02  		& 4.5006E-02  					&  & 20    				& -2.9186E-02    		& 4.5110E-02    \\
$320\times10\times10$      & -2.9186E-02  		& 4.5005E-02					&  & 40    				& -2.9186E-02    		& 4.5109E-02    \\ $320\times20\times20$      & -2.9186E-02  		& 4.5005E-02 					&  & 80    				& -2.9186E-02   		& 4.5109E-02    \\ 
\Xhline{3\arrayrulewidth} 
\end{tabular}
\end{table}
\subsection{Cantilever beam under end moment}
\label{app_sup_end_mnt}
\begin{figure}[htp]	
	\centering
	\begin{subfigure}[b] {0.4875\textwidth} \centering
		\includegraphics[width=\linewidth]{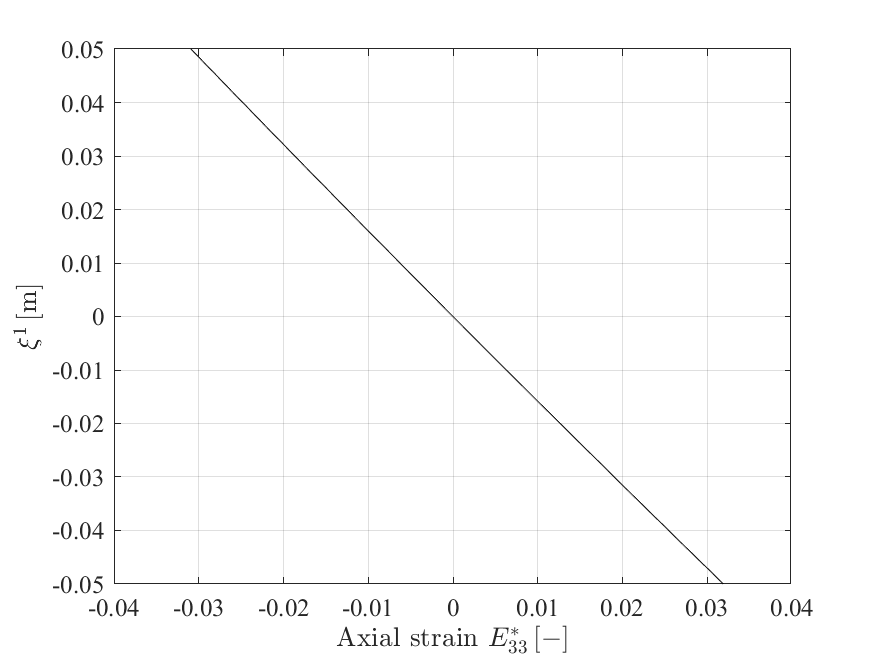}
		\vskip -2pt
		\caption{Initial cross-section height ${h}=0.1\mathrm{m}$}
		\label{pure_bend_analytic_th010}
	\end{subfigure} 
	\begin{subfigure}[b] {0.4875\textwidth} \centering
		\includegraphics[width=\linewidth]{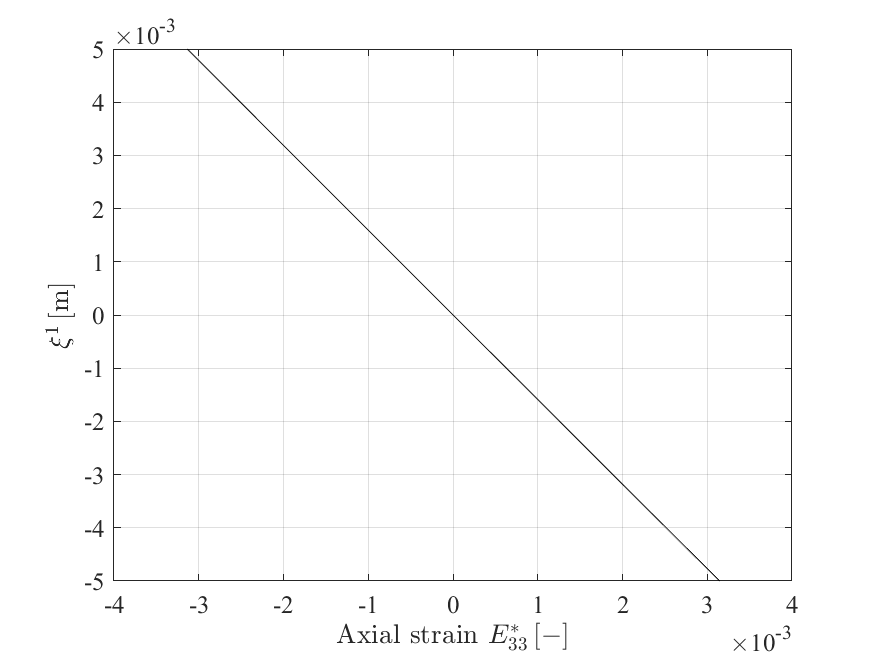}
		\vskip -2pt
		\caption{Initial cross-section height ${h}=0.01\mathrm{m}$}
		\label{pure_bend_analytic_th001}
	\end{subfigure} 
\caption{Cantilever beam under end moment: Analytic solution of axial strain through the cross-section height at the loaded end ($s=L$) under pure bending assumption. Note that, in the analytical solution, the axial strain is zero on the central axis (${\xi^1}=0$).}
\label{pure_bend_analytic_graph}
\end{figure}

Fig.\,\ref{pure_bend_analytic_graph} plots the analytical solution of axial strain in Eq.\,(\ref{ex_end_mnt_axial_comp_GL}). Tables \ref{app_conv_test_xdisp_tip_h010} and \ref{app_conv_test_xdisp_tip_h001} show the convergence results of brick and beam elements.
\begin{table}[]
\small
\centering
\caption{Cantilever beam under end moment: Convergence result of $X$-displacement at the tip of the central axis ($u_\mathrm{tip}$) for the case of ${h}=0.1\mathrm{m}$. All results are obtained by IGA.}
\label{app_conv_test_xdisp_tip_h010}
\begin{tabular}{cccccccc}
\Xhline{3\arrayrulewidth}
\multicolumn{2}{c}{Brick, deg.=(2,1,1)} & & \multicolumn{2}{c}{Brick, deg.=(2,1,2)} &  & \multicolumn{2}{c}{Beam, $p=4$} \\
\cline{1-2} \cline{4-5} \cline{7-8}
\multicolumn{1}{c}{${n_\mathrm{el}}$} & \multicolumn{1}{c}{$u_\mathrm{tip}\,[\mathrm{m}]$} &  & ${n_\mathrm{el}}$       & $u_\mathrm{tip}\,[\mathrm{m}]$         &  & ${n_\mathrm{el}}$     & \multicolumn{1}{c}{$u_\mathrm{tip}\,[\mathrm{m}]$}                                           \\ 
\Xhline{3\arrayrulewidth}
$160\times1\times1$  & -9.9850E+00  &  & $320\times1\times10$  & -1.0010E+01     &  & 5         & -1.0807E+01    \\
$320\times1\times1$  & -9.9849E+00  &  & $640\times1\times10$  & -1.0011E+01     &  & 10        & -9.9899E+00    \\
$640\times1\times1$  & -9.9848E+00  &  & $1280\times1\times10$ & -1.0011E+01     &  & 20        & -9.9849E+00    \\
$1280\times1\times1$ & -9.9848E+00  &  & $2560\times1\times10$ & -1.0011E+01     &  & 40        & -9.9848E+00    \\
$2560\times1\times1$ & -9.9848E+00  &  & $2560\times1\times20$ & -1.0011E+01     &  & 80        & -9.9848E+00    \\
\multicolumn{1}{c}{} & \multicolumn{1}{c}{} &  & \multicolumn{1}{c}{} & \multicolumn{1}{c}{}  &  & 160 & -9.9848E+00 \\ 
\Xhline{3\arrayrulewidth}
\end{tabular}
\end{table}
\begin{table}[]
\small
\centering
\caption{Cantilever beam under end moment: Convergence result of $X$-displacement at the tip of the central axis ($u_\mathrm{tip}$) for the case of ${h}=0.01\mathrm{m}$. All results are obtained by IGA.}
\label{app_conv_test_xdisp_tip_h001}
\begin{tabular}{cccccccc}
\Xhline{3\arrayrulewidth}
\multicolumn{2}{c}{Brick, deg.=(2,1,1)} & & \multicolumn{2}{c}{Brick, deg.=(2,1,2)} &  & \multicolumn{2}{c}{Beam, $p=4$} \\
\cline{1-2} \cline{4-5} \cline{7-8}
${n_\mathrm{el}}$  & \multicolumn{1}{c}{$u_\mathrm{tip}\,[\mathrm{m}]$} &  & ${n_\mathrm{el}}$       & $u_\mathrm{tip}\,[\mathrm{m}]$         &  & ${n_\mathrm{el}}$     & \multicolumn{1}{c}{$u_\mathrm{tip}\,[\mathrm{m}]$}                                           \\ 
\Xhline{3\arrayrulewidth}
$160\times1\times1$  & -1.0020E+01 &  & $320\times1\times10$   & -1.0001E+01   &  & 5      & -1.2414E+01   \\
$320\times1\times1$  & -1.0001E+01 &  & $640\times1\times10$   & -1.0000E+01   &  & 10     & -1.0362E+01     \\
$640\times1\times1$  & -9.9999E+00 &  & $1280\times1\times10$  & -1.0000E+01   &  & 20     & -1.0001E+01       \\
$1280\times1\times1$ & -9.9999E+00 &  & $2560\times1\times10$  & -1.0000E+01   &  & 40     & -9.9999E+00    \\
$2560\times1\times1$ & -9.9998E+00 &  & $2560\times1\times20$  & -1.0000E+01   &  & 80     & -9.9998E+00    \\
\multicolumn{1}{l}{} & \multicolumn{1}{l}{}     &  & \multicolumn{1}{l}{}  & \multicolumn{1}{l}{} &  & 160 & -9.9998E+00   \\ 
\Xhline{3\arrayrulewidth}
\end{tabular}
\end{table}
\subsection{Cantilever beam under end force}
\label{app_sup_end_shear}
Table \ref{app_conv_test_xdisp_tip} shows the convergence test of IGA using brick element to obtain the reference solution.
\begin{table}[]
\small
\begin{center}
\caption{Cantilever beam under end force: Convergence test of displacement at the point A.}
\label{app_conv_test_xdisp_tip}
\begin{tabular}{cccccccc}
\Xhline{3\arrayrulewidth} 
\multicolumn{3}{c}{\begin{tabular}[c]{@{}c@{}}$\nu=0$,\\ Brick, IGA, deg.=(2,1,2)\end{tabular}} & \multicolumn{1}{l}{} & \multicolumn{4}{c}{\begin{tabular}[c]{@{}c@{}}$\nu=0.3$,\\ Brick, IGA, deg.=(3,3,3)\end{tabular}} \\ \cline{1-3} \cline{5-8} 
${n_\mathrm{el}}$                                  & $X$-disp. [m]                         & $Z$-disp. [m] &                      & ${n_\mathrm{el}}$                           & $X$-disp. [m]     & $Y$-disp. [m] & $Z$-disp. [m]               \\ \Xhline{3\arrayrulewidth} 
$80\times1\times10$ & -6.4736E+00   & 7.9951E+00   &  & $80\times10\times10$  & -6.4626E+00           & -1.4513E-03           & 8.0033E+00          \\
$160\times1\times10$ & -6.4737E+00  & 7.9951E+00   &  & $160\times10\times10$ & -6.4629E+00           & -1.4533E-03           & 8.0035E+00          \\
$200\times1\times10$ & -6.4737E+00  & 7.9951E+00   &  & $200\times10\times10$ & -6.4629E+00           & -1.4533E-03           & 8.0035E+00          \\
$200\times1\times15$ & -6.4737E+00  & 7.9951E+00   &  & $200\times15\times15$ & -6.4630E+00           & -1.4528E-03           & 8.0037E+00          \\
 							  &   				 &    				&  & $200\times20\times20$ & -6.4631E+00           & -1.4522E-03           & 8.0037E+00          \\ \Xhline{3\arrayrulewidth} 
\end{tabular}
\end{center}
\end{table}
\subsection{Laterally loaded beam}
\label{app_lat_load}
Table\,\ref{app_lat_load_conv_test_ref_sol} shows the convergence test of IGA using brick elements to obtain the reference solution. We check the convergence of $Z$-displacement at the mid-point of the central axis, i.e., at the point of $s=L/2$ and ${\xi^1}={\xi^2}=0$.
\begin{table}[]
\small
\begin{center}
\caption{{Laterally loaded beam: Convergence test of $Z$-displacement at the mid-point of the central axis in IGA using brick elements with deg.=(3,3,3).}}
\label{app_lat_load_conv_test_ref_sol}
\begin{tabular}{cc}
\Xhline{3\arrayrulewidth} 
${n_\mathrm{el}}$        & $Z$-displacement [m] \\ 
\Xhline{3\arrayrulewidth} 
$80\times5\times5$    & -2.1729E-01 \\
$120\times5\times5$   & -2.1729E-01 \\
$160\times5\times5$   & -2.1729E-01 \\
$160\times15\times15$ & -2.1731E-01 \\
$320\times15\times15$ & -2.1731E-01 \\ 
\Xhline{3\arrayrulewidth}
\end{tabular}
\end{center}
\end{table}
\subsection{45$^{\circ}$-arch cantilever beam under end force}
\label{app_45deg_end_shear}
Table\,\ref{app_45deg_conv_test_ref_sol} shows the convergence test of IGA using brick elements to obtain the reference solution.
\begin{table}[]
\small
\begin{center}
\caption{{45$^{\circ}$-arch cantilever beam: Convergence test of displacements at the point A in IGA using brick elements with deg.=(3,3,3).}}
\label{app_45deg_conv_test_ref_sol}
\begin{tabular}{cccc}
\Xhline{3\arrayrulewidth} 
${n_\mathrm{el}}$     & $X$-displacement [m] & $Y$-displacement [m]	  & $Z$-displacement [m]\\ 
\Xhline{3\arrayrulewidth} 
$200\times10\times10$ & 1.4506E+01 & -2.4987E+01 & 5.2112E+01 \\
$240\times10\times10$ & 1.4507E+01 & -2.4987E+01 & 5.2113E+01 \\
$240\times15\times15$ & 1.4505E+01 & -2.4987E+01 & 5.2116E+01 \\ 
\Xhline{3\arrayrulewidth}
\end{tabular}
\end{center}
\end{table}
\section*{Acknowledgement}
M.-J Choi would like to gratefully acknowledge the financial support of a postdoctoral research fellowship from the Alexander von Humboldt Foundation in Germany.


\bibliography{mybibfile}

\end{document}